# Modelos de interacción de enlace de hidrógeno

## para la simulación del plegamiento

## y la agregación de proteínas

---

# Hydrogen bond models

# for the simulation of protein folding and aggregation

**Marta Enciso Carrasco**

Memoria de Tesis Doctoral

para optar al grado de Doctor

por la Universidad Complutense de Madrid

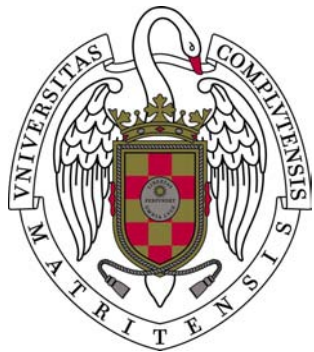

Director: Antonio Rey Gayo

Departamento de Química Física I

Universidad Complutense de Madrid

It was the best of times, it was the worst of times, it was the age of wisdom, it was the age of foolishness, it was the epoch of belief, it was the epoch of incredulity, it was the season of Light, it was the season of Darkness, it was the spring of hope, it was the winter of despair, we had everything before us, we had nothing before us, we were all going direct to heaven, we were all going direct the other way -in short, the period was so far like the present period, that some of its noisiest authorities insisted on its being received, for good or for evil, in the superlative degree of comparison only.

Charles Dickens, *A Tale of Two Cities*

# Agradecimientos

*Y ahora, en primer lugar (cómo no), gracias a Antonio, mi director de tesis. Por ser sin lugar a dudas una referencia personal y profesional.*

*Inmediatamente después vienen mis compañeros del grupo. Desde David, al que he conocido fundamentalmente gracias las historias que Lidia y María contaron mil veces con enorme cariño, a Ramiro, que me ha acompañado en esta recta final. Una mención especial se merece María, dentro y fuera de ese cuarto que siempre estará "al fondo donde el reno".*

*Gracias también a Nacho y al resto de Depar (dejadme que no os nombre uno por uno, necesitaría las listas de Mercedes para cerciorarme de que estáis todos. . . ). Particularmente a Mac y a Jacobo (los grandes maestros del humor suelen ir de dos en dos) y, sobre todo, a $M^a$José, por todo lo que ella sabe.*

*Gracias a los que me acompañaron en Leeds, tanto en la universidad (Emanuele y Sergei; Kostas, James y Piero) como al llegar a casa (Ele, Felix y Li). Gracias a mis chicas Talisa, Giada, Diana y Lily, por sus planes de fin de semana y sus pizzas en los días de nieve.*

*Gracias a "los olímpicos". A Renuncio y $M^a$Carmen por acabar de despertarme la curiosidad por la ciencia. A Patri, Jaime y Vicente con los que tan buenos ratos pasé. Y a todos los chavales que han venido después y que me recuerdan que esto vale la pena.*

*Gracias también a la gente de la Delegación, de Cáritas y de la parroquia (Inma, Jero, Ainhoa, Gema, María, Elena, Dori. . . ), porque más vale tener claro lo que es importante en*

*esta vida, y que el mundo no empieza y acaba en uno mismo.*

*No quiero olvidarme de mis amigos, los de siempre y los de los últimos años. De las amistades forjadas en un autobús, en comedores de facultad y en el Auditorio Nacional. Es decir, gracias a Gloria, Sandra, María, Sonia, Reyes, Isra, Alberto, Félix, Manu; gracias también a las Irenes, los Jaimes y los Javis que me alegran los días.*

*Ignasi se merece un párrafo aparte. Por mi ración diaria de risas descontroladas. Porque Madrid-Barcelona ya no es sólo un partido de fútbol. Porque dos tesis y una boda hacen de 2012 un año apasionante.*

*Las últimas líneas van dedicadas a mi familia. No hay páginas suficientes para explicar lo que les debo. Gracias a mis tíos por estar ahí. Y a mi hermano Alberto, por sus constantes desvelos. Gracias a mis padres por su ejemplo y su apoyo, y por la genética química y docente que me ha hecho disfrutar enormemente de estos años de tesis doctoral.*

*Gracias, en definitiva, a todos por todo.*

# Contents

# 1

---

# Introduction

Protein folding is defined as the process by which a polypeptide chain reaches its tridimensional functional layout. This short definition holds an important challenge within the biophysics realm. Proteins present a unique structure (called *native state*[1]) that is said to be somehow encoded in the peptide sequence. What determines the folded structure? Moreover, the native state is acquired spontaneously under physiological conditions in a relatively fast conformational transition. Which are the factors that rule this process? Can we understand and control them?

The latter statement implies that the proteins' conformational search is not random, but guided by the minimization of the free energy of the system, as the Nobel laureate Christian B. Anfinsen stated in 1973.[2] As the free energy of a given system depends on the different interactions that stabilize it, the so-called *protein folding problem*[3] can be tackled from the point of view of a physical-chemist.[4]

Apart from their many significant roles in living beings,[1] proteins are essential in many applied fields. For instance, they can be used as biomaterials because of their favorable mechanical properties.[5,6] They are fundamental in medicine, too, as they form a new family of peptide-based drugs or targets for the treatment of some types of cancer.[7,8] In addition, some proteins are closely related to many neurodegenerative pathologies (such as Alzheimer or Huntington, frequently called *protein folding diseases*[9]).

These applied aspects of proteins are closely related to the protein folding problem, as they share a common hindrance: the formation of alternative (and non functional) structures under certain conditions (such as the high concentrations that are necessary for the preparation of an economically competitive material or medicine) or in the case of folding diseases. These alternative bodies (frequently called amyloid fibrils) are formed through the interaction of several peptide chains, ending in the formation of aggregates that precipitate in solution.[10] They adopt a very characteristic shape known as cross-$\beta$ structure[11] that is shared by all kind of proteins, regardless their specific sequences.

This double dimension of the protein folding problem (folding itself and aggregation processes) may be explained through thermodynamics: if we know which the fundamental interactions of the system are and how they behave, we will understand the driving rules of protein folding and aggregation.

In this PhD Project we have focused on a specific kind of interaction: the backbone hydrogen bonds, as they are necessary for the formation of the secondary structure elements in the native state and constitute the main stabilizing interaction in protein aggregates. As we shall explain along this Dissertation, we have adopted a computational approach that, starting from the design of a hydrogen bond potential, has allowed us to pose (and answer) a number of questions: which are the distinguishing characteristics of a hydrogen bond? Which is its role in the protein folding problem? How do hydrogen bonds interplay with the rest of interactions so that they lead to folding or aggregation depending on the environment conditions?

## §1.1   Structural organization and protein interactions

The native structure of proteins is stabilized by different types of interactions. They are responsible for the structural levels in which proteins are organized, as represented in Figure 1.1.

The first level is the **primary structure**, i.e. the amino acid sequence itself. It results from the covalent link or *peptide bond* between the amino and carboxylic ends of the different amino acids that form the polypeptidic chain (see Figure 1.2 for details).

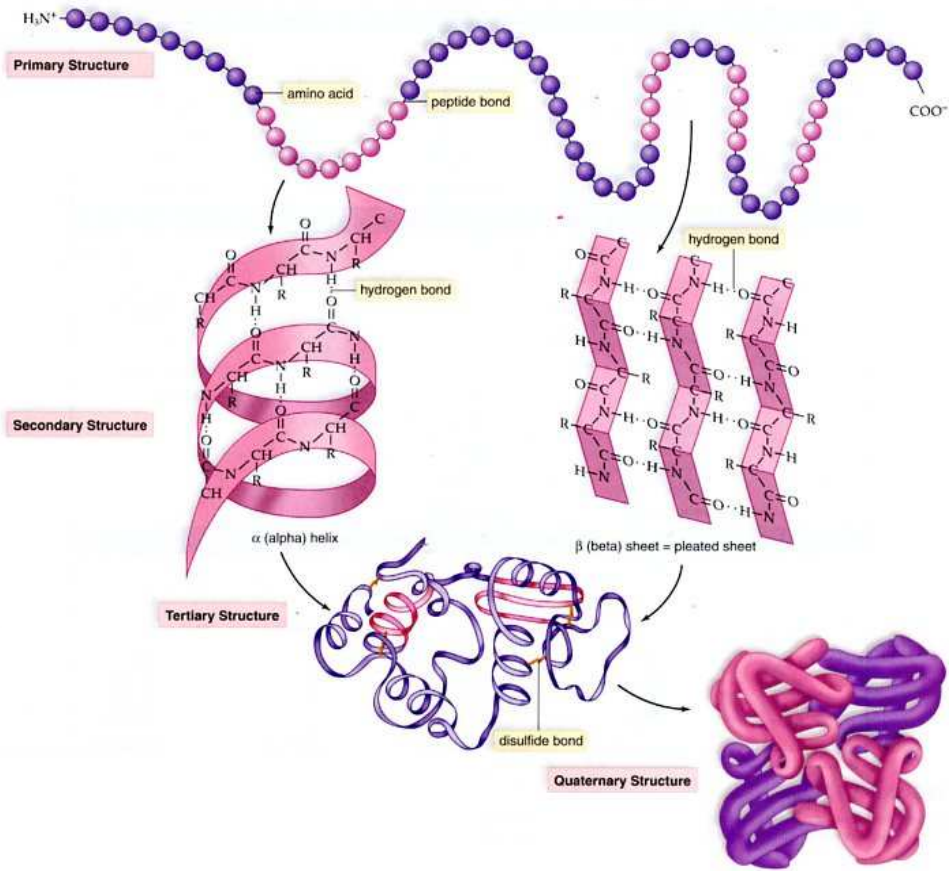

**Figure 1.1:** *Cartoon representation of the structural levels of proteins (primary, secondary, tertiary and quaternary structures). Figure adapted from*[12] *.*

The peptide bond imposes some conformational restrictions to the bonded amino acids due to the double character of the bond. The torsional angles associated to each amino acid $i$ (called $\phi_i$ and $\psi_i$ in the literature) are constrained within certain limits due to steric restrictions, as we have illustrated in Figure 1.3($a$).

The representation of these angles in a $\phi$ versus $\psi$ map is called the Ramachandran plot.[13] Shown in Figure 1.3($b$), it exhibits some highly populated regions. The repetition of these more probable angles along peptide fragments leads to the formation of regular motifs that constitute the second level of organization: the **secondary structure**.

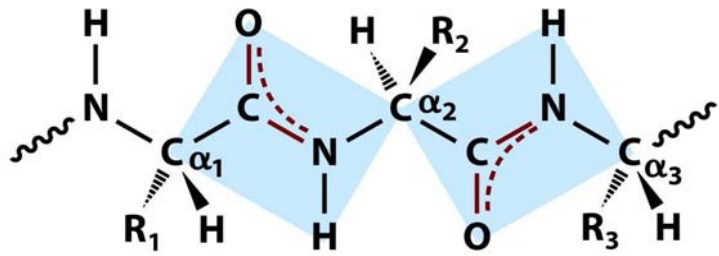

**Figure 1.2:** *Peptide bonds in a polypeptidic chain fragment. Their partial double bond nature creates a planar geometry in its contiguous atoms, as shown by the blue regions. Figure adapted from[1].*

The secondary structure consists, then, in the repetitive arrangements in which different segments of the polypeptide chain are placed. Among other forces, they are chiefly stabilized by hydrogen bonds between the amino and carbonyl groups of the main chain. As these atoms are present in every protein, secondary structure is also ubiquitous in nearly all the proteins.[1,15] They may adopt two different schemes, that have been drawn in Figure 1.1:

**α-helices:** They are formed by the hydrogen bond between the carbonyl group of an amino acid $n$ and the amino group of the $(n-4)$ amino acid, resulting in a right-handed helix of 3.6 amino acids by turn.

**β-sheets:** They consist of the placing of two (or more) extended polypeptide segments in a parallel or antiparallel way, so that the amino and carbonyl groups of one fragment are linked by hydrogen bonds to the complementary groups in the other fragment.

The particular arrangement of these motifs in the complete tridimensional layout is known as **tertiary structure** and is characteristic of each protein. The stabilizing interactions of this structural level essentially take place among the lateral chains of the amino acids so, unlike backbone hydrogen bonds, depend on the specific sequence of the protein. These interactions may have different natures: disulfide bridges, hydrophobic interactions,

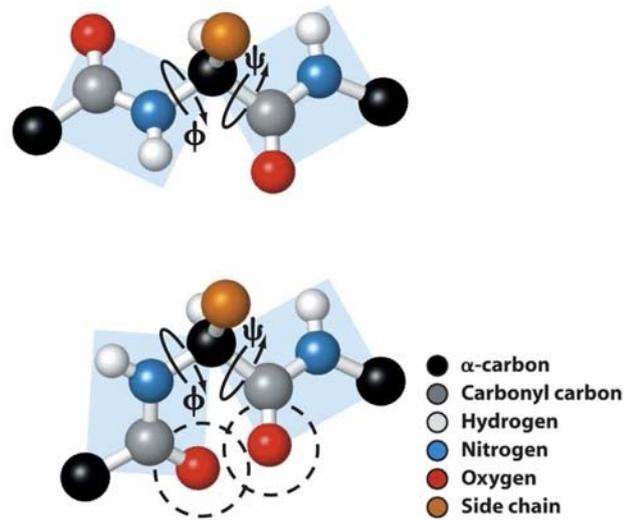

(a) Peptide planes and related torsional angles $\phi$ and $\psi$. The lower cartoon also shows the consequent steric restrictions derived from the peptide bond. Figure adapted from.[1]

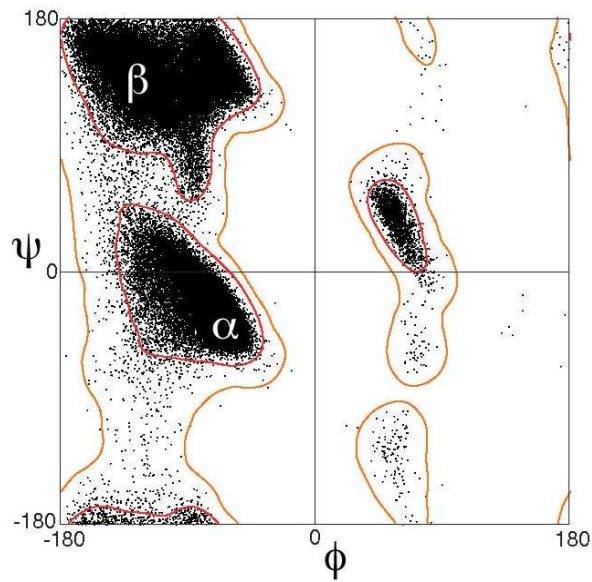

(b) Ramachandran plot for a given dataset of proteins (data taken from[14]). The $\phi$ and $\psi$ values of each amino acid in the dataset are represented by black dots, while the circle lines represent equiprobable values for these variables. The characteristic angles of $\alpha$-helices and $\beta$-sheets are also marked.

**Figure 1.3:** *Torsional constrains derived from the peptide bond.*

electrostatics or hydrogen bonds between lateral chains. Among them, the most common and relevant ones are hydrophobics.[1]

Some functional proteins present more than one polypeptide chain. Each of them has a specific tertiary structure and constitutes an independent covalent unit, but interacts with the rest of the chains forming an oligomer with a given **quaternary structure**. The association between the different chains is stabilized by the same types of interactions as the tertiary structure.

The interactions among polypeptide chains are also a key aspect in aggregation. As we have already stated, aggregates are spontaneously formed by every kind of protein under high concentration conditions. In a concentrated environment, interactions among different polypeptide chains are very frequent and can stabilize a particular structure known as cross-$\beta$ (see Figure 1.4 for a representation).[11,16] The formation of aggregates is a common feature to all proteins regardless their sequence,[17] and the nature of the stabilizing interactions is also shared among all of them: they are the hydrogen bonds of the main chain,[18] similarly to the secondary structure case. Consequently, hydrogen bonds play an important role in the understanding of the competition between protein folding and aggregation.[19]

We have shown that there is a tight connection between the structural organization of a protein and its stabilizing interactions. How do they merge to build the native conformation? How is the native structure achieved? Why do aggregates appear? All of these questions belong to the *protein folding problem*.

## §1.2   The protein folding problem

If protein folding was a random event, the so-called *Levinthal paradox* states that this process would take longer than the age of the universe.[20] As it usually takes from milliseconds to seconds, it is obviously an oriented transition. In fact, Anfinsen showed that proteins do not only fold *in vivo*, but also in a test tube,[2] proving that protein folding is

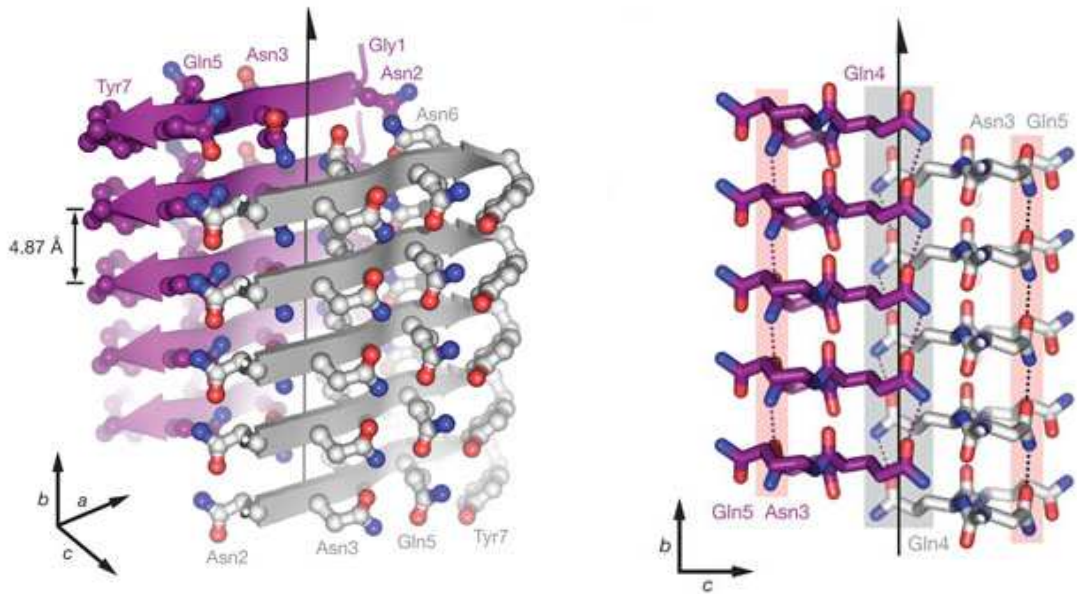

(a) Pair of sheets of the cross-$\beta$ structure. The growth direction is marked with a vertical arrow.

(b) Lateral view of the structure, showing the vertical shift of the sheets.

**Figure 1.4:** *Representations of the typical cross-$\beta$ structure of aggregates, obtained from the GNNQQNY peptide. Images adapted from[11].*

driven by the thermodynamics of the system. In this sense, the more we know about the interactions that rule protein folding, the better we will understand this process.

Different theories have been proposed for nearly a century.[21–24] Currently, the most widely accepted one states that there is not a unique folding pathway, but several ones[25, 26] that depend on the system conditions and are mediated by the free energy landscape of the folding process.[2] This approach is known as the *folding funnel theory*.[25, 27]

## 1.2.1 The folding funnel

The *thermodynamic hypothesis* of Anfinsen derived in an increasing importance of the free energy landscape, mainly due to the effort of Bryngelson and Wolynes.[27–29] They propounded the *principle of minimal frustration*,[30] which asserts that evolution has selected the amino acid sequences of natural proteins so that interactions between amino acids favor

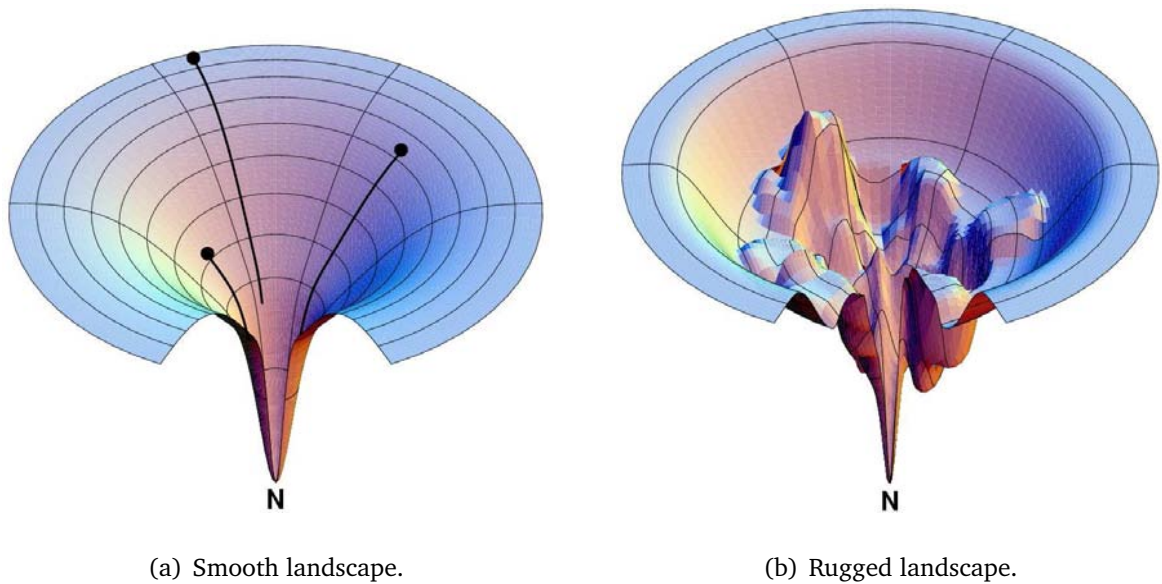

(a) Smooth landscape.                    (b) Rugged landscape.

**Figure 1.5:** *Representations of the folding funnel. Figures adapted from[25] .*

the molecule's acquisition of the folded state.

As a consequence, proteins are generally thought to have funneled energy land-scapes in which the main energetic minimum corresponds to the native state and the dena-tured one is seen as an ensemble of degenerated configurations in which the polypeptide chain presents a large entropy. This folding funnel landscape is unique for a given sequence of amino acids[31] and allows the protein to fold to the native state through any of a large number of pathways and intermediates, rather than being restricted to a single mecha-nism.[25, 26] Each point in the free energy surface represents a possible conformation of the polypeptide chain, while each individual molecule diffuses through the funnel, populating or not this conformation. An extensive exploration of the funnel, accessible by computa-tional means, seems a relevant tool for the understanding of protein thermodynamics and kinetics.

A schematic representation of the folding funnel is shown in Figure 1.5(*a*). Ac-cording to this picture, all the interactions that a protein exhibits in its native state have been evolutionarily designed to contribute to the protein folding in a cooperative way.[26]

Due to the high complexity of proteins, the idea of a smooth funnel is clearly oversimplified. A more realistic scenario is pictured in Figure 1.5(*b*). Although the free energy landscape is still funnel-shaped, the surface presents some ruggedness that results in local minima. Thus, the protein can be transitorily trapped during the folding transition.[32] This may imply the formation of kinetic or thermodynamic intermediates, stabilized by either native or non-native interactions.[26]

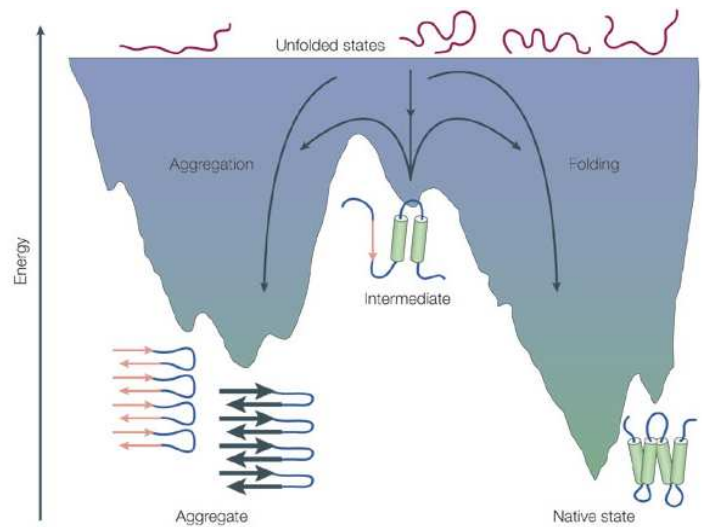

**Figure 1.6:** *Funnel-shaped free energy landscape of protein folding and aggregation. Figure taken from*[35]*.*

The importance of non-native interactions is particularly crucial if aggregation is considered. The presence of aggregates is often represented by an alternative minimum, whose relative depth depends on the environment conditions (see Figure 1.6).[33] The formation of the aggregates takes place through the population of intermediates in which some native or non native secondary structure is observed,[34] highlighting again the implication of hydrogen bonds in the aggregation process.

As a result, a realistic approach to the protein folding problem cannot be limited to the ideally smooth funnel landscape, but must consider explicitly the particular nature of the protein interactions.

## 1.2.2 Understanding individual interactions

In Section 1.1 we showed that the organizing levels of proteins are controlled by different types of interactions that work together to obtain the complete folded structure; in the case of aggregate formation, the specific interactions among chains play also an important role.

In general terms, two of these interactions deserve a particular attention due to their influence in the aforementioned processes. **Hydrophobic interactions** are said to be the main driving force in the formation of folded structures. The so-called hydrophobic effect[3] (related to the entropic cost of ordering the water molecules around a denatured protein) explains the natural tendency of a protein to collapse, burying some residues (depending on their hydrophobicity) inside of the folded structure.

**Backbone hydrogen bonds**, formed between the amino and carbonyl groups of the main chain of the polypeptide, stabilize secondary structure elements[1] and are said to be determinant in aggregation processes.[18] As the principal target of this PhD Project is the interplay between folding and aggregation, we have paid special attention to hydrogen bonds

The role of the latter in the protein folding process is still under debate.[36] Over the last half century, views about the relationship among the different forces of protein folding have changed many times. At first, intramolecular hydrogen bonds were thought to be the driving force of folding,[37] but afterwards this major role was assigned to the hydrophobic effect.[24] Since then, hydrogen bonds have been argued to be either stabilizing[38,39] or destabilizing.[40] In addition, the undoubtfully key role of hydrogen bonds in protein aggregation, where they are the main driving force,[18] has also put this interaction in the limelight.[41,42] How is protein folding achieved? What is the role of hydrogen bonds?

### 1.2.3   Investigating protein folding and hydrogen bonds

Hydrogen bonds have received a remarkable attention from the scientific community since quite a long ago,[1,43] as valuable insight can be obtained by different techniques, either experimental or theoretical. In general terms, hydrogen bonds can be defined as the attractive interaction of a hydrogen atom with an electronegative atom, such as nitrogen, oxygen or fluorine, that comes from another molecule or chemical group. In the case of proteins and protein backbones in particular, a hydrogen bond consists in the partially covalent attraction between the amino group of a given amino acid and the carbonyl group of a different one, mediated by a certain share of the hydrogen that comes from the amino group.

Within the experimental realm, H/D exchange constitutes one of the most powerful methods.[44] Coupled with mass spectrometry[45] or nuclear magnetic spectrometry,[46] it detects hydrogen-bonded pairs. Thanks to the chirality of secondary structure elements, the hydrogen bonded amino acids absorb circularly polarized light, making circular dichroism a suitable technique for the indirect detection of the secondary structure of a protein.[47] The combination of these hydrogen bond analyses with other common techniques for protein folding (such as differential scanning calorimetry[48] or infrared spectroscopy[49]) is extremely useful for extracting thermodynamic and kinetic information of protein folding.

Aggregates, however, present more technical difficulties for their experimental study. As we have already stated, they precipitate in solution in an amorphous way, which means that the common structural tools such as X-ray crystallography and nuclear magnetic resonance cannot be used. Instead, they are usually characterized by indirect techniques such as circular dichroism (that detects secondary structure, no matter the protein state, as already mentioned) or light scattering.[50] Additionally, they have been studied by other methods such as electron microscopy[51] or small angle X-ray scattering methods,[52] but high resolution structures have not been obtained and a comprehensive view of the evolution of aggregates with time is still missing.

The computational strategy has, then, played an important role in the understanding of protein folding and aggregation.[42,53] Regardless the specific characteristics of each simulation, every approach must address two items: an efficient algorithm for the conformational search and an suitable description of the protein and its interactions.[54]

The election of the algorithm mainly depends on two factors: the kind of information we need and the complexity of the task (i.e. the size of the system and the timescale of the process). There are two main methods: molecular dynamics and Monte Carlo algorithms. The former ones are usually employed for the obtention of dynamic properties and were initially limited to short timescales; in contrast, Monte Carlo can be applied to longer processes and gives thermodynamic information, but used to miss the dynamics of the process.[55]

More recently, however, molecular dynamics has extended its accessible timescales thanks to different methods of accelerated dynamics[56,57] (such as umbrella sampling, metadynamics, paradynamics, boxed molecular dynamics, etc.); on the other hand, kinetic Monte Carlo is able to explore a process dynamics keeping the good properties of the traditional method.[58,59]

In relation to the system description, the size of the system limits again the accessible timescale, as the computational time increases with the number of interacting centers. In this way, the resolution ranges from atomistic models, where the protein is described in full detail (but is commonly circumscribed to local processes), to elastic networks, where the different beads are roughly described as springs with a given constant force.[60]

Based on the previous experience of our research group,[61–64] we have focused our attention on an intermediate resolution approach: the so-called coarse-grained models.[42,65–67]

## §1.3   Coarse-grained potentials: state of the art

These potentials are characterized by the description of each amino acid by one or more centers of interaction, reducing the degrees of freedom of the system but still being able to keep the physics of a protein. They are especially appealing for protein folding studies, as they imply a remarkable speed up in the landscape sampling (that must explore both the constrained motions of the folded protein and the large conformational variability of the unfolded state), still describing the system in an accurate way for the structural and energetic properties of proteins processes (if the interacting potential is suitably chosen).

The lack of an atomic definition of the system derives in a purpose-based design for the potential, which must take into account the particular characteristics of the interaction itself. We have mainly analyzed the role of hydrogen bonds, which are essential for the correct modelization of secondary structure elements and aggregation processes. As we will explain later, the use of a single interaction center per amino acid presents some technical difficulties that have derived in the design of our own hydrogen bond potential.

A complete understanding of the folding event also requires an suitable treatment for the tertiary structure of proteins. In this Project, we have considered two different strategies: the use of a structure-based potential and a hydrophobic one.

### 1.3.1 Hydrogen bonds

The backbone hydrogen bond is one of the most common and relevant interactions in proteins, so understanding the internal basis of these interactions has become an active research field in the last few years,[68,69] especially for the computational community.[70–73] Although some efforts have been carried out in the all-atom side,[74,75] folding studies usually require a coarse-grained approach, due to timescale limitations.[72,76–79] Lowering the level of detail of the protein description allows a faster exploration of the conformational space, losing the specific information of each atom, yet trying to provide a realistic description of the behavior of the protein as a whole.

This type of hydrogen bond potentials has been used either combined with other interacting potentials for the complete folding of proteins[80–82] or in isolation, where they are suitable for the exploration of the interplay between helices and sheets or peptide ag-

gregation.[83–85]

These potentials frequently use a mean-field approximation.[86,87] Thus, its design comes after a thorough statistical study of the structural data from an extensive dataset of proteins, usually taken from the Protein Data Bank database.[88] The case of hydrogen bonds using a single bead representation exhibits, however, some peculiarities:

1. It is the only non-bonded protein interaction with a partially covalent character and, consequently, a directional nature. As an example, we show in Figure 1.7(*a*) an all-atom representation of an $\alpha$-helix where the hydrogen bonds and their directions have been marked with dashed black lines. As this directional feature has not been widely considered until very recently,[79,85] most of the hydrogen bond potentials present a spherical symmetry that cannot model hydrogen bonds accurately.[36]

2. The use of a single center of interaction per amino acid implies that neither the amino nor the carbonyl group that form the hydrogen bond are explicitly present. This is illustrated in Figure 1.7(*b*), where only the $\alpha$-carbon trace of the peptide has been drawn. As a result, even the models that take into account the geometrical restrictions of hydrogen bonds usually obtain distorted structures, as we will show along this Dissertation.

The evaluation of coarse-grained hydrogen bond potentials has been an important task for our research group interests during the last years, either in fragment alignment studies[90] or in preliminary folding studies developed at the first stages of this PhD project. We have found that the lack of an appropriate consideration of these distinctive characteristics of hydrogen bonds usually leads to distorted structures and inaccurate descriptions of the hydrogen bond thermodynamics.

One of the alternatives to undertake these drawbacks could be an increment in the number of centers of interaction per amino acid, significantly raising the computational cost of the model. Although this approach has been used by some groups,[78,80] we have kept the single-center approach, looking for relatively fast potentials in consonance with the rest of the potentials developed by our group.[61–63] Therefore, we have designed our own hydrogen bond potential (see Chapter 3),[91] aiming to overcome these aforementioned drawbacks through a deep understanding of how real hydrogen bonds behave.

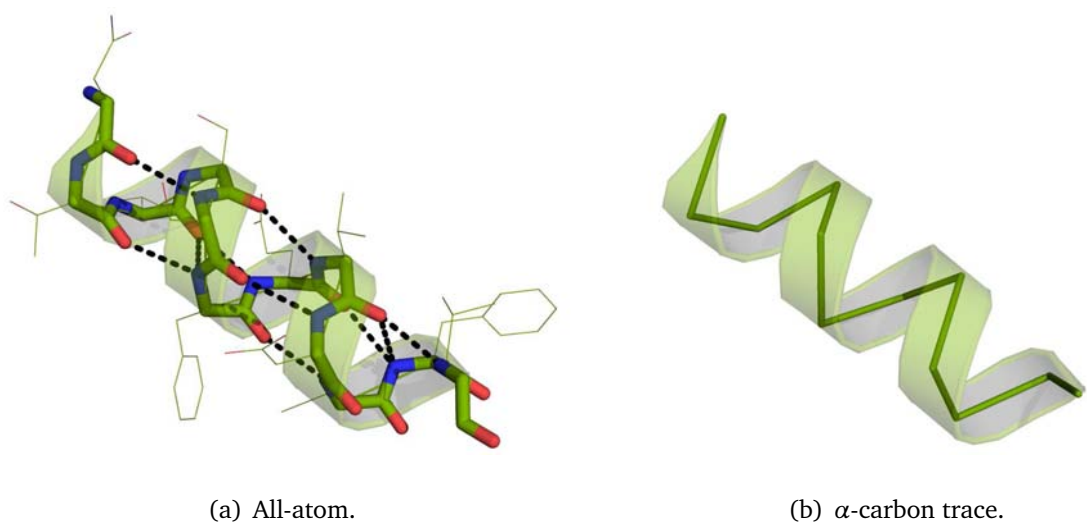

(a) All-atom.                    (b) α-carbon trace.

**Figure 1.7:** *Representation of an α-helix with two different levels of resolution (residues 4-15 of the NS1 protein of the influenza A virus, drawn with PyMOL[89]).*

## 1.3.2  Structure-based potentials

According to the principle of minimal frustration, the interactions that are present in the native state have been somehow evolutionarily chosen to favor the particular native topology of each protein. This view boosted the use of the so-called Gō or structure-based potentials since the 70's.[65,92] Initially proposed by Gō *et al.*, their guiding rule is that only the native interactions (i.e. those between residues that are close in the native tridimensional layout) contribute in a favorable way to the acquisition of the native structure.

In this way, pure Gō interactions are independent on the nature of the amino acids, as the only interacting criterion is spatial proximity.[93] All of them are equally stabilizing, defining therefore smooth funnel-shaped landscapes[26] like the one in Figure 1.5(*a*). Their enormous simplicity makes these potentials computationally efficient, while describing protein folding in a reasonable way.[3,93,94] As a result, they have successfully investigated fold-ing transitions and even on-pathway intermediates.[63, 95–97]

In our group, we developed a structure-based potential some years ago,[62] providing a thorough understanding of coarse-grained modeling as well as a useful insight into folding transitions according to perfect funnel landscapes.[64, 98, 99]

However, Gō models are unsuitable in some cases, as non native interactions are not considered in these models and the unfolded state is poorly defined.[97, 100–102] The investigation of complex folding processes such as the presence of off-pathway intermediates[103] or aggregates[104] implies, then, a more sophisticated view of Gō potentials.[105]

Even if these latter circumstances do not hold, making every interaction equal seems rather crude. Each interaction may have a different strength depending on the chemical nature of the involved amino acids; besides, hydrogen bonds are directional, so the usually spherical Gō potentials cannot grasp their characteristics.

Therefore, a kind of "second generation" Gō models that explicitly consider individual interactions have arisen.[102, 106] The resulting potentials enhance the accuracy of the original Gō models and provide a more detailed tool for protein folding studies. We have also undertaken this task, merging the plain Gō model previously developed in our group[62] with the hydrogen bond one,[91] creating a combined potential. A detailed explanation will be provided in Chapter 4, where some examples will be discussed.

In general terms, the main drawback of Gō potentials lies in the necessity of a native structure. The native state always drives folding, preventing by definition the formation of aggregates with a different structure. Although it maybe suitable for domain swapping *in silico* experiments,[81, 99] it is certainly not enough for broader aggregation studies. This latter purpose requires statistical potentials based on the general behavior of the different interactions and, more specifically, in the most common ones within proteins: hydrogen bonds and hydrophobics.

### 1.3.3   Hydrophobic potentials

The interactions between pairs of residues obey the rules imposed by their chemical natures. Even if we do not know the *rules* (as it happens when coarse-grained descriptions are used and *ab initio* potentials cannot be applied), it is possible to extract them by the statistical analysis of the *results*, i.e. how the residues are placed. This is the essence of mean field potentials, regardless the kind of interaction they aim to describe.[57, 107]

As it also happens in hydrogen bond potentials, one of the main differences among hydrophobic ones is the number of centers of interaction considered in the model and their location.[108, 109] Within this PhD project, we have paid special attention to computational efficiency, so the use of a single center per amino acid is essential. Thus, we have used a hydrophobic potential defined in terms of the same coarse-grained description ($\alpha$-carbons) as the rest of the interactions.[110]

Hydrophobic interactions take place mostly among the lateral chains of the amino acids, being sequence-specific. Locating them in the $\alpha$-carbons implies missing some resolution. For that reason, these simplified hydrophobic potentials generally use highly regular sequences and structures.[110–112] In spite of their simplicity, these potentials can provide useful insight into the folding/aggregation problem, as we shall see in Chapter 5.

## §1.4   Principal aims of this PhD Thesis

In this work we aim to study the role of hydrogen bonds as a common feature in protein folding and aggregation. Due to their chemical peculiarities in terms of strength and directionality, a particular attention must be paid to the definition of the hydrogen bond potential itself, as it should merge simplicity and an accurate description of their geometry and thermodynamics.

This global target has been tackled through a computational approach based on a minimalist description of the protein and the proper design of algorithms. Due to the large sizes of the studied systems (such as the full energetic and structural transitions for complete proteins and aggregation-prone systems), the performance of computationally costly simulations has been an ever-present challenge along this PhD Project. For this reason, we have been particularly careful about efficient programming and code parallelization. Most of the **technical details** will be described in Chapter 2 of this Dissertation.

As the heart of this Project is the study of hydrogen bonds and their role in protein systems through computational means, a proper interaction potential is mandatory. For this reason, we evaluated the properties of many of the previously existing potentials, looking for a hydrogen bond model that fulfilled the requirements of simplicity and accuracy. However, we could not find a model that combined a proper geometric description of the hydrogen bond interaction with a coarse-grained representation. For this reason, the first principal aim of this PhD Thesis has been the obtention of a hydrogen bond interaction potential to carry out our global Project.

Chapter 3 is dedicated to the description of this task, that does not only involve the design of the potential itself, but also its testing. The evaluation of the potential has been performed through a complete study of sequenceless peptide systems under the action of this potential. The lack of tertiary structure in these systems allows the study of the hydrogen bond effects in a full system under different environment conditions, such as temperature and concentration, providing a first glance of **peptide aggregation**.

However, our final aim is not just the understanding of peptide behavior, but of complete proteins. And, for this purpose, we need additional potentials to describe the characteristic tertiary interactions of proteins. Along this Dissertation, we will discuss two different points of view.

The first one consists in the combination of the hydrogen bond potential with a structure-based one, so that hydrogen bonds are treated in terms of their specific potential, and the structure-based potential applies to the rest of the native interactions. Note that this strategy is only possible if the hydrogen bond potential is able to describe in accurate terms the native geometry of secondary structure elements. Therefore, in Chapter 4 we will describe this **combined Gō + hydrogen bond potential**. Thanks to this strategy, we intend to analyze the implications of a proper definition of hydrogen bonds in the thermodynamic and dynamic aspects of protein folding.

All along this PhD Project, we have mainly focused on structural and thermodynamic properties of the studied systems, as the existing algorithms in the research group were based on a traditional Monte Carlo scheme. In the case of Chapter 4, the protein folding studies have involved some kinetic analyses, too. This additional aspect has been motivated by two main reasons. The first one, mainly methodological, is the development of a Kinetic Monte Carlo algorithm to enable the obtention of dynamic information, which was not previously available in the research group. The second reason is the building of a complete view of the folding process, combining thermodynamic and kinetic information.

Going back to aggregation studies, neither hydrogen bonds on their own nor combined with a structure-based potential can provide a complete view of aggregation. For that purpose, we have undertaken a second strategy, based on the use of a generic hydrophobic model in combination with the hydrogen bond one. In this way, the two main factors of folding and aggregation are merged to create a simple but complete **hydrophobics + hydrogen bonds potential** that will be detailed in Chapter 5.

The application of this potential allows the re-study of peptide aggregation, including sequence as an additional factor. However, the core of this Chapter will be dedicated to the obtention and understanding of complete proteins, defined in very basic terms due to the simplicity of the interaction potential. We will define proteins with different folded shapes depending on the used sequence. After that, we will study the competition between folding and aggregation, and how sequence can influence the interplay between these processes, providing therefore a comprehensive view of these two aspects and the role that hydrogen bonds have in them.

Finally, Chapter 6 will consist on a summary of the most relevant results obtained in this PhD, as well as some concluding remarks.

# 2

---

# Methods

In Chapter 1 we have stressed the role of computation in the study of biological processes like protein folding. It constitutes a valuable tool for the extensive and efficient sampling of the free energy landscape. The constant improvement of computers, the efforts in algorithm optimization and the publication of new refined sampling methodologies are fundamental for the undertaking of more and more complex problems like the competition between folding and aggregation.

The design of the simulation methodology is, then, crucial for a suitable description of the process we would like to solve. The simulation technique for the energy landscape sampling must be adapted to the type of information we are interested in. Similarly, the degree of resolution of the system description may vary from very detailed and computationally costly models to low resolution ones, according to the kind of information we are interested in.

In this Chapter we will expose the main features of our simulation algorithms and system definition. As we have already stated, we have chosen a reduced representation that has allowed us to undertake intricate problems at, in principle, an affordable CPU time.

Along the development of this PhD thesis, however, the size of our systems and the complexity of their behavior have continuously increased from single peptide numerical experiments to aggregated proteins. Together with the availability of multi-core computers, this has lead us to parallelize our main simulation programs. For that reason, we have also included in this Chapter some remarks on code parallelization.

Extracting relevant conclusions does not only need sophisticated simulation methodologies, but also a bold analysis of the data. Thus, the last part of this Chapter is dedicated to the discussion of some relevant tools that we have used, or even designed ourselves, for the interpretation of the results.

## §2.1   The Monte Carlo method

Given the broadness and complexity of the folding landscape, we need a highly efficient sampling technique. As we are mainly interested in the thermodynamic aspects of folding, we have chosen the Monte Carlo methodology.[55] Unlike Molecular Dynamics (where Newton laws must be integrated step by step), Monte Carlo does not need differentiable potentials. Thus, this technique is particularly suitable for hydrogen bonding, where all-or-none potentials are frequently used.[78, 113]

Moreover, Monte Carlo is not limited to thermodynamic information, but is also suitable to kinetic studies through the so-called *Kinetic Monte Carlo* method[114–116] (that we will describe in Section 2.4), providing therefore a wide range of possibilities for folding and aggregation studies.

The Monte Carlo method was initially proposed at the end of the 1940's[117] and consists, for our purposes, of the sampling of the energy landscape of a given system[55] in terms of a certain probability that depends on its conditions. During a Monte Carlo simulation, different configurations are generated through non-physical movements (that will be described in the next Section), following the so-called *Markov chain algorithm*, i.e. these configurations are stochastically created and present no correlation with the previous ones, losing in principle the dynamic information of the process. In order to preferably explore the most representative states of the folding funnel, the so-called *importance sampling* methodology is applied by means of the *Metropolis criterium*, which accepts or rejects the new configuration according to the energetic Boltzmann distribution at a given temperature.[118]

A scheme of a Monte Carlo algorithm is sketched in Figure 2.1. It has the following elements:

1. Starting from a given configuration $C_o$ with energy $E_o$, a new one $C_n$ is built through some non-physical movement and its energy $E_n$ is calculated.

2. The new configuration is accepted with a certain probability $p_{acc}$ that depends on the associated Boltzmann factor, $p = \exp(-\beta \Delta E)$:

$$p_{acc} = \min[1, \exp(-\beta \Delta E)] = \min[1, p] \tag{2.1}$$

In this formula, $\Delta E$ is the energy change between the new and old configurations ($\Delta E = E_n - E_o$) and $\beta = (k_B T)^{-1}$, where $k_B$ is the Boltzmann constant and $T$ is the system temperature. To evaluate $p_{acc}$, a random number $\gamma$ between 0 and 1 is uniformly generated and compared to $p$. If $p \geq \gamma$, $C_n$ is accepted and a new configuration is generated starting from it; in this way, $C_n$ becomes $C_o$ for the next Monte Carlo cycle (right branch in the diagram of Figure 2.1). If $p < \gamma$, $C_n$ is rejected and a different configuration is created from $C_o$.

Initially, the Monte Carlo method was applied at constant temperature or under a simulated annealing scheme,[55] in which temperature progressively decreases. In this fashion, it first samples the high energy unfolded states and ends in the native state as the system cools down. The main disadvantage of simulated annealing is the possibility of getting trapped in local minima, something that becomes more probable as the system complexity increases, i.e. the explored folding funnels become more rugged, as shown in Figure 1.5(*b*).

Instead, we have used the *parallel tempering* or replica exchange technique.[119,120] It consists in the building of $N$ replicas of the system that evolve simultaneously, each one at a different temperature. Periodically, we try to exchange the configurations between pairs of replicas $C_a$ and $C_b$ at consecutive temperatures $T_a$ and $T_b$. The acceptance criterium depends again on the linked Boltzmann factors:

$$p_{acc} = \min\left[1, \frac{\exp(-\beta_a E_b - \beta_b E_a)}{\exp(-\beta_a E_a - \beta_b E_b)}\right] = \min[1, \exp(-\Delta\beta\Delta E)] \qquad (2.2)$$

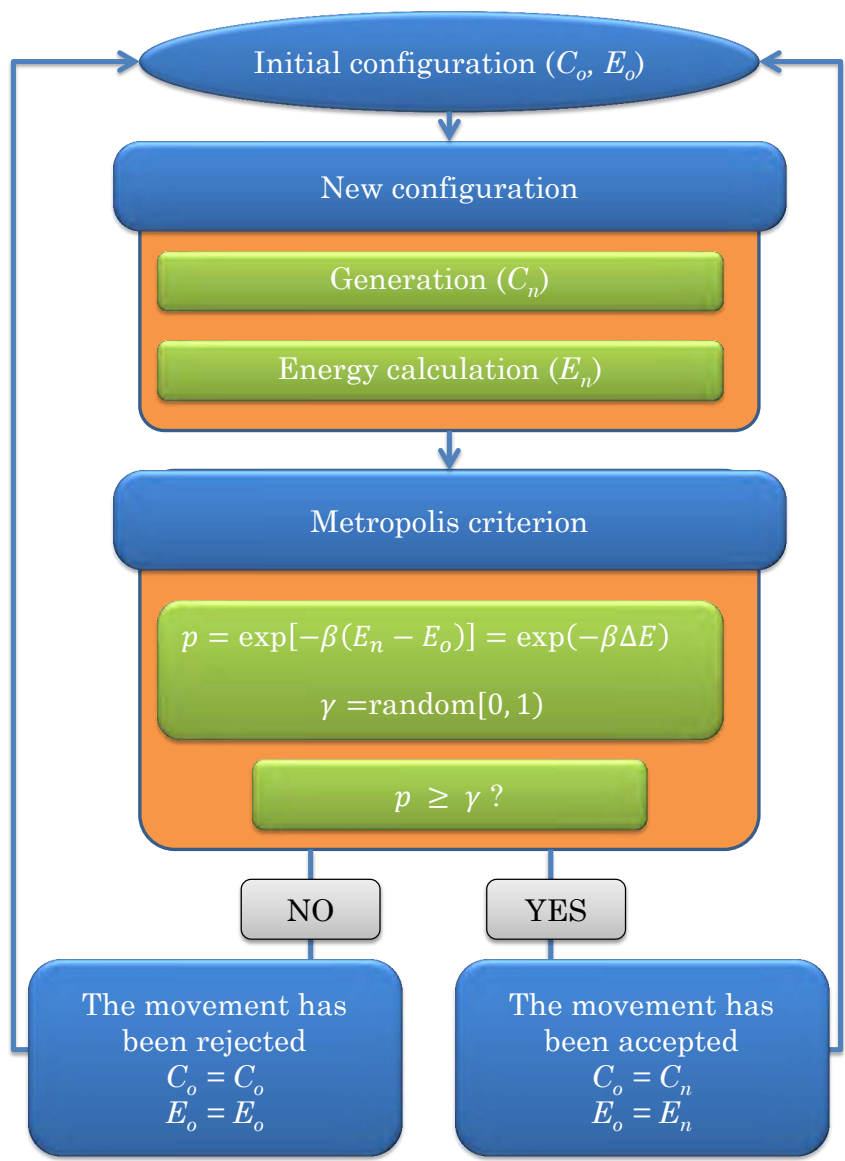

**Figure 2.1:** *Flowchart of the Monte Carlo method.*

In this way, if a configuration $C_a$ at temperature $T_a$ is trapped in a local minimum and can be swapped with a different configuration $C_b$ at a slightly higher temperature $T_b$, $C_a$ will be able to escape from that minimum, while $C_b$ will explore the ruggedness of its accessible configurational space.

The temperatures of the different replicas must be carefully selected to have an efficient swapping rate, that is achieved when there is a significant overlapping of the energy distributions for consecutive temperatures.[121] In this work, the number of replicas varies from 24 to 56 depending on the system complexity and has been chosen according to the system characteristics. The frequency of replica exchange also varies from one simulation to another, usually ranging from $2 \cdot 10^3$ to $10^4$ *Monte Carlo cycles*, where a cycle is defined as the attempt to move each unit of the system for the $N$ replicas.

Each simulation aims to sample the energy landscape as thoroughly as possible. For that purpose, it needs a large number of Monte Carlo cycles. In first place, a given number of *thermalization cycles* are performed (typically, from $3 \cdot 10^6$ to $5 \cdot 10^6$), starting from a completely extended conformation for each chain of the system. Along these cycles, replicas adapt their configurations to their temperatures until they are equilibrated. After that, a large number of additional cycles (from $5 \cdot 10^6$ to $10^7$) is simulated, during which the properties and configurations of the system are periodically registered. In order to improve the statistics, each simulation has been performed at least three times independently.

Up to this point, we have decided the kind of information we want to extract (mainly thermodynamics) and the method to use (Monte Carlo). How will we define the system itself?

# §2.2   System description

An extensive sampling of the free energy surface is essential for a correct understanding of folding. Generally speaking, sampling is tightly linked to the type of representation of the system: atomic descriptions may provide full detail, but are usually constrained to a small portion of the energy surface due to computational limitations; in contrast, the level of resolution allows a faster and larger exploration of the conformational space.

In this work, we have explored the complete folding landscape of complex processes (including several structural transitions within the same simulation) and relatively large systems (up to 200 amino acids distributed in several different peptides or proteins). For that reason, we have chosen an intermediate resolution approach that allows a correct sampling of the full folding process at a reasonable computational cost.[56]

In this Section, we will discuss the definition of the Monte Carlo movements according to our system description, that will help us to explore the energy landscape more efficiently.

## 2.2.1   Protein representation

The protein geometry is defined by one single center of interaction (also referred as *bead* in this context) per amino acid, placed in the $\alpha$-carbon position, as shown in Figure 2.2. It is described by the position vector $\mathbf{r}_i$ associated to each amino acid $i$. Two consecutive centers of interaction $(i-1)$ and $i$ are linked together by the so-called *virtual bond vector*, $\mathbf{v}_i$, of modulus 3.8 Å, corresponding to the average length of a *trans* peptide bond.

The use of an $\alpha$-carbon representation considerably reduces the computational effort. If an average amino acid is formed by about 20 atoms, our coarse grained representation reduces 20 times the number of degrees of freedom to sample; and pair-wise interactions in a system with $N$ atoms decrease from $\frac{1}{2}N^N$ to $\frac{1}{2}(N/20)^{(N/20)}$. This is a huge reduction even for moderate values of $N$. If we take, for instance, the case of a polypeptide with 5 amino acids, a full atom model would sample 300 degrees of freedom and $100^{100}$ interactions, but only 15 degrees of freedom and $5^5$ interactions with our α-carbon representation. For this reason, reduced representations are ideal for folding studies,[55,62,64,93] provided they still keep the physics of the process.

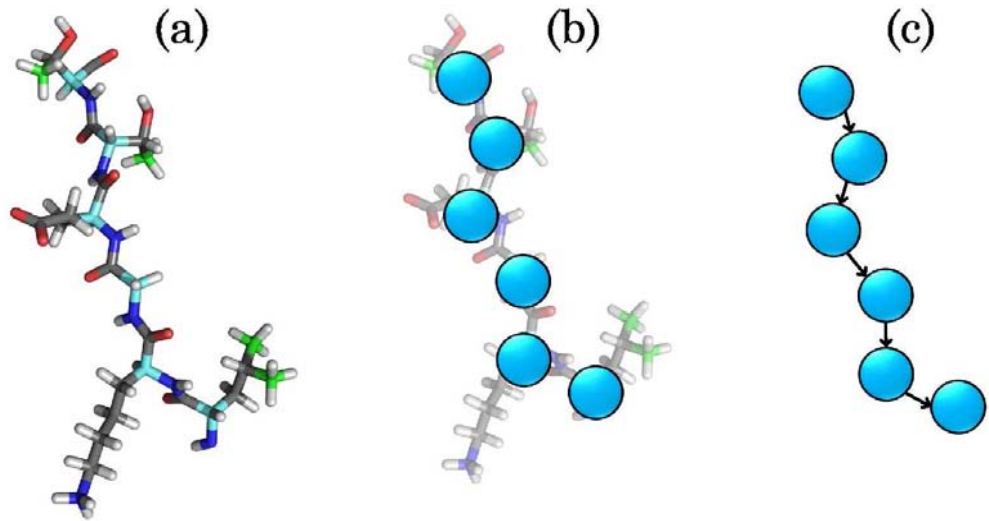

**Figure 2.2:** *Reduction of the degree of resolution of the system for a peptide chain, step by step. (a) All-atom cartoon (drawn with PyMOL[89]). α-carbons are colored in light blue for clarity. (b) Spheres of interaction, placed in the α-carbon positions of the chain. (c) Coarse-grained representation, where the centers of interaction have been linked by virtual bonds.*

Choosing an α-carbon representation, although still advantageous in terms of computational efficiency, may present some technical drawbacks in the accurate description of certain types of interactions such as hydrogen bonds or hydrophobics.[66,96] In the following Chapters we will describe different interaction potentials that have dealt with this challenge. In each specific case, we will explain how we have faced this problem, obtaining precise descriptions through a fine design of the interacting potential.

Once we have decided the protein representation that defines the characteristics of the folding surface, we can tackle our next methodological question: how can we define the *movements* that will drive this search?

## 2.2.2   Improving the conformational search

In the Monte Carlo method, each configuration is built from the previous one without following physical rules. The idea is to generate uncorrelated states of the system that efficiently explore the conformational space. Thus, we have defined a bunch of *movements*. In some cases, they are able to sample the surroundings of a given configuration; in other occasions, they can convert it in a very different one in just one step. The movements we have used can be classified in three different groups:

**Movements linked to a given residue:**  Used in the group's Monte Carlo algorithm since its early stages,[122] they are characterized by the change in the position of *one or more beads* within a certain chain (see Figure 2.3). If we choose to move the first or last residue of a chain, we generate a new virtual bond vector at random, as it is shown in Figure 2.3($a$). If it is an internal bead, we perform one of the three following movements according to their probabilities $p$:

1. Spike ($p = 0.45$): the residue $i$ is rotated around the axis defined by the previous and following residues, as drawn in Figure 2.3($b$).

2. Displacement ($p = 0.45$): equivalent to the movement of the end of the chain, a new bond vector is generated for residue $i$. After that, the rest of the chain is rebuilt accordingly (see Figure 2.3($c$) for details).

3. Pivot ($p = 0.10$): the residues that follow residue $i$ in the same chain are rotated a maximum of $10°$ around a random axis according to the Baker-Watts method,[118] as can be observed in Figure 2.3($d$).

**Movements of a single chain:**  The interest in concentrated peptide systems made us think that it may be suitable that one chain diffuses in the system without losing its internal conformation. For these occasions, we have designed two types of movements that will be chosen with equal probability for every chain (see Figure 2.4):

1. Translation: we generate a tridimensional vector which will move the positions of all the chain residues a maximum distance of 1 Å in a randomly chosen direction, as shown in Figure 2.4(*a*).

2. Rotation: Analogously to the pivot movement, we choose a reference bead inside the chain, and rigidly rotate all the chain residues around it (see Figure 2.4(*b*)).

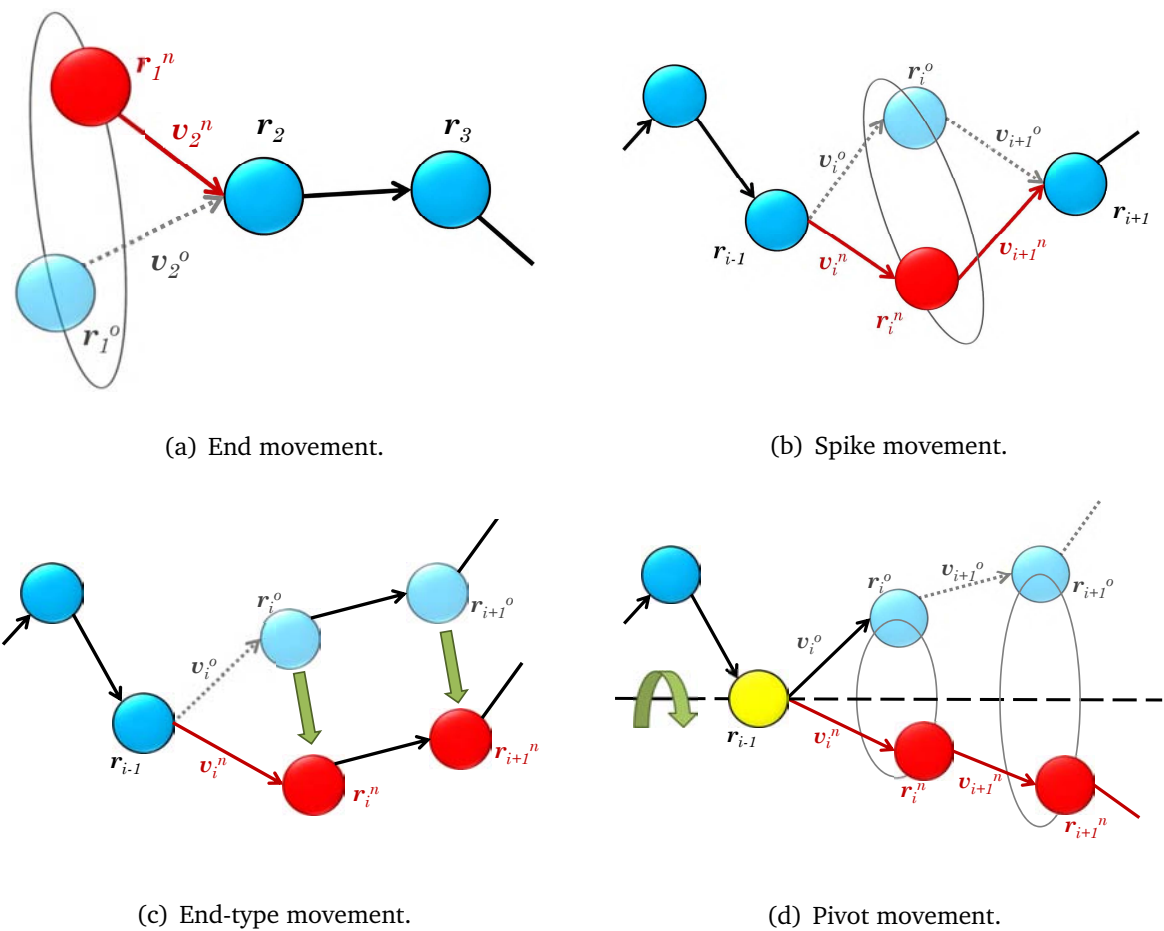

(a) End movement.

(b) Spike movement.

(c) End-type movement.

(d) Pivot movement.

**Figure 2.3:** *Types of movements for a given residue. The residue positions are represented by vector* **r** *and the virtual bonds by the* **v** *vectors. The new vectors and positions are marked with red colors. In (d), the reference bead is colored in yellow.*

**Movements of a group of polypeptide chains:** Other times, several chains are interact-ing (especially in aggregation-oriented simulations). Therefore, we have designed some additional movements where a group of chains can move together without splitting apart. These movements are applied in systems with more than three chains, as shown in Figure 2.5. Two or more of these chains are randomly chosen to do together either a translation movement (called *collective translation*) or a rotation one (*collective rotation*). Similarly to the single chain movements, they are performed with equal probability.

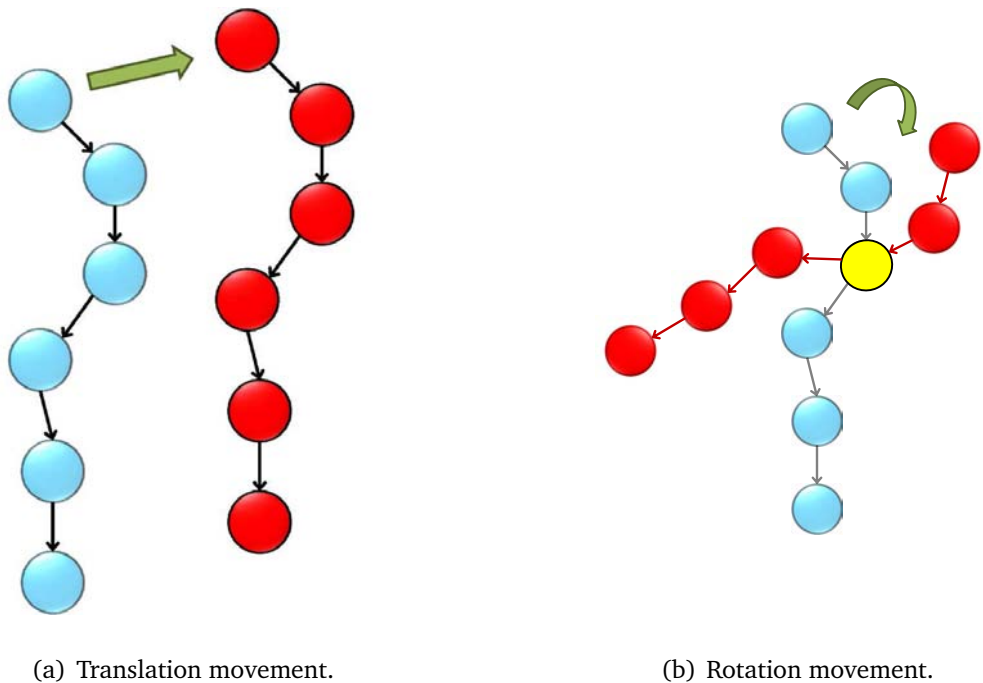

(a) Translation movement.      (b) Rotation movement.

**Figure 2.4:** *Types of movements for a whole chain. Again, the new positions are colored in red.*

Placing several chains together pursues the reproduction of different concentration conditions. How can we simulate them microscopically?

### 2.2.3 Periodic boundary conditions

Simulating different concentration conditions is critical in the context of protein aggregation, as high protein concentrations always derive in the formation of aggregates.

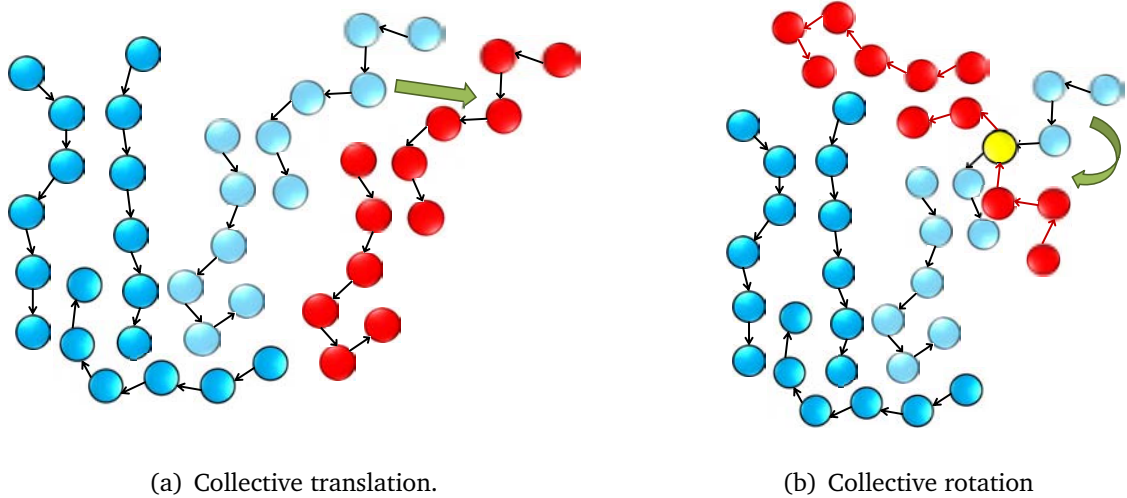

(a) Collective translation.

(b) Collective rotation

**Figure 2.5:** *Types of movements for a set of two or more chains. The new positions are colored in red.*

In this Thesis, we have modified the concentration conditions to model aggregation-prone environments. Real systems are formed by a huge number of particles (similar to the Avogadro number) and a macroscopic volume, so these particles have a *bulk* behavior. Simulating so many particles is unaccessible to numerical simulations, whereas a microscopic confinement would not be appropriate because the results would reflect the presence of the simulation walls instead of presenting the bulk behavior.

The most common strategy is the use of the so-called *periodic boundary conditions*,[55] illustrated in Figure 2.6. In this way, the simulated particles are placed in a reference box of given dimensions. Each particle moves freely according to the previously described movements. If a particle moves out of the original box, a new *mirror particle* gets in through the opposite side of the box.

For example, in Figure 2.6 two particles are moved out from the original box (in a vertical and a horizontal direction), being automatically placed in the opposite side of the reference box. In other words, the real box is surrounded by virtual ones in all directions, removing *de facto* the simulation walls. In this way, the relationship between the number of particles (in this case, polypeptide chains) and the box volume gives a *microscopic*

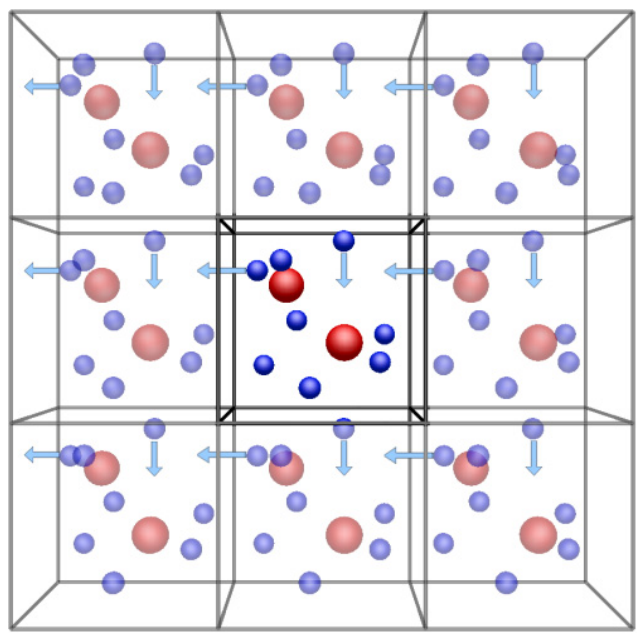

**Figure 2.6:** *Periodic boundary conditions for an ideal system. The reference box is placed in the middle of the Figure, surrounded by mirror boxes.*

*concentration* that qualitatively correlates with a macroscopic one.

The interactions of each particle with the rest of the system are calculated through the *minimum image convention*.[55] Given a residue $i$ and its interaction with residue $j$, the distances with of all the $j'$ mirrors are calculated and the nearest neighbor is identified. Then, we take this minimum image for the calculation of the energy between $i$ and $j$, avoiding in this way to calculate the interaction for each mirror.

To apply this convention, the box must be large enough to avoid the interaction of one particle with its own mirrors. In our case, where the "particles" are chains with multiple beads, not only the single beads must comply with this rule, but the box dimensions also have to prevent the image interactions among beads that belong to the same chain.

## §2.3   Code parallelization

Up to this point, we have spelled out our main simulation technique (Monte Carlo) and the kind of representation (based on the $\alpha$-carbon trace) that we have used in all our simulations. A key factor in the choice of these items has been the computational efficiency, as the aim of this work is the understanding of complex processes such as protein folding and aggregation. Thanks to these methodologies, simulating the folding of a single chain (either a peptide or a small-size protein) can be achieved in no more than tens of hours or a couple of days.

However, the development of this PhD project has implied, from nearly its very early stages, to deal with multiple chain systems (that also derived in the definition of periodic boundary conditions in all our programs) and more and more complex scenarios that required a more extensive sampling of the free energy surface. These new challenges notably increase the simulation times, making these experiments extremely costly with traditional methodologies.

For that reason, we decided to take advantage of the new hardware possibilities, specifically the shared-memory architecture. Using this kind of computers, we have parallelized our codes, making all the single processors (or *cores*) of a machine to work together in a single simulation.

In order to parallelize a program, its different parts need to be evaluated and classified in terms of the relationships among them. If some piece of code does not need the rest of the program, it can be sent to one core that will perform that given task independently. During the execution of the program, each independent part of the code is distributed to the different processors by means of the so-called *threads*. These threads are opened and closed dynamically by the *master thread* of the program, that also executes the *serial regions* (those that are not parallelized).

There are many ways of parallelizing a program, which depend on the computer architecture and the programming language in which the code is written.[123] In our case, our aim was to parallelize a FORTRAN 90 code in multicore machines with shared memory (several cores that have a common memory reservoir and, thus, can easily have shared variables), so the OpenMP methodology[124, 125] seems ideal for our purposes.

In addition, parallel tempering Monte Carlo method is an optimal candidate for this kind of parallelization, as each replica can run independently (i.e. in a different core) except in replica exchange trials. We show a working example of our parallelization strategy in Figure 2.7.

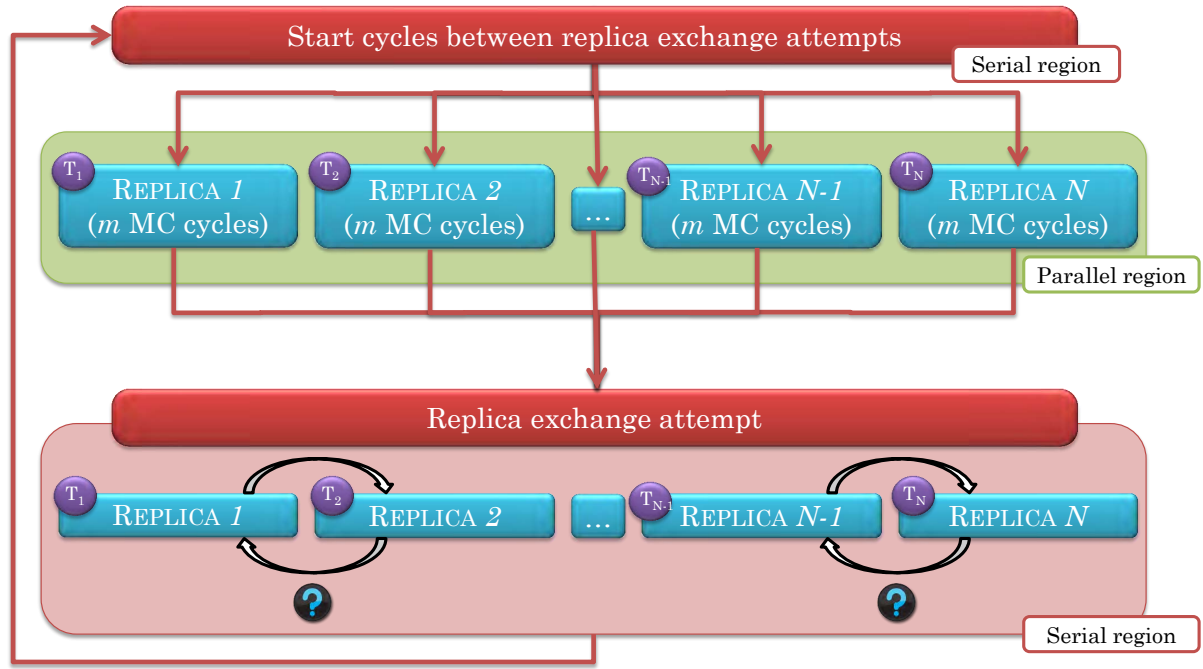

**Figure 2.7:** *Flowchart of our parallelized Monte Carlo code.*

The total number of Monte Carlo cycles, $M$, is divided in $M_p$ packages, according to the replica exchange trial frequency $m$ (number of Monte Carlo cycles between replica exchange attempts): $M_p = M/m$. Then, the following scheme is repeated for each of the $M_p$ packages:

1. The master thread creates the independent threads that will be progressively allocated to the different cores available in the computer (for this reason, the highest efficency is reached when the number of independent threads is a multiple of the number of cores).

2. Each thread performs $m$ Monte Carlo cycles for its replica in a given core.

3. Once all the threads have finished, the master thread receives again all the data and tries the replica exchange between all consecutive pairs of replicas.

Thanks to the particularly suitable characteristics of the parallel tempering Monte Carlo algorithm, we have achieved very high efficiency rates (close to 90%) for computers with 8 to 24 independent cores. This implies that simulations that could last months with the serial program can be performed in just days, allowing complex and ambitious studies that would not have been possible otherwise.

## §2.4   Kinetic Monte Carlo

As we have already explained, Monte Carlo is particularly suitable for the calculation of thermodynamic information and shows a remarkable efficiency, especially if combined with parallelization strategies. Nevertheless, dynamic information is also essential for a correct understanding of folding processes and cannot be obtained through the traditional scheme.

We have used a different technique for this purpose, known as *Kinetic Monte Carlo*.[58,59,114,116] Keeping the Monte Carlo main scheme, it is able to obtain qualitative kinetic information thanks to some modifications that we describe in this Section.

The underlying idea of Kinetic Monte Carlo is that the timescale of the overall process (e.g. the complete folding of a protein) is considerably slower than the single Monte Carlo movements.[59] As a consequence, it is assumed that folding is achieved in a large number of much smaller steps (Monte Carlo cycles), proportional to the real timescale of the process. In this way, the evolution of the Kinetic Monte Carlo system can be compared to the real one.

The underlying scheme is equivalent to the one of Figure 2.1, being driven by the evaluation of the related Boltzmann factors. However, in this case we do not use replica exchange, but a different methodology expressed in Figure 2.8 that is described as follows:

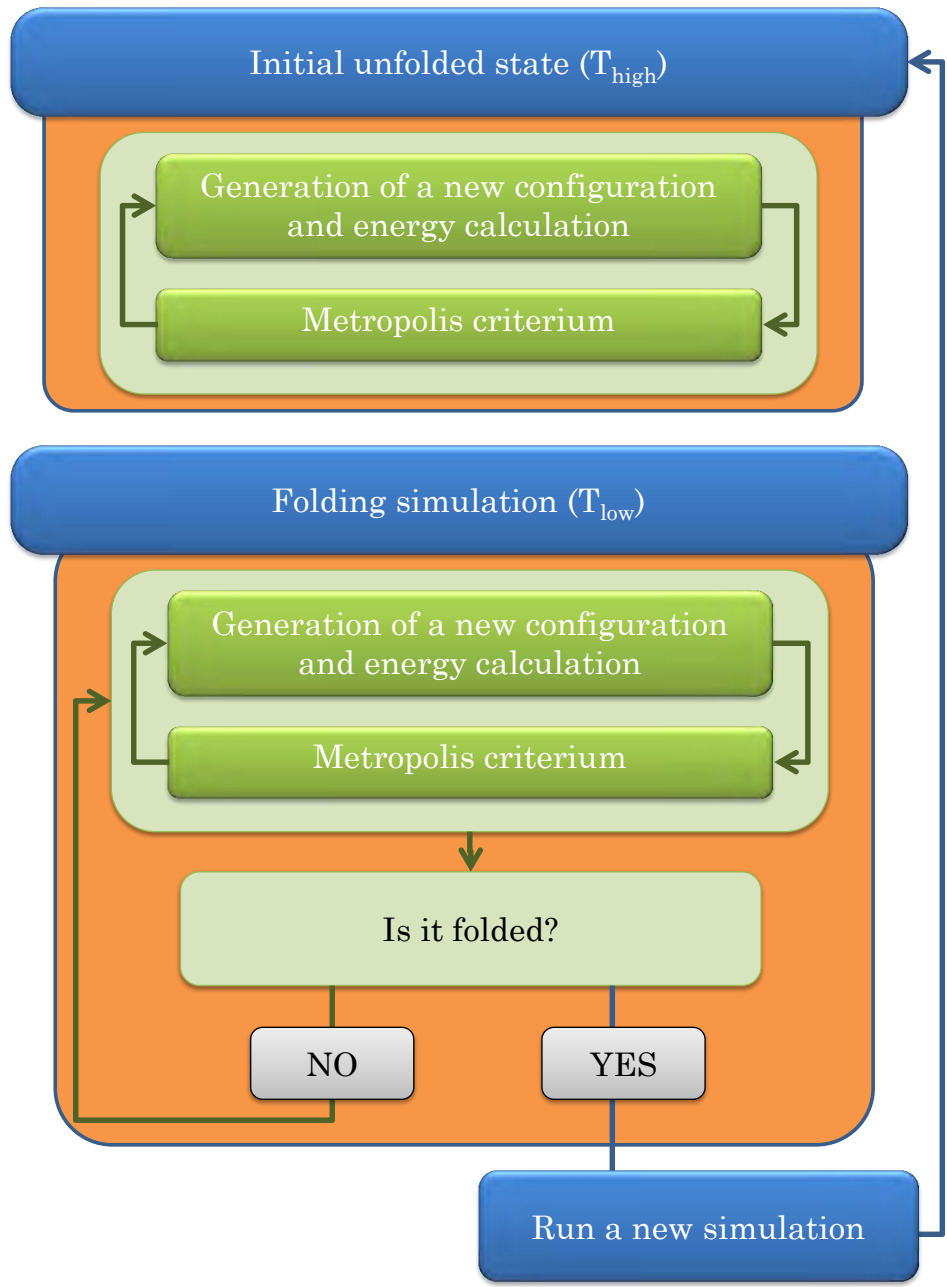

**Figure 2.8:** *Flowchart of the Kinetic Monte Carlo method.*

1. Generation of an unfolded configuration: starting again from a completely extended system, the system is relaxed at a high temperature (typically, 1.2 times the transition temperature).

2. Folding simulation: first, the system is instantaneously cooled down to the desired lower temperature (below the transition one). Then, it is relaxed at constant temperature until the folded state is reached; at this point, we write down the folding time of the simulation. All along the folding, the evolution of the system properties are conveniently registered, as we are not only interested in the final time, but also in how the folded state is achieved.

3. Once the system reaches the folded state, a new unfolded initial state is generated and a new folding simulation starts. Typically, this cycle continues until many folding events have been recorded (2000 on average), guaranteeing significant statistics. The need of so many independent folding events makes parallelization unnecessary in this case, as they can be obtained by several runs (each one of 250 folding cases, for instance) of the serial program, sent to the different cores of a computer.

Regarding the algorithm itself, the main difference with the traditional Monte Carlo scheme, described in Section 2.1, lies in the generation of the new configuration and the used *movements*. As we explained Section 2.2.2, Monte Carlo needs many different types of movements to sample the free energy landscape efficiently. In Kinetic Monte Carlo, however, we are only interested in movements with a small and similar amplitude. In these way, these movements can be thought as microscopic steps much shorter than the overall process we are studying. At a technical level, this results in the elimination of the rotation-related movements (pivot, single chain rotation and collective rotation), which are the widest movements of all.

The obtention of good quality simulations is an important step to obtain meaningful conclusions. But the acquisition of them needs a careful analysis of the raw data.

# §2.5   Data Analysis

In our simulations, some information of the system is periodically registered. We record three types of data along the simulation:

1. Energy data: the different contributions to the total energy of the system are stored.

2. Structural data: some global properties are also computed along the simulation:

- Radius of gyration ($R_g$): it is a measure of the size of the system, being especially relevant in the case of single chain simulations. It is calculated as the root mean square distance of each bead $i$ (characterized by its position $\mathbf{r}_i$) from the chain center of gravity ($\mathbf{r}_{CoM}$), being $N$ the total number of beads:

$$R_g = \sqrt{\frac{1}{N} \sum_{i=1}^{N} (\mathbf{r}_{CoM} - \mathbf{r}_i)^2} \qquad (2.3)$$

- Root mean square deviation ($RMSD$): it measures the difference between two systems, typically a given conformation ($A$) and the native state ($B$). First, the two systems are optimally superimposed.[126] After that, the mean quadratic deviation is computed between the equivalent beads of the two structures:

$$RMSD = \sqrt{\frac{1}{N} \sum_{i=1}^{N} (\mathbf{r}_i^A - \mathbf{r}_i^B)^2} \qquad (2.4)$$

- Bead coordinates of all the residues of the system: in this way, we have a structural representation of the proteins or peptides in different simulation steps and we can develop any further structural analysis on a given configuration (e.g. partial $RMSD$s, contact maps, etc.).

Apart from these quantities, that are automatically generated in our simulations, we have paid special attention to two different kinds of information: thermodynamics and kinetics. In the following Subsections, we will describe the most relevant and frequently used techniques concerning these two items.

## 2.5.1    Thermodynamic analysis

One of the most relevant thermodynamic properties we have calculated is the heat capacity of the system, $C_v$. It constitutes one of the best indicators of the stability of the protein structure, either experimentally or in simulations. It can be defined as a measure of the energetic fluctuations of the system as a funtion of temperature:[48, 127] the higher it becomes, the greater the range of accessible energies is. For that reason, a peak in the heat capacity curve indicates an energetic (and, usually, also structural) transition. At each temperature, it can be calculated as follows:

$$C_v^* = \frac{<E^{*2}> - <E^*>^2}{T^{*2}} \tag{2.5}$$

In the upper Equation, all the variables have an asterisk. It means that they are expressed in *reduced units*, i.e. referred to a certain *reference state* of temperature $T_{ref}$ and energy $E_{ref} = k_B T_{ref}$. We have performed our simulations in terms of these reduced and adimensional units, defined as:

$$T^* = \frac{T}{T_{ref}} \tag{2.6}$$

$$E^* = \frac{E}{k_B T_{ref}} \tag{2.7}$$

Sometimes, especially when simulating idealized system, we have kept the reduced units for the discussion, given that the variations in the properties are similar regardless the specific units. In other cases, such as Chapter 4, where we have dealt with real proteins, we have converted our reduced units to the real ones using the experimental folding transition temperature as a reference.

Besides the heat capacity, we have paid special attention to the free energy of the system. Being a combination of energy and entropy, it is the key variable that defines the folding landscape.

Finding the free energy of a system is a very difficult task because the calculation of entropy needs a thourough sampling of the accessible conformational space. Taking advantage of the information from different temperatures, we have followed a broadly accepted strategy called the *weighted histogram analysis method* (WHAM).[120, 128, 129] Having different temperatures means that each of them explores a different region of this space, obtaining an overall information that enables an accurate estimation of the system entropy.

The WHAM method aims to find out the Helmholtz free energy of a system (*A*) starting from microscopic data obtained from simulations:[130]

$$A(T) = -k_B T \ln z(T) \tag{2.8}$$

According to this Equation, the free energy depends on the system temperature, $T$, and the partition function at that temperature, $z(T)$. This latter variable measures how the system is distributed along its different accessible energetic states, and depends on the *density* of these states:

$$z(T) = \sum_{i=1}^{N_S} \Omega_i \exp\left(-\frac{E_i}{k_B T}\right) \tag{2.9}$$

In this expression, $N_S$ is the total number of accessible states and $\Omega_i$ is the density of each state. Each state is characterized by its internal energy, $E_i$.

The density of states is calculated by the building of energy *histograms*. The energy values that have been recorded along the simulation are classified in $M$ boxes of energy $E_m$ and width $\Delta E$, adding up to $H_m$ cases in each box. After that, the number of configurations at a given temperature $l$ and energy box $m$ is computed ($N_{m,l}$). The density of states is, then, calculated as follows:[130]

$$\Omega_m = \frac{H_m}{\sum_{l=1}^{L} \Delta E \exp\left(\frac{A_l - E_m}{k_b T_l}\right)} \tag{2.10}$$

Equations (2.8), (2.9) and (2.10) are interconnected, as the calculation of the free energy needs to know the density of states; but calculating the density of states requires the value of the free energy at each temperature. As a result, we have employed an iterative method starting from an arbitrary value of the density of states until convergence.[130]

Once the free energy $A$ has been computed for a certain system, we can calculate the free energy profile at every temperature using any other system property (e.g. energy, $RMSD$, etc.) as reaction coordinate.

### 2.5.2   Kinetic analysis

Kinetic studies have constituted an additional challenge for us, as Kinetic Monte Carlo had never been applied in our research group until very recently.[116] As a result, our kinetic analysis has followed a rather fundamental point of view, focusing on methodological questions.

The first one is related to the definition of *folded state*. As we have seen in Section 2.4, the folding time is recorded when the protein reaches the folded state. But how do we decide whether our system is folded? We have looked for a common methodology to perform our numerical experiments and their subsequent data treatment, in order to be consistent in our simulations.

The most appropriate criterion is $RMSD$, as this property reflects the similarity between a given configuration and a reference one, i.e. the native state. We have taken as examples two of the studied proteins in Chapter 4, which illustrate the behavior of two-state folders (the engrailed homeodomain) and downhill ones (the BBL protein). We have looked at the $RMSD$ distributions of our proteins at the transition temperature, taken from equilibrium simulations. They are plotted in Figure 2.9, where we look for a common cut-off for all the simulations, regardless the kind of protein, below which our system unavoidably folds.

The two-state folder in Figure 2.9($a$) presents a bimodal distribution. The peak at low $RMSD$ (below 4.0 Å) matches the native state, while the other one corresponds to the denatured one. This means that in a folding simulation, every configuration that reaches that low $RMSD$ value would end up in the native state in a straightforward way. As a matter of fact, we tried different cut-offs for these proteins, obtaining identical behaviors and similar folding rates for $RMSD$ values between 4.0 and 1.2 Å.

The behavior of downhill proteins is illustrated in Figure 2.9($b$), where the results for 1BBL are plotted. In this case, the $RMSD$ histogram is unimodal, showing the continuous distribution of configurations along folding. Therefore, the election of the cut-off is more problematic: a large $RMSD$ value may not select folded or nearly folded conformations, but a too low value is so infrequent in the simulation conditions at the transition temperature that may not be reached in sensible folding times. As a matter of fact, we performed some simulations using different cut-offs and observed that very low ones lead to flat folding time distributions and a lack of convergence.

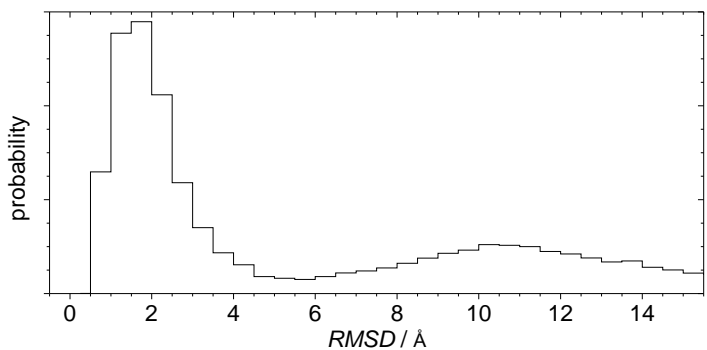

(a) Engrailed homeodomain.

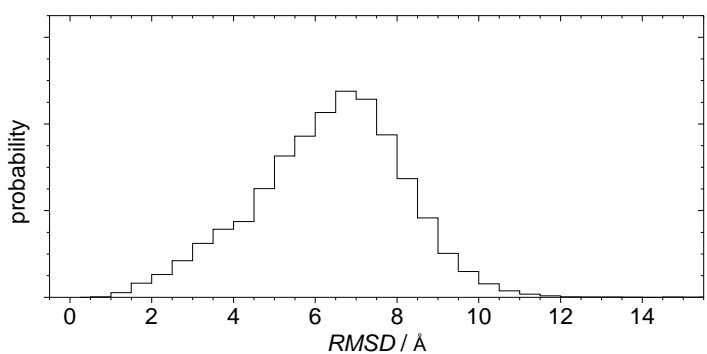

(b) BBL protein.

**Figure 2.9:** *$RMSD$ distributions at the transition temperature for two proteins.*

For this reason, we decided again to base our decision on the experimental approach. We found that experimentalist usually measure folding rates in terms of the dissapearance of some measurable signals of the unfolded state, i.e. they may not be able to detect what is folded, but definitely detect what is not unfolded.[43] In this sense, the choice of $RMSD \leq 4.0$ Å as stop criterion roughly corresponds to the dissapearance of the unfolded state.

Once we have chosen a common stop criterion, we have focused on the specific aspects of folding kinetics, pursuing two different aims: to link kinetics and thermodynamics and to find out information of folding pathways that could add additional insight into our understanding on the folding process.

In first place, we have focused on two-state folders, evaluating the rate of the folding process in terms of the different applied potentials and their related free energy barriers, obtained from the WHAM methodology we have already discussed.

For each folding event, the *first passage time* has been recorded, i.e. the number of Monte Carlo cycles that have been necessary to reach the folded state. We have used them to study the kinetic laws followed by our simulations. Commonly, two-state folders (proteins that present no folding intermediates) are said to fulfill *first order kinetics*:

$$-\frac{\mathrm{d}\,U}{\mathrm{d}\,t} = kU \tag{2.11}$$

In this Equation, $k$ is the rate constant, $U$ is the number of simulations that remain unfolded at a given time (in our case, expressed in Monte Carlo cycles) and $\mathrm{d}\,U/\mathrm{d}\,t$ is the folding rate. Integrating this Equation, we obtain the following expressions, represented in Figure 2.10:

$$\frac{U}{U_0} = \exp(-kt); \quad \ln\frac{U}{U_0} = -kt \tag{2.12}$$

According to these Equations, folding kinetics can be studied by the observation of folding times. Folding kinetics can be characterized by a single value, the *mean first passage time*, known as $MFPT$ or $\tau$ (see Figure 2.10). If a process complies with a first order kinetics, its rate constant is equal to the inverse of the $MFPT$.

If the kinetics are measured at the transition temperature, the rate constant $k$ can be related to the height of the free energy barrier at that temperature $\Delta G^{\#}$ through the transition state theory:[131]

$$k = k_0 \exp\left(\frac{-\Delta G^{\#}}{RT}\right) \tag{2.13}$$

As we shall see in Chapter 4, we have explored the kinetics of some of our systems in order to find out whether our simulations follow these experimental laws. Besides, we have performed time-dependent structural analyses that have allowed us to explore the folding pathway of the studied protein, as well as the building of a coordinated view of folding by an integrative analysis of structural, thermodynamic and kinetic data.

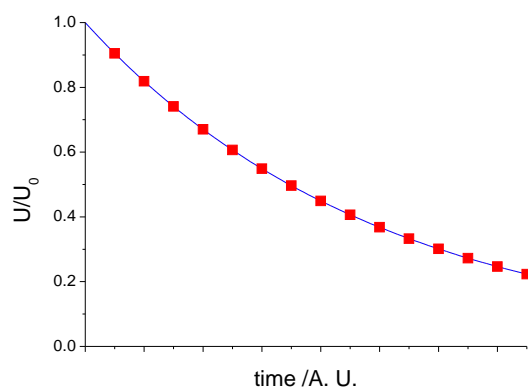

(a) Exponential law.

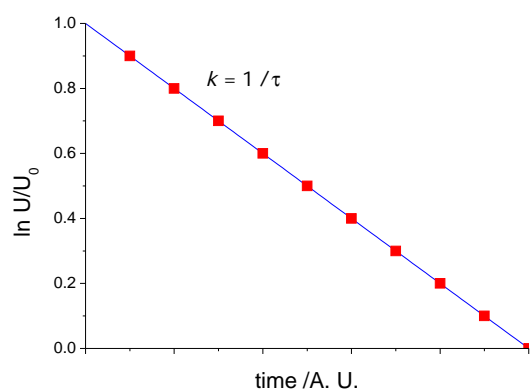

(b) Linearized law.

**Figure 2.10:** *Example of the integrated laws of a first order kinetics.*

Many extra analysis have been carried out along the development of this Thesis. Some of them have been used in preliminary steps of the research, while others are relevant for a coherent explanation of the specific aspects of folding and aggregation processes, and have been included in this Dissertation. They will be explained as they appear in the Results Chapters we are about to begin.

# 3

---

# Modeling hydrogen bonds

In Chapter 1 we explained the importance of hydrogen bonds for the formation of native and aggregated structures such as secondary structure elements and cross-$\beta$ structures, respectively. Regarding the computational field, we also discussed the peculiarities of this kind of interaction for its modeling through coarse-grained potentials, that make the obtention of a *good* hydrogen bond representation a difficult task that has not been solved yet. This has prompted us to design a new hydrogen bond potential based on an accurate geometric description, which constitutes the leitmotiv of this PhD Project.

In this Chapter, we will start with a brief review of some of the most relevant hydrogen bond potentials that had been proposed before the beginning of this work. From previous studies developed in our group[90] and mainly from additional analysis especially designed for this current purpose along this work, we have linked the observation of unwanted effects (such as distorted regular structures or unnatural hydrogen bonds) to particular characteristics of the different potentials. Thanks to an extensive statistical analysis of secondary structure elements as they appear in nature, we have merged the strongest aspects of the studied models to build the refined hydrogen bond potential that will be described afterwards.

Its evaluation has involved an exhaustive analysis of the geometry of the secondary structure elements that the potential generates during folding simulations. In this Chapter we report the simulation of several peptide systems, studying their structural and thermodynamic properties in terms of temperature and concentration. This has resulted in the obtention of a peptide "structural" phase diagram, which will constitute the last part of this Chapter.

## §3.1  Motivation: hydrogen bond potentials

The use of coarse-grained potentials has widely proved its suitability for protein folding and aggregation studies.[132–134] Their simplicity makes accessible longer times and larger systems than the full-atom approach, allowing the evaluation of complex problems using limited resources. In more fundamental terms, they grasp the key aspects of a process from a minimalist starting point.

As we already stated in Chapter 2, we have used an off-lattice protein representation in which each amino acid is described by one center of interaction, placed at the $\alpha$-carbon position. This is a common strategy in different kinds of protein folding studies[93,102,135] that had been also used in our research group in conjunction with structure-based potentials,[62] providing excellent results for the thermodynamic description of protein folding.[63,64,98,136] Coherently, we have sought a hydrogen bond definition in terms of an $\alpha$-carbon representation.

For that purpose, we counted with a previous study from our group in which we had evaluated different hydrogen bond potentials with different levels of resolution,[90] specifically the ones proposed by Irbäck,[137] Chen[77] and Kolinski[78] laboratories. In that work, the consequences of using different levels of detail were properly evaluated, discussing how those diverse definitions were able or not to grasp the properties that, as we stated in Chapter 1, make hydrogen bonds so peculiar.

In this former work, the three different potentials were assessed by means of a methodology that involves the alignment of rigid fragments.[90,138] In the present work, however, our interest does not lie in the packing of rigid fragments, but in the folding of flexible chains. This made us extend the bibliographic search to have a broader view of the currently used hydrogen bond potentials. Although most of the hydrogen bond potentials that will be discussed in this Section have been proposed in combination with other energetic interactions and not on their own (as we aim to use them in this part of the Thesis Project), they can give us a useful insight of the different strategies that are commonly adopted in the definition of hydrogen bonds.

We have found that all of them share an explicit care about the two main features of hydrogen bonds: their geometry and their strength. However, they present significant differences, chiefly in the number of interacting centers and the way the bond directionality is taken into account.

In relation to the former, the more detailed models explicitly represent the atoms involved in the hydrogen bond (i.e. hydrogens, nitrogens and oxygens of the chain backbone).[137, 139] This results in a very accurate representation, but the all-atom representation makes them computationally costly. An intermediate step is the construction of the so-called *virtual atoms* for the acceptors and donors (CO and NH, respectively) along the chain; these atoms are not independent beads, so they do not increase the degrees of freedom of the system, but participate in the energy calculations.[76, 78] Other hydrogen bond potentials present a much simpler definition of the system, just keeping the $\alpha$-carbon trace of the protein either as a chain of spheres[140, 141] or a flexible tube.[79, 142]

Hydrogen bond potentials also differ from one another in the way directionality is described inside the energetic calculation. Some potentials ignore this feature, presenting spherical symmetry.[76] Others insert an angular dependence[137, 143] or a more refined description.[78, 79, 142] Not surprisingly, there is a kind of inverse relationship between the number of centers of interaction and the complexity of the energetic description, i.e. simpler system descriptions need a more detailed representation of the hydrogen bond in order to describe this interaction in a suitable way.

In our case, we are looking for a hydrogen bond potential that, being as accurate as possible, is defined in simple terms at the smallest computational cost. For that reason, we tested in first place the Kolinski potential,[78] implementing it in our simulation algorithm.

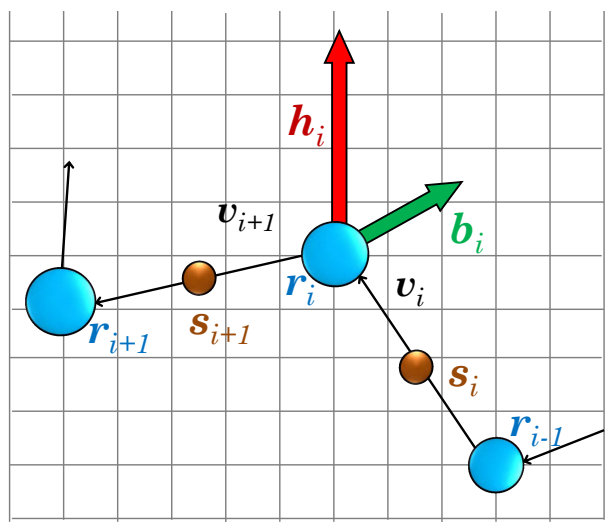

**Figure 3.1:** *Geometric description of the Kolinski potential. The centers of interaction are represented by circles, colored in blue for the α-carbons (defined by their positions **r**) and in brown for the PB positions (**s**). The involved vectors are illustrated by arrows: virtual bond vectors between α-carbons are the black ones (**v**) and the auxiliary vectors are colored in green (unitary bisector vector, **b**) and red (hydrogen bond vector, **h**).*

### 3.1.1   The Kolinski potential

As we have already stated, the Kolinski potential[78] seemed ideal for our purposes in an early stage of this Project, as it is based in an α-carbon representation of the protein, considers in detail the directionality of the hydrogen bond and had previously proved its good characteristics in fragment alignment studies.[90]

The system description of this potential is shown in Figure 3.1. As we can observe, it consists in a lattice representation of the chain where each amino acid is represented by one center of interaction placed at the α-carbon position.

Instead of creating the aforementioned virtual atoms for the donor and acceptor fragments, the authors just place a new center of interaction (called *PB* in the original works[78] and whose position is defined by the **s** vector) in the middle of the virtual bond distance (outside the lattice positions). It roughly corresponds to the —NH–CO— position in the real chain. Having a single center of interaction for both donor and acceptor shifts the numbering of the hydrogen bonds. As a result, the real backbone hydrogen bond between the $i$th and the $j$th residues is replaced by the interaction between the $i$th and the $(j-1)$th beads in the model.

The *PB* center associated to residue $i$ is located between the $\alpha$-carbon of residues $i$ and $(i-1)$. Its position is computed with the following Equation, leading to the brown center of Figure 3.1:

$$\mathbf{s}_i = \frac{\mathbf{r}_i + \mathbf{r}_{i-1}}{2} \tag{3.1}$$

In addition, two auxiliary vectors associated to each residue $i$ have been defined:

- The unitary bisector vector, $\mathbf{b}$: it is placed in the same plane of the adjacent virtual bond vectors:

$$\mathbf{b}_i = \frac{\mathbf{v}_i - \mathbf{v}_{i+1}}{|\mathbf{v}_i - \mathbf{v}_{i+1}|} \tag{3.2}$$

- The hydrogen bond vector, $\mathbf{h}$: this vector, 4.6 Å long, is perpendicular to the latter plane and indicates the direction of the hydrogen bond:

$$\mathbf{h}_i = 4.6 \cdot \frac{\mathbf{v}_i \times \mathbf{v}_{i+1}}{|\mathbf{v}_i \times \mathbf{v}_{i+1}|} \tag{3.3}$$

The hydrogen bond energy of the different configurations, that we call $E^{kol}$, is calculated through a pair-wise individual potential, $e^{kol}$, according to the following expression:

$$E^{kol} = \sum_{i=2}^{N-4} \sum_{\substack{j=i+3 \\ j \neq i+4}}^{N-1} e_{i,j}^{kol} \tag{3.4}$$

In this Equation the $(i, i+4)$ interaction has been removed. According to the definition of the model, it corresponds to the real $(i, i+5)$ one. It stabilizes *thick* helices that do not match any natural structure, so its removal favors the formation of $\alpha$-helices, that are formed by $(i, i+3)$ interactions in the Kolinski model.

Besides, this potential does not allow the formation of a hydrogen bond interaction if one of the involved residues is a terminal one within the polypeptide chain, as neither the *PB* center nor the auxiliary vectors **b** and **h** can be calculated.

The interaction energy between residues $i$ and $j$, $e_{i,j}^{kol}$, is obtained through a two-step process. In first place, the following four *geometric restrictions* are calculated. Based on the author's analysis of real structures, they are fulfilled if their values lie within the indicated acceptable ranges:

**Restriction 1:** The $\alpha$-carbons of two interacting residues must be spatially close (the norm of their connecting vector, $\mathbf{r_{i,j}}$, must be smaller than 6.1 Å):

$$|\mathbf{r}_i - \mathbf{r}_j| = |\mathbf{r}_{i,j}| \leq 6.1\,\text{Å} \tag{3.5}$$

**Restriction 2:** The auxiliary hydrogen bond vectors must adopt a parallel or anti-parallel arrangement (a 40° deviation is allowed):

$$|\mathbf{h}_i \cdot \mathbf{h}_j| = 4.6^2 \cdot \cos(\mathbf{h}_i, \mathbf{h}_j) \geq 16\,\text{Å}^2 \tag{3.6}$$

**Restriction 3:** The backbone fragments must also present a parallel or anti-parallel orientation:

$$\begin{cases} \mathbf{v}_i \cdot \mathbf{v}_{j+1} \leq 0 \quad \text{and} \quad \mathbf{v}_{i+1} \cdot \mathbf{v}_j \leq 0 \\ \qquad\qquad \text{or} \\ \mathbf{v}_i \cdot \mathbf{v}_j \geq 0 \quad \text{and} \quad \mathbf{v}_{i+1} \cdot \mathbf{v}_{j+1} \geq 0 \end{cases} \tag{3.7}$$

**Restriction 4:** The connecting vector $\mathbf{r}_{i,j}$ and the related hydrogen bond vectors must present some spatial and orientational similarities:

$$|\mathbf{r}_{i,j} \pm \mathbf{h}_i| \leq 1.83\,\text{Å} \tag{3.8}$$

$$|\mathbf{r}_{i,j} \pm \mathbf{h}_j| \leq 1.83\,\text{Å} \tag{3.9}$$

These restrictions and their allowed ranges aim to reflect the hydrogen bond directionality by several constrains that consider both distance (restrictions 1 and 4) and angular dependencies (restrictions 2, 3 and, again, 4). After the restriction calculations, the interaction energy is computed through the following expression:

$$e_{i,j}^{kol} = \varepsilon^{\gamma}\, \delta_{i,j}^{\gamma}\, e_{i,j}^{\gamma} + \varepsilon^{h}\, \delta_{i,j}^{h}\, e_{i,j}^{h} \tag{3.10}$$

Its minimum (most stabilizing) value is $-1.00$ in reduced units. The energetic expression consists of two terms that share the same structure: a weighting factor $\varepsilon$, that determines the relative contribution of each addend to the total energy; a $\delta$ factor that depends on the fulfillment of the previous restrictions; and the functional expression itself, $e$.

The first part of the expression, called $\gamma$ term, is a minoritary contribution to the total energy, as its weighting factor is small ($\varepsilon^{\gamma} = -0.07$). According to its $\delta$ factor, it is non-void if the first three restrictions are achieved:

$$\delta_{i,j}^{\gamma} = \begin{cases} 1 & \text{if Equations (3.5), (3.6) and (3.7) are fulfilled} \\ 0 & \text{otherwise} \end{cases} \tag{3.11}$$

The functional form of the $\gamma$ term is defined as follows:

$$e_{i,j}^{\gamma} = 2 - \max\left(\frac{\mathbf{b}_i \cdot \mathbf{r}_{i,j}}{6.1^2}, 0.125\right) - \max\left(\frac{\mathbf{b}_j \cdot \mathbf{r}_{i,j}}{6.1^2}, 0.125\right) \tag{3.12}$$

In this way, the stabilizing energy of this contribution can reach $-0.12$ (in reduced units) if the scalar products of the $\mathbf{b}_i$ or $\mathbf{b}_j$ and $\mathbf{r}_{i,j}$ vectors are small, i.e. if the $\alpha$-carbons of the interacting residues are close and the connecting vector is nearly perpendicular to the unitary bisector ones. In principle, each residue can form an unlimited number of $\gamma$ interactions.

Regarding the second contribution of Equation (3.10), the so-called $h$ term is the major one, having a weighting factor of $\varepsilon^{h} = -0.35$. It is only calculated if the following restrictions are satisfied:

$$\delta_{i,j}^{h} = \begin{cases} 2 & \text{if Equations (3.5), (3.6), (3.7), (3.8) and (3.9) are fulfilled} \\ 1 & \text{if Equations (3.5), (3.6) y (3.7) are satisfied, but only one of (3.8) or (3.9)} \\ 0 & \text{otherwise} \end{cases}$$

$$\tag{3.13}$$

The calculation of the functional form $e_{i,j}^h$ needs two more auxiliary vectors, $\mathbf{r}_{pp}$ and $\mathbf{r}_{qq}$, that represent the distance between the opposite *PB* centers (adjacent to the $\alpha$-carbons involved, see Equation (3.1)):

$$e_{i,j}^h = 0.5 + \left( \frac{4.25}{\max[4.25, \min(6.01, |\mathbf{r}_{qq}|)]} \right)^4 + \left( \frac{4.25}{\max[4.25, \min(6.01, |\mathbf{r}_{pp}|)]} \right)^4 \quad (3.14)$$

The $h$ term is more stabilizing than the $\gamma$ one, reaching $-0.88$ in reduced energy units. This minimum value is achieved if the *PB* centers are placed together (i.e. at a distance shorter than 4.25 Å). Thanks to the restrictions characteristics, each residue can interact through the $h$ term with two other residues at most, being this interacting contribution more specific than the $\gamma$ one and also more similar to what happens in nature, where each amino acid can form one hydrogen bond as donor and a second second one as acceptor.

In the original papers,[78] the system was inserted in a lattice; to include it in our off-lattice representation, we have added a hard sphere potential fixed at 4.0 Å that prevents the overlapping of the beads.

After our implementation, we carried out a number of numerical experiments to prove the ability of this model to generate secondary structure elements. Thus, we built small systems formed by short chains (ten residues long), fast to simulate and straightforward to analyze. We simulated infinite dilution conditions and highly concentrated systems, favoring in principle the obtention of $\alpha$-helices and $\beta$-sheets, respectively. In this part of the study we only cared about the structural characteristics of the stable structures at low temperature, an example of which is shown in Figure 3.2.

The structures of this Figure exhibit a reasonable native-like appearance. Nevertheless, a more detailed analysis reveals the following important differences:

**Helix chirality:** the Kolinski potential does not include any preference towards right-handed helices over left-handed ones. As a consequence, both of them are obtained in a similar proportion, in spite of the fact that every natural $\alpha$-helix is right-handed.[1] Con-cerning our interests, this constitutes a disadvantage of the model because it results in an unsuitable definition of this secondary structure element.

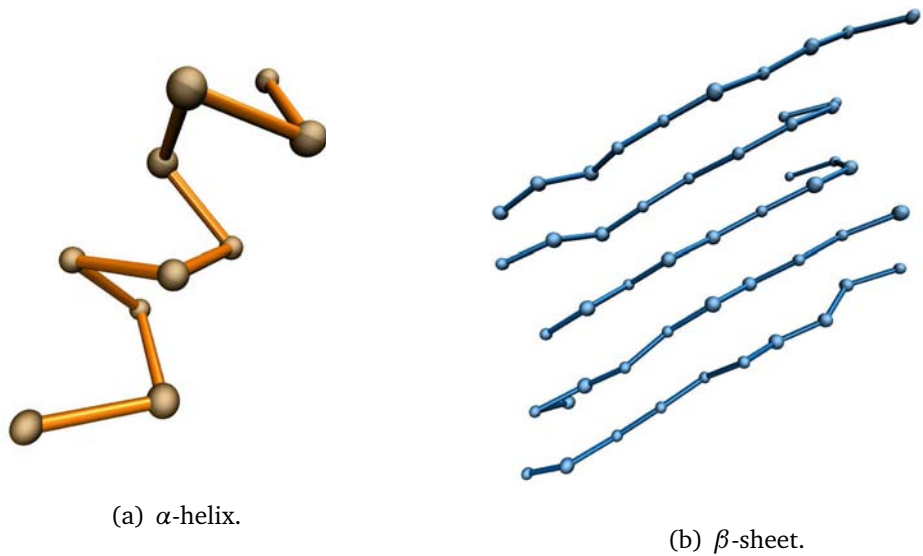

(a) $\alpha$-helix.

(b) $\beta$-sheet.

**Figure 3.2:** *Typical secondary structure elements obtained with the Kolinski hydrogen bond potential in peptide systems of ten amino acids per chain. Images drawn with VMD.[144]*

**Local geometry:** as we stated in Chapter 1, secondary structure elements are characterized by the local arrangement of the peptide chain, as reflected by the Ramachandran angles (see Figure 1.3). In the case of our coarse-grained definition, we cannot compute those angles (as the atomic positions are lost), but the particular geometry of secondary structure elements is also reflected in the local properties of the $\alpha$-carbons. For example, we can calculate the $d_{i,i+3}$ distance associated to the virtual torsional angles of natural proteins, i.e. the one between the $\alpha$-carbons of residues $i$ and $(i+3)$. In Figure 3.3 we show the histograms of these distances in natural proteins (taken from a 1590 protein database generated from the PDB databank[145]) and in the secondary structure elements that have been obtained with the Kolinski potential.

Our results are split in two different histograms ($\alpha$-helices and $\beta$-sheets) while the native data merge both sets. Generally speaking, distances below 6.1 Å correspond to the local geometry of $\alpha$-helices and larger ones correspond to the extended conformations that are typically observed in $\beta$-sheets.

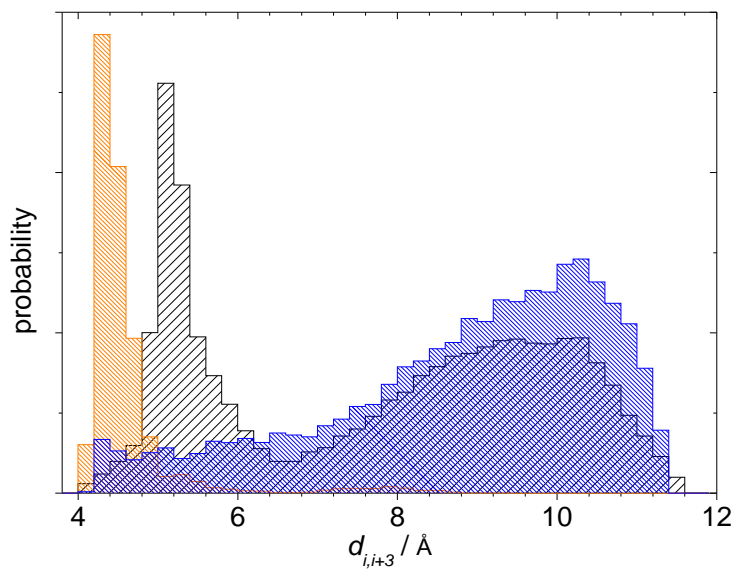

**Figure 3.3:** *Histograms of $d_{i,i+3}$ distances for a dataset of native structures (black) and the secondary structure elements obtained by the Kolinski potential ($\alpha$-helices in orange and $\beta$-sheets in blue).*

This Figure shows that the $\alpha$-helices generated by the studied potential present remarkable distortions. Their histogram shows a maximum around 4.2 Å, shifted to smaller distances in relation to the native peak of helical structures. The distribution of this distance is a critical feature of helices, as the $i$ and $(i+3)$ residues are the ones that form the hydrogen bond interaction. Therefore, we can conclude that Kolinski potential modifies the local geometry of $\alpha$-helices. The case of $\beta$-sheets, however, shows a broad distribution that matches the native one reasonably well.

**The hydrogen bond interactions themselves:** as we commented on while describing the Kolinski potential, the presence of two different terms in the energetic calculation leads to the possibility of having more than two interactions per residue. As an example, we show in Figure 3.4 the hydrogen bond energetic maps for the $\alpha$-helix and

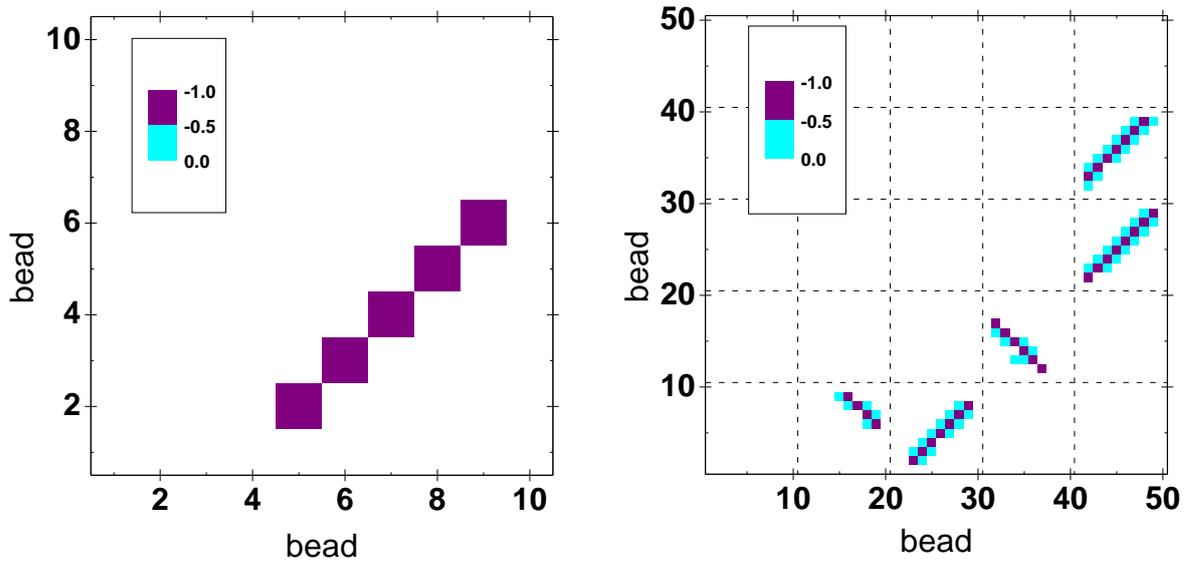

(a) $\alpha$-helix formed in a 10-residue chain.

(b) $\beta$-sheet formed by five chains of 10 residues each.

**Figure 3.4:** *Hydrogen bond energy maps for the secondary structure elements, according to the Kolinski potential. In these representations, a colored spot indicates the presence of a hydrogen bond contact whose energy is shown in the plot legend.*

$\beta$-sheet that are drawn in Figure 3.2, which belong to 10-residue peptides. The axes stand for the residue numbering in the sequence (if there are several chains, this numbering is consecutive and the different chains have been separated by discontinuous lines) and the colored dots indicate the presence of an interaction and its intensity.

The helices, drawn in Figure 3.4($a$), are characterized by a colored diagonal of ($i$, $i + 3$) interactions. As it can be observed, the terminal residues cannot form hydrogen bonds due to the impossibility to define Kolinski auxiliary vectors. Helices just present strong interactions, as the weak ones (that could be formed by ($i$, $i + 4$) interactions) are forbidden by definition (see Equation (3.4)).

In Figure 3.4($b$), $\beta$-sheets exhibit strong interactions among the chains, which can be either parallel (along the principal diagonal of the map, like between the first and third chain in the showed Figure) or antiparallel (see the interaction between the second and fourth chain).

Additionally, there are many weak interactions in the surroundings of a strong interaction, with no physical meaning. They imply that one residue (for instance, residue 45 in the Figure) can form two strong hydrogen bond interactions (in this specific case, with residues 25 and 35) but also many more weak ones (with residues 24, 26, 34 and 36), resulting in an unreasonable value of 6 hydrogen bonds in this residue.

Most of the disadvantages we have discussed are common for most hydrogen bond potentials. Frequently, these considerations are not relevant for their authors, as their models usually have additional energetic contributions that can also modify the structural description of the system, or because they are not interested in strict geometric descriptions of the natural helices and sheets, just needing the approximate features of these elements.

However, our interest in the accurate obtention of native-like structures made us look for other definitions of hydrogen bonds. We decided not to use the Kolinski potential in its original terms, but to find an improved definition of hydrogen bonds by looking into alternative published potentials. For us, one of the most helpful potentials has been the Hoang *et al.*'s one[79] that, based on a tubular description of the chain, describes the hydrogen bond geometry very efficiently.

### 3.1.2   The Hoang *et al.* potential

The Hoang *et al.* potential is defined in terms of a tubular description of the protein, with a tube of radius 2.5 Å, as shown in Figure 3.5. According to the authors, a tubular description is able to reproduce the geometry of a natural polypeptide and mimic an anisotropic interaction such as the hydrogen bond.[79,146] As the edges of the tube are placed at the $\alpha$-carbon positions, the system description of this model is also compatible to ours. In this way, we have not tried to use the Hoang *et al.* model as it was formerly defined, but to take advantage of its characteristics to generate our hydrogen bond potential, based on a stick and bead definition.

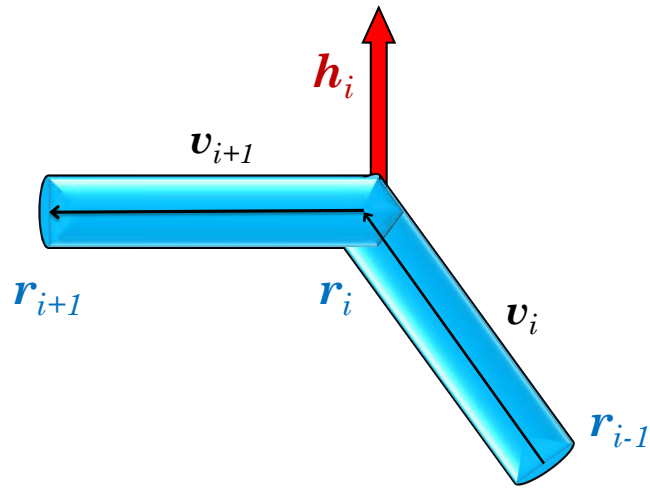

**Figure 3.5:** *Geometric description of the Hoang et al. potential. The polypeptide chain is represented by a tube (colored in blue) formed by rigid fragments whose edges are located at the α-carbon positions,* **r***, and connected by virtual bond vectors* **v***. A unitary virtual bond vector* **h** *(in red) is also defined.*

The system is characterized by the edge positions, **r**, and virtual bond vectors, **v** (see Figure 3.5), similarly to the Kolinski description of Figure 3.1. An auxiliary *hydrogen bond vector*, **h**, is also needed. It is calculated exactly as Kolinski's, but in this case its modulus is 1 Å instead of 4.6 Å (compare the following Equation to Equation (3.3)):

$$\mathbf{h}_i = \frac{\mathbf{v}_i \times \mathbf{v}_{i+1}}{|\mathbf{v}_i \times \mathbf{v}_{i+1}|} \tag{3.15}$$

This potential also allows a large chain flexibility, but limits the virtual bond angles of the system, i.e. those formed by three consecutive residues, from a minimum value of 82° (that prevents residue overlappings) to a maximum value of 148° (avoiding too extended conformations).

The energetic calculation is also based on a number of geometrical restrictions, followed by the energy calculation itself. In this case, the geometrical constrains are split in local (those that form $\alpha$-helices and are located between the $i$ and $(i+3)$ residues) and non-local ones (that are present in $\beta$-sheets).

**Restriction 1:** The $\alpha$-carbons of the interacting residues must lie in a certain range, that differs for local and non-local bonds.

$$\begin{cases} 4.7\,\text{Å} \le |\mathbf{r}_{i,j}| \le 5.6\,\text{Å} \quad \& \quad \kappa > 0 \quad \text{if } j = i+3 \\ |\mathbf{r}_{i,j}| \le 5.3\,\text{Å} \qquad\qquad\qquad\qquad \text{otherwise} \end{cases} \qquad (3.16)$$

In the case of local bonds, the helix chirality is explicitly considered by the variable $\kappa$ that reflects the sign of the virtual torsional angle between the $i$ and $(i+3)$ residues through the triple product of the involved virtual bond vectors (which is computationally more efficient than calculating the torsional angle itself):

$$\kappa = [\mathbf{v}_{i+1}, \mathbf{v}_{i+2}, \mathbf{v}_{i+3}] \qquad (3.17)$$

**Restriction 2:** The auxiliary hydrogen bond vectors must be nearly parallel or antiparallel (allowing a 37° deviation):

$$|\mathbf{h}_i \cdot \mathbf{h}_j| = |\cos(\mathbf{h}_i, \mathbf{h}_j)| \ge 0.8\,\overset{\circ}{\text{A}}{}^2 \qquad (3.18)$$

**Restriction 3:** The hydrogen bond vector and the connecting vector must share a similar orientation. To obtain a restriction that does not depend on the distance but only on the angular dependence, a unitary connecting vector, $\mathbf{m}_{i,j}$, is defined:

$$\mathbf{m}_{i,j} = \frac{\mathbf{r}_{i,j}}{|\mathbf{r}_{i,j}|} \qquad (3.19)$$

Then, Restriction 3 (calculated for the hydrogen bond vectors of both $i$ and $j$ residues) follows this expression, allowing a 20° deviation:

$$|\mathbf{h}_k \cdot \mathbf{m}_{i,j}| \ge 0.94 \qquad \text{for } k = i, j \qquad (3.20)$$

In addition, the Hoang *et al.* potential also considers the formation of a hydrogen bond interaction where one or both involved residues are placed at a chain end. In these cases, the auxiliary hydrogen bond vectors cannot be defined, so their Restrictions 2 and 3 are merged and redefined in terms of the virtual bond vectors. For example, if we consider the interaction between an end residue (whose virtual bond vector will be called $\mathbf{v}_e$) and an internal residue $j$, the restriction is defined in the following way:

**Restriction** $2_e$**:** The ending virtual bond vector ($\mathbf{v}_e$) and the unitary connecting vector must be nearly perpendicular, allowing a 20° deviation:

$$|\cos(\mathbf{v}_e, \mathbf{m}_{i,j})| \leq 0.342 \tag{3.21}$$

Concerning the energy calculation, the Hoang *et al.* potential follows a pair-wise expression:

$$E^{hoang} = \sum_{i=1}^{N-3} \sum_{j=i+3}^{N-1} e_{i,j}^{hoang} \tag{3.22}$$

In contrast to the Kolinski potential of Equation (3.4), the authors do not remove the $(i, i+4)$ interaction, adding up all the possible terms.[79] Each interaction term, $e_{i,j}^{hoang}$, consists of two addends:

$$e_{i,j}^{hoang} = \delta_{i,j} \lambda_{i,j} + \eta \tag{3.23}$$

The first of them depends on two parameters, $\delta_{i,j}$ and $\lambda_{i,j}$. As it happened with the Kolinski potential, the $\delta$ factor reflects the restrictions fulfillment:

$$\delta_{i,j} = \begin{cases} 1 & \text{if Equations (3.16), (3.18) and (3.20) are fulfilled} \\ 0 & \text{otherwise} \end{cases} \tag{3.24}$$

The $\lambda$ factor, i.e. the functional form itself, presents in this case a step-wise functional form. In spite of its simplicity, its all-or-none expression reflects the partially covalent nature of the hydrogen bond. In addition, it distinguishes between local and non-local bonds, as the local bonds are thought to be more stabilizing than the non local ones:[147]

$$\lambda_{ij} = \begin{cases} -1.0 & \text{if } j = i+3 \\ -0.7 & \text{otherwise} \end{cases} \tag{3.25}$$

The last term of Equation (3.23) is the cooperativity factor $\eta$, that explicitly reflects the additional stabilization (here, $\eta = -0.3$ reduced energy units) that is achieved when consecutive hydrogen bonds along the sequence are formed. This strategy, that naturally appears when many-body interactions are considered,[148] forces the cooperative effects that have been found in real proteins.

In relation to the possibility of forming more than two hydrogen bonds per residue, the Hoang *et al.* potential explicitly forbids it during the simulation, including in this way an additional multibody term.

The Hoang *et al.* potential has been used for folding simulations and detailed structural analyses have been carried out by the authors.[79,146] Thanks to these studies, we have evaluated the capability of this potential to obtain native-like secondary structures. The most representative structures obtained with this potential are drawn in Figure 3.6.

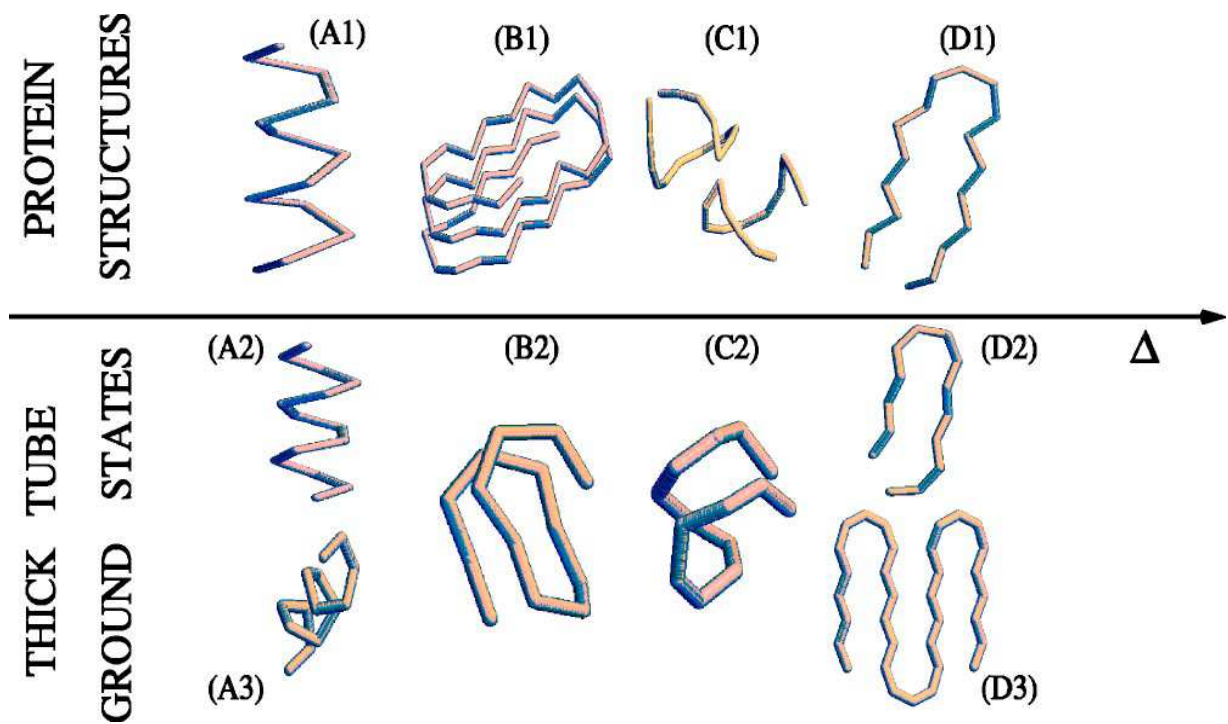

**Figure 3.6:** *Comparison among natural (up) and simulated (down) secondary structure elements according to the Hoang et al. potential. Figure taken from[146] .*

As it can be observed, they resemble the natural ones, but present slightly different characteristics. In some cases (such as the A2 or D2 structures of this Figure), only an enhanced regularity (proper of every hydrogen bond potential) is observed. However, these acceptable structures coexist with clearly unnatural ones (such as A3 or B2) in the relevant temperature range, failing in the obtention of *just* native-like elements, as it was the purpose of our project.

To sum up, the potential of Hoang *et al.* presents many advantageous features, such as the explicit care about the proper chirality of $\alpha$-helices, the regard of terminal bead hydrogen bonds and the constrained flexibility of the chain. Nevertheless, it also presents some drawbacks for the accurate description of native-like helices and sheets.

Up to this point, we have discussed different ways of describing the hydrogen bond interaction. After a thorough bibliographic search, we have found that hydrogen bond potentials frequently present some common features and a certain number of differences among them. As a result, the obtained secondary structures usually resemble the natural ones, but show slight dissimilarities that result, as a consequence, in inacurate descriptions. Interestingly, we have found that the undesirable properties of the structures strongly depend on the specific characteristics of each potential. Can we design a refined potential that reduces these hindrances to a minimum?

## §3.2   Hydrogen bond design

Based on the previous evaluation of several hydrogen bond potentials, we undertook the design of our own model as the natural following step in this Project. Our main hypothesis was that the common drawbacks of other potentials could be overcome by a careful description that merges their good characteristics with an extensive analysis of the natural elements' geometry. As a result, we have built the refined hydrogen bond potential we are about to describe.[91]

First, we have decided the chain description, the needed auxiliary vectors and the conformational chain flexibility. After that, we have devised a set of geometrical constrains to describe the particular geometry of hydrogen bonds in computationally efficient but simple terms. Then, the allowed values for each restriction have been settled according to a detailed analysis of natural structures. As a final step, we have chosen the most appropriate functional form for the potential, tuning the specific values of the different parameters by model optimization.

Our system is described by the $\alpha$-carbons of the different residues, as we show in Figure 3.7, and the following associated vectors for each residue $i$:

- Position vector ($\mathbf{r}_i$): It refers to the location of residue $i$ according to an external reference system.

- Virtual bond vector ($\mathbf{v}_i$): It connects the previous residue in the chain ($i-1$) to residue $i$.

- Unitary hydrogen bond vector ($\mathbf{h}_i$): Identical to the hydrogen bond vector of the Hoang *et al.* potential, it is perpendicular to the plane defined by the preceding ($\mathbf{v}_i$) and following ($\mathbf{v}_{i+1}$) virtual bond vectors (see Equation (3.15)).

Regarding the chain flexibility, we have defined a hard sphere potential among every pair of residues, placed at 4.0 Å, that reflects the excluded volume effects. To avoid too extended unnatural conformations, we have added a maximum limit of 150° to the virtual bond angle between three consecutive residues. In this way, we have described a flexible chain whose possible conformations comply with those of natural proteins.

The definition of the geometry of the model has become an important task during the design process. It has to be precise enough to describe hydrogen bonds in an accurate way. But we also seek intuitive definitions (i.e. geometrical restrictions with a clear physical meaning) and computationally efficient mathematical expressions for them.

We have developed this task in two steps. First, we have designed three geometric variables ($R1$, $R2$ and $R3$) and assigned suitable allowed ranges for each of them. These are our variables for the hydrogen bond interaction between two internal residues:

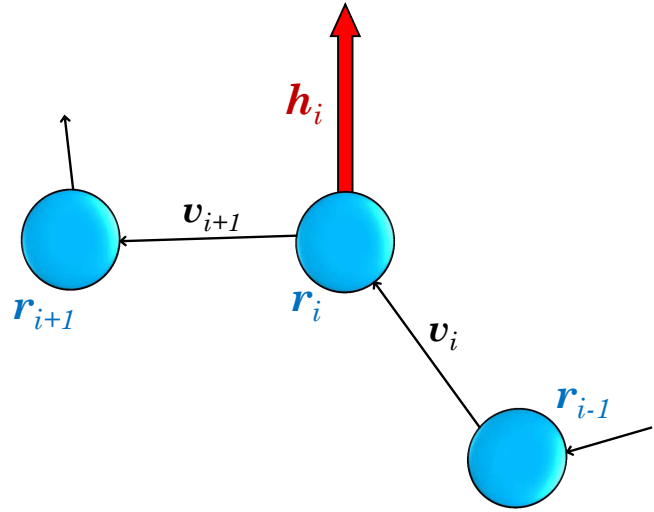

**Figure 3.7:** *Geometric description of our hydrogen bond potential. The polypeptide chain is represented by beads, placed at the α-carbon positions, **r**, and connected by virtual bond vectors **v**. A unitary virtual bond vector **h** (in red) is also defined.*

**Restriction $R1$:** It is a spatial restriction that limits the distance between the two α-carbons of the hydrogen bonded residues.

$$|\mathbf{r}_{i,j}| = |\mathbf{r}_j - \mathbf{r}_i| \qquad (3.26)$$

**Restriction $R2$:** It is an orientational restrain that quantifies the relative alignment between the auxiliary hydrogen bond vectors. It is computed as the cosine of the associated angle:

$$|\mathbf{h}_i \cdot \mathbf{h}_j| = \left|\cos(\mathbf{h}_i, \mathbf{h}_j)\right| \qquad (3.27)$$

**Restriction $R3$:** It is also an orientational quantity that, calculated as the cosine of the angle between the direction of the tentative hydrogen bond in the model and each of the auxiliary vectors, describes the relative orientation between the chain fragments and the hydrogen bonds. $R3$ is independently calculated for both $i$ and $j$ beads ($R3_i$ and $R3_j$):

$$\left|\cos(\mathbf{h}_k, \mathbf{r}_{i,j})\right| \text{ for } k = i, j \qquad (3.28)$$

In the case of having one or both terminal residues involved in the hydrogen bond calculation, the lack of the auxiliary vector prevents the application of the second and third restrictions. In these occasions, we have followed the approach by Hoang *et al.* that adapts the constrains to these specific circumstances. Thus, the $R2$ restriction (i.e. angular dependence between the hydrogen bond vectors) is not calculated. The $R3$ restriction (i.e. relative orientation of the hydrogen bond vector and the connecting vector) is changed to the following one, $R2_e$, for the residue(s) where the hydrogen bond vector cannot be defined. It uses the unitary connecting vector $\mathbf{m}_{i,j}$, previously introduced in Equation (3.19):

**Restriction $R2_e$:**

$$|\cos(\mathbf{v}_k, \mathbf{m}_{i,j})| \qquad \text{for } k = \text{terminal residue} \qquad (3.29)$$

We think that having one spatial restriction and two orientational ones is optimum, as each of them fulfills a specific role in the hydrogen bond definition but still keeps the simplicity of the model.

Once our restrictions have been defined, we have chosen the allowed intervals for each of them. In our experience, this is the critical part of the design. Most hydrogen bond potentials share a similar description of the geometric quantities that define the interaction. However, the specific acceptable ranges present some variability that allows in some cases the unwanted structures we discussed in the previous Section.

The avoidance of these effects in our potential has been achieved thanks to a very extensive study of the natural geometry of $\alpha$-helices and $\beta$-sheets, as stored in the PDB database. We have used for this purpose a dataset of 1590 proteins, previously built in our group,[145] to apply a statistics-based strategy. As we stated in Chapter 1, a useful insight of the rules that govern protein behavior can be obtained by the careful observation of the resulting structures.

In Figures 3.8 and 3.9 we show the histograms we have obtained by the calculation of the aforementioned geometric variables to the pairs of residues of the protein database that form a backbone hydrogen bond. The results have been split into local (Figure 3.8) and non local interactions (Figure 3.9). We have observed that their behavior is slightly different, which justifies the selection of different allowed intervals for local and non-local interactions, in each restriction.

| Restriction | Local range | Non-local range |
|:---:|:---:|:---:|
| $R1$ | $4.7\,\text{Å} \leq R1 \leq 5.6\,\text{Å}$ | $4.0\,\text{Å} \leq R1 \leq 5.6\,\text{Å}$ |
| $R2$ | $0.74 \leq R2 \leq 0.93$ | $0.75 \leq R2 \leq 1.00$ |
| $R3$ | $0.92 \leq R3 \leq 1.00$ | $0.94 \leq R3 \leq 1.00$ |
| $R2_e$ | $0.10 \leq R2_e \leq 0.44$ | $0.00 \leq R2_e \leq 0.34$ |

**Table 3.1:** *Optimal ranges for the three geometrical restrictions chosen in our model for backbone hydrogen bonds.*

The choice of these intervals is a critical aspect of our potential. On one hand, a very large proportion of native hydrogen bonds should be identified by our model with our chosen range (in our case, nearly 80% of the native hydrogen bonds lie in the allowed ranges). On the other hand, the selected intervals should be narrow enough to discriminate between native-like and abnormal backbone hydrogen bonds, since this is precisely the type of result we want to enforce in our interaction model.

The chosen intervals are indicated with shadowed regions in Figures 3.8 and 3.9 and their values are shown in Table 3.1. They have been obtained after an extensive investigation of the effects that each of them has in the overall behavior of the system through numerical experiments (i.e. tendency to form distorted structures, relative stabilization of structural elements, etc.). The interval limits, however, may vary within approximately 5% of the data in Table 3.1 with minor consequences in the model performance.

If we start with the local hydrogen bonds of Figure 3.8, every interaction follows a well defined trend for each restriction, showing therefore relatively narrow peaks in the four plots. The histogram of the spatial restriction $R1$, see Figure 3.8($a$), shows a sharp

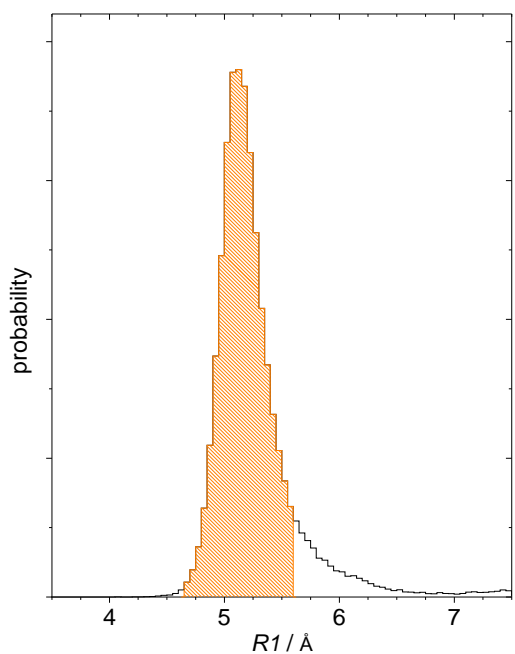

(a) Spatial restriction *R*1.

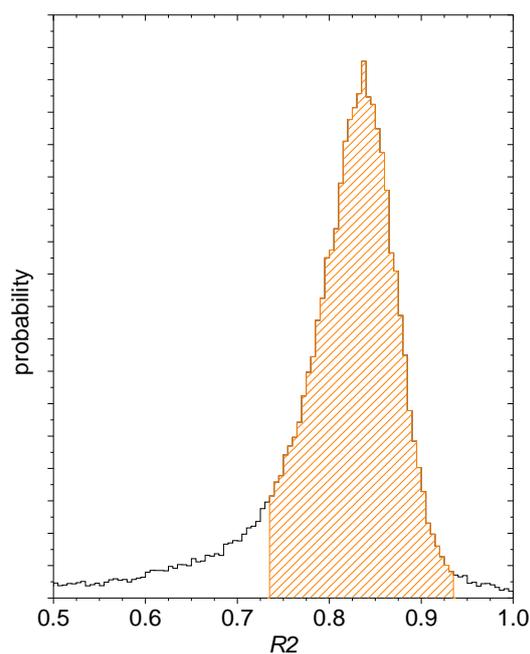

(b) Orientational restriction *R*2.

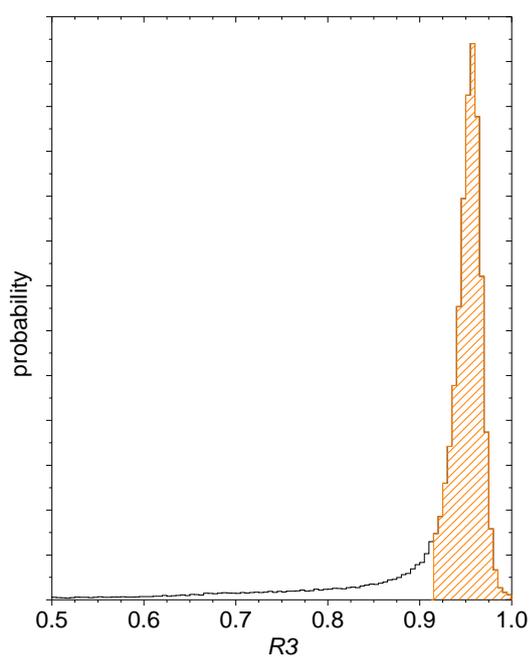

(c) Orientational restriction *R*3.

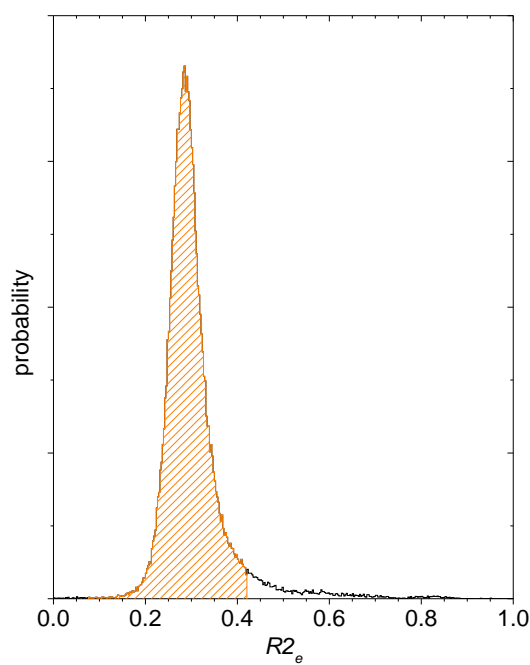

(d) Orientational restriction $R2_e$

**Figure 3.8:** *Statistics over the studied protein database of native local hydrogen bonds for the geometrical restrictions R1, R2, R3 and R2ₑ. The orange stripped regions indicate the selected range of values in our model.*

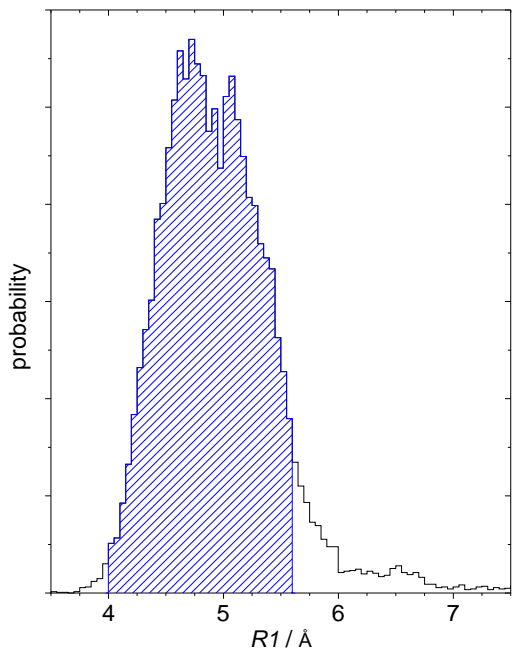

(a) Spatial restriction *R*1.

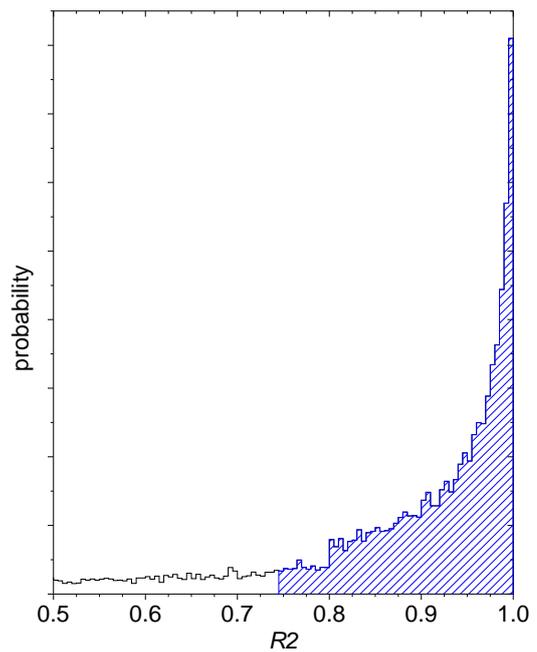

(b) Orientational restriction *R*2.

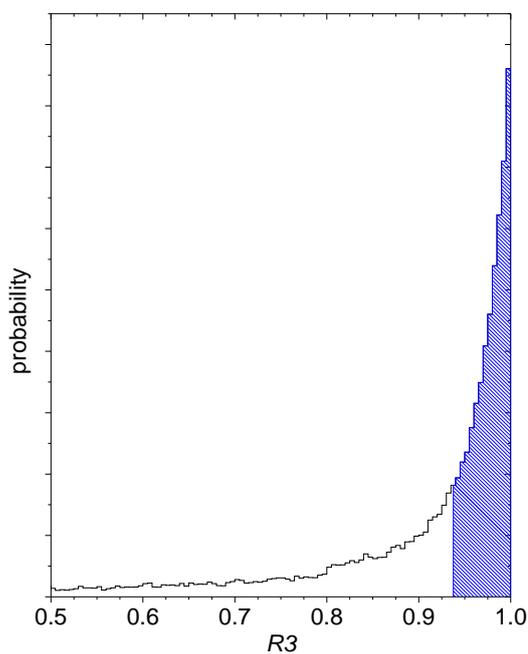

(c) Orientational restriction *R*3.

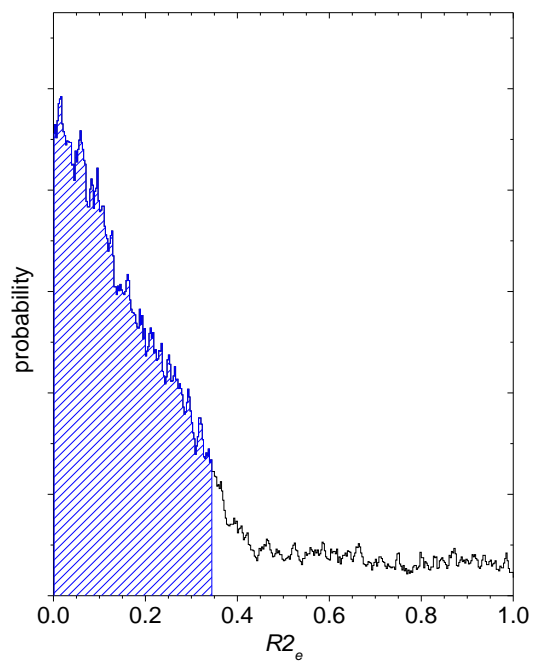

(d) Orientational restriction *R*2$_e$

**Figure 3.9:** *Statistics over the studied protein database of native non-local hydrogen bonds for the geometrical restrictions R1, R2, R3 and R2$_e$. The blue stripped regions indicate the selected range of values in our model.*

peak centered at 5.2 Å. At this point, we have also introduced the helix chirality through the $\kappa$ term we defined in Equation (3.17). Thus, this restriction is only fulfilled if the value of the variable $R1$ lies in the acceptance range and $\kappa > 0$.

Regarding the orientational constrains, clear tendencies are also observed in all of them. In this way, the cosine of the angle formed by auxiliary hydrogen bond vectors (restriction $R2$) exhibits a maximum at 0.84, i.e. the angle between them is approximately 33°, which reflects the advance of the $\alpha$-carbon positions along the helical axes direction. The relative orientation between the auxiliary hydrogen bond vectors and the connecting vector corresponds to an angle of 15° and is linked to the advance in the helix and the fact that the real hydrogen bond is not formed between $\alpha$-carbons but between the amino and carbonyl groups of the amino acids. In relation to the terminal restriction ($R2_e$), the most frequent angle between the connecting vector and the terminal virtual bond vector is around 75°. Around these maximum values, we have allowed some deviations (usually close to 10%) to obtain the acceptable ranges of Table 3.1.

In the case of non local hydrogen bonds (see Figure 3.9 and the right-hand side of Table 3.1), similar trends have been observed, although the histogram distributions are a bit wider in this case. Regarding the spatial restriction $R1$, we find a broad maximum between 4.5 and 5.5 Å, reflecting the larger structural variability of $\beta$-sheets compared to $\alpha$-helices. The orientational restrictions do present a clear maximum, although the probability peak is not as sharp as in local interactions, revealing again the higher permissiveness of $\beta$-sheet structures. They all reflect a marked tendency to parallel or antiparallel arrangements of the peptide fragments ($R2$ and $R3$ near to 1.0, and $R2_e$ close to 0.0).

The interval choice for non local hydrogen bonds has been tougher than in local interactions due to the broader shape of the histograms. As it can be observed in Figure 3.9, our chosen intervals include most of the native cases, altough in some cases (like $R3$ and $R2_e$) we have taken slightly more restrictive ranges in order to destabilize the unnatural structures we will comment in the next Section.

Once the geometric restrictions have been defined and their intervals properly cho-sen, we have designed the functional form of our hydrogen bond potential. We aimed again to obtain a high computational efficiency and the avoidance of undesirable effects such as the additional hydrogen bond interactions that were allowed by the Kolinski potential and the presence of unnatural structures in our simulations. The general expression, $E^{hb}$, is a typical pair-wise potential:

$$E^{HB} = \sum_{i=1}^{N-3} \sum_{\substack{j=i+3 \\ j \neq i+4}}^{N} e_{i,j}^{hb} \qquad (3.30)$$

As it can be observed, the $(i, i+4)$ interactions have been removed from the calculation. This feature, previously observed in the Kolinski potential, destabilizes *thick* helices, formed by these interactions and similar to the B2 structure in Figure 3.6.

In relation to the individual interaction itself, we have chosen a step-like potential, encouraged by the nice results of the Hoang *et al.* potential. We have not considered an explicit cooperative effect, as multi-body terms substantially increase the computational cost and reasonable cooperative behaviors had been obtained without it in other potentials designed by our group.[62] Because of similar reasons, we have not explicitly limited the number of hydrogen bonds per residue, leaving this task to the accurate design of the potential itself.

Our hydrogen bond interaction follows this expression:

$$e_{ij}^{hb} = \begin{cases} \lambda & \text{if R1, R2, R3}i \text{ and R3}j \text{ are fulfilled} \\ 0.25\,\lambda & \text{if R1, R2 and only one of R3}i \text{ or R3}j \text{ are fulfilled} \\ 0 & \text{otherwise} \end{cases} \qquad (3.31)$$

As it can be observed, it is divided in three steps and defined in terms of a factor $\lambda$ that depends on the kind of interaction ($\gamma_{int}$), either local or non local, and the kind of residue ($\gamma_{res}$), either internal or terminal:

$$\lambda = \gamma_{int} \cdot \gamma_{res} \qquad (3.32)$$

The interaction-dependent term distinguishes between local and non-local hydrogen bonds, as it happened in the Hoang *et al.* potential. We have considered the energetic difference as a parameter of the model, optimized to give a sensible competition between $\alpha$-helix and $\beta$-sheet structures. The best results are obtained with:

$$\gamma_{int} = \begin{cases} -1.00 & \text{if it is a local bond} \\ -0.92 & \text{otherwise} \end{cases} \tag{3.33}$$

The residue-dependent term has been included due to the acquired knowledge of secondary structures during our statistical exploration of native elements. We have found that the terminal residues present more variability in their behavior compared to the internal ones (also reflected in the histograms of Figures 3.8($d$) and 3.9($d$)). So, we have reproduced this fact by establishing a kind of energetic penalty that makes terminal hydrogen bonds easier to form and break:

$$\gamma_{res} = \begin{cases} 1.00 & \text{if it is an internal residue} \\ 0.25 & \text{otherwise} \end{cases} \tag{3.34}$$

In the previous paragraphs, we have thoroughly explained the most important features of our hydrogen bond potential, such as the different treatment of local and non local bonds, both for the geometric restrictions' ranges and the energetic interaction. This strategy has also been applied to the terminal residues, tackling in this way the lack of a full description for this type of residues. Will this strategy overcome the drawbacks of other available potentials?

## §3.3 Results

Along the current Chapter, we have discussed several hydrogen bond potentials and their drawbacks, concerning our main purpose of generating a hydrogen bond potential able to describe secondary structure elements as accurately as possible, using an $\alpha$-carbon representation. As a result, we have proposed a new one that, in principle, would solve these disadvantages. Therefore, in this Section we will carry out an extensive analysis of the model in order to evaluate its applicability. In first place, will it show the desired structural characteristics we are seeking?

The most common evaluation method for hydrogen bond potentials (and usually their first use, too) is the study of the thermal stability of the different secondary structure elements. In this way, there are many studies of the folding and unfolding of $\alpha$-helices[143, 149–151] and $\beta$-sheets[152] and the competition between them.[79, 153–155]

This kind of analysis, which has provided the ultimate tune of the model parameterization, has been divided in two main parts. In first place, we have studied the helix-coil transition with single chain experiments, focusing on the variation of the thermodynamic properties with the chain length. Secondly, we have addressed the multi-chain problem: we have simulated several peptides in a simulation box, varying their concentration. As a result, we have studied the different structures that arise from the application of our potential and the impact of the unnatural ones in the overall behavior of peptides. Finally, we have performed a combined analysis of temperature and concentration, culminating into a two-dimensional phase diagram for peptide systems under the effect of our hydrogen bond potential.

### 3.3.1 Geometric analysis

For this Section, we have followed the same approach as the one in Section 3.1.1, just caring about the geometry of native-like features obtained with our potential, again using 10-residue long peptides in infinite-dilution (for obtaining $\alpha$-helices) and concentrated (for $\beta$-sheets) conditions. An example of our simulated structures is shown in Figure 3.10. Again, they present a native-like look at first glance. Will they still show these suitable characteristics after a detailed analysis? For this purpose, we have carried out the same kind of analysis we previously did, centered in three major items:

**Helix chirality:** As our refined model explicitly defines the chirality of the local bond through Equation (3.17), every formed $\alpha$-helix presents the characteristic right-hand

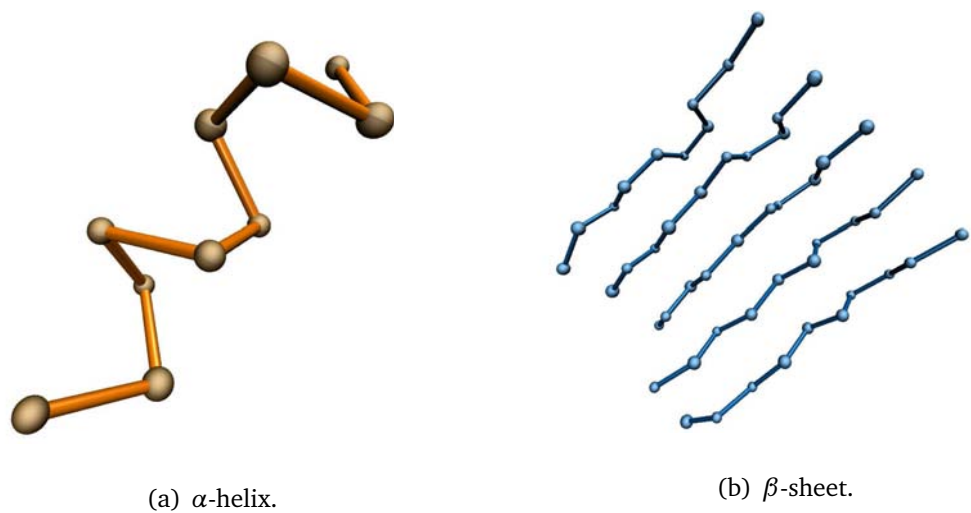

(a) *α*-helix.

(b) *β*-sheet.

**Figure 3.10:** *Typical secondary structure elements obtained with our refined hydrogen bond potential in peptide systems of ten amino acids per chain. Images drawn with VMD[144].*

arrangement.

**Local geometry:** The short-range organization of peptides is usually characterized, when dealing with *α*-carbon representations, by the $d_{i,i+3}$ distances, drawn in Figure 3.11. We can observe that our helical histogram nicely overlaps with the native one. In the case of *β*-sheets, the $d_{i,i+3}$ distance is slightly shorter than the native one, motivated by the control of the model over too extended conformations. Anyhow, it presents a broad peak that mostly overlaps with the native distance distribution, so we have considered that the geometry is successfully reproduced.

**Hydrogen bond contact map:** The main problems of Kolinski potential regarding the contact maps were *(i)* the absence of hydrogen bonds for the terminal residues and *(ii)* the appearance of unnatural non-specific hydrogen bonds around the real ones. In our model, these two drawbacks had been considered from the beginning, so our typical hydrogen bond contact maps (drawn in Figure 3.12) just show the desired interactions, allowing them when the terminal residues are involved –see, for instance, the local interaction between beads 1 and 4 in the $\alpha$-helix of Figure 3.12(*a*)– and just presenting two hydrogen bond interactions per residue at most.

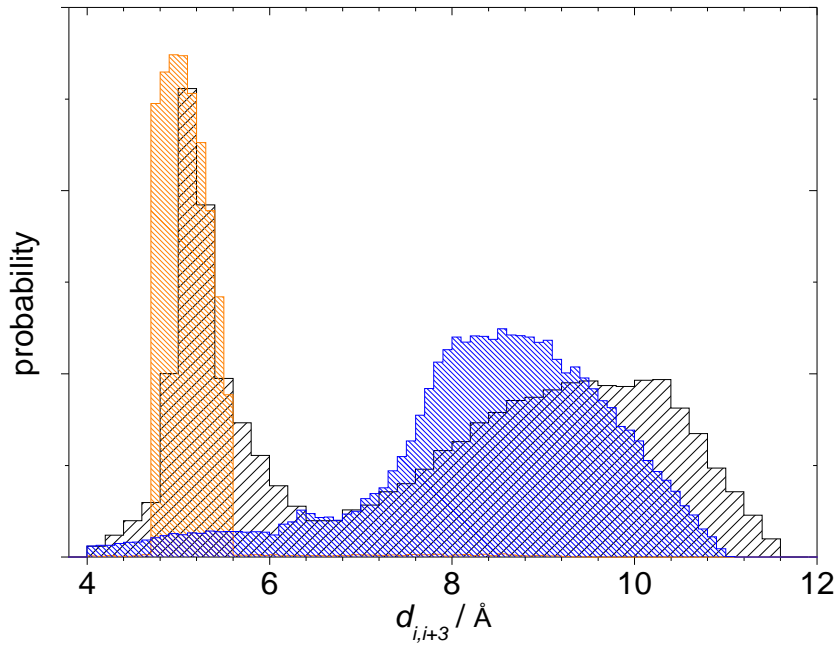

**Figure 3.11:** *Histograms of $d_{i,i+3}$ distances for a dataset of native structures (black) and the secondary structure elements obtained by our refined potential ($\alpha$-helices in orange and $\beta$-sheets in blue).*

Then, we can conclude that our hydrogen bond potential successfully describes secondary structure elements with a native-like geometry, as we have checked in a model system of short peptides. Will their properties depend on the chain length? How will they behave in complete folding experiments? Let's start with the most simple case, the helix-coil transition.

### 3.3.2 Single chain experiments

The structural and thermodynamic study of hydrogen bonds has been carried out in different stages of increasing complexity. The first one has been the performance of single chain numerical experiments. In this kind of simulations, interchain interactions are not possible, reproducing therefore infinite dilution conditions. Apart from being the most simple folding simulations that can be carried out (and, for that reason, always useful for model testing), they also have an impact in real systems, as certain homopeptides fold into helices in diluted conditions.[156]

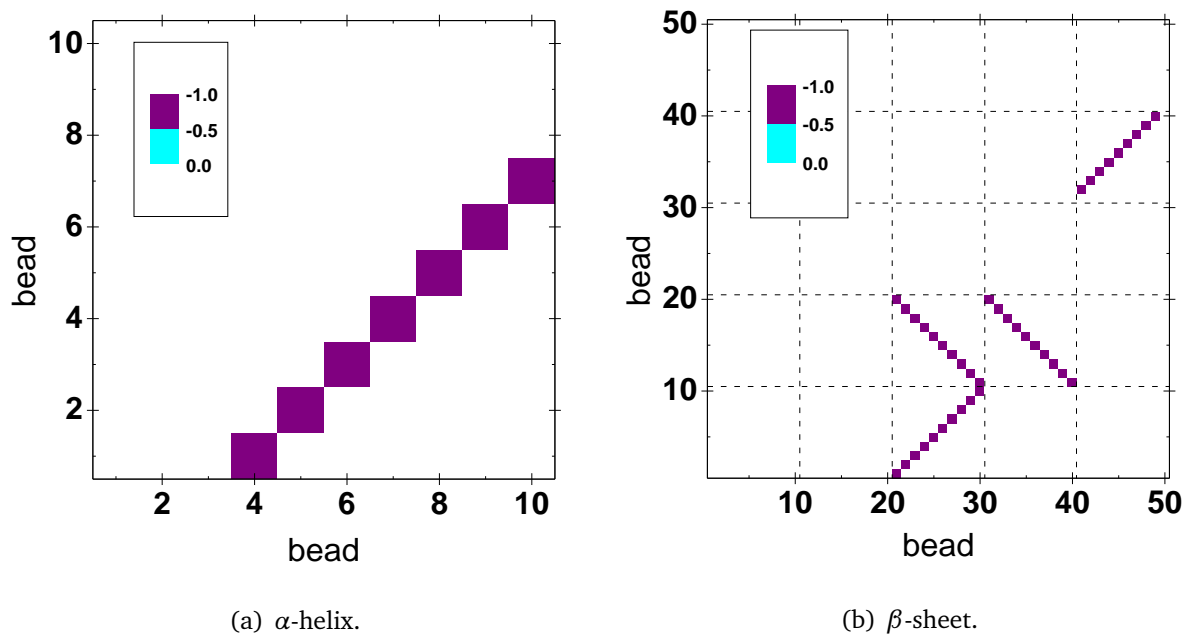

(a) α-helix.                    (b) β-sheet.

**Figure 3.12:** *Hydrogen bond energy maps for the secondary structure elements, according to our refined potential.*

For us, these numerical experiments pursue two aims: to check whether our potential favors native helical states in diluted systems and to determine if other alternative structures are present in our simulations, allowed by the geometrical features of the defined interaction but lacking any physical meaning in a real polypeptide chain. Particularly, we have wondered if there is any sort of *applicability range* for our systems, i.e. if we can simulate secondary structures of any size and where is the limit for the obtention of native-like elements.

We have performed equilibrium folding/unfolding simulations for systems of one chain of different length, *L*, from 10 to 25 residues. In these simulations we have used reduced energy units, as we are not interested in the specific values of temperature or energy, but in the general behavior of the system. Due to technical details, we have increased the energy units ten times, assuming therefore that the maximum stabilization of a hydrogen bond is $-10.0$ energy units instead of $-1.00$.

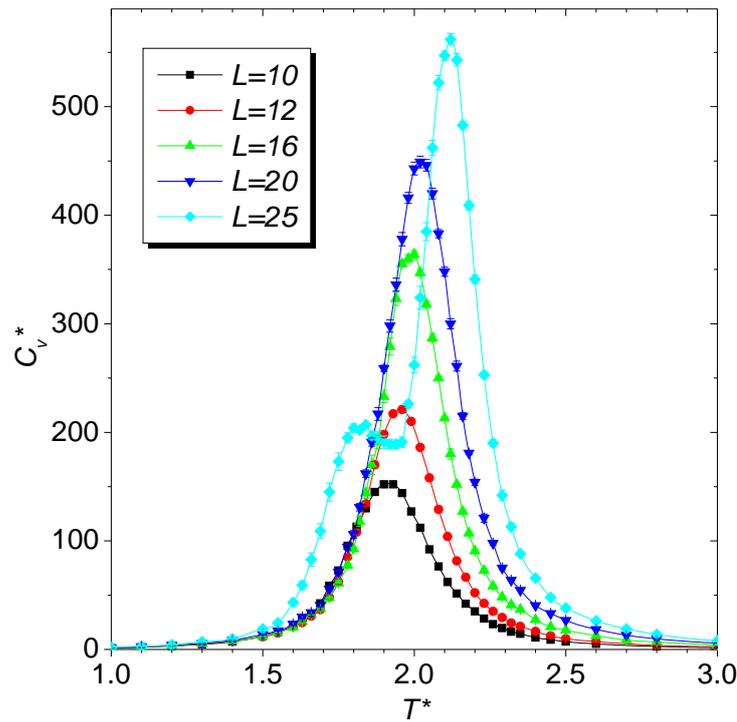

**Figure 3.13:** *Heat capacity curves vs temperature for infinite-dilution systems of different chain lengths L. Note the use of reduced units in both axes.*

We show in Figure 3.13 the heat capacity curves versus temperature, previously introduced in Chapter 2. In these curves, a peak indicates an energetic transition that, as we shall see afterwards, matches a structural one.

Short chains show a common behavior with only one peak in this curve, which corresponds to the transition from a folded state to a denatured (unfolded) one. On the contrary, the longest chain exhibits an unexpected behavior, with a double peak in the heat capacity curve (see Figure 3.13 for $L = 25$).

Regarding the structural evolution with temperature, we have found three main types of structures: $\alpha$-helices, formed by $(i, i + 3)$ interactions; distorted *thick* helices, formed by long range interactions, mostly $(i, i + 5)$; and unfolded structures with just sporadic interactions. For each system, we have performed a population analysis, shown in Figure 3.14. A schematic plot of each kind of structure (drawn with VMD[144]) has also been inserted in this Figure.

Again, the modification of the transition characteristics with the chain length is observed. At low temperatures, every chain folds into an $\alpha$-helix. The shortest chains, $L = 10$ and $L = 12$, unfold at the transition temperature without any intermediates, as seen in Figures 3.14($a$) and 3.14($b$). At intermediate temperatures, the increase in the chain length (shown in Figures 3.14($c$) and 3.14($d$)) results in the growth of the population of thick helices. This distorted structure, built thanks to nonlocal hydrogen bonds, becomes dominant at intermediate temperatures for the 25-residue chain, explaining therefore the double peak in the heat capacity curve of Figure 3.13.

According to our results, the population of thick helices is null or almost negligible for chains shorter than 20 residues (it adds up to less than 5% of the registered configurations). As the average length of a native $\alpha$-helix in globular proteins is about 12 residues,[1] we can conclude that our model succeeds in the obtention of native-like helices in infinite dilution conditions for realistic chain lengths.

We have performed additional simulations including $(i, i + 4)$ interactions in the energy calculations (data not shown). They highlight the importance of this detail: the removal of these interactions widens the thermal stability range for the $(i, i + 3)$ ones, obtaining unique native-like structures in the significant chain length range commented on above.

We have deepened in the thermodynamic analysis of these systems, computing their free energy in terms of the internal energy, $E = E^{hb}$, through the WHAM method we explained in Chapter 2. We have observed that the behavior of this property is modified with the chain length. It is illustrated in Figure 3.15 by the free energy profiles at different temperatures for a short chain ($L = 12$) and a long one ($L = 25$). In both cases, we have plotted the free energy profile at three different temperatures: a low one, $T^* = 1.50$ (where the folded structure is observed); a high one, $T^* = 2.50$ (to show the denatured state), and also the high temperature transition, $T^* = 2.00$. Note that the graphs show the typical step-like look of discrete potentials.

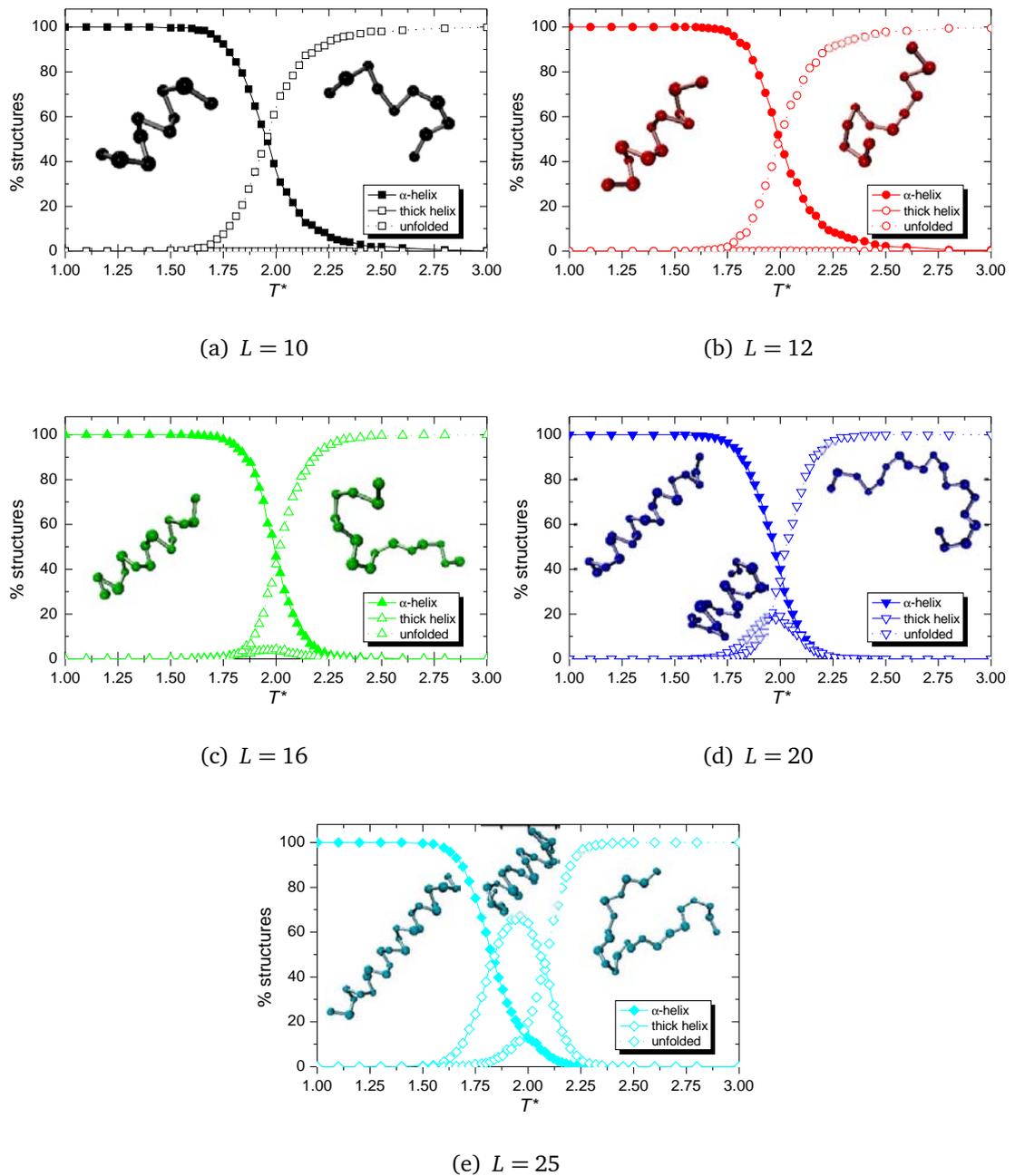

(a) $L = 10$

(b) $L = 12$

(c) $L = 16$

(d) $L = 20$

(e) $L = 25$

**Figure 3.14:** *Variation in the population of each type of structure, i.e. native-like helices stabilized by (i, i + 3) hydrogen bonds (solid symbol and solid line), thick helices stabilized by nonlocal hydrogen bonds (open symbol and solid line) or unfolded structures (open symbol and dashed line), with temperature for different chain lengths. Structures represented using VMD.[144]*

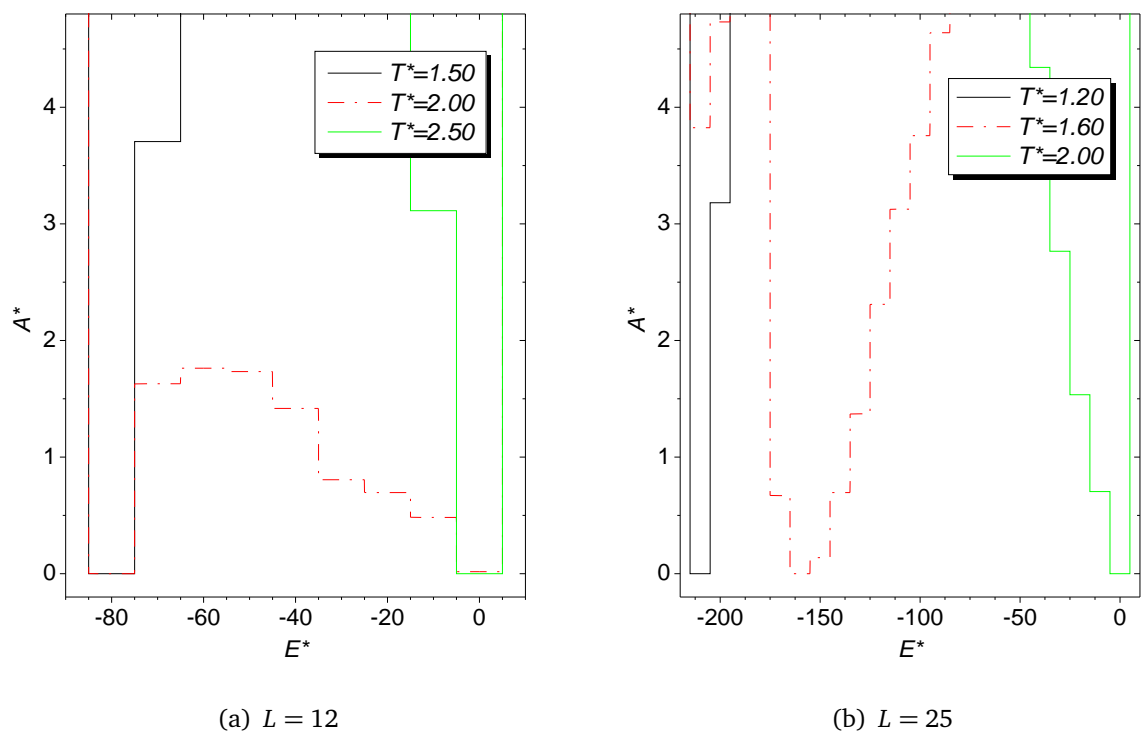

(a) $L = 12$            (b) $L = 25$

**Figure 3.15:** *Free energy profiles at different temperatures for two single chain simulations of different chain length.*

The extreme temperatures show similar characteristics in both cases: the free energy presents a sharp minimum, centered at the hydrogen bond energy of an $\alpha$-helix (according to the chain length of each system) in the case of the low temperature, and close to 0.00 energy units for the high temperature one, reflecting the complete disappearance of hydrogen bond interactions (as it should happen in the fully denatured state). Regarding the transition temperature, the short chain presents a bimodal curve, mainly populating the folded and the unfolded states. The longer chain, on the contrary, presents a broader minimum at an intermediate energy, reflecting the presence of thick helices that populate an intermediate state along this transition.

In this way, systems of a reasonable chain length present a helix-coil transition with a free energy barrier that could be interpreted as a sign of the system cooperativity. It is a desirable property in a hydrogen bond potential that, remarkably, we have achieved without an explicit cooperativity term and just relying on a careful selection of the meaningful variables of the defined potential.

### 3.3.3 Multichain systems

After the study of the helix-coil transition, we have investigated the properties of multichain systems, i.e. different concentration conditions. For us, concentration is a critical factor, as concentrated conditions favor interchain interactions and, thus, the possible formation of $\beta$-sheet structures that can be linked to the secondary structure element itself but also to aggregates. In this way, the study of multichain systems under the effect of a hydrogen bond potential is a first step in the comprehension of the interplay between folding and aggregation.

We have simulated different concentrations within the same order of magnitude (either varying the number of chains or the size of the simulation box). Showing similar properties, we shall expose here the results of a system with five chains of twelve residues each (comparable to the typical length of secondary structure elements),[1] under four relevant concentrations, ranging from 0.01 to 0.06 chain moles/L. Note that the numerical values of the simulated concentration are just representative of the variation analyzed in this work, but they do not try to reflect a real experimental concentration.

In the investigation of these systems, we have found four most populated types of structures, as well as essentially random structures that have been classified as "unstructured". These structures are shown in Figure 3.16, where a hydrogen bond map and a representative cartoon are included for each of them. Two of them (structures A and B) match the native-like helices and sheets, respectively. But we have also found two additional non-natural structures, labelled C and D:

**Structure A:** It corresponds to the typical native-like helices, formed by local $(i, i + 3)$

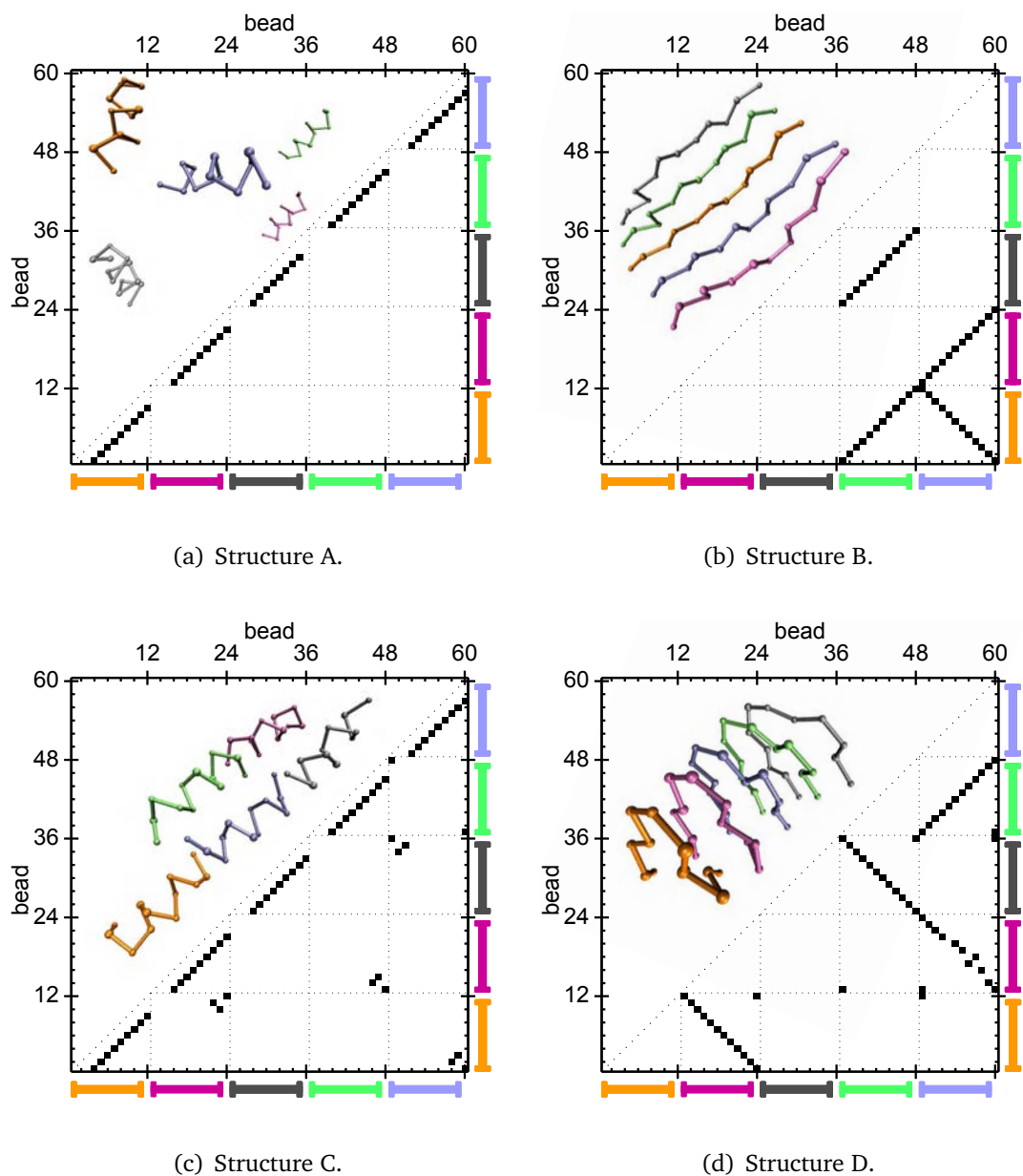

(a) Structure A.

(b) Structure B.

(c) Structure C.

(d) Structure D.

**Figure 3.16:** *Hydrogen bond maps for the different structures obtained in the multichain simulations. The thin dotted lines indicate the end of a chain and the beginning of another one; the black dots indicate a hydrogen bond contact. Structures represented using VMD.[144]*

hydrogen bond interactions.

**Structure B:** It is a $\beta$-sheet formed by the conjunction of both parallel and antiparallel interactions among the five chains of the system, showing diverse arrangements in our simulations such as the one we show in Figure 3.16(*b*). They are stabilized by long range interchain hydrogen bond interactions, easily detected by the black spots on the block diagonals of the energy map of this Figure.

**Structure C:** Similarly to Structure A, the internal conformation of each chain is stabilized by local interactions, presenting an essentially helical structure. The main difference lies in the presence of interchain associations between the terminal residues of a pair of chains. This results in an oligomeric superstructure joined by the chain ends that finds more hydrogen bonds than those initially expected, propagating them as if the system was a large single chain, instead of multiple independent ones. This structure drastically minimizes the system entropy through a strong reduction of the system mobility.

**Structure D:** It is stabilized by interchain interactions that create a sort of distorted $\beta$-type structure, wrapped into itself by the formation of extra hydrogen bonds among terminal residues. This additional energetic stabilization illustrates, as in Structure C, a violation of a property of real proteins, at a high entropic cost.

As we have already stated, structures C and D do not match any relevant natural one, constituting artifacts of our model. They highlight an important feature of our potential: the number of hydrogen bonds for a given residue is not explicitly restricted in the potential definition. For inner residues, this fact does not imply any significant drawback, as the careful design of the restrictions and parameters of the model naturally limits the number of hydrogen bonds per bead. However, the laxer definition of the interactions involving terminal residues fosters the formation of these abnormal structures. Since we have penalized the energy of the hydrogen bonds involving terminal residues in our model, as shown in Equation (3.34), the significant population of these structures is restricted to very low temperatures, as we show in the next paragraphs.

The relevance of the different structures along the whole free energy landscape has been investigated through a combined analysis of thermodynamic and structural properties, analogously to the former Section. Our main results are shown in Figures 3.17 and 3.18, where we have plotted the heat capacity curves versus temperature and the temperature evolution of the population of the different structures for the different concentration conditions, respectively.

Starting with the most diluted system (0.01 chain moles/L), it exhibits two well separated maxima in the heat capacity curve (see Figure 3.17(*a*)). These two energetic transitions define temperature ranges where a certain structure population is dominant. In Figure 3.18(*a*) we observe that Structure C, the helical oligomeric structure, is preponderant at very low temperatures. The formation of its interchain interactions imply a large entropic cost, so a slight increment in the temperature of the system breaks these associations, resulting in a large stability region (from $T^* = 1.30$ to its unfolding temperature, $T_m^* = 1.95$) for Structure A (isolated $\alpha$-helices) as the unique stable structure within this range. In this sense, we have recovered the helix-coil transition studied in the previous section for infinite dilution conditions —see Figure 3.14(*b*). The presence of multiple chains in the current system is reflected by an extra transition at very low temperatures between the artifactual and natural helical structures (C and A, respectively), but without any unwanted non-natural effects in the not-frozen temperature range.

Following our discussion with the 0.02 chain moles/L system, the corresponding heat capacity curve of Figure 3.17(*b*) shows an additional peak in the low temperature region in comparison to the more diluted system we have already discussed. At the lowest temperatures, Structure C is again the most populated (see Figure 3.18(*b*)), but a small temperature increase within this almost-frozen range leads to an energetic and structural transition to Structure D, as interchain interactions are not so infrequent as before. It is also stable only at very low temperatures, as the entropic cost of blocking the terminal residues is still very high. For this reason, the stability region of these structures is small, essentially disappearing above $T^* = 1.50$.

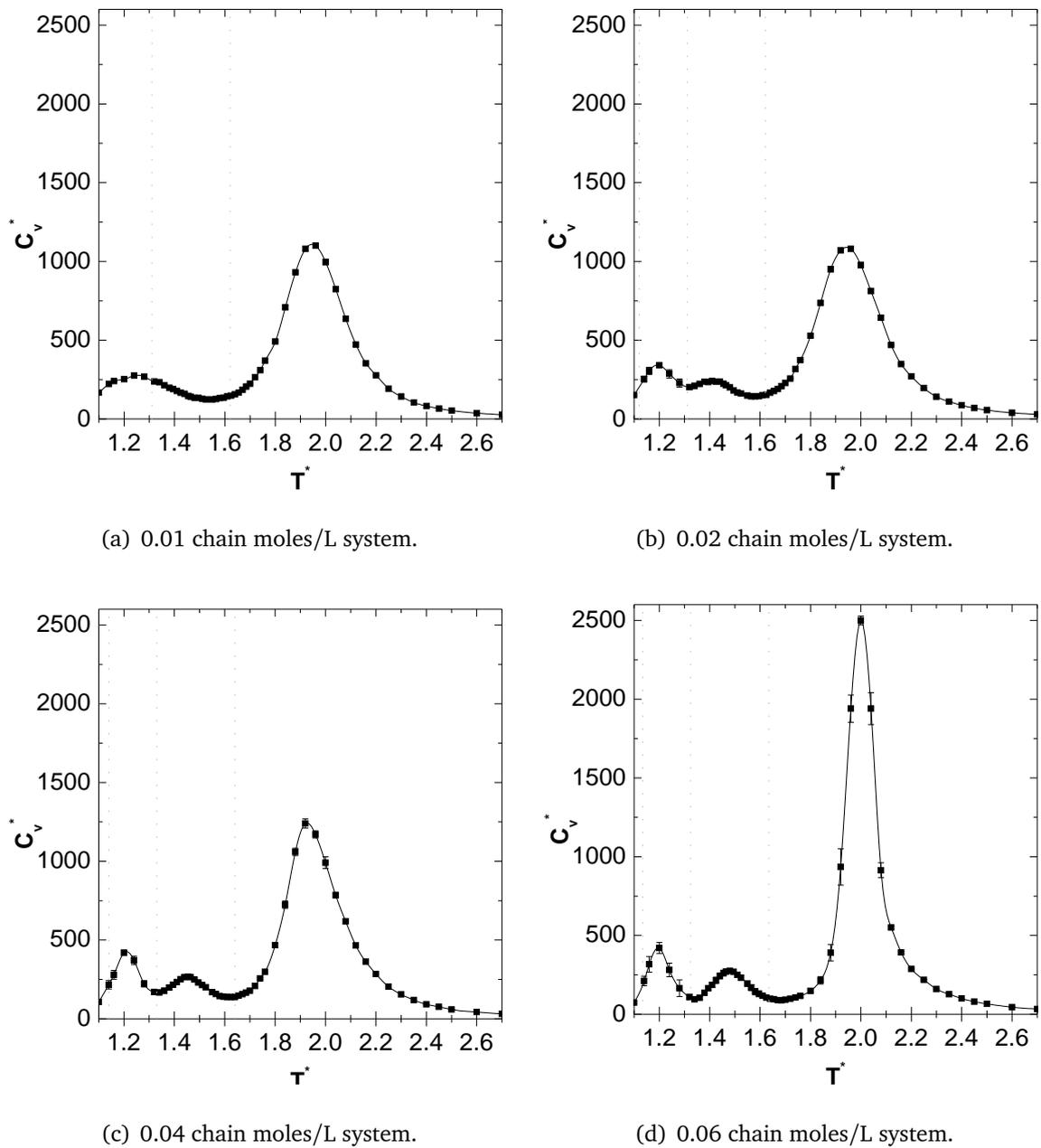

(a) 0.01 chain moles/L system.

(b) 0.02 chain moles/L system.

(c) 0.04 chain moles/L system.

(d) 0.06 chain moles/L system.

**Figure 3.17:** *Thermal evolution of the heat capacity for each multichain simulated system.*

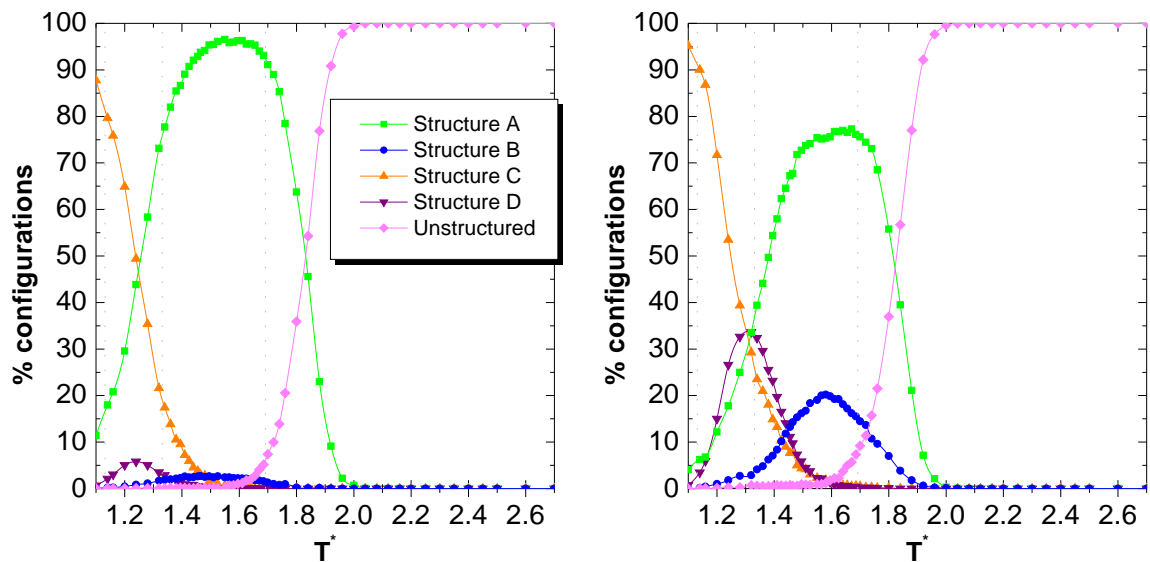

(a) 0.01 chain moles/L system.  (b) 0.02 chain moles/L system.

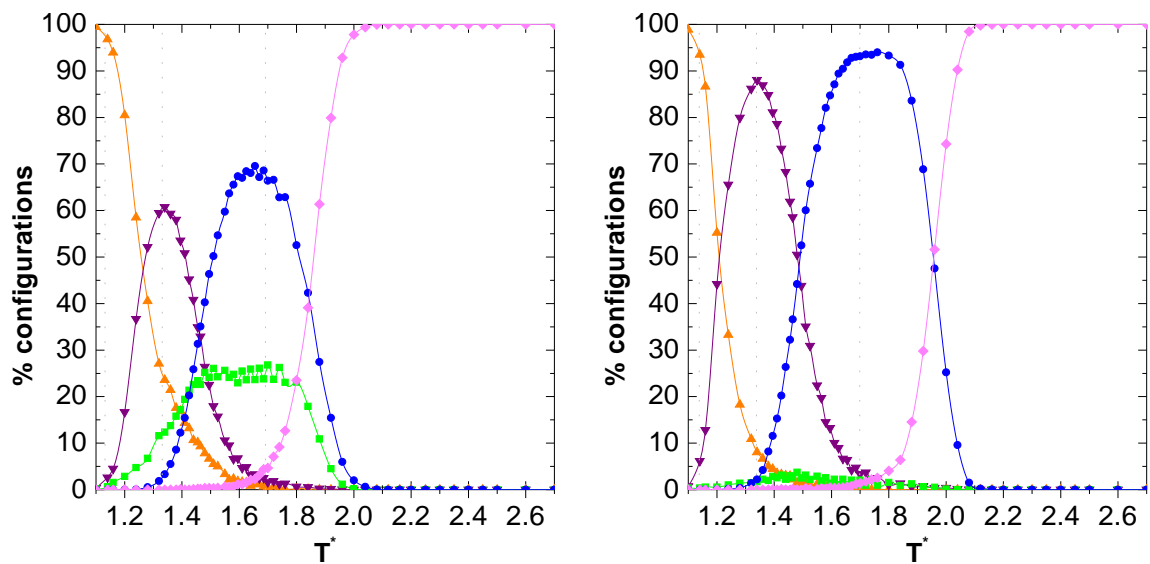

(c) 0.04 chain moles/L system.  (d) 0.06 chain moles/L system.

**Figure 3.18:** *Thermal evolution of the population of the observed structures A, B, C and D (plotted in Figure 3.16) for each multichain simulated system.*

In the subsequent temperature interval, Structure A (isolated $\alpha$-helices) is the predominant feature, although a smaller population of Structure B ($\beta$-sheet) has also been detected. Thus, a concentration increment has revealed the emerging competition between the populations of the two native-like secondary structure elements, mediated by this factor.

It is also remarkable that they are the only stable structures within this relevant temperature range, so our results in the vicinity of the unfolding transition are not marred by the undesired presence of alternative structures. Then, we can argue that the presence of unnatural structures is an artifact of the model that is observed in energetic minimizations (i.e. frozen conditions). But they have no impact at all in the biologically relevant temperature range, that is close to the unfolding transition temperature.

The main results for the simulated system of 0.04 chain moles/L are shown in Figures 3.17($c$) and 3.18($c$). This concentration displays a very similar behavior to the 0.02 chain moles/L system, with the same three energetic and structural transitions we have previously discussed. Importantly, the concentration increase has modified the relative population of helices and sheets (Structures A and B) in the intermediate temperature region, predominating $\beta$-sheets in this case. This shows that concentration really modulates the competition between these two structures in our simulations.

Finally, Figures 3.17($d$) and 3.18($d$) illustrate our results for the 0.06 chain moles/L system. In this case, the low temperature region exhibits the behavior previously described. However, the high temperature transition ($T_m^* = 2.05$ in Figure 3.17($d$)) presents different characteristics, as it is higher and narrower than the transitions observed at lower concentrations. This can be linked to the absence of a significant population of Structure A at immediately lower temperatures (see Figure 3.18($d$), where the population of Structure A is nearly negligible), as interactions among chains (due to the high concentration of the system) are so common that finding isolated helices is very rare. In this concentration conditions, we have essentially obtained the sheet-coil transition.

The origin of the differences in the heat capacity curves seems to be connected to the relative population of the secondary structure elements at the unfolding temperature or, more specifically, to the lack of a stable population of $\alpha$-helices in the most concentrated system. A deeper insight into this fact can be obtained through the calculation of the related free energy profiles.

Consequently, we show in Figure 3.19 the free energy profiles for the two most concentrated systems at the different transition temperatures (matching with the peaks in the related heat capacity curves). These systems mainly present the same kind of transitions (i.e. Structure C to Structure D, Structure D to mainly Structure B and Structure B to unfolded, starting from low temperatures). Nevertheless, the presence of a moderate (although minoritary) population of Structure A in the 0.04 chain moles/L system seems to be linked to the very different shape of the heat capacity curve, that is more similar to the more diluted systems, that essentially present a helix-coil transition with a small population of $\beta$-sheets in some cases.

Regarding the free energy profiles themselves, the low temperature transitions present similar multimodal free energy distributions with very small barriers. However, the unfolding transitions, colored in green, show a significant difference: while the most concentrated system presents a bimodal curve, the more diluted one just has a single and broader peak centered at intermediate energies.

We have analyzed the structures found at the unfolding transition temperature, comparing both systems. In the case of the most concentrated one (0.06 chain moles/L), we have splitted the registered configurations according to its internal hydrogen bond energy. We have inspected the low energy basin (states with $E^{hb} \leq -350$), that corresponds to the stable structure at low temperature, and also intermediate energy configurations (those with $-350 \leq E^{hb} \leq -250$). We have built their hydrogen bond average bidimensional energy maps, shown in Figure 3.20, that tell us, on average, how the hydrogen bonds are distributed in the analyzed configurations.

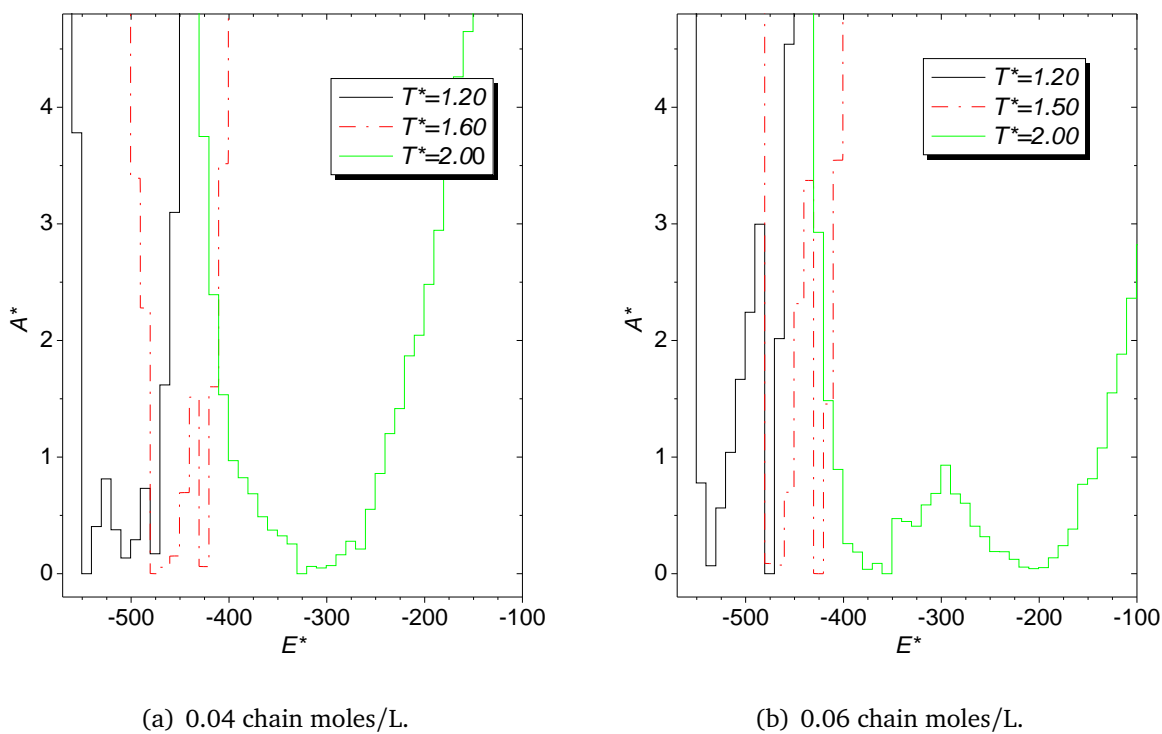

(a)  0.04 chain moles/L.                (b)  0.06 chain moles/L.

**Figure 3.19:** *Free energy profiles at several temperatures for five-chain simulations at two different concentrations.*

In the left part of Figure 3.20 we show the low energy configurations, whose hydrogen bonds are mainly located at the block diagonals, reflecting in this way the presence of $\beta$-sheets in many different arrangements. The right hand map, on the contrary, shows a clear high population of local hydrogen bonds, proving the existence of a helical intermediate in our simulations. Interestingly, a similar property has been observed experimentally, where a helical intermediate is said to play a role in aggregation.[34]

In the case of the diluted system (0.04 chain moles/L), no energetic division is possible according to the free energy profile, as the energetic distribution is barrierless. If we structurally analyze the complete set of configurations, we observe that the chains that separate from the complete $\beta$-sheet present a partially helical structure. They have a wide structural and energetic variability that explains the lack of a free energy bimodal distribution (data not shown). In some sense, the mainly helical intermediates of the most concentrated system have now become the most frequent situation, instead of a minority

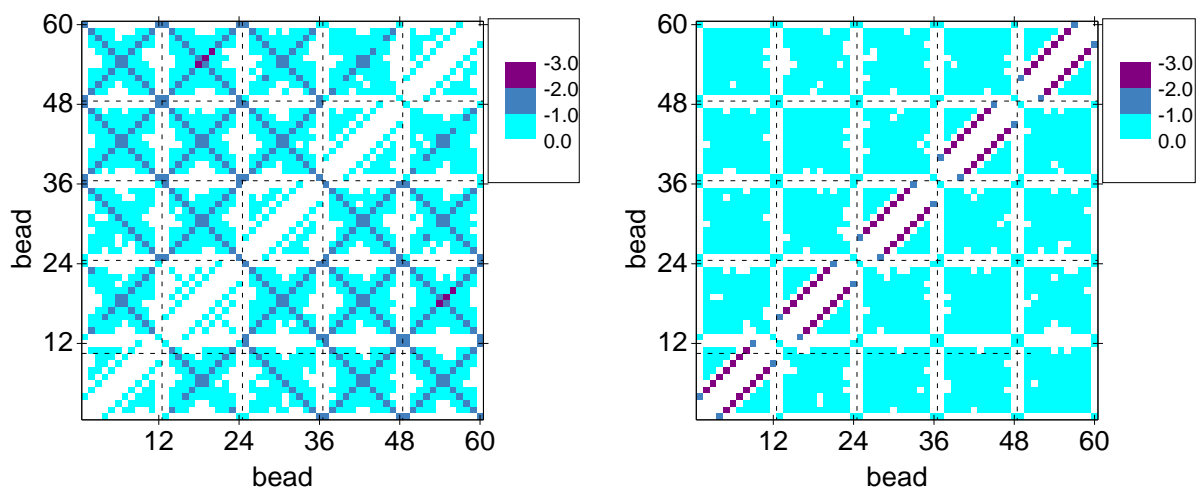

(a) Low energy configurations.          (b) Intermediate energy configurations.

**Figure 3.20:** *Maps for the average hydrogen bond energy obtained at the unfolding transition temperature $T^* = 2.00$ for the most concentrated system (0.06 chain moles/L), in two different energetic regions according to the free energy profile of Figure 3.19.*

one.

Along this Section, we have presented a set of simulations that have allowed us to explore the most relevant situations we would expect in systems driven by the formation of hydrogen bonds, from the structural helix-coil transition in diluted conditions to the sheet-coil one in concentrated systems. This information is condensed in Figure 3.21, a sketched phase diagram where we have roughly matched the simulated structures to their stability regions as a function of concentration and temperature.

At very low temperatures, every system finds Structure C, as it has the lowest energy due to its high helical content and some additional hydrogen bonds among chain ends. It is entropically disfavored, so a small temperature raise drastically destabilizes this structure. In very diluted systems, interchain interactions are scarce, so the disappearance of Structure C releases the chains, keeping the helical configuration of each independent chain (i.e. Structure A).

If the system concentration is moderate or high, our model produces a high popula-tion of Structure D, thanks to the formation of interchain bonds. Its entropic cost, although lower than Structure C's, is still high. For this reason, Structure D also becomes unstable at relatively low temperatures. This way, the abnormal structures' interval ends at temperatures that, being too low, could be easily ignored in a numerical study of the simulation model around the folding/unfolding temperature.

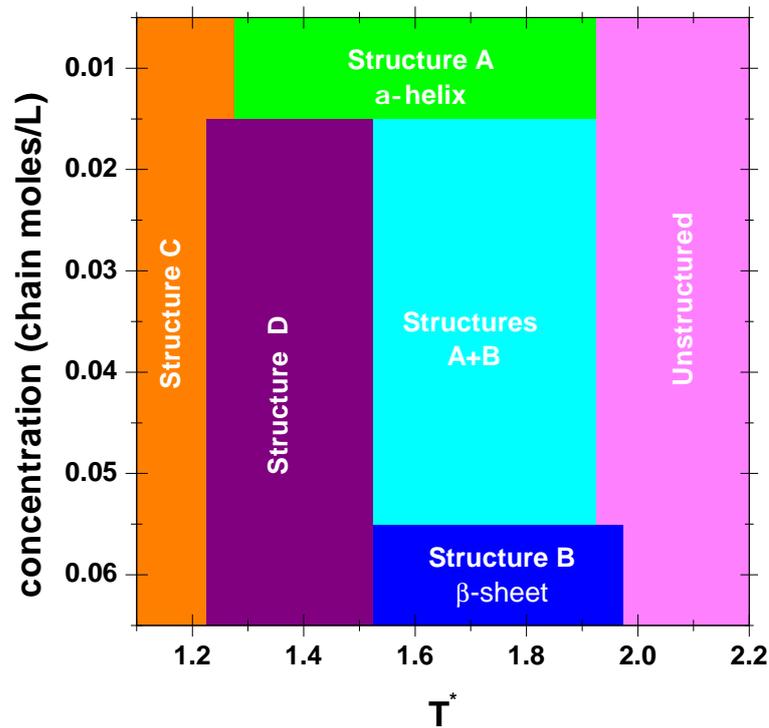

**Figure 3.21:** *Schematic phase diagram for multichain systems according to our simulation model.*

Within the relevant temperature range, the structural situation depends on the system concentration. If concentration is low, isolated $\alpha$-helices (Structure A) are the stable feature until their complete denaturalization. In highly concentrated systems, $\beta$-sheets (Structure B) are stable until unfolding, as crowded systems promote interchain hydrogen bonds. For moderate concentrations, Structures A and B coexist within the same temperature range.

## §3.4   Summary and conclusions of this Chapter

In this Chapter we have tackled the design of a hydrogen bond potential, as a first step for the final aim of this PhD Project: the investigation of the role of hydrogen bonds in the folding and aggregation of proteins and peptides. Bearing this in mind, we have defined a set of properties that had to be fulfilled: *(i)* use of an $\alpha$-carbon reduced representation of the chain to allow extensive simulations at a reasonable computational cost, *(ii)* obtention of secondary structure elements with accurate native-like geometries, *(iii)* reasonable thermodynamic behavior upon thermal folding and unfolding simulations, and *(iv)* sensitivity under different concentration conditions.

We started with a thorough investigation of many hydrogen bond potentials published in the literature, that have provided the necessary background to undertake this task. Particularly inspired by two of them,[78, 79] we have built a hydrogen bond potential where the observation of a small number of geometric requirements has to be fulfilled in order to apply a simple step-like potential.

Concerning the geometrical restrictions, we have taken a knowledge-based approach that consists in the statistical analysis of backbone hydrogen bonded pairs in native proteins. Thanks to it, we have settled the restriction themselves, as well as the acceptable ranges for each of them. They reflect the native tendencies of hydrogen bonds in relation to the distance between $\alpha$-carbons of bonded residues ($R1$) and the relative orientation between auxiliary vectors in the model ($R2$) and between the auxiliary vector of each residue and the hydrogen bond direction in the model ($R3$).

We have calculated these values for each pair of hydrogen bonded residues in a large set of protein structures,[145] extracting recognizable tendencies for both local and nonlocal bonds. In the case of terminal residues, as auxiliary vectors cannot be built, the hydrogen bond definition is laxer. Selecting the best interval for each restriction and case, we have chosen a range of values where each condition has to be fulfilled for the hydrogen bond of our model to be considered.

The simplicity of a step-like potential exhibits a high computational efficiency while reflecting the discrete nature of real hydrogen bonds. It has been adapted to the different situations that take place when considering this kind of interaction. In this way, the energetic stabilization depends on the fulfillment of the geometrical restrictions. Besides, it depends on the type of hydrogen bond interaction (either local or non local), reflecting their experimentally observed different stabilization, and is also tuned if terminal residues are involved, due to the higher structural variability we have observed.

The hydrogen bond potential design has been followed by an extensive investigation of the structural and thermodynamic properties that arise under folding/unfolding simulations of single chain and multichain systems. In this way, we have studied the helix-coil transition and the sheet-helix-coil one, depending on concentration.

Firstly, we have simulated the whole folding transition of a polypeptide system in infinite dilution conditions. We have obtained the helix-coil transition for different chain lengths, reproducing native-like energetic and structural properties for chains shorter than twenty residues, clearly above the average $\alpha$-helix length in globular proteins (twelve residues).[1]

We have also studied the effect of concentration. For that purpose, in this work we report the results of a system composed of five polypeptide chains of twelve residues each, whose concentration has been modified by changes in the size of the simulation box. In this case, the structural and energetic scenarios become more complex, as the number of interacting possibilities increases. Apart from detailed considerations for each individual system, a complete phase diagram has been obtained.

Four different types of structures have been observed. In the very low temperature region, where our system is nearly frozen, we observe two abnormal structures with more hydrogen bonds than those allowed by nature, being the result of a too intense energy minimization and the inherent (and always commented on) model limitations. The population of these structures, anyhow, is negligible at relevant temperatures, and therefore does not imply any serious pitfall of the model used in this work.

Thus, above this extremely low temperature region, we only observe native-like structures and their unfolding processes. The system behavior depends on its concentration. Modifying this parameter within sensible values, we have observed every expected relevant situation. Highly concentrated systems show the typical sheet-coil transition, whereas more diluted ones reflect a tough competition between the population of helices and sheets, the sheet-helix-coil transition. It is also remarkable that our model predicts a helical thermodynamic intermediate in the denaturalization of $\beta$-sheets when the sheet-helix-coil transition is observed.

Altogether, we can say that the use of a very simple representation of the polypeptide chain makes essentially impossible to design a hydrogen bond potential which only produces the natural secondary structures (i.e. found in real proteins) at any simulation condition. However, we have shown that our model is able indeed to clearly penalize other abnormal structures in the relevant temperature regime, and therefore they are only significantly populated at extremely low temperatures. This we have got with a definition of the interaction potential which is based on a careful choice of a proper set of geometrical restrictions, and their corresponding ranges.

In conclusion, we have provided a relevant insight in the simulation of peptide systems with a simple hydrogen bond potential, successfully dealing with the particular geometry of secondary structure elements, as well as the related thermodynamic properties. Now, we are prepared to face the study of complete proteins.

# 4

# Influence of hydrogen bonds in a structure-based potential

In the previous Chapter we have described the refined hydrogen bond model that is guiding us along this PhD Dissertation. As we have explained, its accuracy and simplicity lead to the obtention of secondary structures with a precise native-like geometry and a reasonable thermodynamic behavior in folding and aggregation studies of peptide systems.

We deal now with the effect of hydrogen bonds in the folding of complete proteins. As we discussed in Chapter 1, this task needs to combine the hydrogen bond potential (that can stabilize the secondary structure elements of the protein) with at least another one that simulates the rest of the interactions that take place in a protein.

We have employed in this Chapter a structure-based (or Gō) potential, i.e. one whose interactions depend on the native topology of each particular protein. We will start with a brief overview of this kind of potentials and its philosophy. As we shall expose, the lack of explicit information of each kind of interaction sometimes results in a quite innacurate description of the folding process. Therefore, the aim of this part of the Project is to adapt the Gō strategy so that hydrogen bonds are correctly modeled.

One of the critical aspects of this strategy is to find compatible potentials. Gō potentials build their interactions in terms of the tridimensional arrangement of the studied protein, as it is deposited in the PDB database.[88] For this reason, it is critical that the hydrogen bond potential also reproduces native secondary structure elements in an accurate way. If this property is not fulfilled (as it happens with most hydrogen bond potentials), this approach is incompatible from the beginning. Nevertheless, we have undertaken this task, encouraged by the promising properties of the hydrogen bond potential of Chapter 3.

In first place, we have evaluated the ability of the hydrogen bond potential to detect secondary structure elements in native proteins. As we have already stated, this is a fundamental aspect for the good performance of the combined potential in folding studies. Besides, it can be used as a detection method for hydrogen bonds from PDB structures, as we shall explain in Section 4.2.2.

After checking the suitability of the combined potential for a good description of proteins, we have carried out a number of simulations to investigate the properties of the model itself and to analyze the folding process of different proteins of interest. In Section 4.3.1, we will present a representative set of proteins whose folding has been investigated by thermodynamic means. We have firstly analyzed some proteins that have been widely studied both experimental and theoretically. They present different native structures and folding properties, and therefore they have let us obtain a general overview of our combined potential characteristics. In second term, we have focused on proteins where our previous experience shows that a simple Gō model like ours fails in their description (e.g. results depending on the experimental structure[157] or downhill folders predicted as two-state ones with Gō models).

Finally, we have analyzed the kinetic properties of some of the proteins that have been formerly studied on a thermodynamic basis. Thanks to them, we have tackled three main questions: can our kinetic Monte Carlo method properly describe the protein folding rates?; how does the change in the potential modify the kinetic properties of a given simulated system?; regarding the folding pathway of a protein, can our methodology (i.e. simulation method and applied potential) improve its description?

Let's start, then, with the introduction of structure-based potentials, focusing on their aims and characteristics. How are they defined? Why should hydrogen bonds be explicitly included?

## §4.1  Motivation: structure-based potentials

According to the folding funnel theory, the native structure is a minimum in the free energy surface.[25,27] Based on this principle, structure-based potentials simplify the protein folding landscape by assuming that the only relevant interactions are those present in the native structure, which implies that they completely drive the folding process. In the simplest form of these potentials, every interaction presents the same strength and functional form regardless its chemical nature.[26,62,65]

Apart from their strong theoretical base, related to the principle of minimal frustration,[30] structure-based potentials are computationally very efficient due to their simplicity. For that reason, they have been widely applied by the biophysical computational community in the last decades. Either on a lattice[158,159] or in the continuous space,[62,160] they have led to fruitful studies of the folding properties of proteins, such as the investigation of folding pathways[95,96] or the presence of on-pathway intermediates.[63,161]

Their appealing properties have also extended their use to other fields farther from the natural behavior of proteins, such as force probes[162,163] or aggregation.[81,99] In the former case, the use of structure-based models is widely accepted, as native interactions determine how the native structure is lost under pulling experiments. In the latter case, however, structure-based potentials present remarkable limitations due to their strong bias towards native arrangements. This proneness is crucial when considering additional stabilized structures such as aggregates, but also in the case of off-pathway intermediates.[103]

Even if none of these situations occur, a smooth folding funnel (like the ones that "traditional" structure-based potentials describe) seems oversimplified. The ruggedness of the funnel that one expects in a real system increases when taking into consideration the particular nature of the protein interactions, as well as the introduction of non-native terms into the original structure-based scheme. As a result, a kind of "second generation" Gō models that explicitly consider individual interactions have arisen.[102, 106, 164] The resulting potentials frequently enhance the accuracy of the original Gō models and provide a more precise tool for protein folding studies.

Among the individual interactions, we have focused on hydrogen bonds because of their particular properties, that we have highlighted several times along this PhD Dissertation. In this case, we are particularly concerned about their partially covalent nature, and how it makes hydrogen bonds stronger than other interactions (such as hydrophobics) and also directional, resulting in an extra difficulty for their accurate inclusion within structure-based potentials (as we have also remarked when considering this interaction on its own).

For that reason, seldom has this combination been reported. Instead, hydrogen bonds have sometimes been inserted as a directional modifier of the native interaction strength,[165] although this alternative misdescribes the directionality of hydrogen bonds in strands.[166] In other examples, more centers of interaction per amino acid for either hydrogen bond[80] or Gō interactions[81] are necessary, leading to more complex and time-consuming potentials.

In our case, we count on an accurate hydrogen bond potential based on an $\alpha$-carbon representation that has inspired us in the tackling of this task. How have we dealt with it?

## §4.2  Building the combined potential

Merging two potentials needs a deep understanding of each of them individually, as it is critical to know which properties must be maintained and which should be improved. In this Section, we describe the main technical details that we have considered in the design of the combined potential and that will help us in the discussion of the Results Section in this Chapter.

In the current case, we apply two potentials that have been designed in our research group and previously used independently, having therefore a deep understanding of both of them individually. On the side of the structure-based potential, it was initially optimized by Prieto *et al.* in 2005.[62,122] It has been successfully applied to many cases, displaying good agreement with thermodynamic experimental data and a remarkable co-operativity by itself.[63,98,136,157]

The hydrogen bond potential, already introduced in the previous Chapter, also presents acceptable thermodynamic properties, as well as an accurate representation of the native topology of secondary structure elements.[91]

Along the next paragraphs, we will center our attention in a short description of the structure-based potential and in the most suitable way to add the hydrogen bond contribution.

## 4.2.1 The Prieto *et al.* structure-based potential

As most structure-based potentials, it is based on a coarse-grained representation such as the one drawn in Figure 2.2(*c*), where only the $\alpha$-carbon positions (represented by their position vectors, $\mathbf{r}$) are used. The overlapping between two beads is forbidden by means of a hard sphere potential centered at 4.0 Å.

The particular characteristics of the Gō potential are related to two main aspects: the functional form of the potential and the contact definition.[62] In relation to the first one, Prieto *et al.* calculate the Gō (or structure-based) energy as a sum of terms, where the contiguous pairs have been excluded:

$$E^{go} = \sum_{i=1}^{N-2} \sum_{j=i+2}^{N} e_{i,j}^{go} \tag{4.1}$$

The individual interaction, $e_{i,j}^{go}$, is described by a parabola centered at the native distance, $d_{i,j}^{nat}$, and truncated at $\pm a$, according to the following functional form (also outlined in Figure 4.1):

$$e_{i,j}^{go} = \begin{cases} \delta_{i,j}^{go} \left[ \left( |\mathbf{r}_{i,j}| - d_{i,j}^{nat} \right)^2 - a^2 \right] / a^2 & \text{if} \quad d_{i,j}^{nat} - a < |\mathbf{r}_{i,j}| < d_{i,j}^{nat} + a \\ 0 & \text{otherwise} \end{cases} \tag{4.2}$$

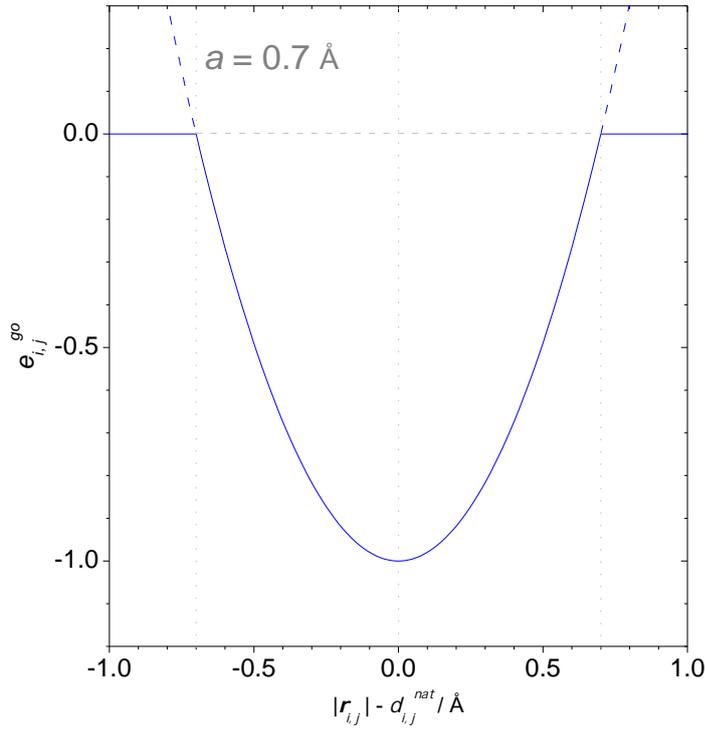

**Figure 4.1:** *Structure-based interaction potential according to the Prieto et al. definition. The dashed blue lines indicate the shape of the untruncated parabola.*

As it is shown in Figure 4.1, $a = 0.7$ Å. This value has been optimized to accurately reproduce the main thermodynamic properties of folding.[62, 122] In this way, the Prieto *et al.* potential presents a simple and computationally efficient definition whose energetic minimum is defined in the native configuration.

The value of $\delta_{i,j}$ tells us if the considered pairs are in contact in the native state or not, which gives an energy $e_{i,j}^{go} = -1.00$ (in reduced energy units) at the well minimum for all the individual contributions. It is defined as a Kroneker's delta function:

$$\delta_{i,j}^{go} = \begin{cases} 1 & \text{if } i \text{ and } j \text{ form a native contact} \\ 0 & \text{otherwise} \end{cases} \tag{4.3}$$

These individual contributions form a a certain *native contact map* for each protein, as we have exemplified in Figure 4.2. But, how do we define a *native contact*? This potential considers two types of interactions: short range and long range ones. In first place, local interactions are those between residues that are near in the sequence, where "near" refers either to $(i, i+2)$ interactions (also known as *virtual angle interactions*) or $(i, i+3)$ ones (called *torsional interactions*). They reflect the chain stiffness and, for that reason, are always present.[136]

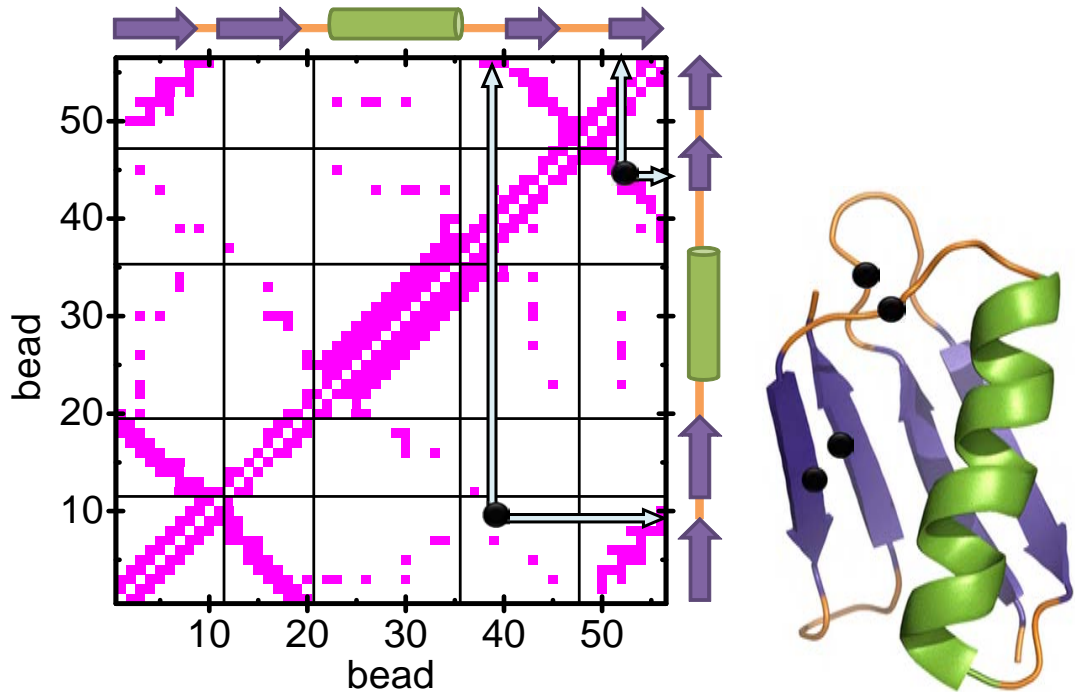

**Figure 4.2:** *Contact map of the B domain of protein G (also drawn in the Figure), according to the Prieto et al. definition of native contact. As an example, we have marked two long range contacts with black circles.*

Long range (or tertiary) interactions between residues $i$ and $j$, on their side, are only observed if these residues are in contact in the native state, i.e. at a distance shorter than $d_{i,j}^{cut}$. To compute the distance between two residues, the full atomic detail of the protein is needed, as it is stored in the PDB.[88] In this way, we calculate the distance between any pair of heavy atoms of the interacting residues and select its minimum, $\min[d_{i,j}]$; a pair of residues are said to be in contact in the native state if $\min[d_{i,j}] \leq d_{i,j}^{cut} = 4.5$ Å.

Our combined potential is also characterized by an accurate description of hydrogen bonds, according to the potential we have described in the previous Chapter. The mixing of the two potentials must be done in a careful way to avoid the overstabilization of those contacts that, being present in the native state (and, thus, affected by the structure-based interaction), also form a hydrogen bond in this native state (and would have an additional hydrogen bond energetic contribution). How can we detect native hydrogen bond contacts? Are hydrogen bonds and structure-based interactions equally stabilizing?

## 4.2.2 Merging hydrogen bonds and the structure-based potentials. Detection of native hydrogen bonds.

In our combined potential definition, native hydrogen bonds need to be suitably considered (i.e. removed from the Gō calculation and just stabilized by the hydrogen bond interaction), while some frustration is allowed by the possibility of forming non native hydrogen bonds between any pair of residues.

Our first task is, then, to detect the native contacts that are forming a hydrogen bond in a PDB structure. The detection of hydrogen bonds is an aim on its own, as their explicit identification is very helpful for structure and function considerations and also marks secondary structure regions. This information is indirectly provided by the PDB headers through the secondary structure experimental assignments, but the precise hydrogen-bonded pairs are not included.

In addition, there are many tools that aim to identify these hydrogen bonded pairs through the analysis of a given structure with atomic detail (such as the PDB ones). Bioinformatic tools such as DSSP[167] or STRIDE[168] have traditionally been used for this purpose, as well as more recent algorithms, frequently inserted in visualization software such as VMD[144] or PyMOL.[89]

In Figure 4.3 we show the percentage of miscorrectly identified hydrogen bonds using some of the most common computational tools. As a reference, we have taken the indirect definition of hydrogen bonds that comes in the PDB file headers, and whose number of hydrogen bonds (approximately identified by the secondary structure definition) can be listed by any visualization program (do not confuse this listing with the identification algorithm itself). In this plot, a positive value refers to the hydrogen bonds that have been identified by a given method, but have not been experimentally detected; a negative value corresponds to the percentage of real hydrogen bonds that has been missed in the identification step.

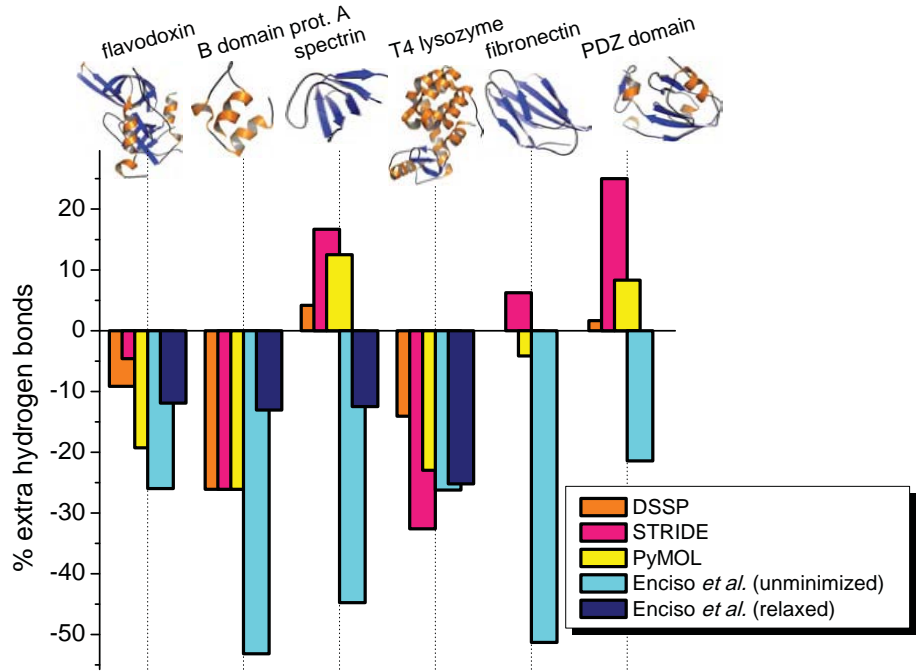

**Figure 4.3:** *Identification of hydrogen bonds in a set of proteins (relative to the experimentally observed number, as it is assigned through the secondary structure regions of the PDB header of the different proteins), according to different identification methods. Cartoons have been drawn with PyMOL[89] and colored by their secondary structure elements (orange for α-helices and blue for β-sheets).*

We have chosen eight proteins that, having different structural motifs and contents of secondary structure (also shown in the Figure), are taken as representative examples. In spite of taking advantage of the atomic detail of PDB structures, we have found that all these methods present some differences in relation to the experimental number of hydrogen bonds. In some cases (like T4 lysozyme or the B domain of protein A, for instance), all of them fail in the identification of many real native hydrogen bonds (up to 30%), while in other examples (like spectrin or the PDZ domain) they detect hydrogen bonds that have not been observed experimentally. For us, the latter case is particularly undesirable, as any pair of residues can form a hydrogen bond during folding in our model, even if they are not hydrogen bonded in the native state, but false positives remove some native structural information that cannot be recovered during the simulation.

For those reasons, we wondered if we could identify real hydrogen bonds using the definition of our hydrogen bond potential. Regarding its geometrical restrictions (see Figures 3.8 and 3.9), nearly 80% of natural hydrogen bonds should lie in the acceptable ranges. Regrettably, we found that this approach also leads to some innacuracies between our results and the protein PDB headers (see the cyan bars in Figure 4.3), reaching an unacceptable deviation of up to 50% in the worst cases (although false positives have never been observed).

But having a complete potential instead of just a detection method prompted us to improve these results by letting the proteins relax under the effect of both the structure-based and the hydrogen bond potentials. As a result, we have obtained relaxed structures where all or nearly all the native hydrogen bonds are identified (see the dark blue bars in Figure 4.3) and without additional hydrogen bonds. It is remarkable that our results are comparable to the other common techniques, but just using the $\alpha$-carbon trace of the protein for the identification instead of the full-atom information that other methods need.

We have also checked that the minimized structures are fully native, showing no remarkable distortions (their root mean square deviation compared to the original PDB structure is less than 0.5 Å, computed from $\alpha$-carbons). Therefore, we have taken these relaxed structures as reference, as they fulfill our major requirements: reasonable matching between identified number of native hydrogen bonds and the original ones, absence of extra hydrogen bonds and nearly identical initial and relaxed structures.

Once all the native hydrogen bonds in a protein have been detected, they are removed from the Gō potential definition, resulting in a decrease in the number of Gō contacts (compared to the pure Gō potential) when the combined potential is built. The removal of these contacts needs special care in the case of helical hydrogen bonds. According to their representation, hydrogen bonds are nominally assigned to the $\alpha$-carbon positions instead of the real amino and carbonyl ones. As we have already discussed in Chapter 3, this requires a "renumbering" of the interacting residues because the real backbone hydrogen bond between the $i$th and the $j$th residues is replaced by the interaction between the $i$th and the $(j-1)$th beads in our model. Thus, helical $(i, i+3)$ hydrogen bonds imply the removal of $(i, i+4)$ Gō interactions. As an example, we show in Figure 4.4 the redrawn Gō contact map we previously exhibited in Figure 4.2. Pink squares correspond to the native contacts that will be considered through the structure-based part of the combined potential; blue squares are Gō interactions that, forming a native hydrogen bond, have been removed from the structure-based calculation.

The next step in the construction of our combined potential is the selection of the optimal relative weight of structure-based and hydrogen bond interactions. According to our definition, the formation of hydrogen bonds is ruled by the hydrogen bond potential; concerning the structure-based potential, it mainly represents the interactions that stabilize the tertiary structure, which principally have a hydrophobic nature. In this way, it seems plausible to assign a more stabilizing role to hydrogen bond interactions, as they are commonly thought to be stronger than hydrophobics.[36]

For us, the relative weighting of the interactions has been taken as a parameter of our model, establishing its optimal value by testing a full set of them in our simulations. The general expression is the following:

$$E = \omega^{hb} E^{hb} + \omega^{go} E^{go} \qquad (4.4)$$

Keeping $\omega^{go} = 1.0$ as a reference, we have tried different values of the hydrogen bond weighting factor, $\omega^{hb} = [0.5, 4.0]$. To choose the best one, we have performed folding simulations for three proteins: the B domain of protein G, the BBL protein and fibronectin. From now on, we will refer to each of them by its PDB code, which concerns to its structure as it is deposited in the PDB. In this way, we will take the X-ray resolved structure of fibronectin (of PDB code 1TEN[169]) and the NMR resolved structures of the G domain and the BBL protein, denoted as 2GB1[170] and 1BBL,[171] respectively.

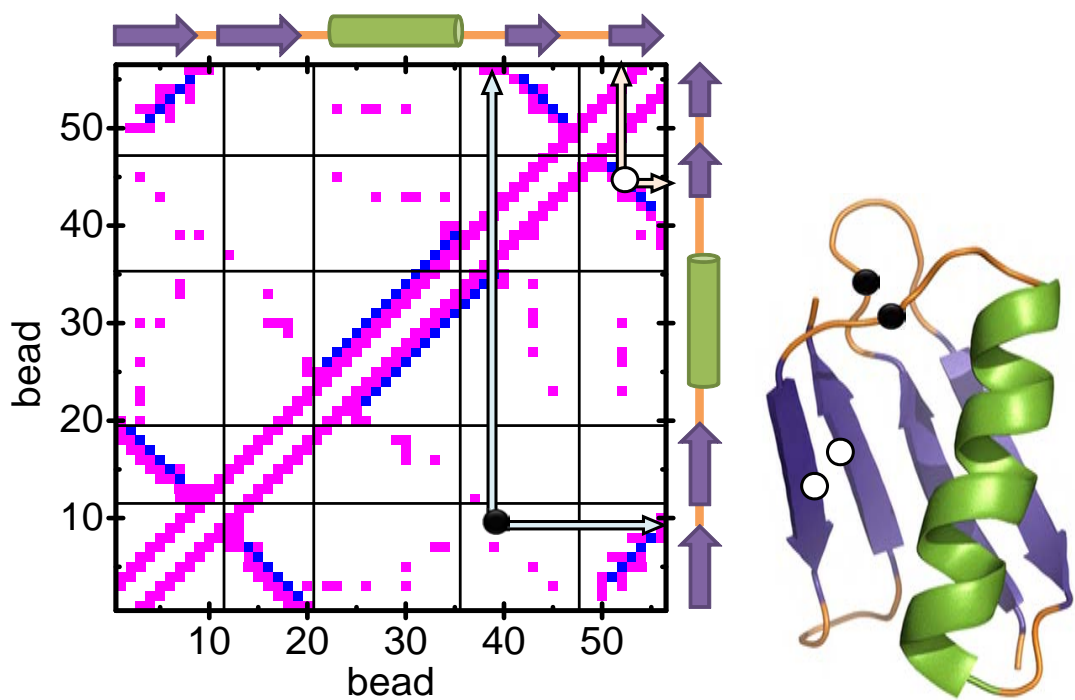

**Figure 4.4:** *Contact map of the B domain of protein G (also drawn in the Figure) for the Gō part of our combined potential (in pink) where the native hydrogen bonds (removed from the structure-based calculation) have been colored in blue. Regarding the two long range contacts we marked in Figure 4.2, one of them is still maintained in the combined potential (black circle), while the other one has been removed because it belongs to a hydrogen bond in the β-sheet (white circle).*

These proteins have been selected due to their different structural motifs and well-established folding properties, both experimentally and in computer simulations (some of them already published using our structure-based potential[62,98]). In this way, 2GB1 is an $\alpha + \beta$ protein that shows a typical two-state folding,[172,173] 1TEN is an all-$\beta$ protein that also shares the same kind of folding process[174,175] and 1BBL is a small all-$\alpha$ protein that shows a downhill folding transition.[176] To clarify the discussion, we will just show the results of 2GB1, keeping in mind that similar conclusions have been obtained from the other two cases.

We show in Figure 4.5 the heat capacity curve versus temperature (both in reduced units) for 2GB1 using different weighting factors (see legend for details). The original structure-based potential ($\omega^{hb} = 0.0$), that will be used as reference, is marked with a black line. It corresponds to a two-state transition, shown by a relatively narrow and high peak in the heat capacity plot.[62,157]

The increase in the hydrogen bond weighting factor shifts the transition towards higher temperatures, as it modifies the system energy. More interestingly, it also changes the peak properties: it gets increasingly narrower (i.e. more two-state-like) until $\omega^{hb} = 2.0$ and after that it gets blunter and more asymmetric, ending in a kind of high temperature plateau for $\omega^{hb} = 4.0$. That tells us that the folding transition, as described by our simulations, is modified by the effect of hydrogen bonds.

Thanks to the structural analysis of our results, we have linked these peak characteristics to the role of the hydrogen bonds in each simulation. This can be illustrated by the histogram of the number of hydrogen bonds in each configuration, computed from all the snapshots registered at a given temperature. It is shown in Figure 4.6, where we have taken the histograms at two different temperatures (a low one, where the folded structure is stable, and the transition one) for three sets of parameters.

If the weighting factor is low ($\omega^{hb} < 2.0$), such as in Figure 4.6($a$), the stable structure at low temperatures nearly finds the total number of native hydrogen bonds for this protein, which is 34. However, some of them are lost before the melting temperature and even the folded structure at this temperature (represented by the peak at a higher number of hydrogen bonds) presents a wide variability in the number of hydrogen bond contacts, that is also reflected by a number of configurations with an intermediate number of hydrogen bonds. This means that the hydrogen bond weight is too low to have a significant effect in the folding process and some native hydrogen bonds are already lost at low temperatures.

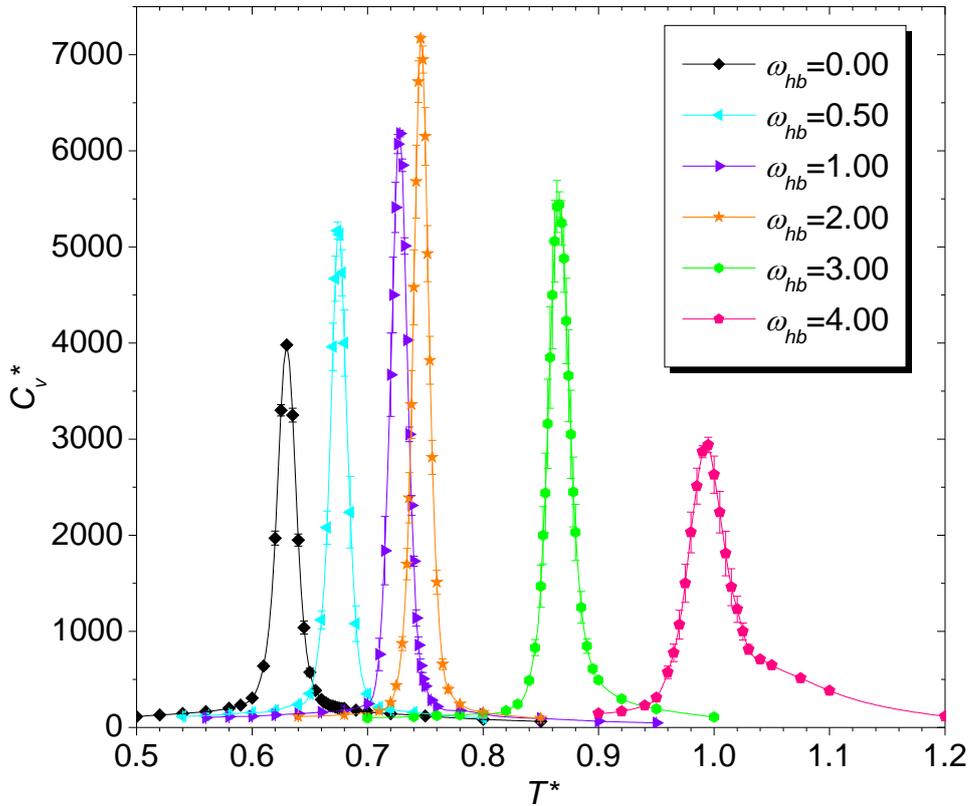

**Figure 4.5:** *Heat capacity curves vs temperature in infinite-dilution conditions for the protein 2GB1 using different weighting factors for the hydrogen bond term, as shown in the legend.*

However, if the hydrogen bond weight is too high ($\omega^{hb} > 2.0$), such as in Figure 4.6(*c*), the stable structure at low temperatures is better defined (look at the narrower peak in the histogram) but the evolution of the number of hydrogen bonds at the transition temperature is also unsatisfactory. In this case, the unfolded structure still maintains around 20 hydrogen bonds (native and non native), leading again to a continuous loss of hydrogen bonds, incompatible to the desired all-or-none transition typical of two-state

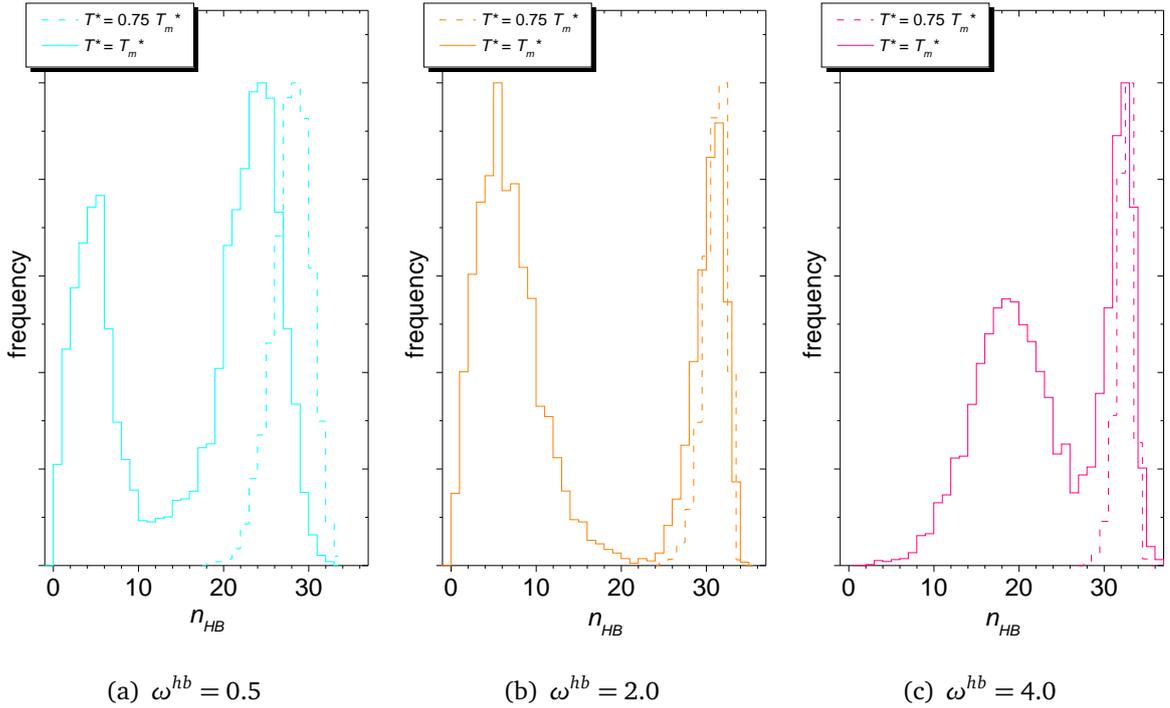

(a) $\omega^{hb} = 0.5$      (b) $\omega^{hb} = 2.0$      (c) $\omega^{hb} = 4.0$

**Figure 4.6:** *Histograms of the number of hydrogen bond contacts, $n_{HB}$, found at two different temperatures (a low one, $T^* = 0.75T_m^*$, and the transition one, $T^* = T_m^*$) for three hydrogen bond weighting factors.*

folders like 2GB1.

In the case of 1BBL, where the proportion of secondary structure is low and mainly helical, the increase in the hydrogen bond weighting factor does not only worsen the transition itself, but also creates an alternative energetic minimum where the protein forms a single $\alpha$-helix using all the protein residues (data not shown). This indicates that an excessive weight makes hydrogen bonds rule the folding process, ignoring the native topology of the protein at the transition temperature and, in some cases, in the low temperature region as well.

It seems clear, then, that the best weight will be reached when structure-based interactions and hydrogen bonds are balanced, so that the native structure is well defined even at the transition temperature and the hydrogen bond energy contributes to its stabi-lization, instead of being lost either at lower or higher temperatures.

In our case, this situation is found when $\omega^{hb} = 2.0$, shown in Figure 4.6(*b*). At low temperatures, we find a single narrow peak, which corresponds to a well defined native state ($RMSD < 0.9$ Å). At the transition temperature, the histogram of the number of hydrogen bonds exhibits a clean bimodal distribution, where the maximum at a higher number of hydrogen bond contacts perfectly matches the low temperature peak (indicating the nativeness of the structure) and the one at a low number of hydrogen bonds (unfolded state) presents its maximum value below 10 hydrogen bonds, which corresponds to sporadic hydrogen bond contacts. This situation increases the cooperativity of the transition in relation to the original structure-based potential, as it can be observed in the heat capacity curves of Figure 4.5 by the higher and narrower peak of the chosen weight.

Similar conclusions have been derived from the analysis of the folding of 1TEN and 1BBL (data not shown), so we can conclude that $\omega^{hb} = 2.0$ is the most appropriate value for the relative stabilization of native interactions in relation to hydrogen bonds. Thus, we are now prepared to start the folding studies that will conform the following Section.

## §4.3  Results

As we have explained, we have built a combined potential that aims to improve the description of protein folding of a pure Gō model by the accurate treatment of hydrogen bonds. In this part of the Project, we have combined the thermodynamic approach (similarly to Chapter 3) with additional kinetic analysis.

### 4.3.1  Thermodynamics of folding processes

We have evaluated the impact of an explicit consideration of hydrogen bonds in combination with classical Gō models in the thermodynamic properties of folding. We aim to know if this approach improves aspects like cooperativity and free energy barriers along the folding process. Furthermore, we want to estimate if this effect is a kind of *a priori* bias or provides a correct modulation of the free energy landscape of the transition. For that purpose, we have focused in a representative set of eight globular proteins, where each of them exemplifies a particular aspect we want to tackle in this study.

It is also important to mention that we have mainly focused in proteins that had been thoroughly investigated by experimental techniques, particularly their thermal folding and unfolding. This has prompted us to use real units instead of the reduced ones (for the thermodynamic properties) in order to obtain semi-quantitative results and compare them with experimental values when possible. Thus, we have correlated the experimental denaturation temperature of each protein to our transition temperature (computed as the maximum of the calculated heat capacity curve). Once the temperature is re-scaled through Equation (2.6), the energy conversion is straightforward, as explained in Equation (2.7).

Let's start, then, with the first question: do hydrogen bonds have any effect on folding for a single coarse-grained model like ours? In order to check this, we have started with a more detailed study of 2GB1 (that we previously introduced when optimizing the hydrogen bond weighting factor). The B domain of protein G is a typical paradigm of a two-state folder that has been carefully studied not only in experiments,[172,173,177] but also in computer simulation studies.[178–180]

In Figure 4.7, we show a cartoon representation of 2GB1 (colored in orange and blue according to the presence of $\alpha$-helices and $\beta$-sheets, respectively), as well as its heat capacity curve versus temperature (Figure 4.7($a$)) and free energy profile versus internal energy ($E = E^{hb} + E^{go}$) at the transition temperature (Figure 4.7($b$)). As we have already stated, sharp maxima in heat capacity curves usually indicate cooperative foldings, while free energy profiles provide the energy distributions of the native and denatured states at a given temperature (in this case, the transition one), as well as the related free energy barrier. Regarding the semi-quantitativeness of our results, two-state folders have free energy barriers greater than 5 kJ·mol$^{-1}$;[176] smaller barriers lie below or very close to thermal fluctuations and indicate a downhill folding.

In each graph of Figure 4.7, the black curve represents the results computed using

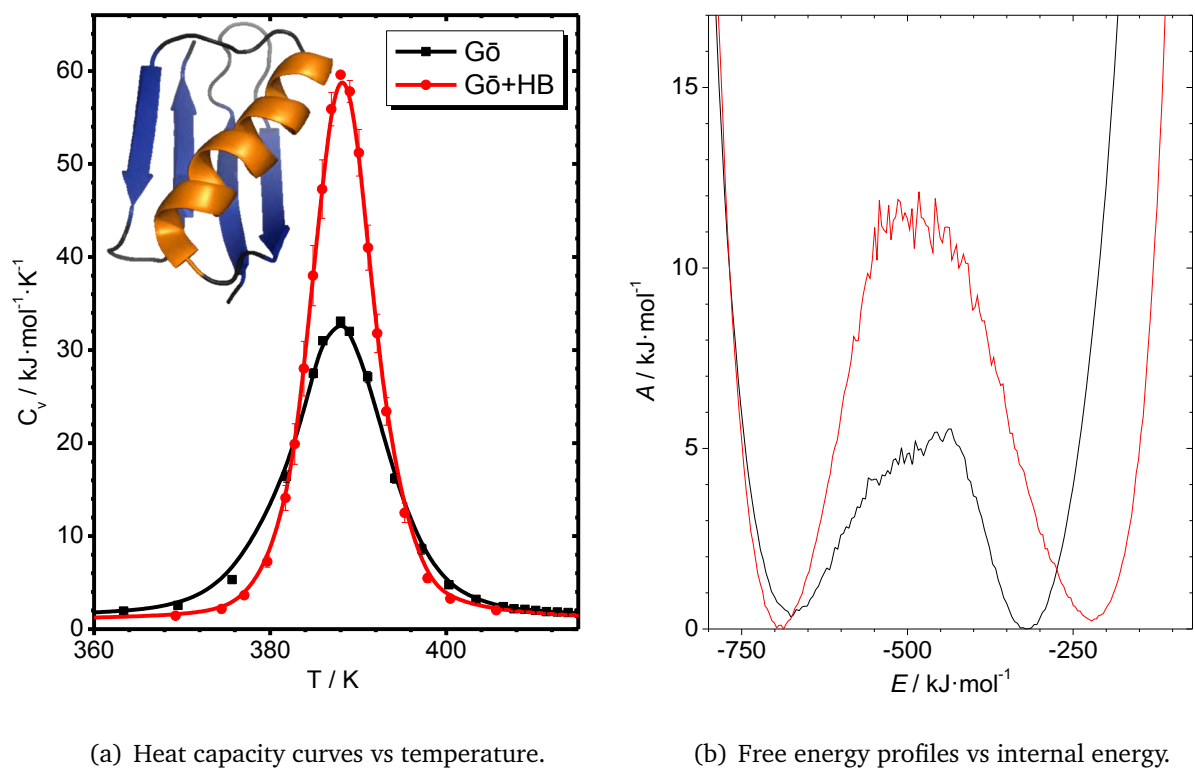

(a) Heat capacity curves vs temperature.

(b) Free energy profiles vs internal energy.

**Figure 4.7:** *Thermodynamic results of 2GB1 for the plain Gō model (black) and the Gō model with hydrogen bonds (red).*

the pure Gō potential, used as reference, while the red one comes from the combination of Gō and hydrogen bond interactions introduced in this work. As we can see in Figure 4.7(*a*), both heat capacity curves present a sharp peak, typical of cooperative transitions such as the 2GB1 one.

The effect of the hydrogen bond interaction is observed as an enhancement in the transition cooperativity, also increasing the free energy barrier, as shown in Figure 4.7(*b*). This latter plot lets us connect these changes to a better definition of the native state (mediated by the specific definition of hydrogen bonds in the native structure) merged with a destabilization of the denatured one, where the removal of the Gō contacts that form hydrogen bonds reduces the residual native structure in the denatured state. Therefore, this first example reflects the kind of effect that an explicit description of hydrogen bonds has in the folding process, mainly connected to a better description of the native state and some

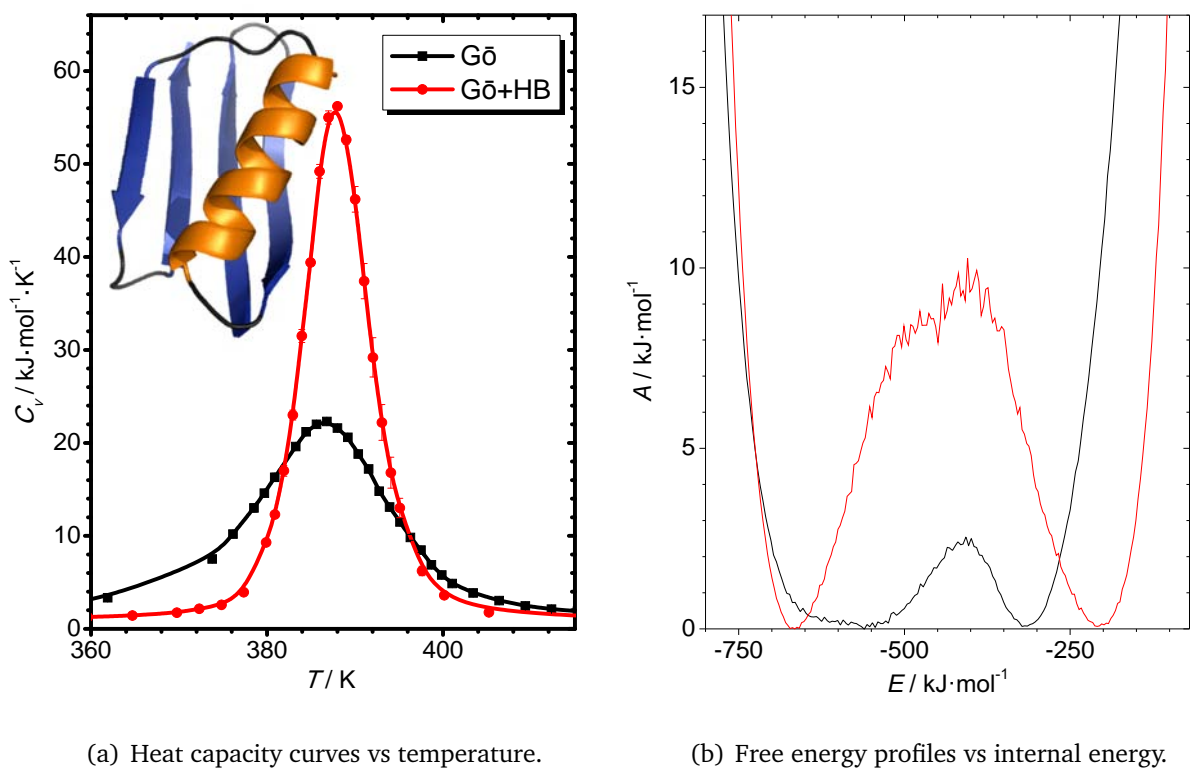

(a) Heat capacity curves vs temperature.

(b) Free energy profiles vs internal energy.

**Figure 4.8:** *Thermodynamic results of 1PGB for the plain Gō model (black) and the Gō model with hydrogen bonds (red).*

modifications in the energetics of the denatured state. We have observed a desirable tune of the folding characteristics that has inspired us to continue with this thermodynamic study and to investigate to what extent hydrogen bonds may improve the results of the original model.

Gō models strongly depend on the specific experimental structure, leading to different results for the same protein depending on the experimental structure considered, when more than one is available for the same protein.[64,98] Encouraged by the good results on 2GB1, we can raise our second question: can hydrogen bonds correct this logical but undesirable feature of plain Gō models? The case of the B domain of protein G is particularly exemplifying, as it has been reported that the previously studied NMR structure (2GB1) leads to different results compared with the X-ray one (1PGB[181]);[98] it can be observed by comparing the thermodynamic data in Figure 4.7 (2GB1) and Figure 4.8 (1PGB).

If we start our discussion with the black curves, we can clearly appreciate the differences between 2GB1 and 1PGB, where the latter one presents an incorrect barrierless folding (free energy barrier less than 3 kJ·mol$^{-1}$) and a poorly defined native state at the transition temperature. The inclusion of hydrogen bonds (red curves) has a clear effect, sharpening the description of the energetics of the native state. As a result, we recover for 1PGB an almost identical transition to the 2GB1 case, proving that the explicit consideration of hydrogen bonds provides fully comparable results for this protein, independently on the experimental structure used for the definition of the native interactions.

After these results, we could think that the explicit consideration of hydrogen bonds in the model always leads to an increase in the free energy barrier, independently of the protein itself. As this statement has been suggested for all-$\alpha$ proteins,[80] this will be our following question: do hydrogen bonds insert a sort of systematic bias in the free energy surface?

To find it out, we have analyzed the folding process for some all-$\alpha$ proteins. As an example, we show in Figure 4.9 the amino terminal domain of phage 434 repressor (PDB code 1R69[182, 183]), an all-$\alpha$ protein that presents a cooperative folding.[184] As we shall explain along the following paragraphs, differences between the former model and the new one are rather subtle, but meaningful.

In both cases we obtain a sharp heat capacity peak, distinctive of two-state folders. In the free energy curves, the combined potential shows a slight reduction in the barrier that refutes the hypothesis of a systematic barrier increase. There are no changes in the native basin between both models but the denatured state presents the already observed shift towards higher energies when hydrogen bonds are considered, linked to the removal of some Gō contacts in our combined potential. As it is accompanied by a widening of this basin, the net effect is a slight reduction in the free energy barrier, as already mentioned.

The broadening of the denatured basin has a structural source: it is due to the presence of some residual helical structure that, interestingly, has been claimed to be essential for the folding of this protein.[184] In Figure 4.10 we show the frequency of hydrogen bonds

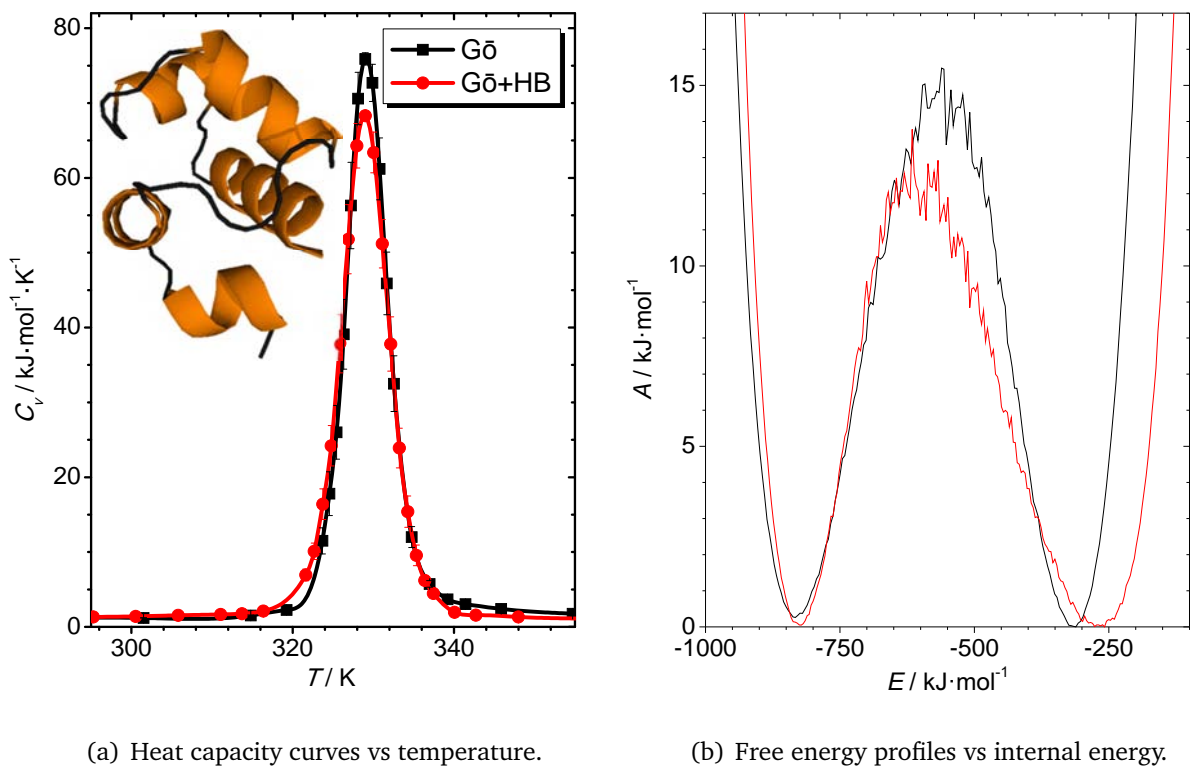

(a) Heat capacity curves vs temperature.      (b) Free energy profiles vs internal energy.

**Figure 4.9:** *Thermodynamic results of 1R69 for the plain Gō model (black) and the Gō model with hydrogen bonds (red).*

that we have detected in our simulations for the denatured state of 1R69 at the transition temperature. It presents a low proportion of helical contacts that are not circumscribed to those present in the native state.

As a matter of fact, a remaining helical content has been stated to be a general feature of the denatured state of helical proteins.[185] The lack of an explicit consideration of hydrogen bonds in Gō models results in a vague representation of helicity in the denatured state. Our approach surmounts this lack of specificity, dealing with these interactions in a more rigorous way.

As an additional example, we show in Figure 4.11 the engrailed homeodomain protein (PDB code 1ENH[186, 187]). Numerous experimental and computational studies state that 1ENH is a two-state folder that nevertheless presents a high folding rate. In this protein, the residual structure is also thought to play an important role in the folding transition.[188] The pure Gō results show a relatively low heat capacity peak and a free energy barrier of approximately 5 kJ·mol$^{-1}$, edging the downhill limit. Adding hydrogen bonds leads to a small increase in the folding cooperativity that can also be observed by the growth of the free energy barrier up to 7 kJ·mol$^{-1}$, a typical value for two-state folders. The widening in the denatured state with the combined model (also observed here) is partly counterbalanced in this case by a better defined native state. In conclusion, the introduction of hydrogen bonds in this kind of proteins implies the retention of some residual helical structure, providing a general broadening of the denatured basin but without forcing any systematic bias on the thermodynamic folding properties, that depend on the intrinsic characteristics of the analyzed protein.

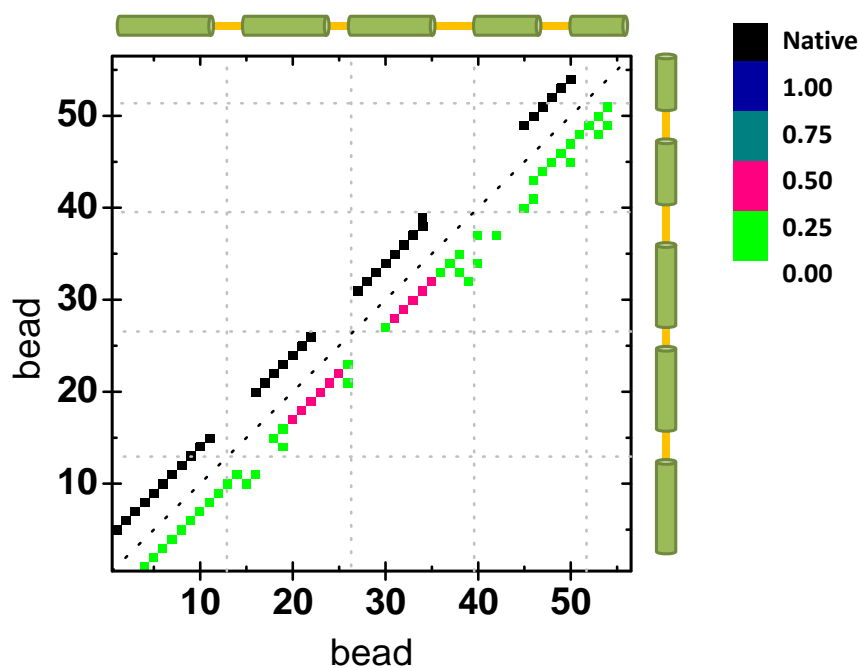

**Figure 4.10:** *Map of hydrogen bond frequency in the denatured state of 1R69 at the transition state. The upper triangle shows the hydrogen bonds that are detected in the native state. The lower one shows the frequency of appearance of hydrogen bonds in the denatured state of the protein, according to the legend. Grey dotted lines frame secondary structure elements.*

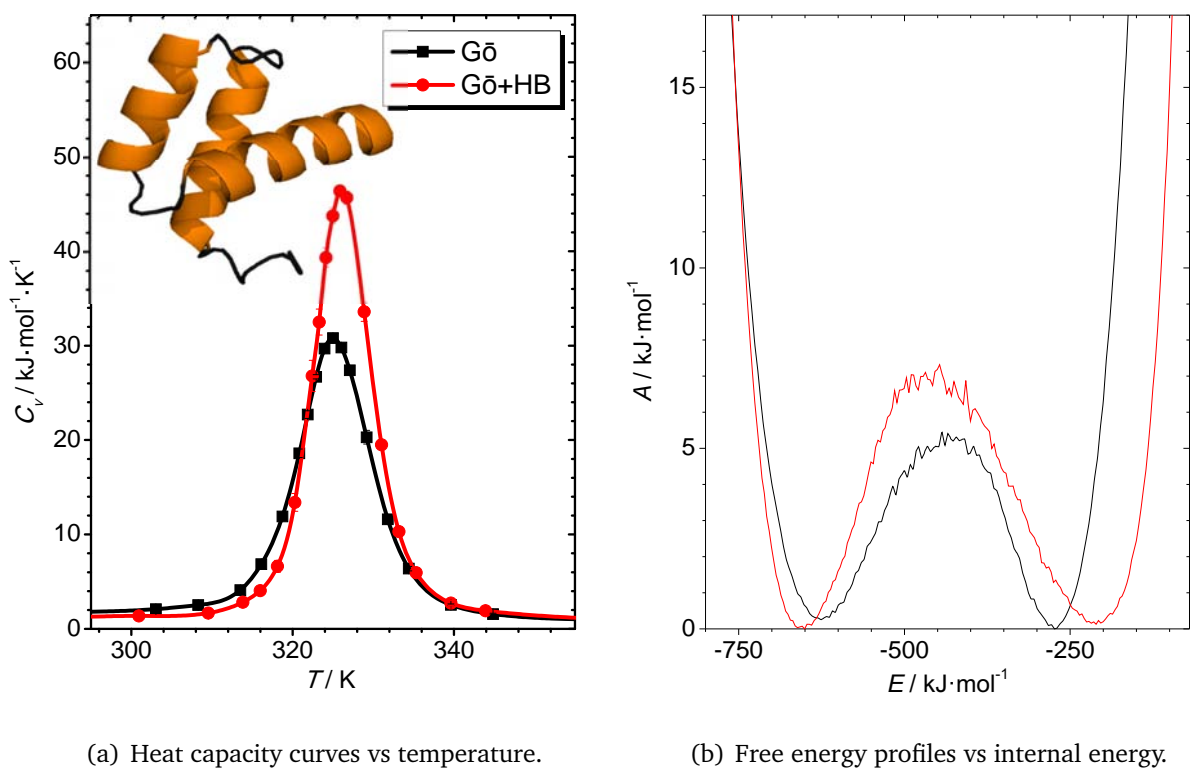

(a) Heat capacity curves vs temperature.

(b) Free energy profiles vs internal energy.

**Figure 4.11:** *Thermodynamic results of 1ENH for the plain Gō model (black) and the Gō model with hydrogen bonds (red).*

Once the impact of hydrogen bonds in helical proteins is correctly understood, how do hydrogen bonds affect β-proteins? In order to answer this question, we have started with the study of the fibronectin type III domain from human tenascin (PDB code 1TEN[169, 189]), a typical example of two-state folder whose folding pathway has been thoroughly studied.[174, 175] This protein is also familiar to us thanks to the preliminary simulations that we performed in Section 4.2.2 to determine the optimal hydrogen bond weighting factor.

Our results are shown in Figure 4.12 and only reveal minor changes when hydrogen bonds are applied. The free energy profile indicates that the native state is slightly better defined (narrower), while the denatured one shows no broadening, evidencing that hydrogen bonds do not induce any particular residual structure in this case (hydrogen bond map not shown).

Within β-proteins, we have also performed numerical simulations on the SH3 do-

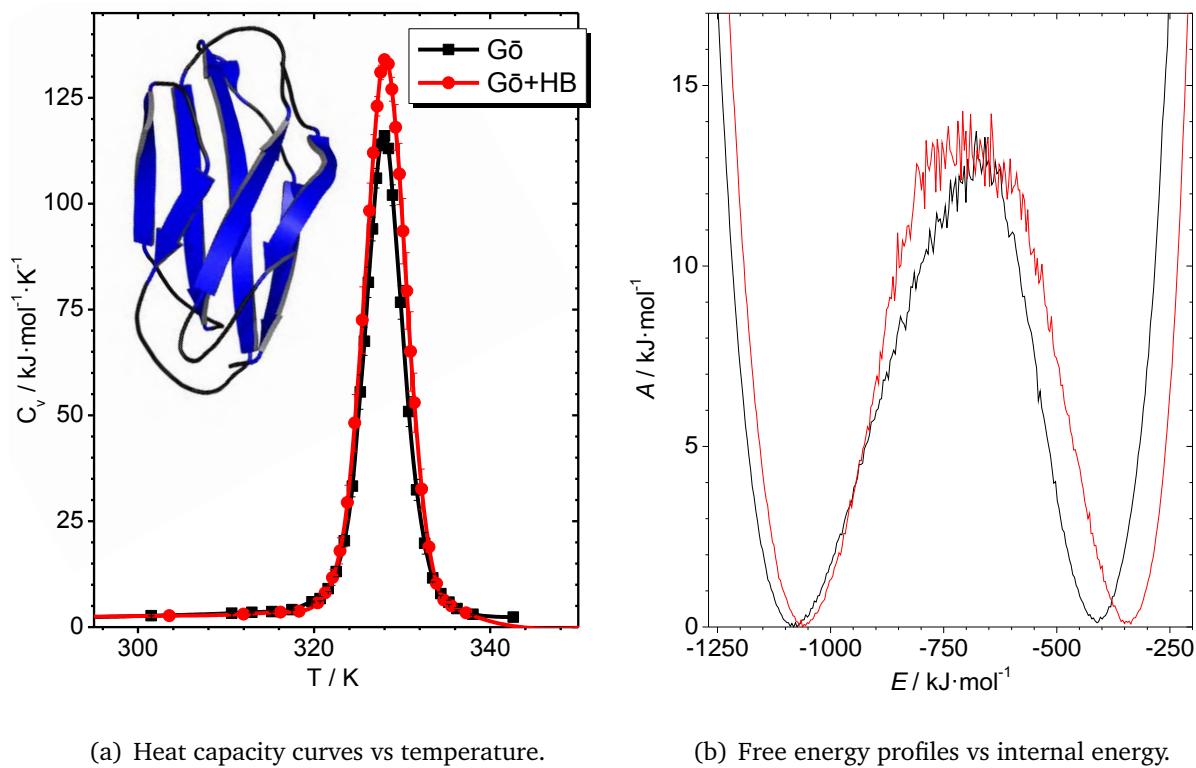

(a) Heat capacity curves vs temperature.

(b) Free energy profiles vs internal energy.

**Figure 4.12:** *Thermodynamic results of 1TEN for the plain Gō model (black) and the Gō model with hydrogen bonds (red).*

main (PDB code 1NLO[190]), due to its biological relevance in aggregation studies and its well-established properties (two-state folder with a relatively high folding rate, comparable to helical proteins[191]). Our results, plotted in Figure 4.13, show that the heat capacity curves are typical from two-state folders, either including or omitting hydrogen bonds. However, the free energy barrier (tightly linked to the folding rate) gets considerably lower if hydrogen bonds are considered.

The origin of this change lies in the presence of a very broad denatured state, where many configurations within this basin share some residual hydrogen bond interactions. The hydrogen bond frequency map, shown in Figure 4.14, mostly locates them in the distal loop of the protein (residues 39–44). Interestingly, this region is part of the structure of the transition state of this protein and is thought to be experimentally stabilized by a hydrogen bond interaction.[192] Moreover, it has been found that the distal loop plays a de-

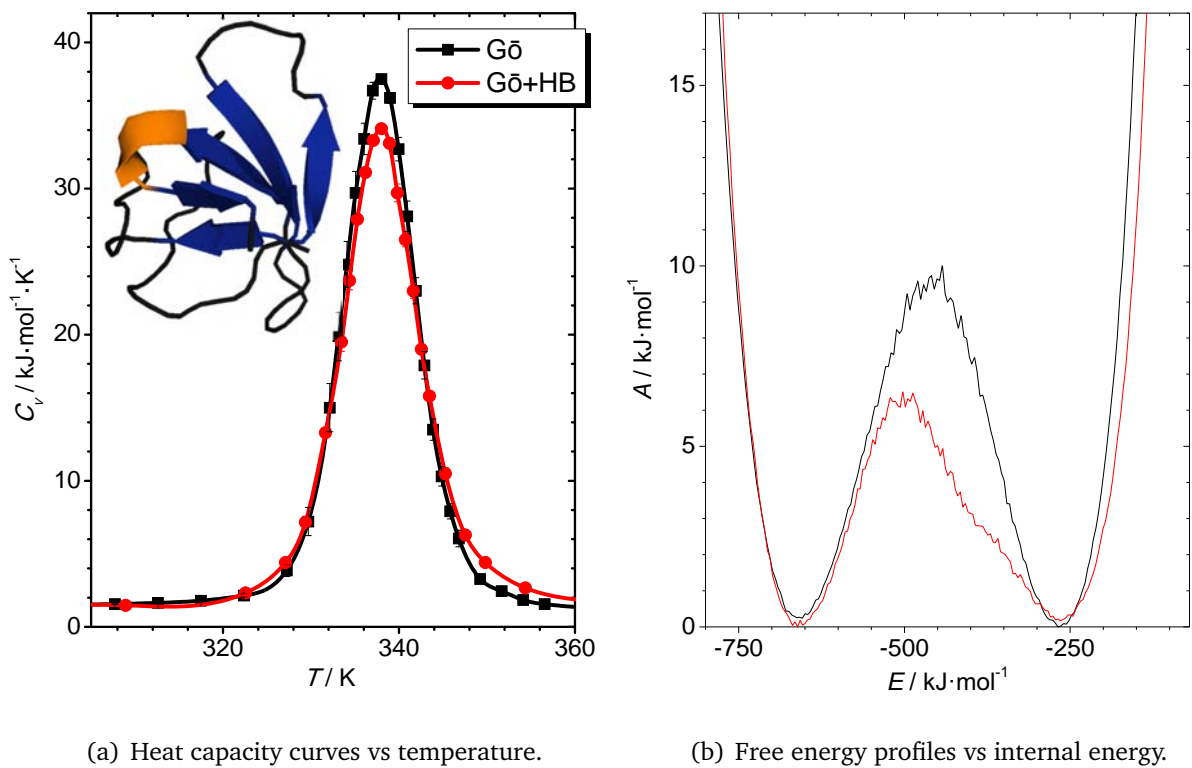

(a) Heat capacity curves vs temperature.

(b) Free energy profiles vs internal energy.

**Figure 4.13:** *Thermodynamic results of 1NLO for the plain Gō model (black) and the Gō model with hydrogen bonds (red).*

terminant role in the dimerization of the SH3 domain,[17] showing therefore the importance of hydrogen bond interactions in the folding and aggregation of this protein.

Up to this point, our systematic study has focused on two-state folders, which are thought to be the most common ones within living organisms.[193] However, some small proteins can also exhibit a downhill folding where the free energy barrier is negligible.[176] Can hydrogen bonds contribute to a better description of barrierless folders?

Firstly, we have simulated the best characterized example of downhill folder, the E3 binding domain of the dihydrolipoamide succinyltransferase core from the 2-oxoglutarate dehydrogenase multienzyme complex of *E. coli*, also known as BBL protein (PDB code 1BBL[171]), that has been extensively investigated.[176,194] This protein was previously studied in our group with the plain structure-based potential, showing that downhill folding processes can be adequately simulated with this approach.[98]

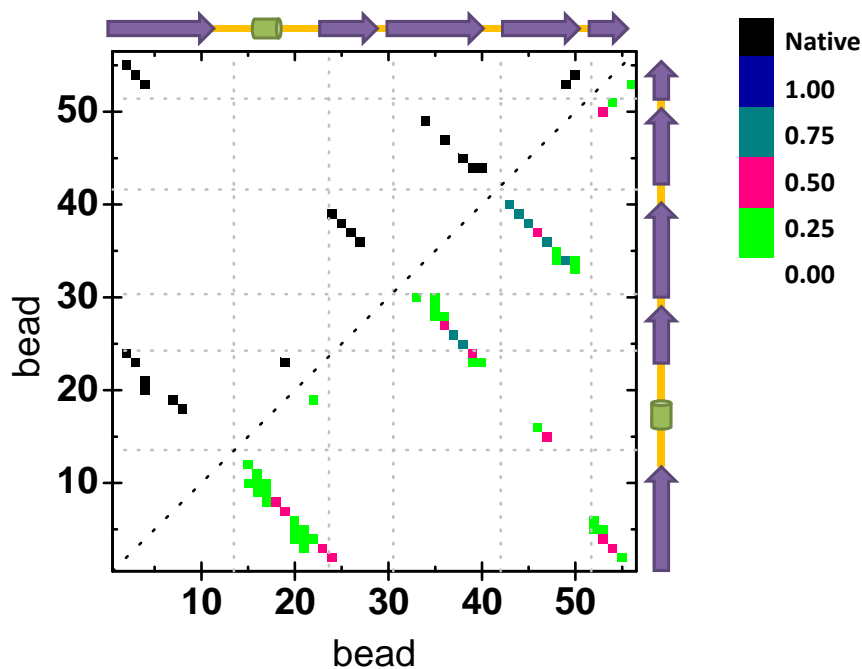

**Figure 4.14:** *Map of hydrogen bond frequency in the denatured state of 1NLO. Similarly to Figure 4.10, the upper triangle shows the hydrogen bonds that are detected in the native state and the lower one shows the frequency of appearance of hydrogen bonds in the denatured state of the protein at the transition temperature.*

Our results, shown in Figure 4.15, are characteristic of downhill folders, with a very broad and shallow heat capacity curve and without any barrier in the free energy profile at the transition temperature. This also proves that the inclusion of hydrogen bonds in our simulations does not induce free energy barriers if they are not present in the real system.

The success in the previous case fostered our study of the gpW protein (PDB code 1HYW[195]). This $\alpha + \beta$ protein experimentally presents a very small folding barrier, showing an overall downhill folding.[196,197] Knowledge-based potentials have given quite elusive results,[61] and so does our plain Gō model.

Shown in the black curves on Figure 4.16, this protein exhibits a clear two-state behavior with the plain structure-based potential. The inclusion of hydrogen bonds (red

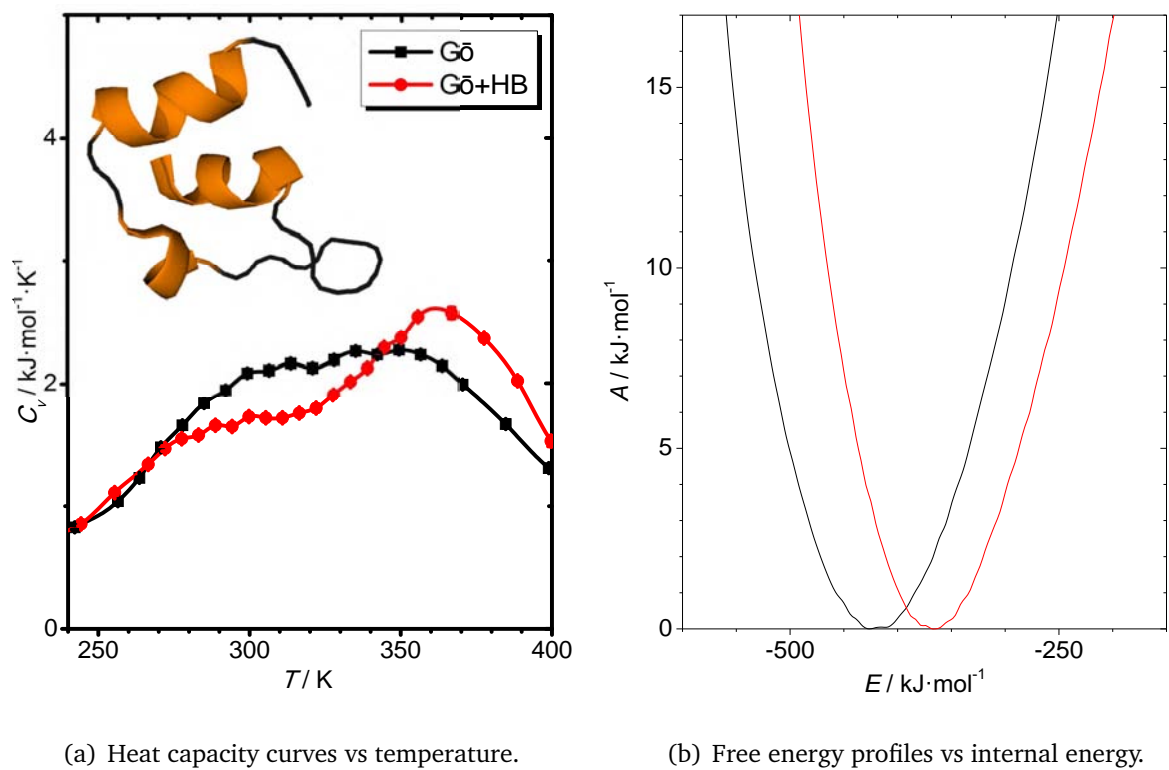

(a) Heat capacity curves vs temperature.        (b) Free energy profiles vs internal energy.

**Figure 4.15:** *Thermodynamic results of 1BBL for the plain Gō model (black) and the Gō model with hydrogen bonds (red).*

curves) flattens the heat capacity curve and considerably reduces the free energy barrier down to about 4 kJ·mol$^{-1}$, a more typical result of a downhill protein with a barrier smaller than the thermal fluctuations.

In Figure 4.16(*b*) we can observe some hints for this change, that are mainly connected to a change in the energetics of the denatured state. In this case, we note the usual broadening of this basin, that is also shifted towards a more stabilized region: the long helices of 1HYW still present some population in the denatured state at the transition temperature.

As a general view, we can conclude that all these profitable results of the combined model are mainly based on three issues: a better definition of the native state in most cases, a certain removal of Gō-like residual structure in the denatured state and a partial retention of helical structure in the denatured state (if this kind of structure is preponderant in the

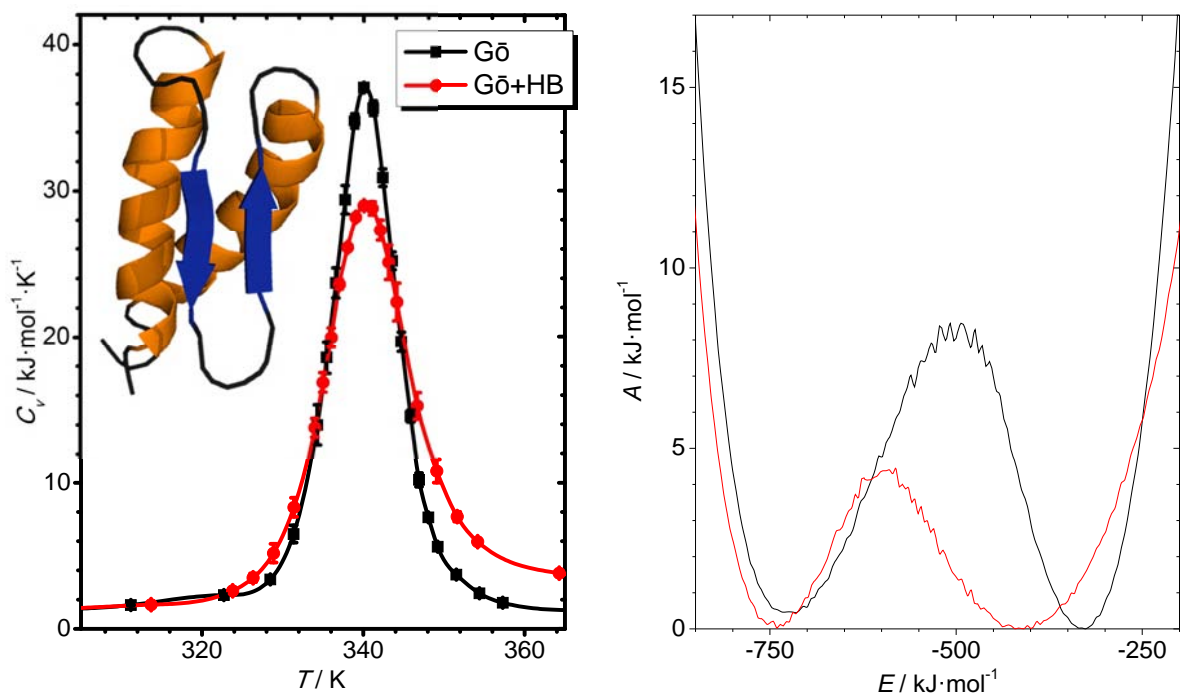

(a) Heat capacity curves vs temperature.　　　(b) Free energy profiles vs internal energy.

**Figure 4.16:** *Thermodynamic results of 1HYW for the plain Gō model (black) and the Gō model with hydrogen bonds (red).*

native conformation). As a result, the protein thermodynamics is not *biased* but adequately *modulated* according to the intrinsic characteristics of the protein itself. Will it also work for the kinetic description of folding?

## 4.3.2   Folding kinetics

In this Section, we will pose some questions about the kinetic properties of our simulations. We have explored the performance of our Kinetic Monte Carlo algorithm (see Section 2.4), as well as the impact of adding hydrogen bonds to the original definition of a structure-based potential. In this way, the aim here is not to perform a thorough investigation of the specific properties of a set of proteins (as we did in the former Section), but to evaluate the different effects that the inclusion of hydrogen bonds may have in the kinetic aspects of protein folding, and link them to the observed thermodynamic properties and the available experimental data.

This Section is divided in two main blocks. The first one is devoted to the analysis of two-state folders, and the comparison between the results of our kinetic simulations and what could be expected for this kind of proteins. In second place, we have taken a couple of examples where we have deepened into the dynamic performance of the different applied potentials (either the original structure-based model or the one combined with hydrogen bonds).

We start with the kinetic characteristics of two-state folders, aiming to find out if our simulations reproduce the expected first law kinetics and to relate their folding rates with the free energy barriers we have calculated in the previous Section, using for these purposes Equations (2.12) and (2.13). For this latter reason, we have carried out kinetic simulations at the transition temperature of each protein, choosing the two-state folders we have already studied (2GB1, 1PGB, 1R69, 1ENH, 1TEN and 1NLO).

As mentioned, we are especially interested in the connection between the height of the free energy barriers and the folding time. It is important to note, then, that we have expressed "time" in terms of the number of Monte Carlo cycles, as we discussed in Chapter 2. The movements associated to each Monte Carlo cycle are accepted or rejected in terms of the relative modification of the system energy. During our simulations, energy has been expressed in reduced units and, therefore, the acceptance or rejection of the system movements (according to the kinetic Monte Carlo scheme) are tightly related to the reduced $k_B T^*$ energetic unit ($\beta$ in Equation (2.1)). For this reason, in this Section we have not converted the free energy barriers into real units, but have used the reduced values that are shown in Table 4.1.

One of our main interests lies on the effect of the different potentials. Therefore, we have paid special attention to two of these proteins, 2GB1 and 1NLO, because their free energy barriers are largely modified by the change in the applied potential. According to Table 4.1, the barrier of 2GB1 gets considerably higher if hydrogen bonds are included in the simulation and the contrary happens in the case of 1NLO.

| | $\Delta A^*$ | |
|---|---|---|
| Protein | Structure-based potential | Combined potential |
| 2GB1 | $1.14 \pm 0.06$ | $3.10 \pm 0.07$ |
| 1PGB | $0.66 \pm 0.02$ | $2.8 \pm 0.1$ |
| 1R69 | $3.2 \pm 0.1$ | $3.42 \pm 0.06$ |
| 1ENH | $1.10 \pm 0.02$ | $1.75 \pm 0.08$ |
| 1TEN | $3.1 \pm 0.1$ | $3.5 \pm 0.2$ |
| 1NLO | $2.10 \pm 0.04$ | $1.48 \pm 0.08$ |

**Table 4.1:** *Free energy barriers (in reduced units) for the simulated two-state folders.*

In first place, we have looked at the distribution of folding times or first passage times ($FPT$), represented in Figure 4.17. In the case of 2GB1, the combined potential shows a broader distribution that is centered at larger folding times. In 1NLO, however, the $FPT$ distributions behave the other way round, presenting larger times with the original structure-based potential.

This trend matches the modification of the free energy barrier, stressing therefore the connection between free energy barriers and folding times. The same tendency is observed for the rest of the proteins, as it is shown in Table 4.2, where we have collected their mean first passage times (called $MFPT$).

The results in this Table validate the kinetic Monte Carlo method that we have used, as the folding times are much larger than the individual Monte Carlo steps (see Chapter 2). Besides, the obtained data show that the impact of hydrogen bonds in structure-based potentials modifies not only the thermodynamics of folding, but also its kinetics, in a coherent way.

The relationship between the free energy barrier and its folding kinetics can be quantified by the calculation of the rate constant. In Chapter 2 we showed that it can be computed (for two-state folders) through the fitting of the data to a first order kinetics or estimated through the inverse of the $MFPT$.

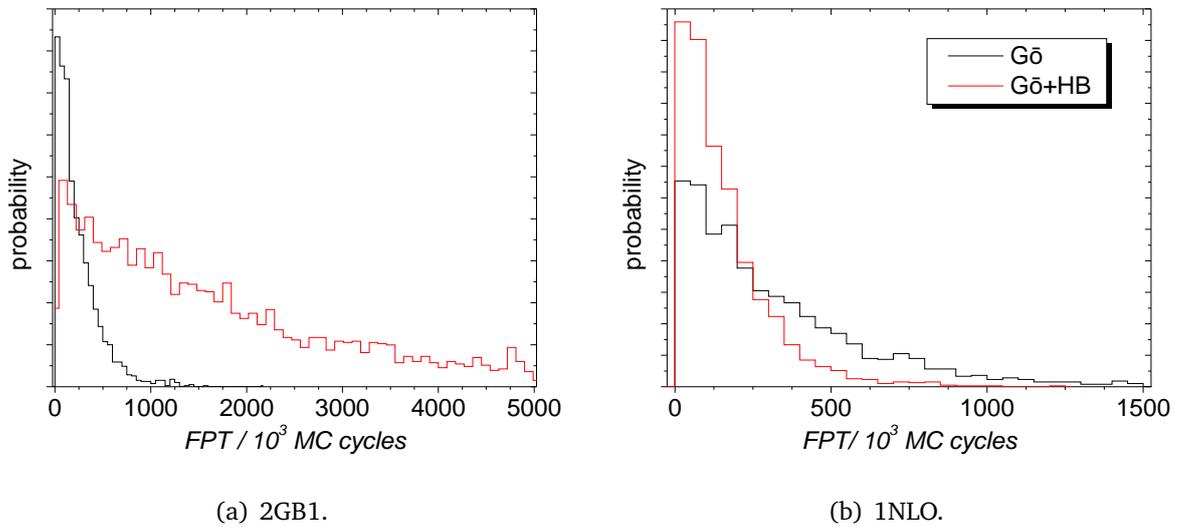

(a) 2GB1.

(b) 1NLO.

**Figure 4.17:** *Distributions of folding times (FPT) for two two-state folders using the original structure-based potential (black) and the combined one (red).*

|  | $MFPT/10^5\ MC\ cycles$ | |
| --- | --- | --- |
| Protein | Structure-based potential | Combined potential |
| 2GB1 | $2.37 \pm 0.05$ | $15.7 \pm 0.3$ |
| 1PGB | $1.27 \pm 0.03$ | $16.5 \pm 0.5$ |
| 1R69 | $10.0 \pm 0.2$ | $12.4 \pm 0.2$ |
| 1ENH | $0.39 \pm 0.09$ | $1.2 \pm 0.02$ |
| 1TEN | $33.3 \pm 0.5$ | $32.5 \pm 0.8$ |
| 1NLO | $3.24 \pm 0.07$ | $1.56 \pm 0.03$ |

**Table 4.2:** *Mean first passage times (MFPT) of the selected two-state proteins at the transition temperature.*

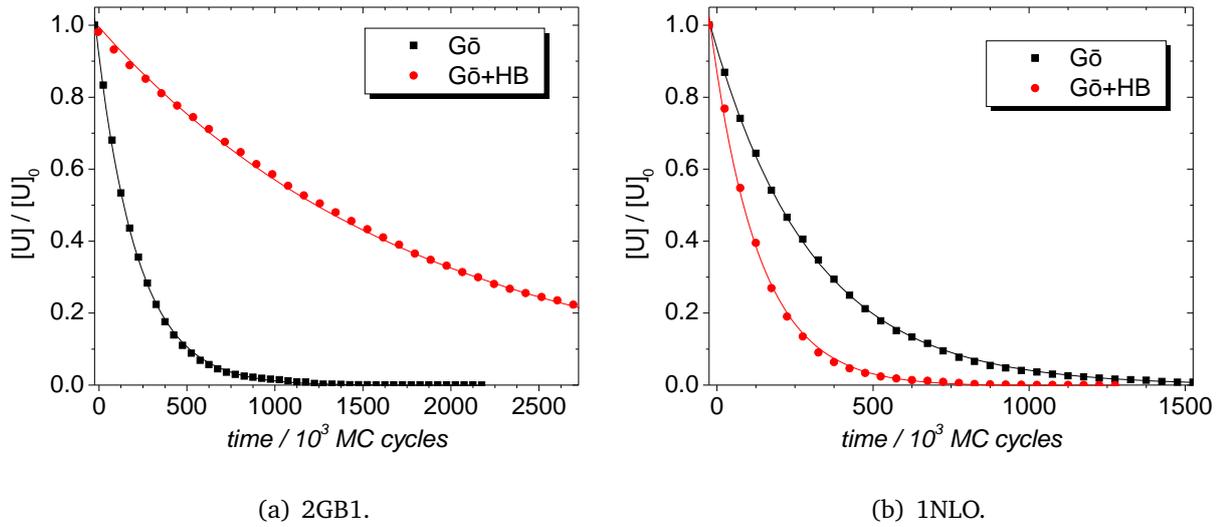

(a) 2GB1.

(b) 1NLO.

**Figure 4.18:** *Exponential fitting of the relative population of unfolded proteins vs time (in number of Monte Carlo cycles) of the selected proteins at their transition temperatures.*

For each protein and kind of potential, we have fitted the unfolded population as a function of "time" to a first order law. We have obtained a proper fitting in all cases, as illustrated in Figure 4.18 for 2GB1 and 1NLO, and a rate constant that matches the inverse of the $MFPT$, differing in less than a 5% for the six studied proteins with the two different potentials. This confirms that our simulations of two-state folders comply with a first order kinetics, as it is predicted by theory.

The connection between rate constant and free energy barrier is illustrated in Figure 4.19. Although our data present some dispersion, we can observe that the logarithm of the rate constant at the transition temperature correlates with the free energy barrier, confirming in this way the agreement between thermodynamic and kinetic information.

Up to this point, we have focused on the relationship between kinetics and thermodynamics, validating the consistency of our results. However, kinetic simulations can add extra information on the folding pathway of proteins. In this second part of the Section, we have studied the folding dynamics of two test proteins. As we aim to compare our results with experimental data, we have performed these simulations at room temperature ($T = 298$ K),[198,199] obtaining the $MFPT$ of Table 4.3 for two test proteins at this new temperature. Note that the $MFPT$ of 1ENH are smaller than those in Table 4.2, as the reduction in temperature speeds up the folding process.

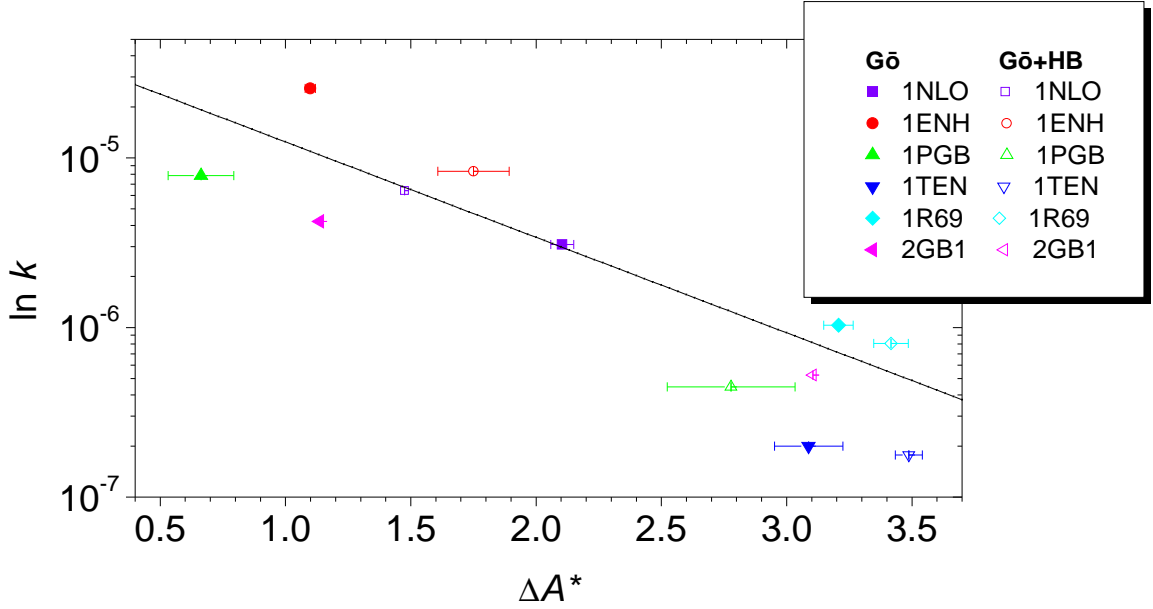

**Figure 4.19:** *Relationship between the constant rate (in logarithmic scale) and the free energy barrier.*

| | $MFPT/10^5 \ MC \ cycles$ | |
|---|---|---|
| Protein | Structure-based potential | Combined potential |
| 1ENH | $0.248 \pm 0.006$ | $0.63 \pm 0.01$ |
| 1BBL | $0.162 \pm 0.006$ | $0.29 \pm 0.01$ |

**Table 4.3:** *Mean first passage times ($MFPT$) of the selected proteins at 298 K.*

As we can see in the Table, we have chosen 1ENH and 1BBL. Having a similar structure (they belong to the all-$\alpha$ family), they present very different folding characteristics, as the first one is a fast two-state folder[186, 188, 200] and the other one presents a downhill behavior.[201, 202]

We have started by evaluating the differences between independent folding trajectories for each protein. It is reasonable to expect a smaller variability for two-state folders, as the presence of a free energy barrier is responsible for a cooperative folding transition and the lack of it results in a large number of states that fast interconvert among them.

For each protein, we have represented the evolution of the $RMSD$ with time for some independent trajectories. We have selected trajectories whose folding time is similar to the mean first passage time. We have observed a similar dispersion among the individual simulations, regardless the applied potential. As an example, we show in Figure 4.20 the evolution of $RMSD$ with time for four independent trajectories of 1ENH, in Figure 4.20($a$), and 1BBL, in Figure 4.20($b$), using the combined potential.

In this Figure, 1ENH spends most of the simulation time far from the native state (with $RMSD$ ranging from 8 to 12 Å) and very quickly acquires its native state in the last steps, showing a typical cooperative behavior. 1BBL, on the contrary, presents a broad distribution of $RMSD$ all along the simulation, having values near 5 Å (matching partially folded configurations) that interconvert with other configurations of larger $RMSD$ values. As a matter of fact, the native state (where the stop criterion is $RMSD < 4$ Å) does not seem to be particularly favored in the simulated conditions, but to constitute just one of the possible states of the protein.

A comprehensive study of the folding kinetics, however, cannot be circumscribed to the individual study of some folding events: it must involve some statistical significance. In fact, kinetic experiments (excluding single probe ones) also measure the average behavior of the different individual protein molecules that are present in the observed sample.[198] Therefore, the take-home message for downhill proteins like 1BBL is that their downhill nature makes them have a quite random behavior, presenting a high variability in which each independent folding event can show different structural features.

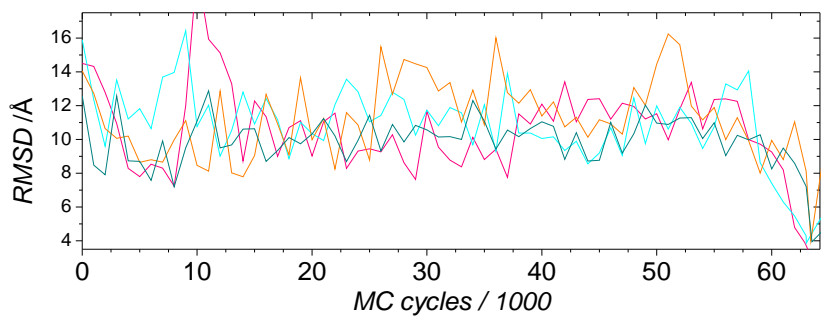

(a) 1ENH.

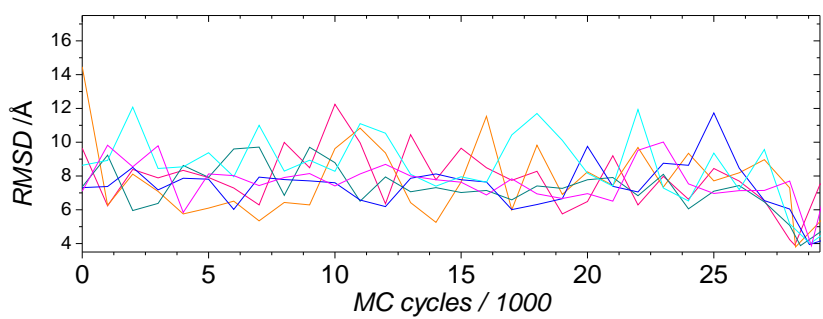

(b) 1BBL.

**Figure 4.20:** *Evolution of the RMSD with time for four different trajectories of 1ENH and 1BBL, using the combined potential.*

Nevertheless, a statistical treatment is valid for two-state folders like 1ENH, where the small data dispersion allows the averaging of the folding events and, therefore, a meaningful study of the dynamic properties of the folding process and the impact that the different potentials make in the simulation results. For a better understanding of the folding process, we have started digging into the available bibliography for experimental and simulation data obtained with higher resolution potentials.[187,200,203,204]

The folding pathway of 1ENH is thought to start with a hydrophobic collapse, followed by a fast formation of the secondary structure elements (three $\alpha$-helices).[200] Among them, the first $\alpha$-helix ($\alpha_I$, that comprises residues 10 to 22) is more stable in isolation and remains undocked during most of the folding process; the second helix ($\alpha_{II}$, from residue

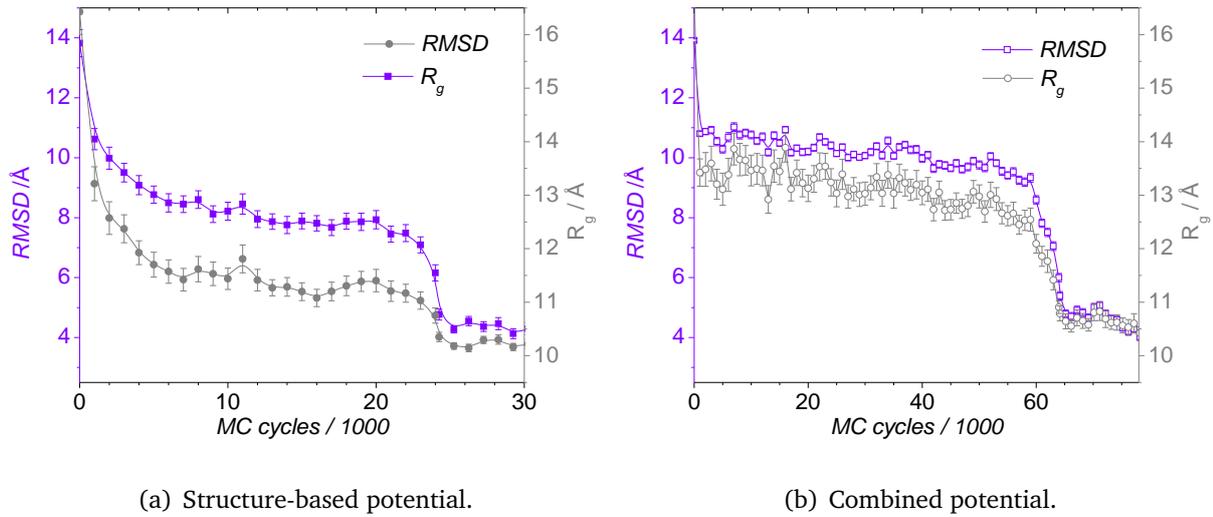

(a) Structure-based potential.                    (b) Combined potential.

**Figure 4.21:** *Evolution of the global RMSD (violet) and radius of gyration (grey) with time, for folding simulations of 1ENH using the structure-based potential (left, filled symbols) and the combined one (right, open symbols), at 298 K.*

28 to 37) and the third one ($\alpha_{III}$, from residue 42 to 54) establish some contacts between them, creating a kind of helix hairpin.[204] The final step in this process is the docking of the three helices together, that occurs through the breaking of some non native contacts in $\alpha_I$ (more specifically, an additional helix turn around residue 24) before its docking to the rest of the structure.[203] Shall we find something similar in our simulations? Will hydrogen bonds help us in a correct description of this folding process?

For this study, we have averaged the simulated folding events whose first passage time differs in less than a 5% from the $MFPT$ (although we have also checked that similar trends are found for $FPT$ values five times larger and smaller). Firstly, we have compared the evolution of the average $RMSD$ and average radius of gyration along the folding time for the original structure-based potential and the combined one. The results are shown in Figure 4.21.

In this Figure, we observe a similar trend regardless the applied potential: starting from a completely denatured state, the trajectories reach a kind of plateau of intermediate $RMSD$ and radius of gyration (that can be linked to a certain hydrophobic collapse of the structure). At the folding time, we observe a final decrease of the radius of gyration, accompanied by the reaching of the cut-off for the *RMSD*.

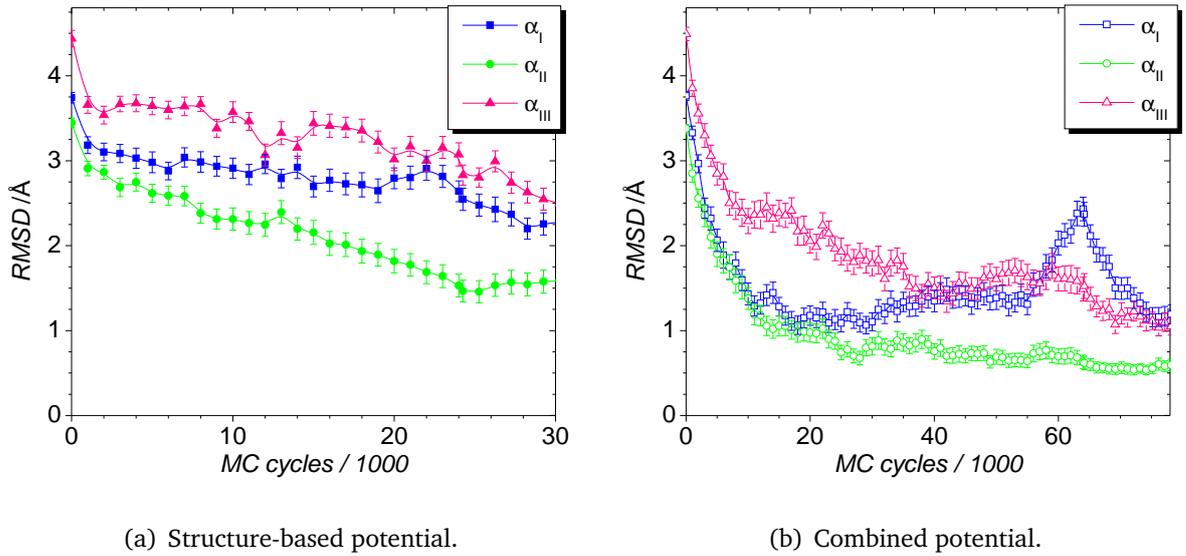

(a) Structure-based potential.   (b) Combined potential.

**Figure 4.22:** *Evolution of the RMSD of the three helices of 1ENH ($\alpha_I$, $\alpha_{II}$ and $\alpha_{III}$) using the structure-based potential (left, filled spots) and the combined one (right, open spots)*

These plots, however, differ in the characteristics of the intermediate step. In the case of the pure structure-based potential, the *RMSD* value smoothly decreases along the folding simulation, while the combined potential exhibits a more abrupt transit that can be linked to the higher cooperativity we have already discussed through previous results. The evolution of the radius of gyration also shows a change in the folding characteristics, presenting more compact configurations in the original structure-based potential. As a result, we can conclude that the inclusion of hydrogen bonds does indeed modify the dynamic characteristics of the transition.

Inspired by experimental data, we have sought the origin of these differences by analyzing the individual secondary structure elements (whose evolution is shown in Figure 4.22) and the formation of the helix hairpins (see Figure 4.23).

If we start with Figure 4.22, we observe that including hydrogen bonds in the potential improves the description of the helices at the early stages of folding (lowering

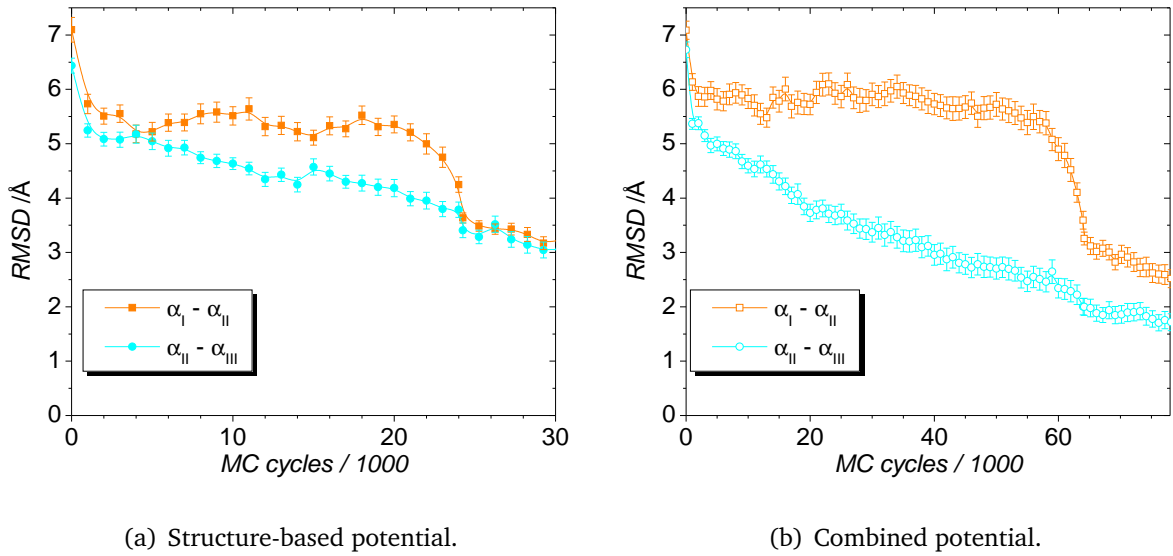

(a) Structure-based potential.     (b) Combined potential.

**Figure 4.23:** *Evolution of the RMSD of the two possible helix-hairpins of 1ENH ($\alpha_I - \alpha_{II}$ and $\alpha_{II} - \alpha_{III}$) using the structure-based potential (left, filled dots) and the combined one (right, open dots)*

the $RMSD$ values for all of them). The case of $\alpha_I$ is particularly interesting because of its partial unfold just before the complete folding of the global structure.

According to the structure-based potential, this helix does not present any particular feature, folding in a similar way to the other two helices of the protein. If the combined potential is applied, however, its folding occurs faster, being almost completely folded from very early stages. Besides, it is the only one of the three $\alpha$-helices of the system that partially unfolds during the whole process, and this happens just before the complete folding of 1ENH. If we observe Figure 4.23, this fact seems to be related to the complete docking of the structure: $\alpha_{II}$ and $\alpha_{III}$ get together in a quite continuous way, but $\alpha_I$ needs that partial unfolding to achieve the global folded structure.

In the case of the combined potential, we have looked for a more detailed view of the folding process. We have computed the frequency maps for Gō and hydrogen bond contacts (shown in Figure 4.24) at two different folding times: before the partial unfolding of $\alpha_I$ ($t = 54000\ MC\ cycles$) and when this rearrangement occurs ($t = 60000\ MC\ cycles$).

Besides, we show in Figure 4.25 some snapshots (from a representative trajectory) that illustrate the behavior of 1ENH under the combined potential at different stages of the folding process.

The first snapshot, in Figure 4.25($a$), corresponds to the situation before the partial unfolding of $\alpha_I$. Helices $\alpha_{II}$ and $\alpha_{III}$ (in green and yellow, respectively) are forming an $\alpha$-hairpin while $\alpha_I$ (in purple) is completely formed, but is not docked to the rest of the structure. The energetic description of this intermediate, shown in Figure 4.24($a$), confirms the presence of the helical hairpin formed by $\alpha_{II}$ and $\alpha_{III}$, as we can observe high populations of local and non-local interactions. Besides, the helical interactions of $\alpha_I$ are mainly formed at this stage (more than 50% population for local hydrogen bond and Gō contacts), while long-range interactions are infrequent.

Figure 4.25($b$) shows the partial unfolding of $\alpha_I$. In the related frequency map, in Figure 4.24($b$), we can observe that $\alpha_I$ looses some local contacts around the C-terminus of the helix while increasing the population of long range interactions.

The formation of these native long range contacts enables the docking of $\alpha_I$, that in Figure 4.25($c$) is refolded again. In this case, it has the correct native orientation with respect to the rest of the system, which allows the complete folding of the full structure.

Therefore, we can conclude that the introduction of hydrogen bonds greatly improves the description of the folding pathway of 1ENH, obtaining a remarkable agreement with experimental data. With the use of the combined potential we have reproduced the distinct behavior of $\alpha_I$, as well as the on-pathway intermediate where $\alpha_{II}$ and $\alpha_{III}$ are placed together.

## §4.4   Summary and conclusions of this Chapter

In this Chapter, we have found out to what extent our hydrogen bond potential is able to improve a structure-based one, in protein folding studies. On one hand, the combination of these potentials is a very relevant test for the accurate obtention of native-like

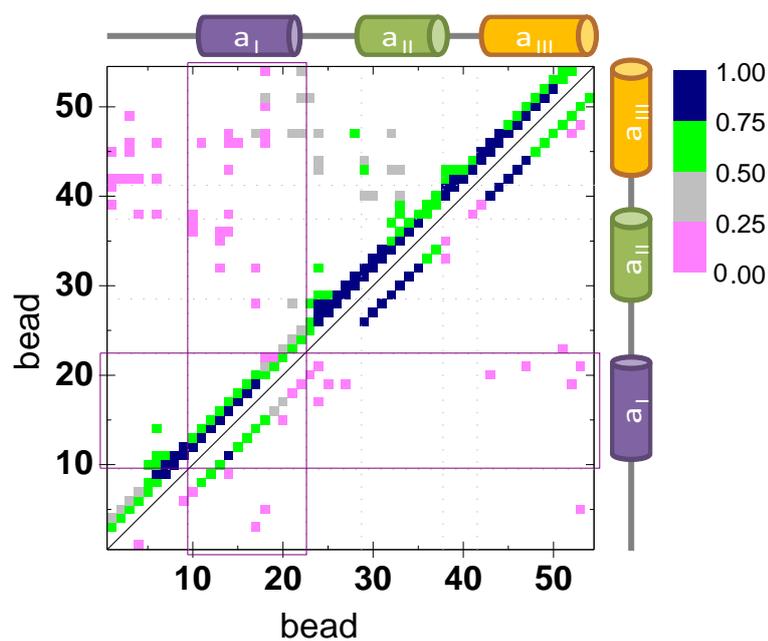

(a) *t = 54000 MC cycles.*

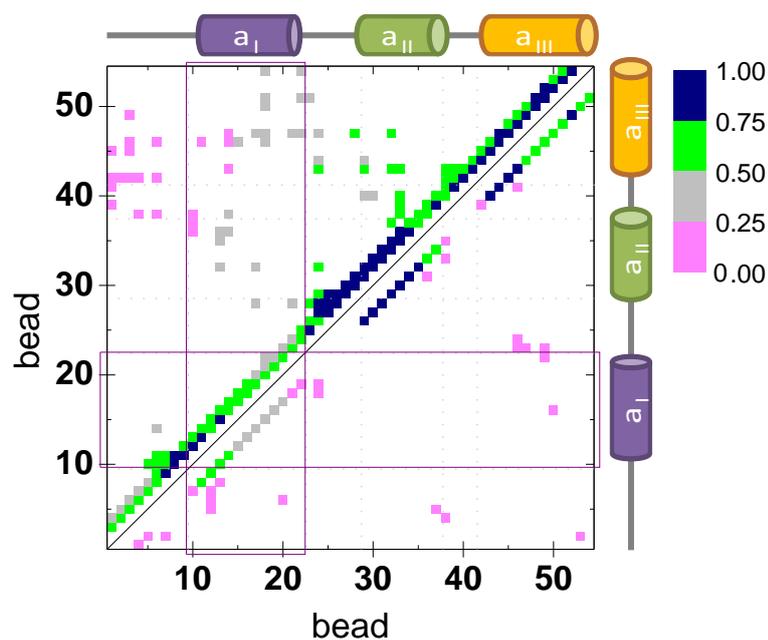

(b) *t = 60000 MC cycles.*

**Figure 4.24:** *Frequency contact map of 1ENH using the combined potential at two different times (averaged from independent trajectories whose FPT matches the MFPT). In these maps, the upper diagonal shows the frequency of Gō contacts and the lower one, the frequency of hydrogen bond ones. The fragment that corresponds to $\alpha_I$ has been boxed in purple.*

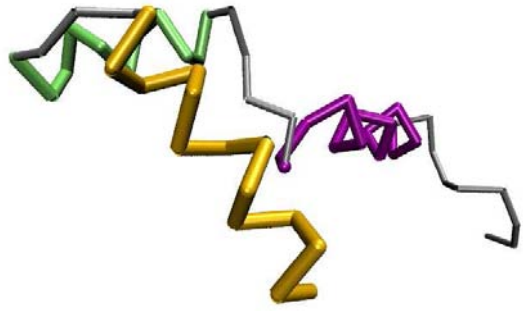

(a) $t = 54000\ MC\ cycles.$

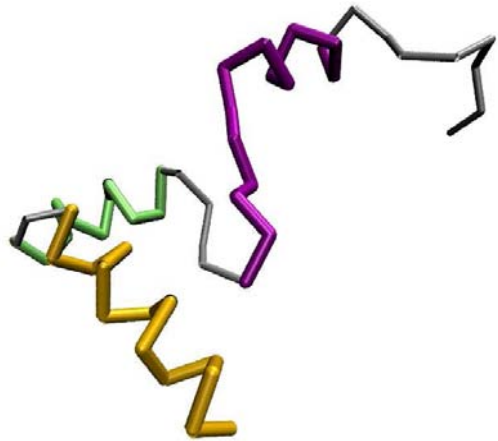

(b) $t = 60000\ MC\ cycles.$

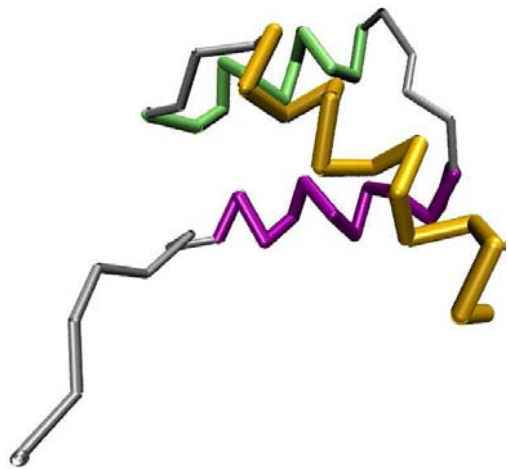

(c) $t = 63000\ MC\ cycles.$

**Figure 4.25:** *Representative snapshots during a folding simulation of 1ENH, at different simulation times. In these cartoons, $\alpha_I$ has been colored in purple, $\alpha_{II}$ is represented in green and $\alpha_{III}$ is yellow. Images drawn with VMD.[144]*

secondary structure elements. Actually, that fact constituted one of the main points during the design of the hydrogen bond model (see Chapter 3) and is essential for a successful merging of both potentials. We have checked that, under a slight minimization of the PDB structures, our hydrogen bond potential is able to identify all the native hydrogen bonds from a native protein structure, obtaining results similar to the usual algorithms for hydrogen bond detection, such as STRIDE[168] or DSSP,[167] but using just the $\alpha$-carbon trace of the protein. On the other hand, the inclusion of hydrogen bonds in a structure-based potential has resulted in a remarkable improvement of the thermodynamic and kinetic folding properties for a representative set of proteins.

We have carried out extensive equilibrium simulations to analyze the thermodynamic properties of eight proteins, each of them representing a particular feature of the structure-based potential that could be improved. In first place, we have studied 2GB1, a typical two-state folder of well-known properties. We have checked that the inclusion of hydrogen bonds does not imply any kind of undesirable effect, maintaining the general two-state scenario and even enhancing the free energy barrier of folding (without any term explicitly favoring cooperativity in the potential definition).

After that, we have simulated the folding of 1PGB. Being the same protein as 2GB1 (the B domain of protein G) but whose structure has been solved by a different experimental technique, this protein exhibits a downhill folding when the original Gō model is applied, in contrast with the two-state folding that 2GB1 shows in the same conditions.[157] In this case, our combined potential is able to correct this Gō artifact, presenting again a two-state folding essentially identical to 2GB1.

Once our approach has been found valid for general cases, we have carried out a systematic analysis of folding events using proteins with different types of native structures. To illustrate this, we have shown a couple of examples of all-$\alpha$ (1R69 and 1ENH) and all-$\beta$ (1TEN and 1NLO) proteins.

The role of hydrogen bonds in all-$\alpha$ proteins is especially relevant, as a residual helical content has been experimentally observed in the denatured state of many of them.

This effect, which the pure structure-based model cannot properly reproduce, is correctly modeled when hydrogen bonds are considered. Interestingly, they do not force changes in the height of the free energy barriers, but correctly modulate them in agreement with experimental results.

In relation to all-$\beta$ proteins, no systematic residual structure is observed in their unfolded state from our simulations, like in 1TEN. In 1NLO, however, some residual structure is observed. Interestingly, these partial interactions are not a model artifact, but they seem to have a biological meaning: they are mainly placed in regions experimentally stabilized by hydrogen bonds in the transition state and with a crucial role in the folding of this protein. Thus, the folding of this protein according to our model constitutes a valuable example of the synergistic effect of the two contributions of this combined potential and its applicability to protein folding studies.

Although we have mainly focused on two-state folders, the combined potential seems to work well in downhill folders too. We have studied 1BBL, a typical example where the structure-based potential had already given correct results.[98] We have checked that the introduction of hydrogen bonds does not have any undesirable effect, and downhill behavior is conserved. Our last challenge has been to determine whether the combined potential can simulate downhill folders even when the structure-based potential cannot detect them. For that purpose, we have studied 1HYW, a downhill protein with a small folding barrier where the Gō model predicts a two-state transition. Interestingly, the insertion of hydrogen bonds considerably reduces the barrier, reaching the downhill range.

After this thorough description of the folding thermodynamics, we have tackled a kinetic study of these proteins. In first place, we have checked the validity of our kinetic Monte Carlo methodology, using for this purpose some two-state folders. We have confirmed that their folding follows a first order kinetics and their folding rates are correlated with the free energy barriers we obtained by thermodynamics.

We have also studied the folding pathways of two all-$\alpha$ proteins, highlighting the different performance of downhill folders (1BBL) in relation to two-state ones (1ENH). In the latter case, we have been able to develop a more extensive analysis of the folding pathways and to compare the results of the original structure-based potential and the combined one. Again, we have found that an explicit treatment of hydrogen bonds notably improves the description of folding, reproducing the main features of the folding of 1ENH along time, as it has been reported by experiments and higher resolution simulations.[200, 204]

Then, we can conclude that our hydrogen bond model has succeeded in the challenges that we posed at the beginning of this Chapter, proving the relevance of this kind of interaction in an accurate representation of protein folding and merging in this way kinetics and thermodynamics to build a global view of the process.

# 5

---

# Hydrogen bonds and hydrophobic potentials: peptides and proteins upon folding and aggregation

Previous Chapters have dealt with some results related to the effect of hydrogen bonds upon folding and aggregation. In this way, we have studied in Chapter 3 the behavior of a peptide system under the hydrogen bond potential especially designed for this Project, and without any additional interactions.[91] This has allowed us to build a "structural" phase diagram that, although very roughly, reflects the role of temperature and concentration in the formation of secondary structure elements. However, this strategy is limited by the lack of an appropriate way to describe tertiary interactions and is, therefore, constrained to sequenceless peptides.

Our next step, then, has been the combination of the hydrogen bond potential with another one able to reproduce the rest of the protein interactions. Thus, in Chapter 4 we have merged our hydrogen bond model with a structure-based potential, previously developed in our group.[62] This kind of potentials is suitable for the study of most folding processes, but prevents a realistic view of aggregation processes (being limited to domain swapping experiments[99]). Even if hydrogen bonds are included, the influence of the structure-based potential is too biased towards the native state to expect a proper competition between folding and aggregation.

Therefore, in the current Chapter we take an alternative strategy to tackle the competition between folding and aggregation in proteins. We have combined our initial hydrogen bond potential with a very simple hydrophobic one that does not depend on the native structure.

Similarly to the previous Chapters, we will start with a brief discussion on simple hydrophobic potentials and their suitability for our purposes. After that, we will expose the main features of the particular hydrophobic potential that we have used in this part of the Project, as well as the special care to make it fully compatible with our hydrogen bond potential.

After these technical details, we will expose the main results concerning the interplay between folding and aggregation in sequence-dependent peptides and proteins. It must be kept in mind that the simplicity of the hydrophobic potential hampers the use of realistic sequences. Consequently, the used ones do not match any real protein. Instead, they have been particularly designed for this Project in a very regular way, reproducing the most relevant structural families of real proteins and peptides.

In first place, we will compute the structural phase diagrams of peptides with different sequences (helical and $\beta$-prone ones). Their comparison with the phase diagram of Chapter 3 is a valuable tool to understand the role of hydrophobic interactions in the general hydrogen bond scheme.

The last part of this Chapter will be dedicated to the study of complete proteins with different structural topologies: an all-$\alpha$ protein, a $\beta$-type one and an $\alpha+\beta$. After their design, we have simulated them in different concentration conditions and observed their thermodynamic and structural properties, successfully reproducing the competition between folding and aggregation and completing therefore the principal aim of this PhD Thesis.

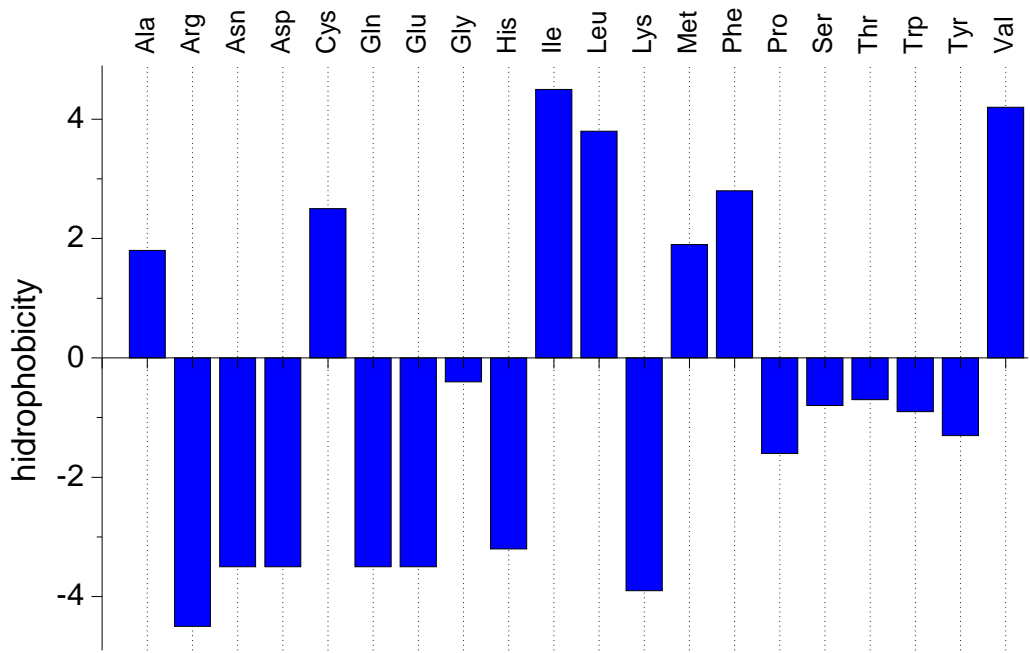

**Figure 5.1:** *Hydrophobicity scale according to Kyte et al. definition.[205]*

## §5.1   Motivation: hydrophobic potentials

As we explained in Chapter 1, hydrophobic interactions play an important role in the folding process through the so-called hydrophobic collapse we mentioned in Chapter 1. Each amino acid presents a different tendency to collapse, which allows its ranking into a hydrophobicity scale,[205] similar to the one showed in Figure 5.1. Three main tendencies can be identified: some residues present a large hydrophobicity (like valine) and tend to be buried inside the protein structure; others (like arginine) are polar (or hydrophilic) and, therefore, more likely to be found at the protein surface; finally, other amino acids like glycine show a mild hydrophobic tendency, being commonly considered as neutral.

This fact is very relevant when considering the modelization of hydrophobic interactions, as some potentials assign a different hydrophobicity to each amino acid[206,207] while others just consider these main tendencies.[110,208,209] The latter are commonly known as *HP potentials*, as they usually implement a two or three-letter alphabet based on *H*ydrophobic and *P*olar residues.

Not surprisingly, these approaches also differ in the level of detail for the protein description. As we explained in Chapter 1, hydrophobic interactions mostly take place between the side chains of the different amino acids. Therefore, potentials that take into account the different interactions which can be formed among the twenty types of amino acids (and, in principle, pursue a detailed representation of the hydrophobic interaction) commonly place their interacting centers somewhere in the side chain, usually in its geometric center[206, 210] or $\beta$-carbon.[78, 211, 212]

In the case of HP potentials, some models still place the interaction centers at the side chains.[213] However, most of them combine this simple description of the protein interaction with a more basic representation of the protein itself, usually defining the hydrophobic interaction between the $\alpha$-carbon positions of the residues.[106, 110, 111, 214, 215]

It is remarkable that HP potentials can provide qualitatively comparable results to more detailed potentials in a number of cases,[110, 216] supporting the validity of their use in folding and aggregation studies if simplified sequences are used. Besides, they reduce the complexity of the system and its conformational space, so they allow long and complex simulations at a moderate computational cost.

## §5.2   Building the combined potential

As we have exposed in the previous Chapter, building a combined potential from two independent ones needs a full compatibility between them. Therefore, we have looked for a hydrophobic potential that fulfills the two main characteristics of our hydrogen bond one: a simple $\alpha$-carbon representation and the accurate reproduction of the native geometry of secondary structures.

The first requirement automatically impedes the use of hydrophobic potentials with a high degree of resolution, that locate their interacting centers somewhere in their side chains.[78, 138, 206, 207, 211] This kind of potentials can reproduce the geometry of native proteins in an accurate way, but are accompanied by a remarkable increase in the complexity of the simulated system and its conformational space, resulting in an undesirable raise in the computing time.

As we do not consider the side chains of the residues in our model, we have chosen an HP potential for the hydrophobic interactions. Among them, there are three main strategies. Some of them are based on a lattice representation of the protein, so they cannot reproduce the precise off-lattice geometry of real systems.[217,218] Others adopt a discrete description of the interaction by means of a step-like potential;[219,220] although we used a discrete functional form in the case of the hydrogen bond potential (see Chapter 3), this approach is not so valid in this case, as the individual hydrophobic interactions do not exhibit a discrete nature. The third strategy is the use of a Lennard-Jones expression, suitably modulated depending on the kind of interacting residues.[110,209,221]

We have chosen the Brown *et al.*[110] potential, that belongs to the latter family. It is based on the work of Thirumalai and Head-Gordon laboratories, and was originally proposed on a lattice scheme.[214,221] The description we used here, however, has been adapted by the authors to off-lattice representations of the polypeptide chain with reported excellent results in folding simulations of designed proteins.[110,111]

## 5.2.1 The Brown *et al.* hydrophobic potential

This HP potential has been designed to be used in isolation, being able to fold a complete simple protein by itself.[110,111] It defines the protein at two different levels: the hydrophobicity of each residue and its local geometry.

Therefore, each residue $i$ is classified in terms of its hydrophobicity into one of three groups, either polar, hydrophobic or neutral, designated in the original work as $L$, $B$ and $N$, respectively.[110]

Its local geometry is also imposed, being defined by the hydrophobicity of the surrounding residues (from $i-1$ to $i+2$), as it is shown in Table 5.1. The underlying idea is that $\alpha$ or $\beta$ propensity is encoded to some extent in the sequence, as a particular combination of hydrophobic and hydrophilic residues can favor the formation of secondary structure elements (see Figure 5.2 for details).[222] In the Brown *et al.* potential, this characteristic is not only a realistic trend, but it inserts an additional *a priori* selection of the type of local structure (i.e. favored secondary structure element for each residue).

| Sequence | Local geometry of the internal residue |
|:---:|:---:|
| *XNNX* | Turn (*T*) |
| *XLLX* | Helical (*H*) |
| *LBBL* | Helical (*H*) |
| Other | Extended (*E*) |

**Table 5.1:** *Assignment of the local geometry of the amino acid i in terms of the hydrophobicity of the surrounding residues i − 1, i, i + 1 and i + 2, according to the Brown et al. potential.[110] In this Table, X refers to any kind of amino acid.*

As the Brown *et al.* potential is applied alone, its energy function, $E_{brown}$, includes both local interactions (to control the local geometry of the peptide and the chain stiffness) and long range ones (the hydrophobic terms themselves). Therefore, it has three different terms, as can be observed in Equation (5.1), taken from the original paper:[110]

$$
\begin{aligned}
E_{brown} = & E^{\theta}_{brown} + E^{\phi}_{brown} + E^{HP}_{brown} = \\
& \sum_{\theta}^{N} \frac{1}{2} k_{\theta} (\theta - \theta_0)^2 + \\
& \sum_{\phi}^{N} \left[ A(1 + \cos \phi) + B(1 - \cos \phi) + C(1 + \cos 3\phi) + D \left( 1 + \cos \left[ \phi + \frac{\pi}{4} \right] \right) \right] + \\
& \sum_{i}^{N-3} \sum_{j \geq i+3}^{N} 4\varepsilon_H S_1 \left[ \left( \frac{\sigma}{r_{i,j}} \right)^{12} - S_2 \left( \frac{\sigma}{r_{i,j}} \right)^6 \right]
\end{aligned}
$$

(5.1)

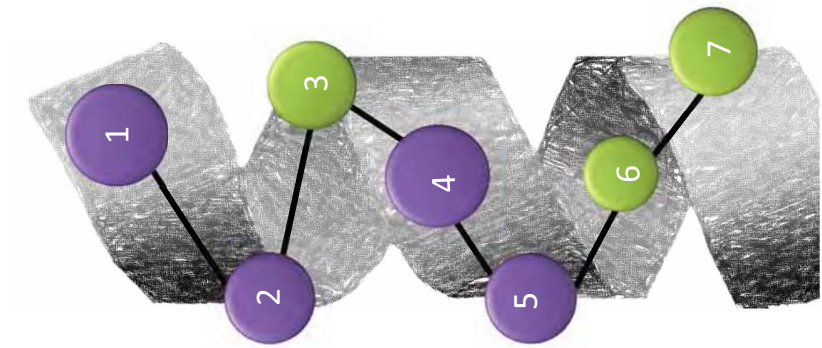

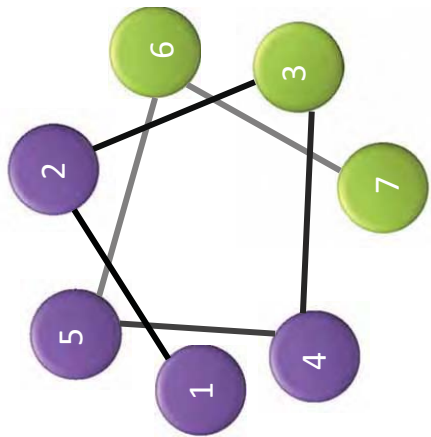

(a) $\alpha$-helix, $L_2BL_2B_2$.

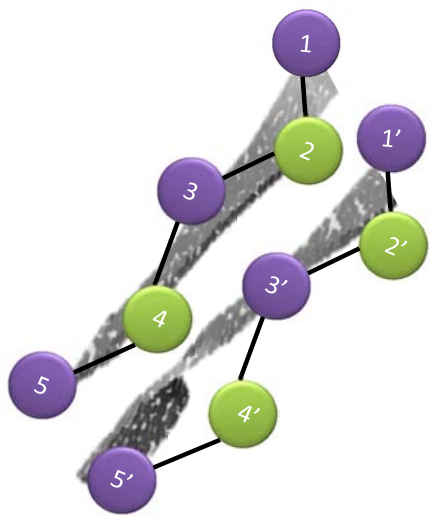

The two first terms, $E_{brown}^{\theta}$ and $E_{brown}^{\phi}$, act on the local geometry of the chain either through a harmonic well potential in the case of the virtual bond angles, $\theta$, or a more complex angular expression for the virtual torsional angles, $\phi$. The last term, $E_{brown}^{HP}$, corresponds to the hydrophobic potential itself and is only applied in long range interactions, as it is shown in its mathematical expression, where a Lennard-Jones shape can be observed.

In this potential, all energetic variables take reduced units and have been set so that $\varepsilon_H = 1.00$. As it was formerly based on a lattice representation, $\sigma = 1.00$, which corresponds to the distance between beads in the original lattice.

A representation of the local terms is shown with the dark green curves of Figure 5.3. The first one, shown in Figure 5.3($a$), is common to all kinds of residues, being centered at $\theta_0 = 105°$ and having a force constant of $k_\theta = 20\ \varepsilon_H/\text{rad}^2$.

Regarding the other local term, $E_{brown}^{\phi}$, the values of the constants $A$, $B$, $C$ and $D$ depend on the kind of residue that is involved. We have collected them in Table 5.2 and the resulting potential is shown in Figures 5.3($b$) to ($d$). Note the different treatment of the $T$, $H$ and $E$ residues. Turns are highly flexible (i.e. their torsional potential provides only a small stabilization of several preferred conformations) and their three minima are equally stabilized.

In the case of helical residues, in Figure 5.3($c$), the energetic difference between minima and maxima are larger. Besides, the three energetic minima present different relative stabilizations, driving in this way the obtention of angular values of $+60°$. In the case of extended ones, shown in Figure 5.3($d$), the relative stabilization of extended conformations ($\pm180°$) is also remarkable. In this way, the local energetic terms unavoidably induce the formation of secondary structure elements.

Regarding the long range terms, they also depend on the particular nature of the involved residues, as we can see in the values of $S_1$ and $S_2$ in Table 5.2. If two hydrophobic residues interact ($B - B$), the positive value of $S_2$ indicates an attractive minimum in the Lennard-Jones potential; if hydrophilic residues interact among them ($L - L$) or with hydrophobic beads ($L - B$), $S_2$ becomes positive and the interaction is repulsive; if a neutral

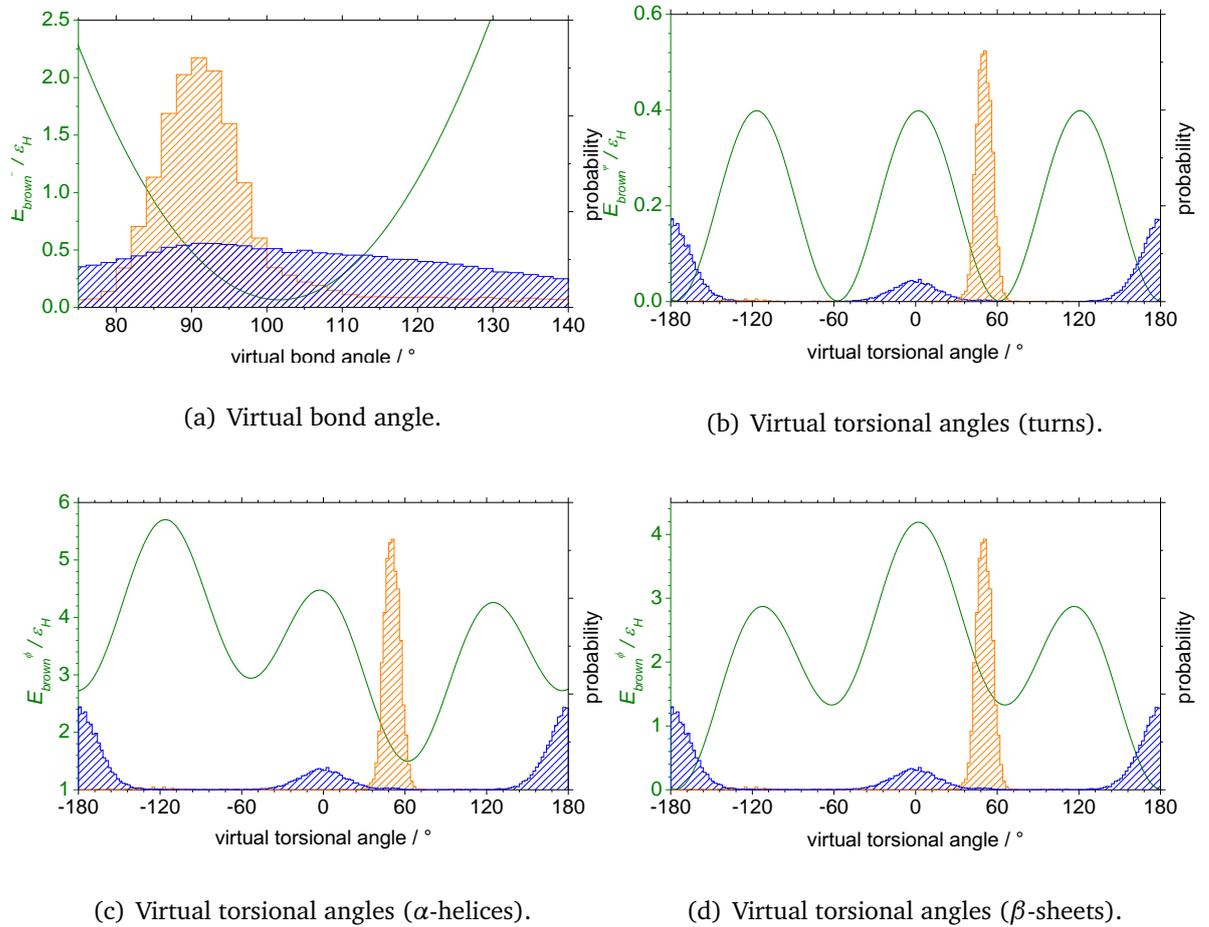

(a) Virtual bond angle.

(b) Virtual torsional angles (turns).

(c) Virtual torsional angles (α-helices).

(d) Virtual torsional angles (β-sheets).

**Figure 5.3:** *Comparison of the local geometries favored by the Brown et al. potential (expressed as the minimum of the appropriate energetic terms in Equation (5.1) and colored in dark green) and the geometries of the secondary structure elements generated with our hydrogen bond potential (colored in orange and blue for α-helices and β-sheets, respectively).*

| Local geometry | $A$ | $B$ | $C$ | $D$ | Type of interaction | $S_1$ | $S_2$ |
|:---:|:---:|:---:|:---:|:---:|:---:|:---:|:---:|
| Turn ($T$) | 0 | 0 | 0.2 | 0 | $B-B$ | 1 | 1 |
| Helical ($H$) | 0 | 1.2 | 1.2 | 1.2 | $L-L$ or $L-B$ | 1/3 | -1 |
| Extended ($E$) | 0.9 | 0 | 1.2 | 0 | $N-L, N-B$ or $N-N$ | 1 | 0 |

**Table 5.2:** *Values of the parameters of the sequence-dependent terms of $E_{brown}$. In the left hand side, we show the values of A, B, C and D in $\varepsilon_H$ units (see Equation (5.1)), according to the local geometry of the residue. The right hand side contains the values that $S_1$ and $S_2$ (needed for the calculation of $E_{brown}^{HP}$) take depending on the kind of non-local interaction (note that these latter quantities are adimensional).*

residue is involved, there is also some short-range repulsion among beads.

In order to evaluate the suitability of this potential for our purposes, we must highlight the fact that it was originally based on a lattice representation. Will it be compatible to our hydrogen bond potential? The answer to this question is illustrated in Figure 5.3, where we show (in addition to the local terms of Equation (5.1), already discussed) the angular distributions of typical secondary structure elements obtained with the hydrogen bond potential.

If we start with Figure 5.3($a$), where the energetic term for the virial bond angle is represented, we observe that $\beta$-sheets present a very broad distribution, while $\alpha$-helices show a narrower one that does not match the minimum of $E_{brown}^{\theta}$. In this way, the inclusion of this term would regularize $\beta$-sheets, while distorting the helical structures.

Regarding the torsional terms, we observe Figure 5.3($c$) that the maximum at $+60°$ (stabilized in $E_{brown}^{\phi}$ for helical residues) matches the angular distribution of $\alpha$-helices. The maximum at $\pm180°$ corresponds to the totally extended conformations found in natural $\beta$-sheets. However, we also find that the $\beta$-sheet distribution presents a smaller maximum around $0°$, that does not match natural structures, but whose presence is allowed by the geometrical restrictions of the hydrogen bond definition. It explains the higher structural variability of these structures, although it does not have a relevant impact on their overall behavior. In any case, the combination with the $E_{brown}^{\phi}$ potential automatically penalizes them, removing any unwanted effect.

The differential treatment of the local geometries (depending on the *a priori* selection of the secondary structure that forms the involved residue) is a distinguishable characteristic of the Brown *et al.* potential. According to the author's definition, it is essential in order to obtain reasonably folded structures, where secondary structure fragments are stable and present the desired local geometry.

However, we pursue a different aim: to model complete proteins (and their secondary structure elements) through the careful combination of hydrophobic interactions and hydrogen bonds. In this sense, we want to reproduce the natural trend of proteins, where the particular combination of hydrophobic and polar residues favors the formation of $\alpha$-helices or $\beta$-sheets (stabilized by the formation of hydrogen bonds) without imposing the local geometry in any additional way.

Besides, we aim to study the interconversion of folded structures (that may have any kind of shape) into aggregated ones (that mostly exhibit a $\beta$-type structure[11]) by the effect of concentration. Changes in this parameter must then be able to modify the balance between the two types of local arrangements and, therefore, the observed stable structure. Obviously, if the local geometry is selected from the beginning, this structural change is impossible. For this reason, we aim to control the chain stiffness of the system, but without imposing any secondary structure from local geometric considerations.

After the discussion of the potential itself, we have analyzed the geometric characteristics of the structures that are stabilized by the Brown *et al.* potential.[111] As we show in Figure 5.4, these structures are unavoidably based on regular sequences, which allow the acquisition of reasonable folded lattice structures in spite of the simplicity of the model. This highlights the fact that this kind of simplified HP potentials exhibit qualitatively good properties, but cannot aim to reproduce the features of a real protein with a high accuracy.

Besides, the structures in Figure 5.4 are highly compact, as a result of the lack of side chains in the model and the use of the underlying lattice.

Regarding the sequences themselves, the secondary structure fragments usually have the same local arrangement of those of Figure 5.2, with slight modifications to favor the packing of the complete structure. As we can observe in Figure 5.2, the formation of secondary structure elements creates hydrophobic faces (that tend to be buried in the global structure) and hydrophilic ones that repel each other. In this way, the positions of the *B* and *L* residues have been chosen so that the burying of the hydrophobic faces places the ends of the fragments as close as possible.

Up to this point, we have exposed the main features and limitations of the Brown *et al.* hydrophobic potential. Some of them are inherent to all HP potentials and we will have to cope with them and sacrifice some accuracy in the protein representation. Others, however, can be improved if we adapt the original potential.

## 5.2.2 The adapted hydrophobic potential and its combination with the hydrogen bond one

We have performed many simulations analyzing the relative weights of the Brown *et al.* and the hydrogen bond potential. We have followed Equation (5.2), simulating different weighting factors for the hydrogen bond energy term, as described by Equation (3.30), a term that keeps the chain stiffness (and can be ascribed to the local terms $E_{brown}^{\theta}$ and $E_{brown}^{\phi}$ of the Brown *et al.* potential, defined in Equation (5.1)) and one term for the long range hydrophobic interactions (similarly to the $E_{brown}^{HP}$ addend in Equation (5.1)):

$$E = \omega^{hb} E^{hb} + \omega^{stiff} E^{stiff} + \omega^{HP} E^{HP} \qquad (5.2)$$

Thanks to these simulations, we have noted some useful observations about the behavior of the potential in equilibrium simulations. For instance, it is important to note the role of the chain stiffness when the non-local hydrophobic interactions are included; in this way, we avoid a too strong non-specific collapse that would result in a "sticky potential". These two terms ($E^{stiff}$ and $E^{HP}$) work synergistically, so any modification in one of them must be followed by the corresponding check of its effect in the other.

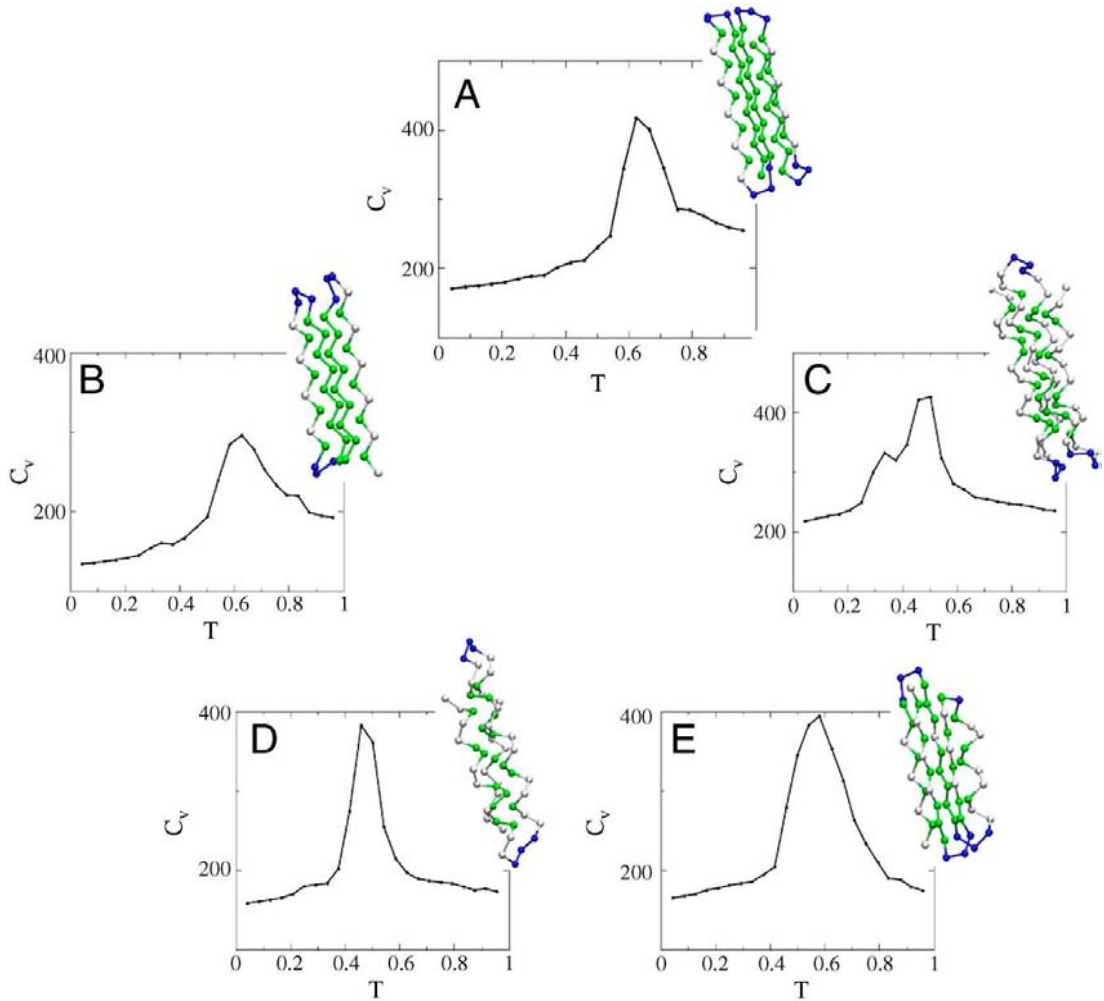

**Figure 5.4:** *Minimum energy structure and heat capacity for five sequences A–E, computed with the Brown et al. (A) $B_9N_3(LB)_4N_3B_9N_3(LB)_5N_3B$; (B) $B_9N_3(LB)_4N_3B_9N_3(LB)_5L$; (C) $((L_2BL_2B_2)_2L_2N_3)_3(L_2BL_2B_2)_2L_2$; (D) $((L_2BL_2B_2)_2L_2N_3)_2(L_2BL_2B_2)_2L_2$; (E) $(LB)_4BN_3B_3LB_5N_3(L_2B)_2BL_2BN(BL)_3B_2N_3B_2(BL)_4$. Figure taken from[111].*

The chain stiffness is maintained in the original potential by the $E^{\theta}_{brown}$ and $E^{\phi}_{brown}$ terms. But we have already discussed that the $E^{\theta}_{brown}$ one is not compatible to our helical structures, while $E^{\phi}_{brown}$ imposes *a priori* a given local geometry of the fragments. Instead, we have removed the $E^{\theta}_{brown}$ interaction, keeping the chain stiffness by just using a modified $E^{\phi}_{brown}$ term.

Its functional shape is similar to the turn-type one that, as already discussed in the previous Section, equally favors $\alpha$-helices and $\beta$-sheets. We have kept the distinction between turn residues and secondary structure fragments by a different energetic stabilization: $\omega^{\phi}_{turn} = 0.25\ \omega^{\phi}$. In addition, we have strongly penalized conformations whose torsional angle lies near $0°$, that would lead to unnatural and too compact conformations. Therefore, the final functional form –plotted in Figure 5.5($a$)– is as follows:

$$E^{stiff} = \sum_i^N E_i^{\phi};\ E_i^{\phi} = \begin{cases} 0.5\ (1 + \cos 3\phi_i) - 1 & \text{if } |\phi_i| \geq 40° \\ 10.0 & \text{otherwise} \end{cases} \quad (5.3)$$

As we have already stated, the change in the local stiffness of the chain affects the performance of the hydrophobic long range term. To avoid the formation of too collapsed structures, we have sharpened the *HP* term with respect to the original one, as shown in Equation (5.4) and Figures 5.5($b$) to 5.5($d$). The modified values for the potential parameters can be found in Table 5.3.

$$E^{HP} = \sum_i^{N-4} \sum_{j>i+4}^N E_{i,j}^{HP};\ E_{i,j}^{HP} = S_1 \left[ \left( \frac{\sigma_{int}}{r_{i,j}} \right)^{12} - S_2 \left( \frac{\sigma_{int}}{r_{i,j}} \right)^{10} \right] \quad (5.4)$$

The relative values of the three weighting factors, $\omega^{hb}$, $\omega^{\phi}$ and $\omega^{HP}$, have been optimized through extensive simulations of peptide systems and complete proteins under different conditions. We have analyzed different sets of parameters looking for optimal

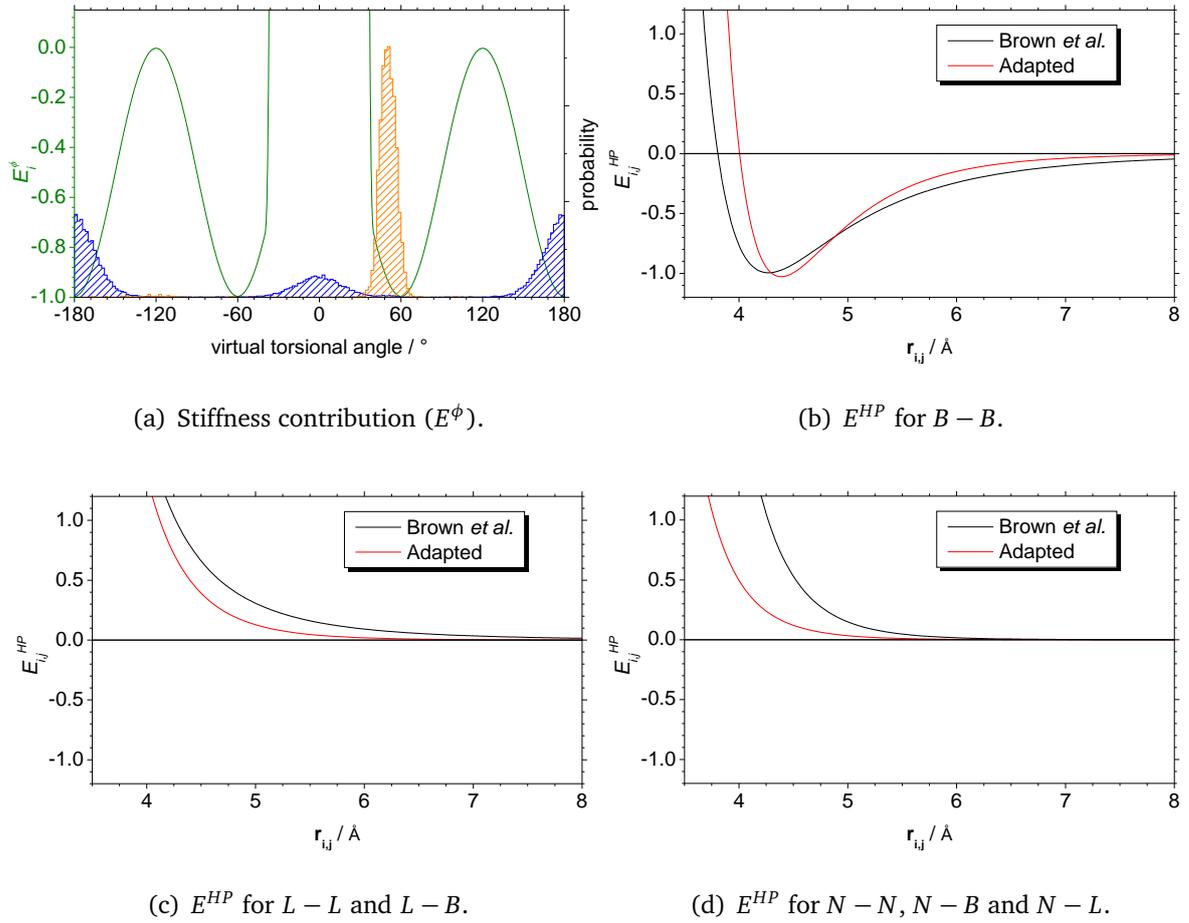

(a) Stiffness contribution ($E^\phi$).

(b) $E^{HP}$ for $B - B$.

(c) $E^{HP}$ for $L - L$ and $L - B$.

(d) $E^{HP}$ for $N - N$, $N - B$ and $N - L$.

**Figure 5.5:** *Representation of the energetic contributions of the adapted hydrophobic interaction (stiffness contribution, related to the virtual torsional angle $\phi$, and sequence-dependent long-range terms).*

| Type of interaction | $\sigma_{int}/\text{Å}$ | $S_1$ | $S_2$ |
|:---:|:---:|:---:|:---:|
| $B - B$ | 4.00 | 15.45 | 1 |
| $L - L$ or $L - B$ | 3.00 | 15.45 | -1 |
| $N - L, N - B$ or $N - N$ | 3.00 | 15.45 | 0 |

**Table 5.3:** *Values of the parameters of the sequence-dependent terms of $E^{HP}$. The right hand side contains the values that $\sigma_{int}$, $S_1$ and $S_2$ take depending of the kind of non-local interaction (note that the two latter quantities are adimensional).*

structural features and thermodynamic properties, such as a proper obtention of helical and $\beta$-type structures in terms of concentration, the acquirement of reasonable folded structures for protein sequences, etc. We have found that the most suitable weights are $\omega^{hb} = 9.5$, $\omega^{\phi} = 7.0$ and $\omega^{HP} = 6.5$. Now, let's start with the discussion of the Results.

## §5.3   Results

In first place, we must stress again the implications of using a simplified hydrophobic potential. On one hand, it shares the $\alpha$-carbon representation that characterizes our hydrogen bond potential and allows extensive simulations at a moderate computational cost; on the other hand, it reduces the detail of real hydrophobic interactions, using a three-letter code and lacking information about the side chains of the residues. Therefore, we lose some detail but can still undertake a qualitative analysis on peptides and proteins with very regular designed sequences.

Once we have set up the scope of our approach, we have tackled the competition between folding and aggregation using a two-step strategy: first, we have studied the behavior of peptides with different sequences (either $\alpha$-prone or $\beta$-prone). This has let us analyze the impact of sequence in our results by comparison with the study of sequenceless peptides in Section 3.3.3.

In second place, we have dealt with the study of complete proteins having different (designed) sequences and, therefore, different folded shapes (i.e. all-$\alpha$, all-$\beta$ and $\alpha + \beta$). Thanks again to the evaluation of peptide behavior, we have identified the optimal number of independent peptides that, in our model, are stable together for a given sequence, providing therefore a valuable hint for the design of complete proteins. After that, we have carried out a thermodynamic analysis of these designed proteins in infinite dilution conditions (necessary to guarantee the appropriate stability of the protein itself). Finally, we have performed multichain simulations, evaluating the propensity to aggregate in terms of concentration, temperature and sequence.

### 5.3.1 Peptide systems

The study of peptides in terms of simple potentials (such as the one we are using here, just based on hydrogen bonds and hydrophobics) has received a remarkable attention from the scientific community.[79,85,223,224] In this Section, we have followed a very similar strategy to Section 3.3, analyzing the structural and thermodynamic performance of our systems (that, in this case, contain six polypeptide chains) in terms of the temperature of the system and its concentration.

Our main interest in the current study is, then, the evaluation of the effect of sequence in the competition between folding and aggregation, given we already know the underlying effect of the hydrogen bond interactions for these systems. For this reason, we will start with the study of peptides in different concentration conditions, aiming to build a structural phase diagram analogous to the one in Figure 3.21. In this case, we have expressed the system concentration in residue moles per liter, as the use of different chain lengths impedes a straightforward comparison in terms of chain moles per liter. In any case, the concentrations simulated in this Chapter are similar to the ones of Chapter 3.

We have selected two different types of sequences to simulate $\alpha$-forming and $\beta$-forming peptides, inspired by previous efforts found in the literature.[110,111] Note that the used sequences are very simple and extremely regular (due to the three-letter code of the hydrophobic definition), as the desired structures will be achieved by a suitable combination of residues "that tend to be buried in the structure" ($B$) and residues "that repel each other" ($L$). We have also adapted the peptide lengths to the type of sequence. Therefore, helical-prone peptides are 13 residues long, and $\beta$-prone peptides have 9 beads.[12] In this way, our helical sequence is $L_2B(BL_2)_2B_2L_2$ and the $\beta$-type one is $(LB)_4L$. Note that both types of peptides present an odd number of beads in order to have symmetric sequences where the first and last residues have a polar ($L$) nature.

**Helical-prone systems**

In first place, we show the results of our simulated systems with helical-prone sequences. Similarly to Section 3.3, we have performed thermodynamic and structural analyses on our simulations. The former are summarized in Figure 5.6, where we show the heat capacity curves versus temperature (in reduced units) at different concentrations.

At first sight, they all present a similar shape, with an energetic transition around $T^* = 2.3$; in the highest concentration case, we find an additional low temperature peak that disappears with the concentration decrease. Does it mean that concentration has little effect in helical systems?

We have looked into the structural characteristics of the different systems, classifying the observed structures in terms of the number and type of hydrogen bond interactions, the radius of gyration of the independent chains and the distances between their centers of mass. We have found three main types of structures (aside from the essentially unstructured configurations at high temperature), that are represented in Figure 5.7. For the sake of clarity, we have colored each type of residue in a different way, using purple for hydrophilic (or polar) residues ($L$) and green for hydrophobic ones ($B$).

The first shown structure, in Figure 5.7($a$), is a snapshot of two three helix bundles. As it can be observed, the six chains of the simulated system are forming $\alpha$-helices thanks to the hydrogen bond interaction. They interact among them three by three, due to the hydrophobic potential. Note, besides, that hydrophobic residues are completely buried in the structure, while polar ones are exposed.

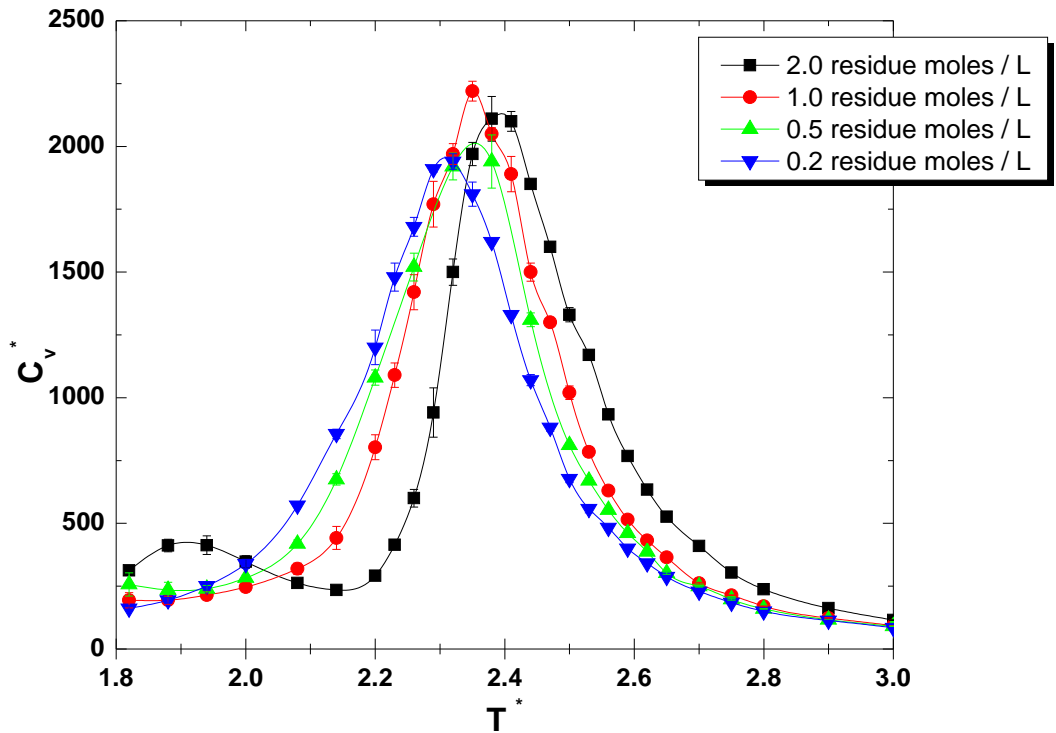

**Figure 5.6:** *Heat capacity curves vs temperature for systems with helical-prone sequences at different concentrations (see legend for details) using the combined potential from hydrogen bond and hydrophobic interactions.*

In Figure 5.7(*b*), we show a kind of $\beta$-type structure, where the main structural item is a $\beta$-sheet. If we look carefully at this structure, we observe that the hydrogen bonds are correctly formed between the different strands. The sequence of these peptides creates a certain "accumulation" of exposed hydrophobic residues, that are somewhat "protected" in this particular configuration by their covering with two $\alpha$-helices.

The last observed structure, in Figure 5.7(*c*), corresponds to the individual $\alpha$-helices, similarly to the ones found in diluted systems in Chapter 3.

After the identification of the structural types that are present in our simulations,

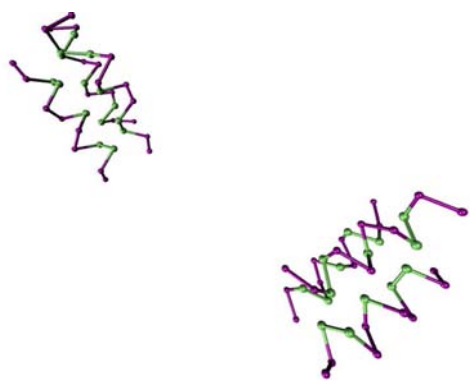

(a) α-helix bundles.

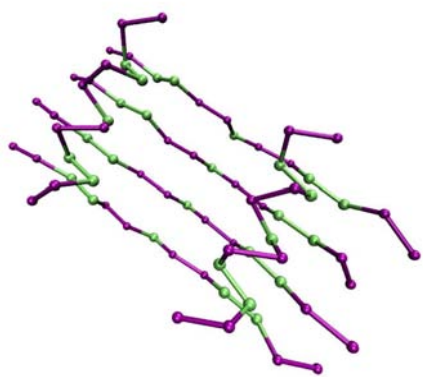

(b) β-type structure

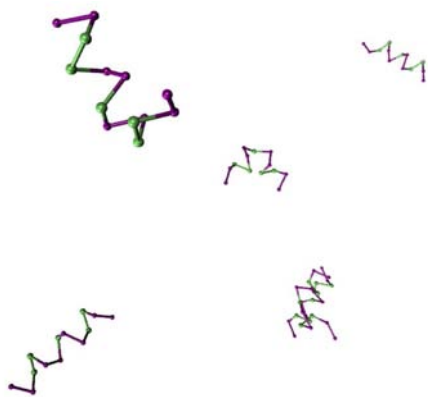

(c) Independent α-helices.

**Figure 5.7:** *Cartoons of the different types of structures found in systems with helical-prone sequences. Structures represented using VMD.[144] In these drawings, B residues have been colored in green and L ones in purple.*

we have evaluated their population as a function of the system temperature, for each of our computed concentrations, as it is shown in Figure 5.8.

If we start with the most diluted system, in Figure 5.8(*a*), we observe that the stable structure at low temperatures is the three helix bundle. When reaching the unfolding temperature bundles are lost, populating an intermediate state where the isolated helices remain stable in a narrow temperature range. If we compare this situation with the diluted system of sequenceless peptides –see Figure 3.18(*a*)–, we conclude that the inclusion of hydrophobic interactions favors the interactions among chains at low and moderate temperatures, forming one of the characteristic arrangements of helical fragments in real proteins, such as three helix bundles.[1]

When we increase the concentration of the system, as it happens from Figure 5.8(*b*) to Figure 5.8(*d*), we observe two effects. Firstly, the reduction of the population of isolated helices at the transition temperature. In second place, the increase in the population of $\beta$-type structures. It confirms that, even in the case of helical-prone sequences, high concentration conditions promote the formation of multichain structures, stabilized by interchain hydrogen bond interactions. This feature is tightly linked to the formation of aggregates in real systems.[11] Compared to sequenceless peptides, the presence of a helical sequence is clearly reflected in high concentration conditions by the different features of the $\beta$-type structure and the fact that there is still some helical structure in the surroundings of the transition temperature.

As we did in sequenceless peptides, we have completed the study of systems with helical-prone sequences with the calculation of a structural phase diagram, shown in Figure 5.9. In order to decide the predominant features at each temperature and concentration, we have considered that there is a significant population of a given structure if its value is greater than a 20%.

The structural analysis has revealed, then, that temperature and concentration changes do affect the thermal equilibrium among the different observed structures in these systems. Therefore, the use of a very simple interaction potential is able to reproduce the main features of more complex descriptions. We have kept the sensitivity of the original hydrogen bond model towards temperature and concentration modifications, while the hydrophobic interaction has successfully modulated the response of the system according to its sequence.

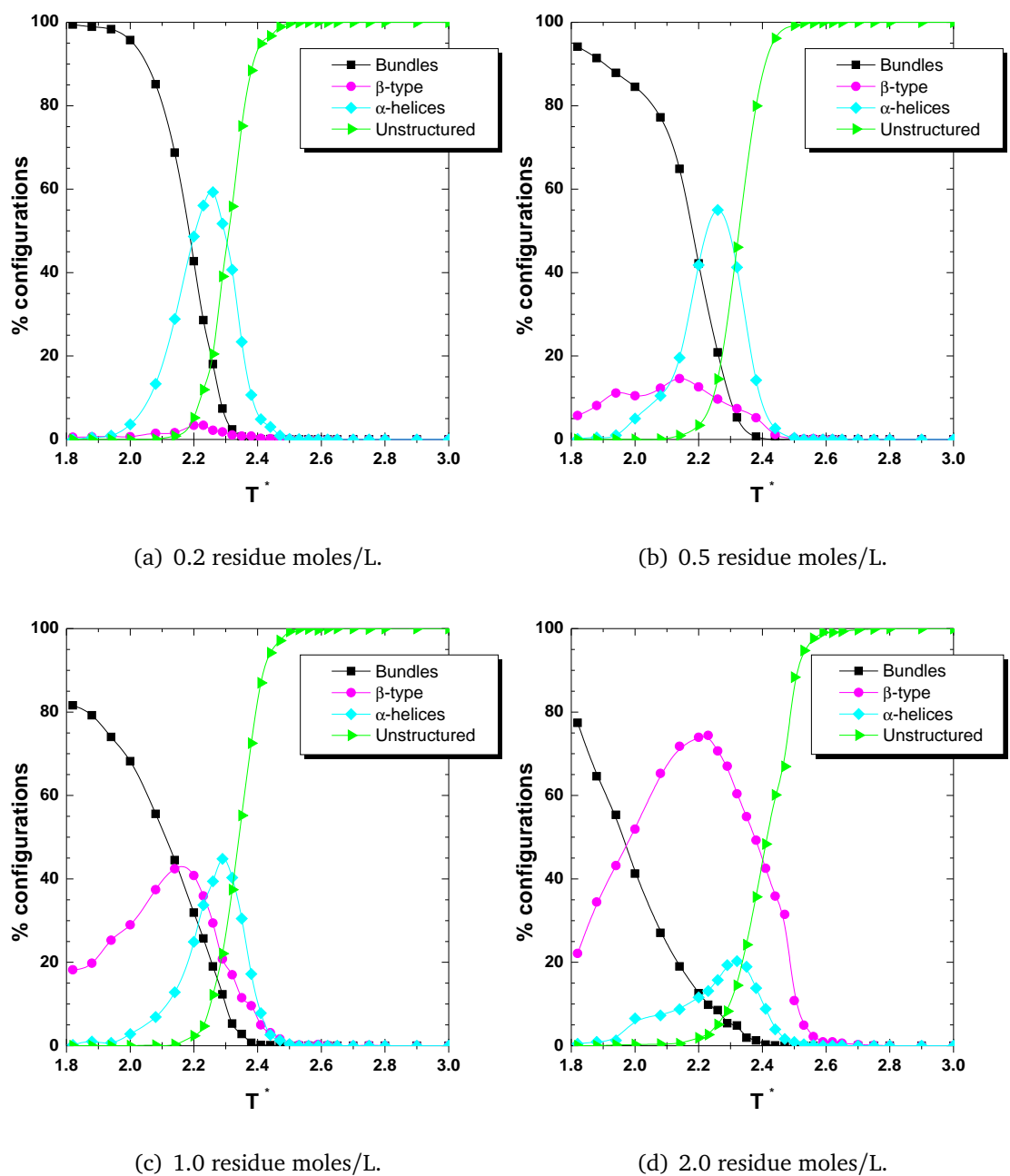

(a) 0.2 residue moles/L.

(b) 0.5 residue moles/L.

(c) 1.0 residue moles/L.

(d) 2.0 residue moles/L.

**Figure 5.8:** *Thermal evolution of the population of the observed structures in systems with helical-prone sequences of different concentrations. The referred structures (β-type structures, α-helix bundles and independent α-helices) are plotted in Figure 5.7.*

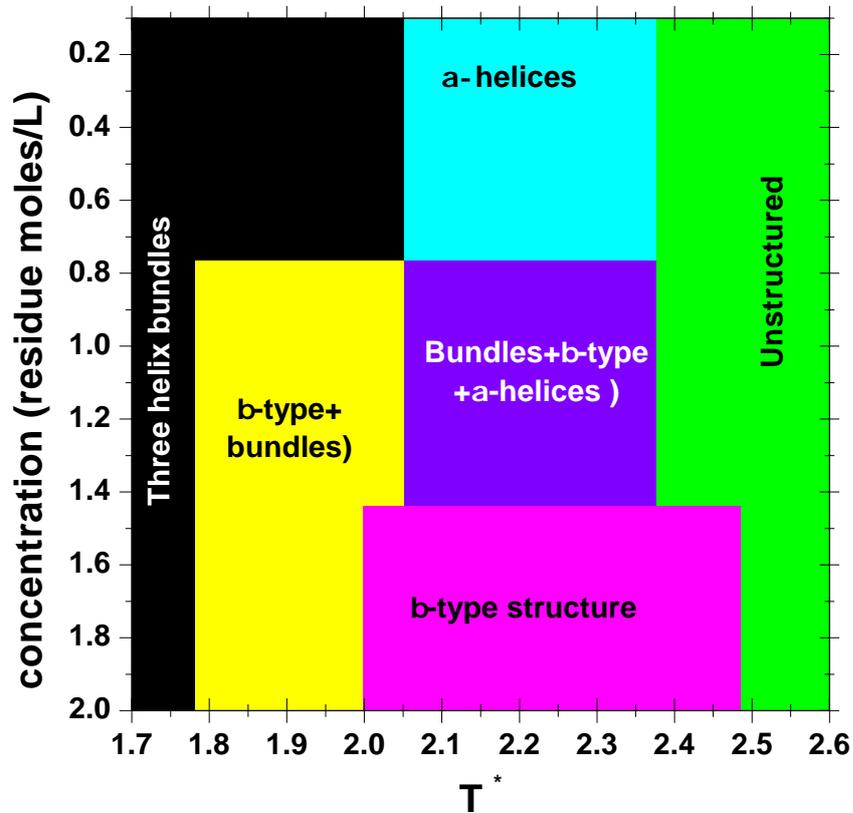

**Figure 5.9:** *Schematic phase diagram for multichain peptides with helical-prone sequences according to our simulation model.*

### $\beta$-prone systems

In second term, we have studied $\beta$-prone systems, performing similar structural and thermodynamic analyses at four concentrations. In this case, we would expect that the tendency towards $\beta$-type structures would result in significant changes in comparison to sequenceless peptides and the former helical systems.

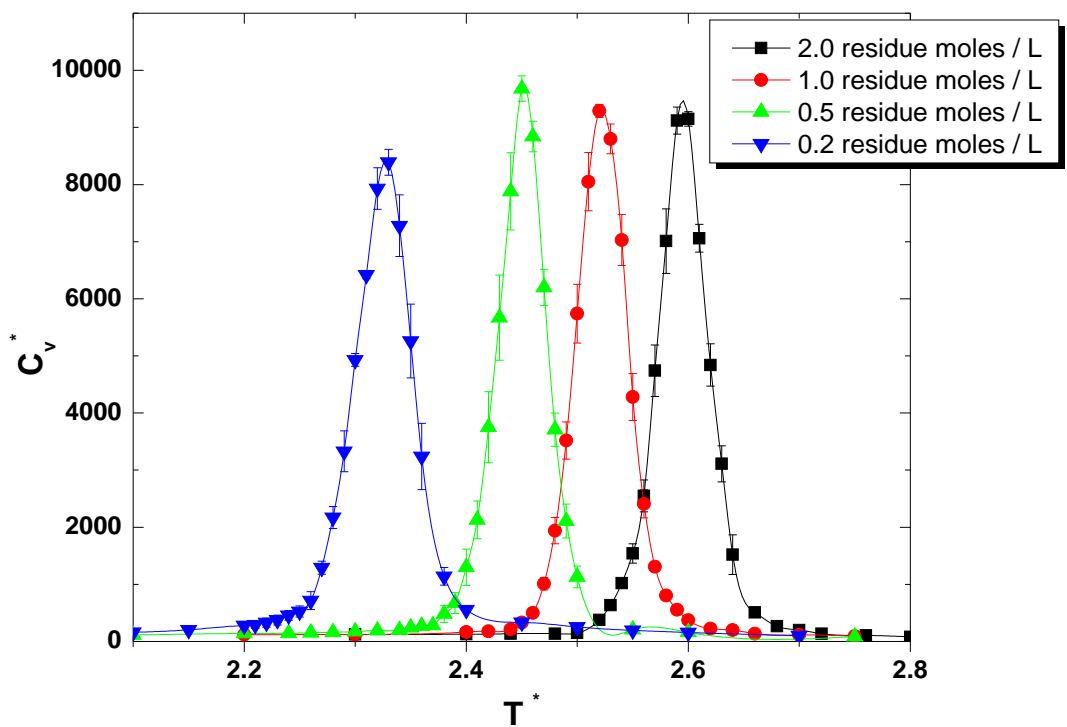

**Figure 5.10:** *Heat capacity curves vs temperature for β-sequence systems of different concentrations (see legend for details) using the combined potential from hydrogen bond and hydrophobic interactions.*

In Figure 5.10, we show the general thermodynamic behavior by means of the evolution of the heat capacity as a function of the system temperature. We observe that the raise in the concentration conditions has a clear effect in these curves, shifting them towards higher temperatures. In addition, the three more concentrated systems seem to have higher and slightly narrower peaks. This indicates from the very beginning that β systems are also sensitive to concentration changes.

The macroscopic source of these changes can be explained through the structural analysis of our simulations, based again in the interactions that are present in each configuration and other general properties like radius of gyration and distances among centers of mass. In Figure 5.11, we show the main types of structural motifs that we have identified in our simulations.

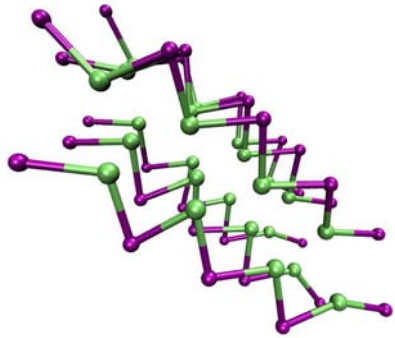

(a) $\beta$-sandwich.

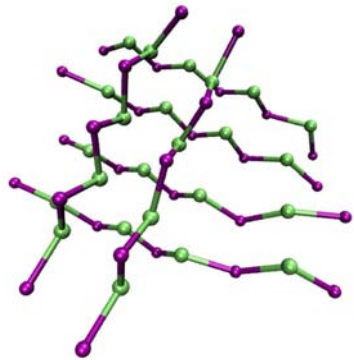

(b) Alternative $\beta$-structure.

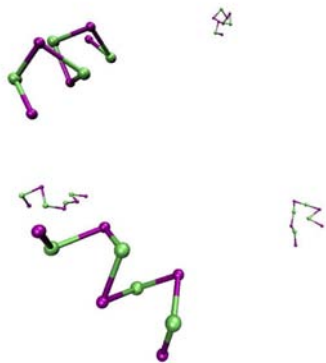

(c) Independent $\alpha$-helices.

**Figure 5.11:** *Cartoons of the different types of structures found in $\beta$-type systems. Structures represented using VMD.[144] Similarly to Figure 5.7, L residues have been colored in purple and B ones in green.*

The first of them, presented in Figure 5.11(*a*), is a $\beta$-sandwich structure. Each sheet of the sandwich, formed by three chains, is stabilized by hydrogen bonds while the two sheets are brought together thanks to the hydrophobic interactions among *B* beads.

In addition, we have found a population of another kind of $\beta$ structure where a four-stranded $\beta$-sheet blocks some of its exposed hydrophobic residues with the other two chains, in a perpendicular conformation. This structure, shown in Figure 5.11(*b*), is energetically very similar to the former one and is a consequence of the simplicity of the applied potential. In our simulations, these two alternatives differ in their structure, but are indistinguishable either by their energy or by their thermal behavior.

Finally, we have found some independent $\alpha$-helices, see Figure 5.11(*c*), that reflect the trend of isolated peptides to get stabilized by the formation of intrachain hydrogen bonds, as it happened in sequenceless peptides.

Regarding the identification of these structures depending on temperature and concentration, we show the related plots in Figure 5.12. Starting again with the most diluted system, we can confirm that the slightly broader heat capacity peak is due to a structural reason: the $\beta$ structures stable at low temperatures (either $\beta$-sandwiches or the alternative $\beta$-type conformations shown in Figure 5.11(*b*), less populated) are lost at the transition temperature, populating a helical intermediate. This situation closely resembles the intermediate concentration systems of sequenceless peptides, where a high population of helical intermediates could be found (see Figure 3.18).

In more concentrated systems, shown in Figures 5.12(*b*), 5.12(*c*) and 5.12(*d*), the helical population is considerably reduced, obtaining a sandwich-coil transition. In addition, it is remarkable that the relative population of the two kinds of $\beta$ structures seems to be independent on the system concentration, predominating the $\beta$-sandwich in all cases.

We have finished our study of $\beta$-forming peptides with its phase diagram, shown in Figure 5.13. We can conclude that the effect of concentration in these systems clearly affects the transition temperature. However, the structural scenario is very similar in all cases, just obtaining a relevant population of independent helices when the system is very diluted.

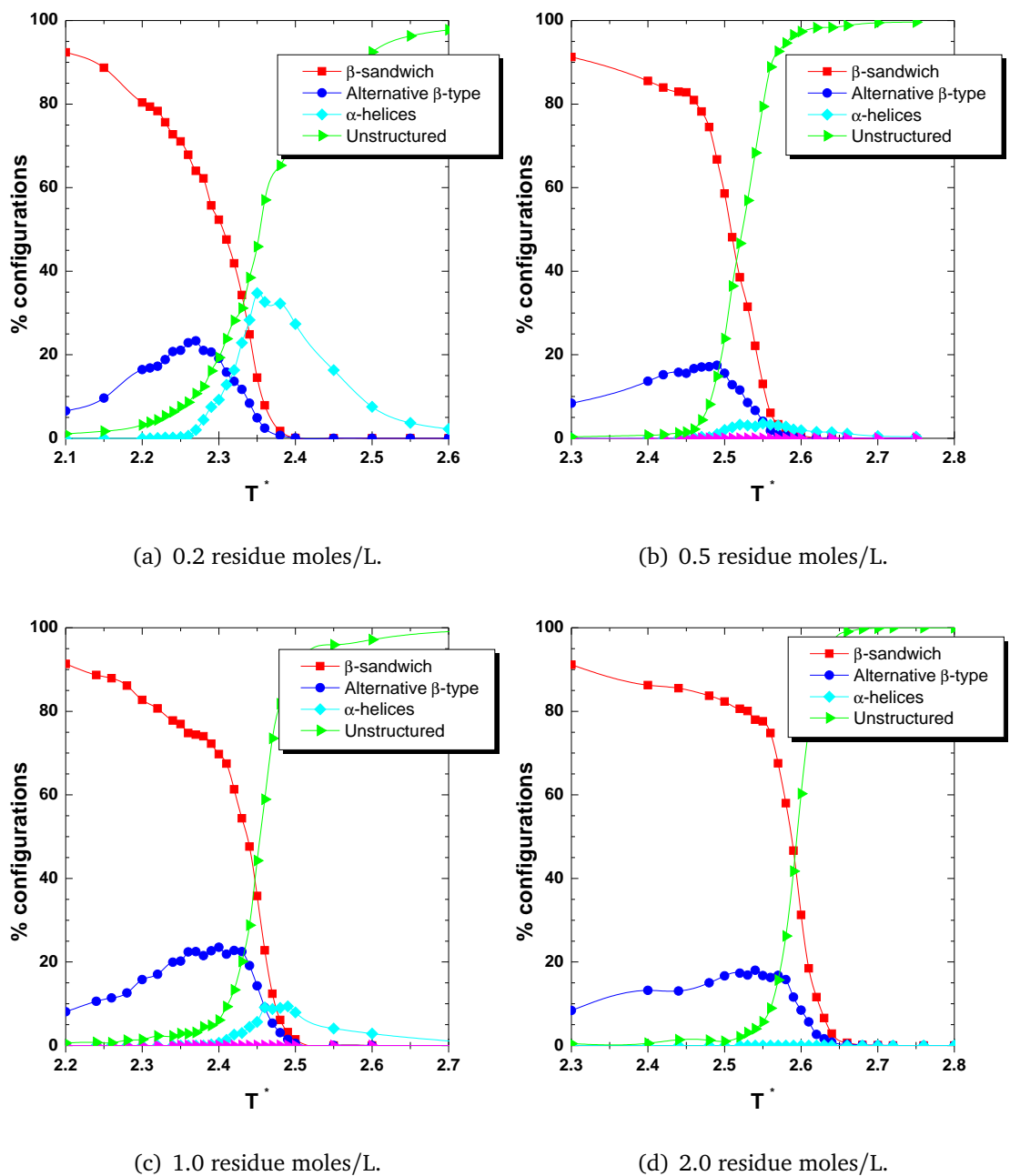

(a) 0.2 residue moles/L.

(b) 0.5 residue moles/L.

(c) 1.0 residue moles/L.

(d) 2.0 residue moles/L.

**Figure 5.12:** *Thermal evolution of the population of the observed structures in β-sequence systems of different concentrations. The referred structures (β-sandwich, alternative β-structure and independent α-helices) are plotted in Figure 5.11.*

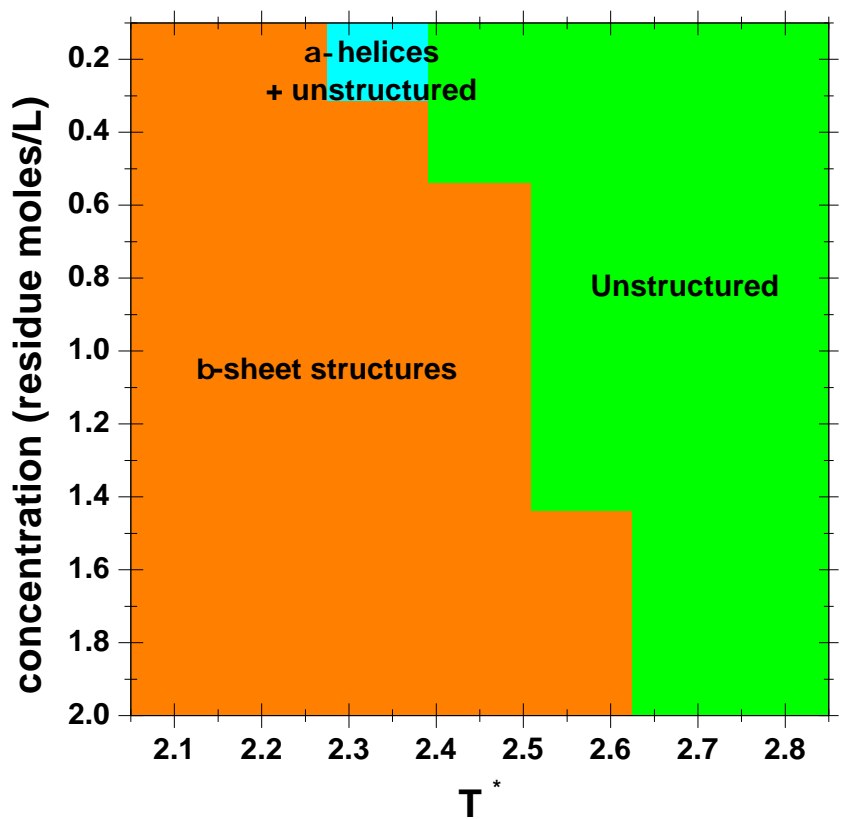

**Figure 5.13:** *Schematic phase diagram for multichain β-sequence systems according to our simulation model.*

At this point, we have shown the differential aspects of sequenced peptides (either α or β-prone) in relation to the sequenceless peptides of Chapter 3. Therefore, if we compare Figures 3.21, 5.9 and 5.13, we can note the same underlying behavior (i.e. helices can convert into sheets at high concentration and vice versa at low ones), nicely modified by the effect of hydrophobic interactions. Once we have checked that our hydrophobic interaction properly modulates the system behavior, we can tackle the study of complete proteins.

## 5.3.2   Complete proteins

The study of full proteins using minimalist potentials like ours must be preceeded by a careful design of the protein itself. We must decide the general kind of structure we want to obtain (i.e. all-$\alpha$, all-$\beta$ or $\alpha + \beta$). After that, we must find a sequence that folds into this structure, taking into account the limitations imposed by the simplicity of the model.

The design of the sequence, then, is divided into two tasks: the choice of the secondary structure elements and the optimization of the packing of the complete protein, in order to bring together these secondary structure elements. Each element is designed in a similar way as in peptide systems (see the previous Section and Figure 5.2), but the exact sequences have been slightly tuned so that their hydrophobic faces can be placed together. Note that in this case the number of possibilities grows, as the covalent linking of the different secondary structure fragments may lead to different possibilities in the way they pack.

As the obtention of unique structures is critical in order to relate the behavior of the model to realistic proteins, we have started analyzing how independent fragments interact among them, finding the most stable "clusters" of secondary structure elements according to our potential definition and the used sequences. After that, we have connected them by short linkers that keep the observed packing of independent fragments.

Finally, we have carried out some simulations varying the system concentration, aiming to explore the interplay between folding and aggregation in these symplified systems. We have especially focused on the conditions for finding aggregates in terms of the system concentration. Thanks to the simulation of three systems with different structural properties, we have analyzed the effect of sequence in the formation of these structures (e.g. type of multichain association, relationship between $\beta$-propensity and sequence, etc.).

For the sake of clarity, we have split this Section in three blocks depending on the structural family of the designed protein.

**All-$\alpha$ proteins**

Thanks to the previous study of helical peptides, discussed in Section 5.3.1, the design of our helical protein has been relatively straightforward, as we already have some information of how helical fragments associate. We have found that the most stable association of $\alpha$-helices within our model is the three helix bundle, as we have confirmed in simulations with eight and twelve chains in a simulation box (data not shown). It is also a very general feature in real proteins.[1]

We have linked the helical fragments through turns (formed by four residues) that have neutral ($N$) nature according to our three-letter hydrophobic code. The inclusion of turns has forced a slight modification in the sequence of the helical fragments. They all follow the same pattern (see Figure 5.2), but the positions of the $B$ and $L$ residues have been shifted so that the burying of the hydrophobic ones ($B$) place the ends of the helical fragments as close as possible. Therefore, our resulting sequence is $L_2BL_2B(BL_2)_2B$ $N_4LBL_2B(BL_2)_2B_2$ $N_4L_2B(BL_2)_2B_2L_2$, whose stable structure is shown in Figure 5.14.

Regarding its thermodynamic stability, we have represented its heat capacity curve in Figure 5.14. As it can be observed, we find a single and relatively narrow peak that indicates that the all-$\alpha$ protein has a quite cooperative transition. We have also performed a structural analysis for these simulations, concluding that the three helix bundle loses its structure at the transition temperature without any perceptible intermediates (data not shown). Therefore, we have succeeded in the design of this protein, finding a single folded structure that experiments a cooperative folding process.

Afterwards, we have simulated two-chain systems under different concentration conditions. We have obtained the heat capacity curves of Figure 5.15. Similarly to helical-prone peptide systems, these curves show slight changes in shape in comparison to the infinitely diluted system of Figure 5.14.

The structural analysis reveals that these systems are sensitive to changes in concentration. Apart from individually folded structures and unstructured configurations, we have found very frequent multichain structures along our simulations. These interchain associations lead to two main types of situations. In first place, we can find structures where a domain swapping has occurred, such as the one represented in Figure 5.16(*a*). In this cartoon, one of the α-helices of each chain has separated from its compact shape, interchanging its position with the analogous helix of the other chain.

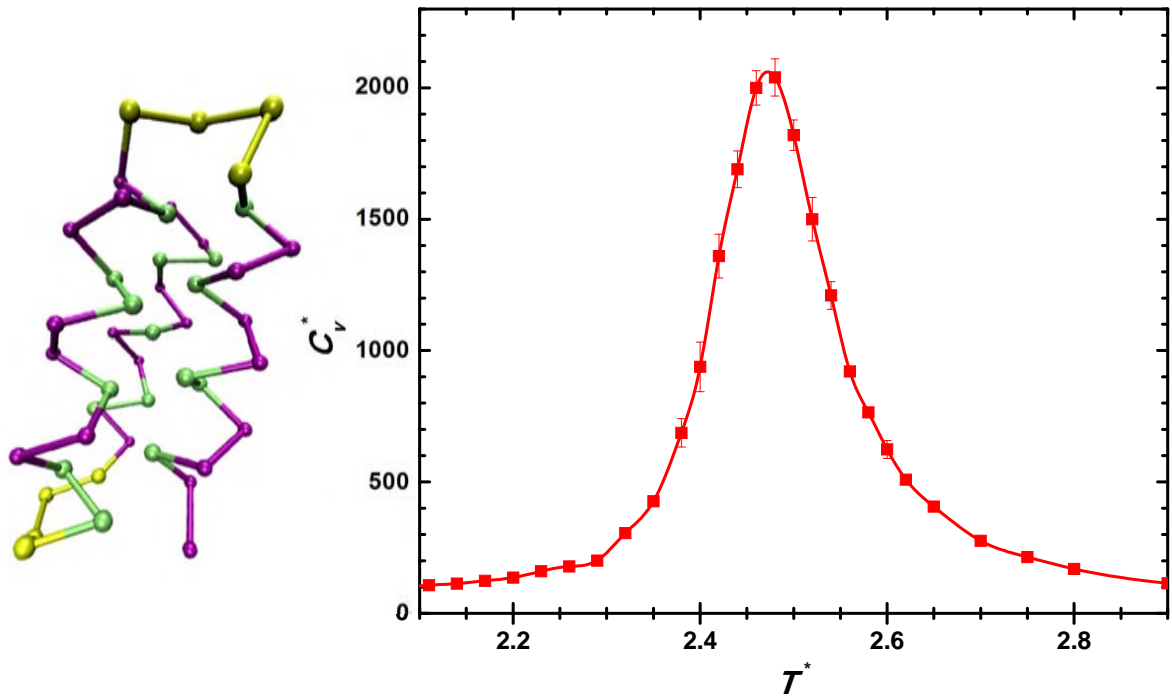

**Figure 5.14:** *Heat capacity curve vs temperature for our helical protein in infinite dilution conditions. A cartoon representation of the folded structure has been included, drawn with VMD.[144] The color code is kept constant during all this Chapter: B residues are colored in green and L ones are purple; N ones are represented by yellow spheres.*

In second term, we find an alternative multichain possibility, labeled as "aggregate" in Figure 5.16(*b*). As we can see, each chain keeps some of its helical structure but also forms interchain interactions that can be ascribed to the formation of long range hydrogen bonds (see the small β-sheet in the middle of the drawn structure) and hydrophobic interactions. These aggregates are characterized by their high mobility, as their interactions are broken and created again many times along the simulation, leading to an energetically

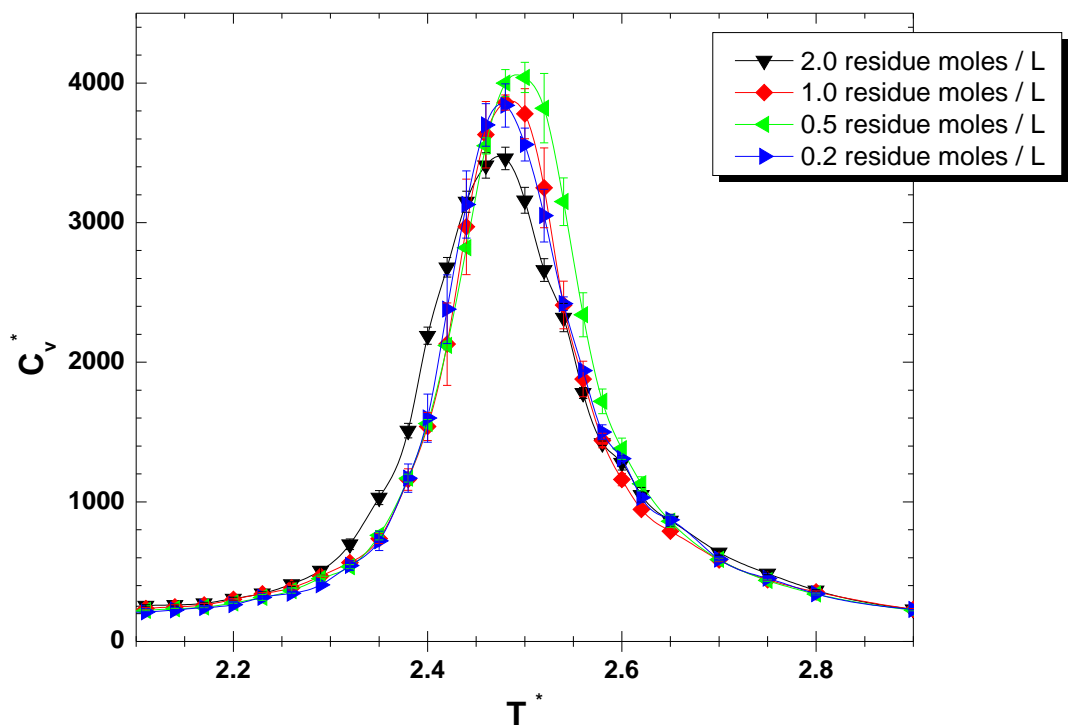

**Figure 5.15:** *Heat capacity curve vs temperature for systems of two all-α proteins under different concentration conditions.*

degenerated set of states without a well-defined structure.

Both structures are said to be involved in aggregation processes and to play a role in the formation of non-functional structures, through different mechanisms.[17,225] This highlights the fact that, even using a very simple model, we can tackle situations that resemble the real ones and allow us to evaluate them in terms of the particular properties of our systems.

We have evaluated the population of the found structures with temperature and concentration, and the results are represented in Figure 5.17. Starting with the two most diluted systems, plotted in Figures 5.17(*a*) and 5.17(*b*), we observe that the independent chains of the system seldom interact between them, leading to the typical folding/unfolding transition of the three helix bundle, similarly to infinite dilution simulations. This means that in both systems the frequency of the interactions among chains is so low that, in general

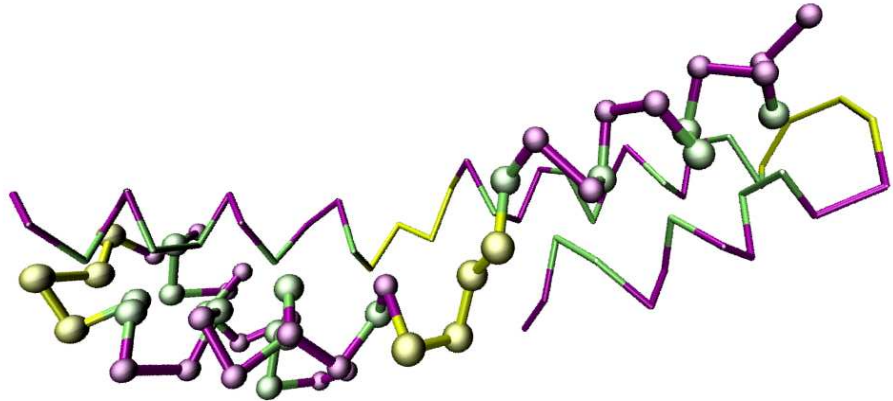

(a) Swapped domains.

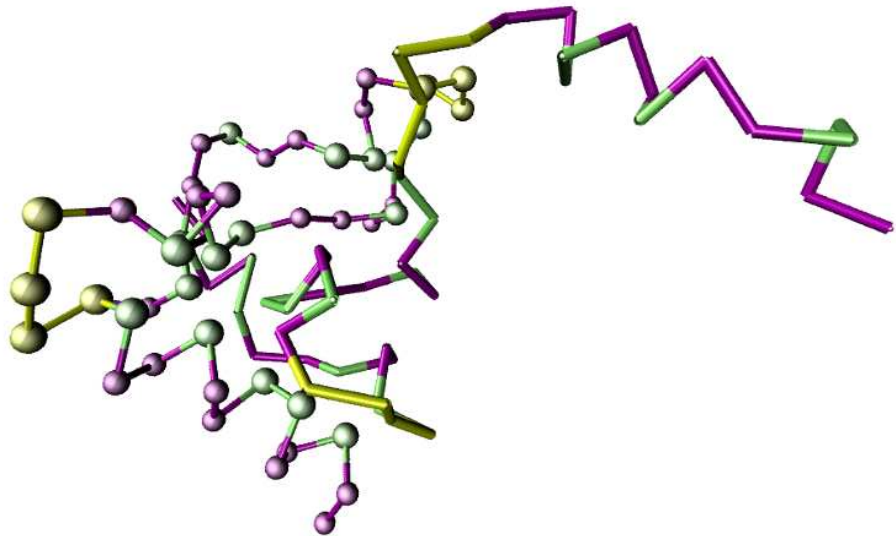

(b) "Aggregate".

**Figure 5.16:** *Cartoons of the different multi-chain structures found in our all-α protein system, where one chain is represented by sticks and the other one with beads and sticks. Structures represented using VMD.[144]*

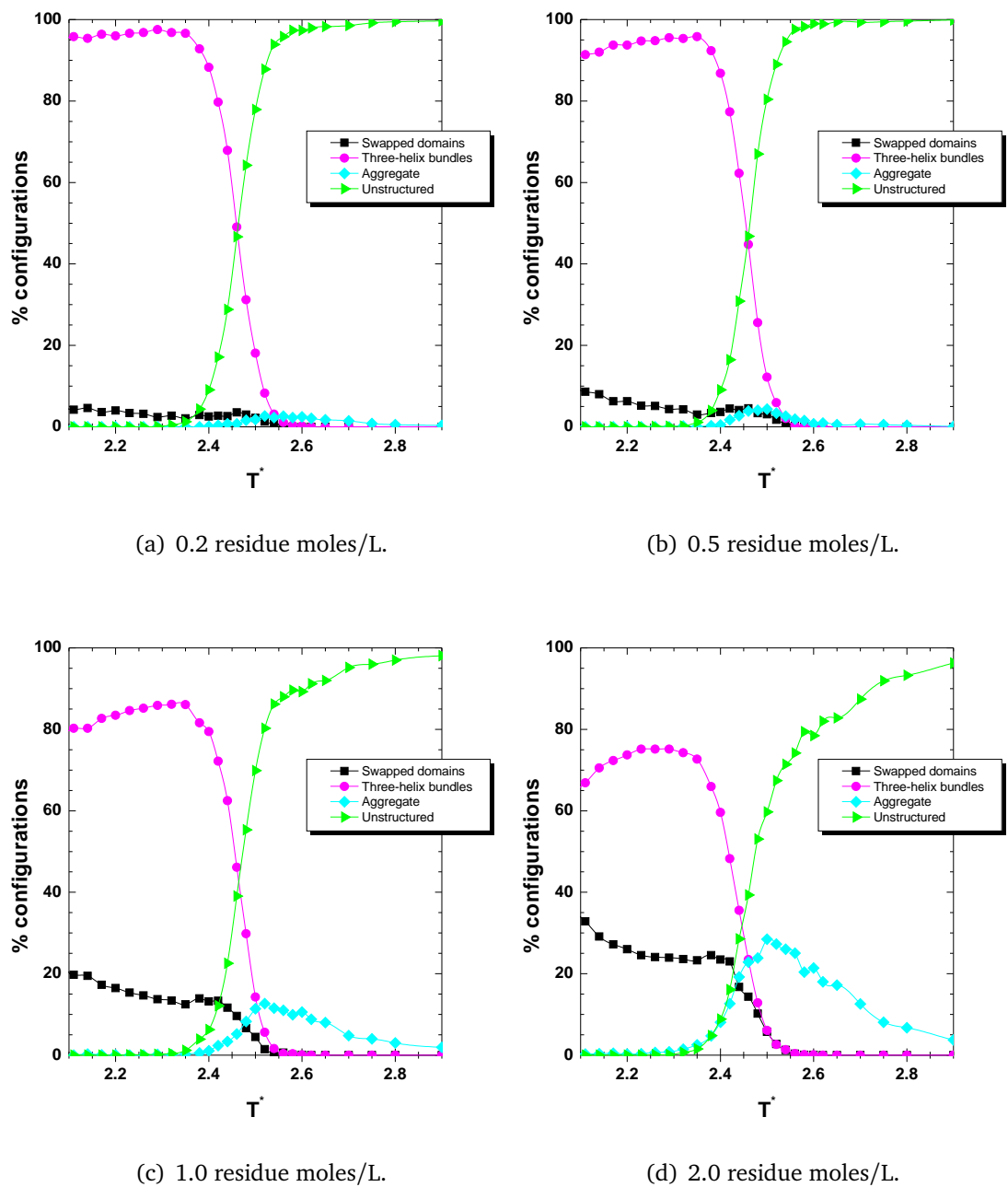

(a) 0.2 residue moles/L.

(b) 0.5 residue moles/L.

(c) 1.0 residue moles/L.

(d) 2.0 residue moles/L.

**Figure 5.17:** *Thermal evolution of the population of the observed structures in all-α proteins at four different concentrations.*

terms, each chain folds independently.

In contrast, the two most concentrated systems, in Figures 5.17($c$) and 5.17($d$), present a higher population of the two multichain structures we commented on in Figure 5.16. In this way, a perceptible population of swapped structures is present below the transition temperature; just above it, we find the "aggregated" $\beta$-type structures competing with unfolded structures. Obviously, the presence of these structures gets higher when the concentration is increased.

These non-specific or partially $\beta$-type collapsed structures can be linked to the natural trend of any kind of protein to aggregate at the transition temperature through partially folded intermediates if the concentration is high enough.[17] Besides, a reduction in temperature lets us observe a different kind of multichain structures (also related to aggregation processes), probably mediated by the very regular helical sequence.

As a result, we can conclude that, in spite of the simplicity of our interaction model and the regularity of our sequences, we have been able to reproduce some important features of helical proteins through our simplified three-helix bundle protein, such as the independent folding in diluted conditions and the presence of a kind of "aggregates" in highly concentrated systems, formed by the interaction between $B$ beads and the formation of interchain hydrogen bonds or due to domain swapping.

After our all-$\alpha$ protein, let's start a different structural family, also formed by a single type of secondary structure element.

**All-$\beta$ proteins**

In this case, we can also rely on our previous peptide simulations of $\beta$-prone chains. We have observed a remarkable stability for $\beta$-sandwiches where each sheet is formed by three independent chains, being therefore a good candidate for the building of a complete protein.

In first place, we have checked if systems with a larger number of chains (eight and twelve chains per system) lead to the same or different stable sandwiches. At low temperatures, we have found that sandwiches with more strands per layer are stabilized; however, in the naturally relevant temperature range these larger sandwiches split, mainly populating the $(\beta_3 + \beta_3)$ one.

We have kept this small structural feature for the design of the all-$\beta$ protein, linking the independent fragments by short neutral turns (formed by three beads each). Therefore, we have used the following sequence, where each $\beta$-strand is equal to the rest: $[(LB)_4LN_3]_5(BL)_4L$.

Similarly to the all-$\alpha$ protein case, we have started by checking the thermal stability of the complete protein. We have evaluated the impact of having this extremely regular sequence, as well as the conformational constrain that turns insert in the protein compared to independent fragments. We have found that our potential does not find a unique folded structure for this sequence, but two alternative ones (shown in Figure 5.18) in proportion 85:15. This fact is not surprising, as two different $\beta$-type structures are also found in $\beta$-peptides, as shown in Figures 5.11($a$) and 5.11($b$). Nevertheless, the heat capacity curve presents good thermodynamic properties, exhibiting a single and relatively sharp peak at the transition temperature, which means that these two structures are energetically indistinguishable.

Therefore, we have continued with this all-$\beta$ protein, performing two-chain simulations at four different concentrations, obtaining the heat capacity curves of Figure 5.19 that, as always in this work, will inform about the stability of the systems and their energetic (and probably also structural) transitions. We observe that the heat capacity peaks slightly vary with concentration, shifting towards higher temperatures when the concentration is increased. Therefore, the concentration seems to have some effect on all-$\beta$ proteins.

The explanation of these changes is linked to the different structural situations that have been detected in these systems. They can be divided in unstructured configurations, independent folded chains (with one of the two aforementioned structures) or interchain associations. We have found two different kinds of associations, represented in Figure 5.20.

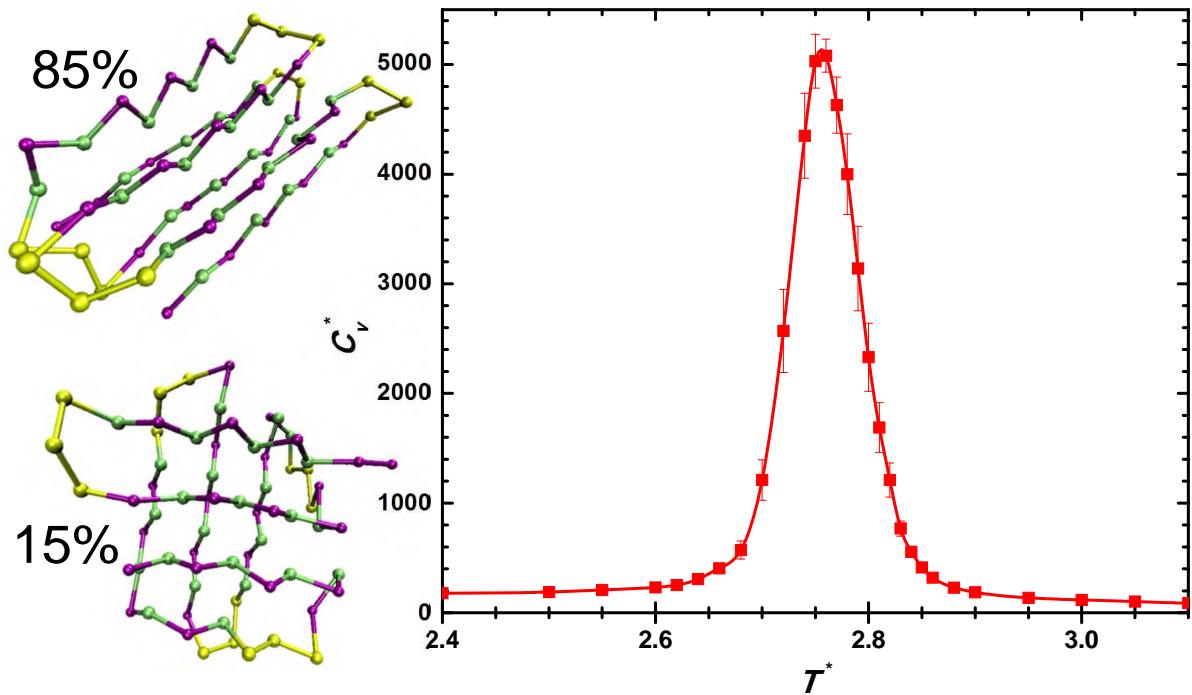

**Figure 5.18:** *Heat capacity curve vs temperature for our all-β protein in infinite dilution conditions. A cartoon representation of the folded structures has been included, drawn with VMD.[144] The percentages indicate their relative populations at low temperatures.*

They basically reflect the dual nature already commented on. Some structures keep the β-sandwich scheme, as exemplified in Figure 5.20(*a*). Others tend to cross some strands in relation to the others, creating a sort of "honeycomb" similar to the one in Figure 5.20(*b*). In both cases, the structures are stabilized by long range (both inter and intra chain) hydrogen bonds and hydrophobic interactions. In this case, it is not possible to distinguish between domain swapped structures and "aggregated" ones, as the individual folded structure also presents some structural variability.

The evolution of these structures with temperature, shown in Figure 5.21 for the different concentrations, explains the thermal behavior of the simulated systems. In these plots, we have just distinguished between independently folded structures (regardless the structure itself), those with associated chains (regardless its β-sandwich or honeycomb interaction) and unstructured configurations.

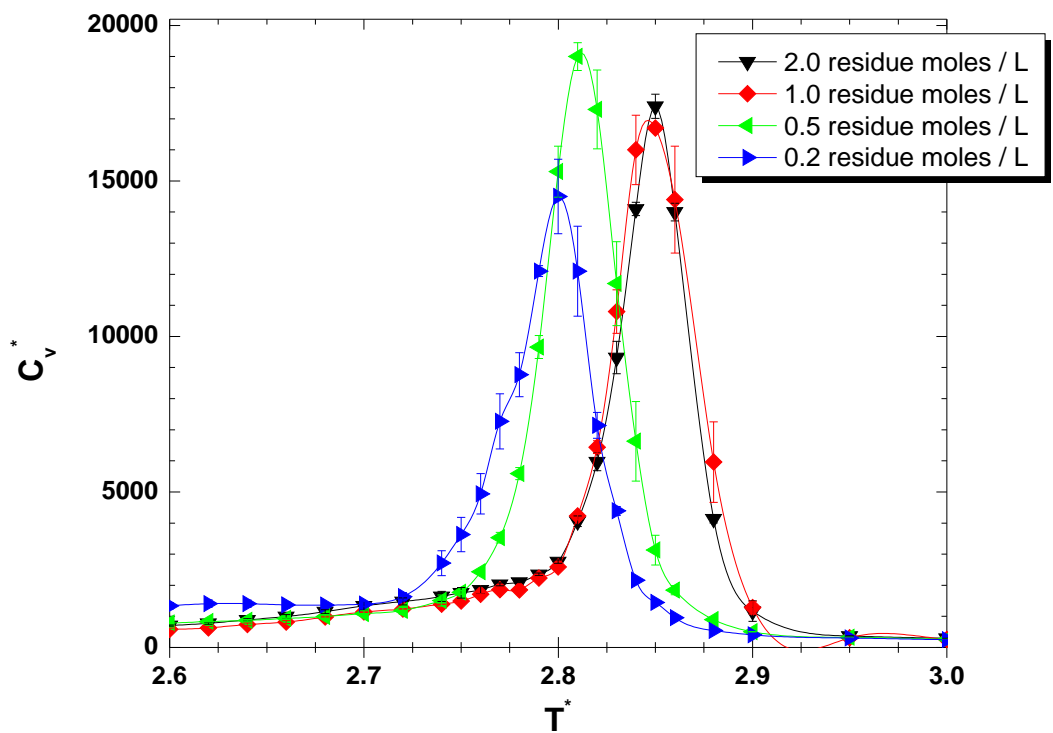

**Figure 5.19:** *Heat capacity curve vs temperature for systems of two all-β proteins under different concentration conditions.*

In all these systems, we observe that the most stable structure at low temperature is the associated one. A low concentrations (see Figure 5.21(*a*)), however, there is a drastic decrease of its population in the surroundings of the transition temperature, predominating the independent folded proteins below the transition temperature (i.e. in the relevant temperature range).

In the other systems, the population of independent proteins get remarkably lower at any temperature. In these cases, the *β*-type associated structures predominate in the full temperature range. Again, we have obtained a different behavior of the system in terms of its concentration. In this case, we have found a kind of "aggregates" that are stabilized by interchain hydrogen bonds and hydrophobic interactions.

As a result, we can say that our all-*β* protein presents a higher tendency to aggregate in comparison to the all-*α* one. As a matter of fact, the propensity to aggregate

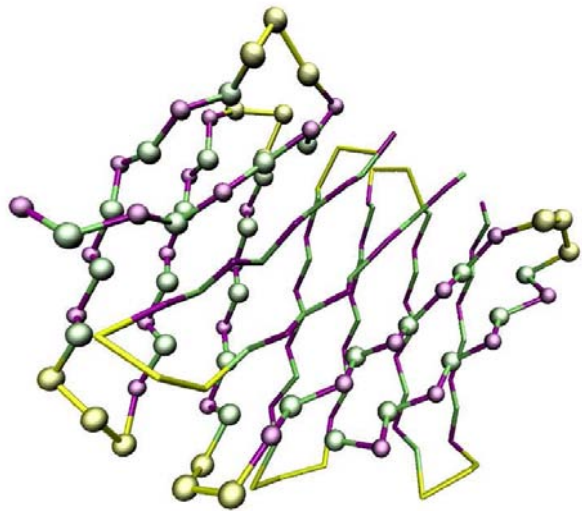

(a) Mixed $\beta$-sandwich.

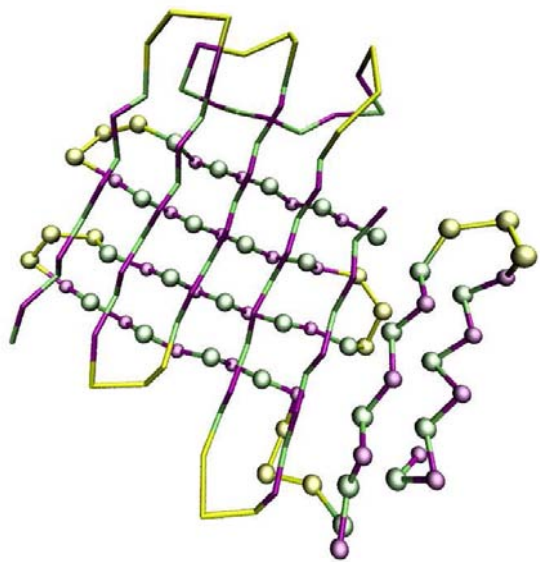

(b) "Honeycomb".

**Figure 5.20:** *Cartoons of the different multi-chain structures found in $\beta$-type proteins. Structures represented using VMD.[144]*

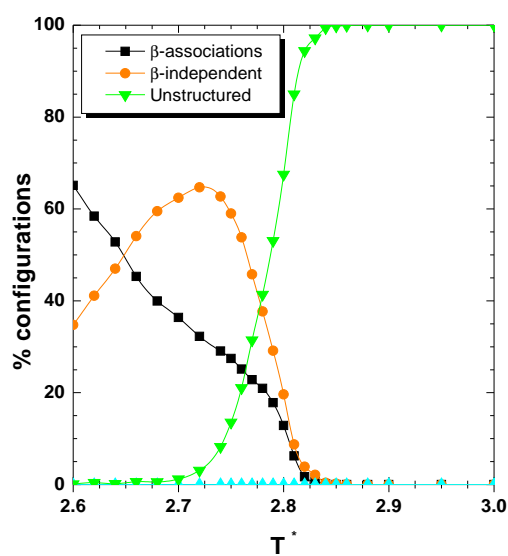

(a) 0.2 residue moles/L.

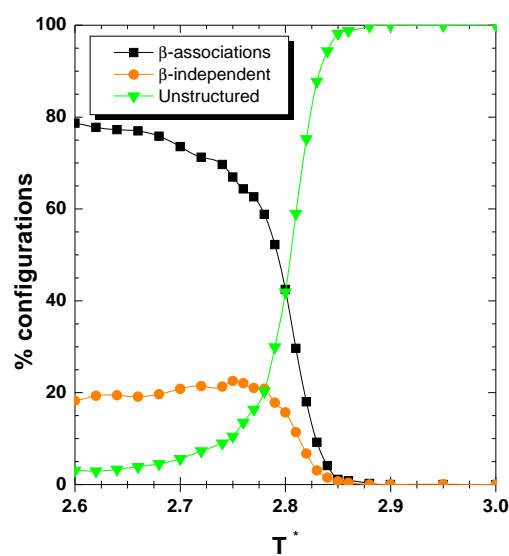

(b) 0.5 residue moles/L.

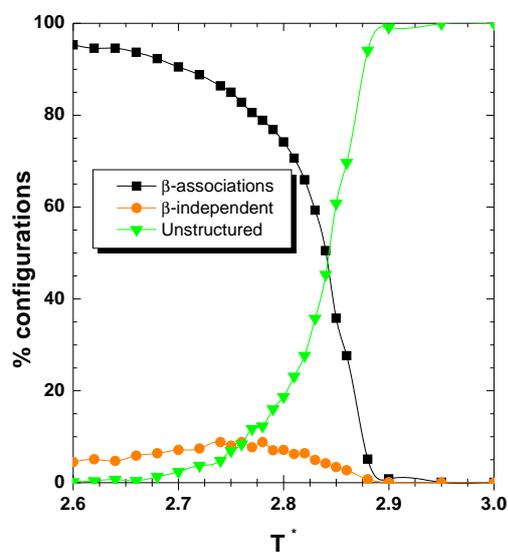

(c) 1.0 residue moles/L.

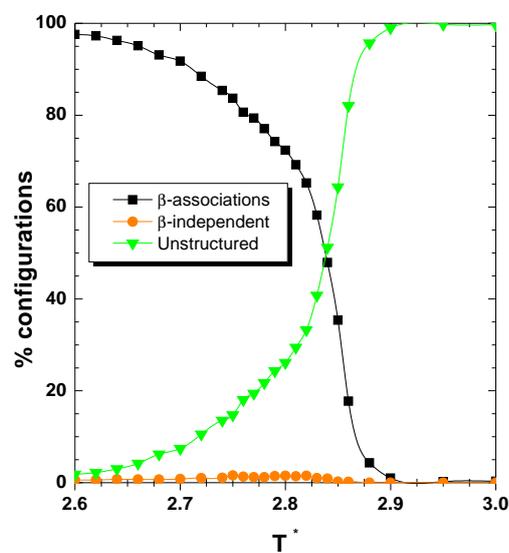

(d) 2.0 residue moles/L.

**Figure 5.21:** *Thermal evolution of the population of the observed structures in all-β proteins at four different concentrations.*

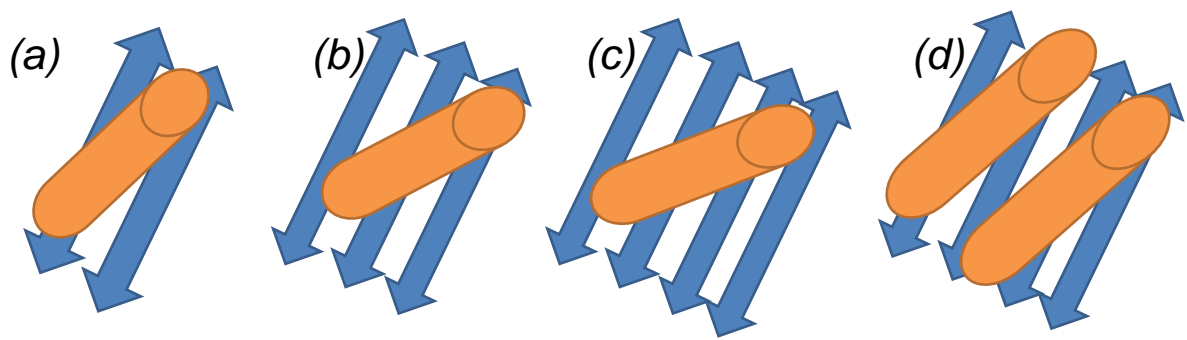

**Figure 5.22:** *Most common combinations of helical and $\beta$-type fragments found in nature, represented by orange cylinders and blue double-headed arrows, respectively. The double arrows indicate the possibility of forming either parallel or antiparallel associations. (a) $\alpha + \beta_2$, (b) $\alpha + \beta_3$, (c) $\alpha + \beta_4$, (d) $\alpha_2 + \beta_4$.*

has been linked to the proportion of alternating hydrophobic ($B$) and hydrophilic ($L$) residues.[226] If we evaluate this proportion in our simulated sequences, it is just 42% in our helical protein, but raises up to 78% in the case of the all-$\beta$ sequence. What would happen if we combine both types of structures into the same protein?

### $\alpha + \beta$ **proteins**

In this case, we do not have previous peptide simulations for $\alpha + \beta$ peptides. Therefore, the first question is: which arrangements are stable according to our model? Although it may seem quite naïve, this question is very relevant for our purposes, as having a stable multipeptide arrangement is the first requirement to undertake the design of a complete protein, at least according to our interaction definition.

We have looked for the most common motifs found in nature,[12] schematically plotted in Figure 5.22, and we have checked if a combination of the independent fragments would be stable in solution according to our model. We have used relatively high concentration conditions in order to favor multichain structures. Our results are summarized in Figure 5.23.

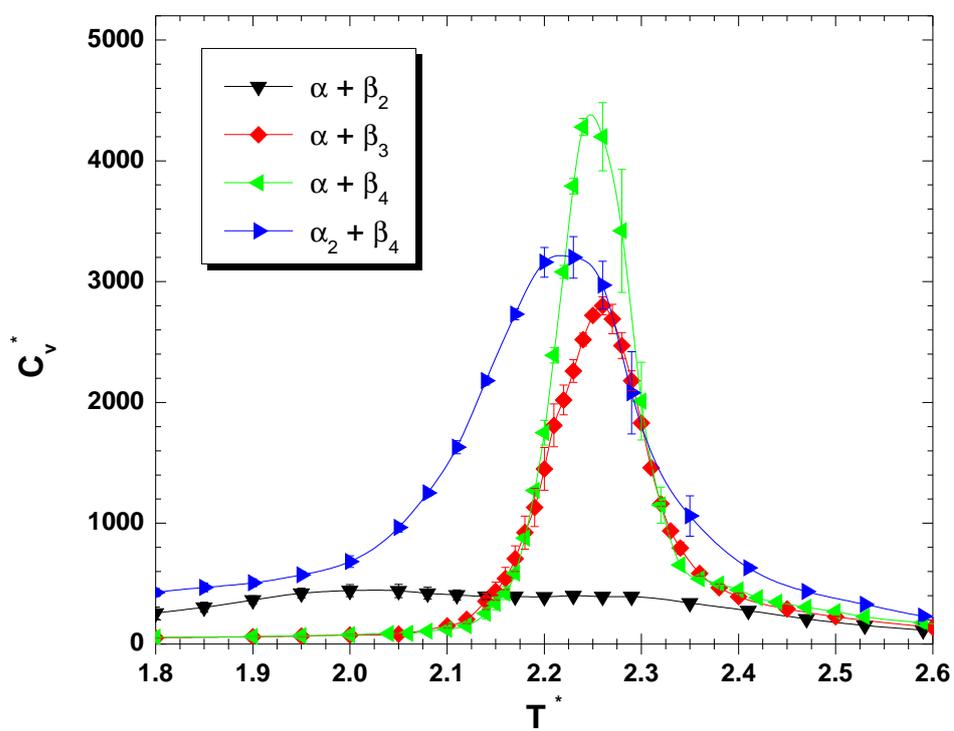

**Figure 5.23:** *Heat capacity curve vs temperature for systems of different combinations of helical and β-peptides.*

The most simple peptide arrangement (one $\alpha$-helix and two $\beta$-strands) is not stable at all, presenting a broad transition where the $(\alpha + \beta_2)$ associated structure is seldom found together (data not shown). The systems with a narrower heat capacity curve present a clearer transition, finding the associated structure below the transition temperature and unfolded structures above it. Therefore, we have concluded that the most stable combination is composed by a 4-strand $\beta$-sheet and one $\alpha$-helix. Linking the different fragments by short neutral linkers, we followed one of the most common arrangements, composed by two $\beta$-hairpins with one $\alpha$-helix in the middle of the sequence. We have used, then, the following sequence: $[(LB)_4LN_3]_2L_2B(BL_2)_2B_2L_2[N_3(LB)_4L]_2$.

The last step in the protein design is the analysis of the thermal stability of our complete protein, that we have illustrated through the heat capacity curve of Figure 5.24. We have found that our desired $\alpha + \beta$ protein is stable below the transition temperature. The height and shape of the heat capacity peak correspond to a quite cooperative folding/unfolding of the structure.

We have continued with the study of our $\alpha + \beta$ protein under different concentrations, using the same four conditions we previously used in our peptide and protein simulations. We show the corresponding heat capacity curves as a function of temperature in Figure 5.25. In these plots, we observe that the heat capacity peak is slightly modified by the concentration conditions. What is structurally happening?

The analysis of the obtained configurations reveals, aside from the independent folded proteins and unstructured configurations, the presence of two different multichain arrangements. The first of them, shown in Figure 5.26(*a*), is a domain swapped structure, where part of the $\beta$-strands of the two proteins have been interchanged. The other multichain situation, presented in Figure 5.26(*b*), corresponds to a completely $\beta$-type aggregate that resembles the one obtained with our all-$\beta$ protein and is far more ordered than the one obtained in the all-$\alpha$ structure.

Therefore, in this case we recover the two different kinds of "aggregates" that we observed in our all-$\alpha$ protein. How are these structures distributed along our simulations?

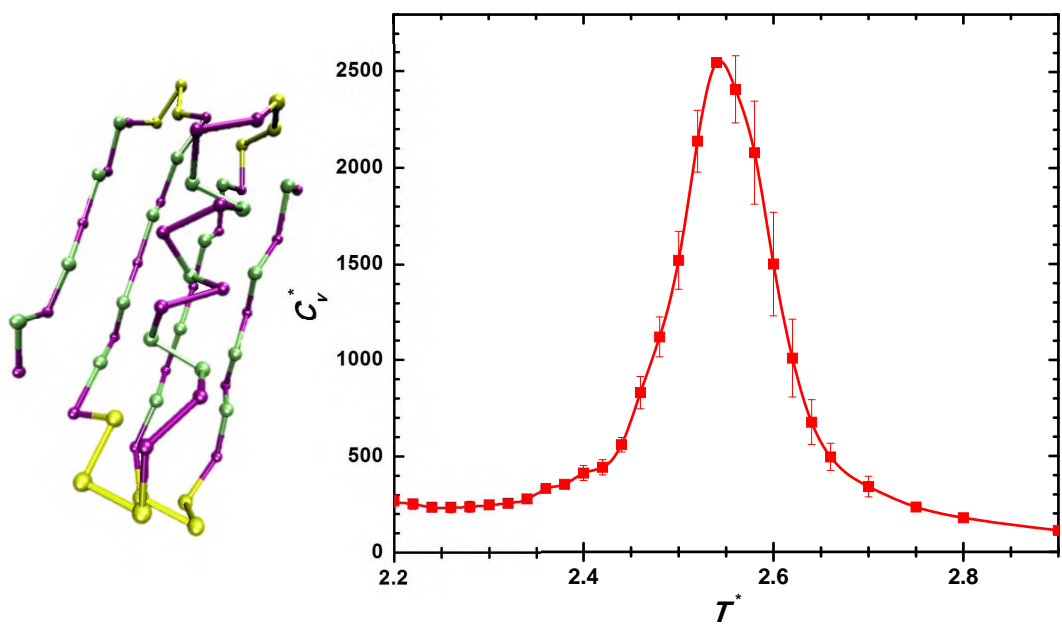

**Figure 5.24:** *Heat capacity curve vs temperature for our designed $\alpha + \beta$ protein in infinite dilution conditions. A cartoon representation of the stable structure has also been included.*

These results are shown in Figure 5.27, where we can see that the most stable structure at low temperatures in all our systems is the swapped domain (regardless the concentration) and also a small proportion of $\beta$-aggregate is found. If we look at the most diluted system, in Figure 5.27($a$), we can see that these two multichain structures become less populated near the transition temperature, predominating the independently folded structures in this temperature range.

The raise in the concentration conditions modifies the competition between independently folded structures (with the proper $\alpha + \beta$ fold) and aggregated ones, favoring aggregated conditions when concentration increases. The current system presents intermediate properties between the all-$\alpha$ protein and the all-$\beta$ one in relation to its propensity to aggregate. If we compute the proportion of alternating hydrophobic and hydrophilic residues in the sequence, we also observe that the $\alpha + \beta$ one has a 54% of alternating residues, between the all-$\alpha$ sequence (48%) and the all-$\beta$ case (78%).

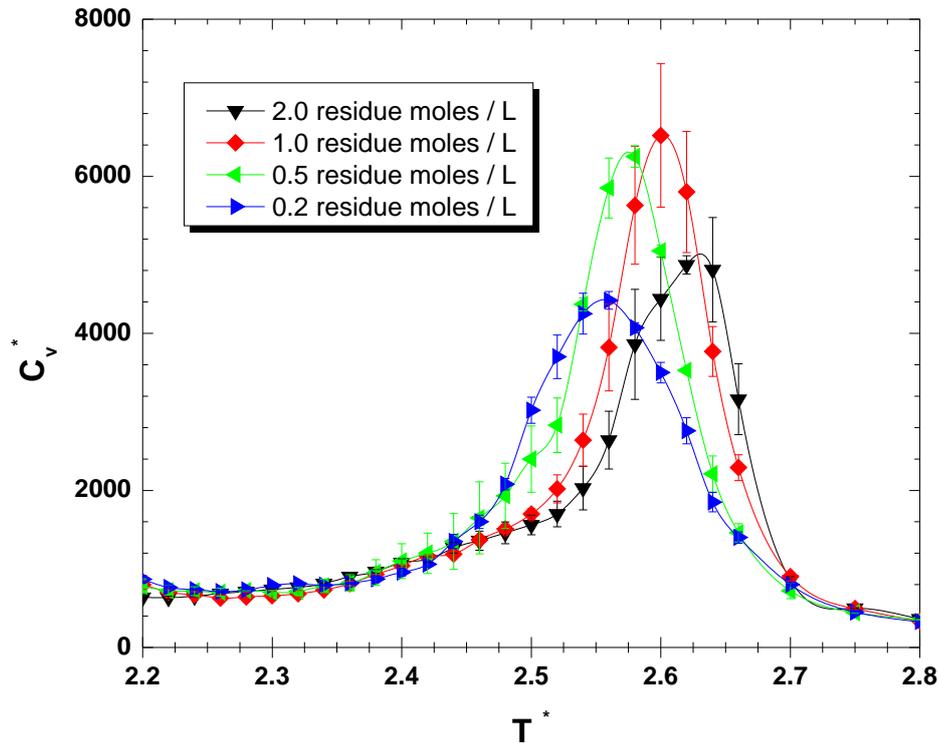

**Figure 5.25:** *Heat capacity curves vs temperature for systems of two $\alpha + \beta$ proteins under different concentration conditions.*

Regarding the properties of the aggregates themselves, we find again the two kinds of aggregated structures, either by domain swapping or $\beta$-type structures, prevailing the latter one at intermediate temperatures. We can conclude that this system nicely exemplifies the interplay between folding and aggregation, proving that every protein aggregates under high concentration conditions.

As we have discussed, this general rule is maintained regardless the particular sequence of the simulated protein, in agreement to experimental findings.[17] The different kinds of proteins, however, differ in the characteristics of this competition and the particular structural properties of the detected situations.

According to our results, our all-$\alpha$ protein forms aggregates with less internal structure and mainly stabilized by hydrophobic interactions; our all-$\beta$ and $\alpha + \beta$ systems present a larger number of interchain hydrogen bonds, that are related to the characteristics of their sequences. Regarding the formation of domain-swapped structures, they are more frequent in the case of the all-$\alpha$ protein, suggesting that, according to our interaction model, there could be different ways of forming these aggregates. As a matter of fact, the amount of alternating hydrophobic/hydrophilic residues is said to be proportional to the propensity to form $\beta$-type aggregates,[226] which agrees with the observations of our simulated systems.

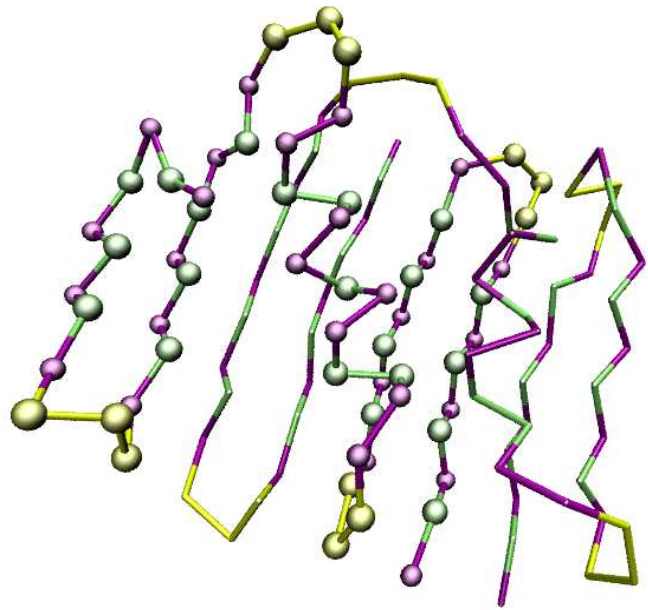

(a) Swapped domains.

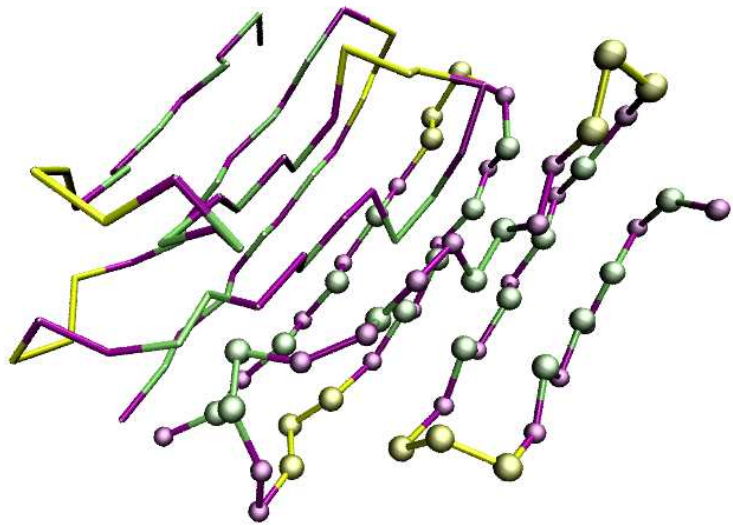

(b) $\beta$-type aggregate.

**Figure 5.26:** *Cartoons of the different multi-chain structures found in $\alpha + \beta$-type proteins. Structures represented using VMD.*[144]

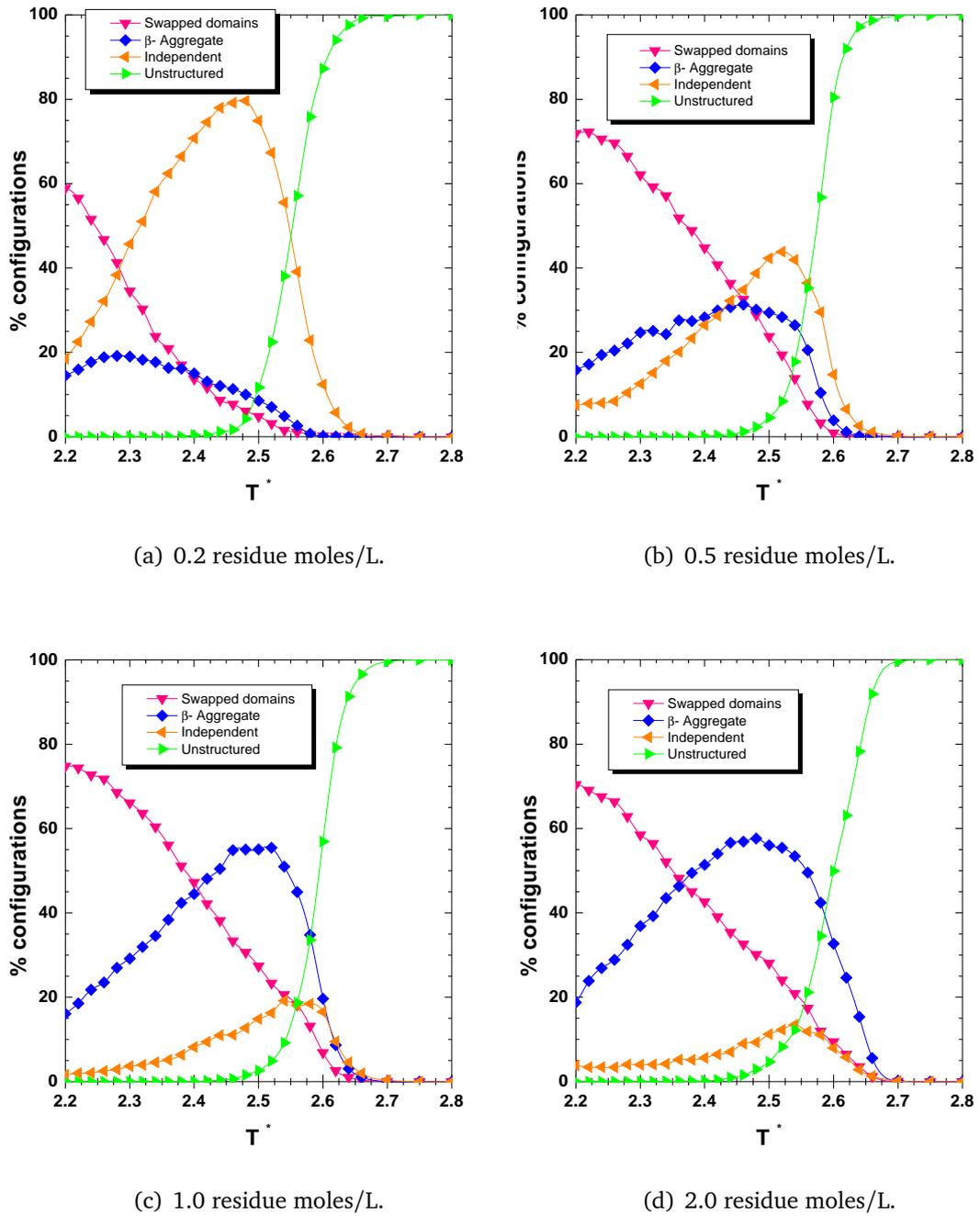

(a) 0.2 residue moles/L.

(b) 0.5 residue moles/L.

(c) 1.0 residue moles/L.

(d) 2.0 residue moles/L.

**Figure 5.27:** *Thermal evolution of the population of the observed structures in our $\alpha + \beta$ protein at four different concentrations.*

## §5.4   Summary and conclusions of this Chapter

In this Chapter, we have tackled an ambitious aim (in terms of complexity and size of the numerical experiments): the understanding of folding and aggregation processes in peptides and proteins using a minimalist model with just a simple representation of hydrogen bonds and hydrophobic interactions. As we had designed a suitable hydrogen bond potential (whose performance has been checked and discussed in many different circumstances along this PhD Dissertation), our first task has been the selection of a hydrophobic potential, that had to fulfill the requirements of simplicity (i.e. $\alpha$-carbon representation of the polypeptide chain) and geometric consistence with the hydrogen bond model.

We have chosen the Brown *et al.* potential,[110] based on an $\alpha$-carbon representation and a three-letter code for the hydrophobic classification of the amino acids. The combination of two independent potentials to build a fully compatible combined one has resulted in some modifications of the original hydrophobic model, necessary to our purposes. Therefore, we have modified the geometric description of the interactions (originally based on a lattice model[214]) to match the native-like geometric definition of the hydrogen bond potential.

In addition, the Brown *et al.* potential includes some terms related to the chain stiffness and its local geometry, imposing an associated secondary structure for each residue. This feature has been also modified in the combined potential. We have kept some control of the chain stiffness, but we have not enforced the local geometry of the chain towards helices or strands according to the sequence, leaving the stabilization of a specific kind of structure to the merged effect of hydrogen bonds and hydrophobics. In this sense, it is important to remark that the hydrogen bond contribution plays a determinant role in these simulations, as it drives the formation of secondary structure elements without any additional bias, as it was needed in the original Brown *et al.* potential.

Thanks to the simplicity of our potential, we have been able to carry out extensive simulations (several months each in a single processor) for peptides and full proteins un-der different concentration conditions. In first place, we have checked the performance of this new combined model in peptide systems. We have compared the behavior of helical peptides and $\beta$-prone ones (according to its sequence), also having in mind the results obtained for sequenceless peptides in Chapter 3. We have concluded that all peptide systems are sensitive to the concentration conditions. The natural tendency of peptides to form $\alpha$-helices or $\beta$-sheets is also observed in our simulations, resulting in different structural scenarios depending on the system concentration that reflect the trends that the sequence imposes in peptide systems.

Apart from their role in peptides, hydrophobic interactions are crucial for the study of complete proteins, as these interactions are responsible for the formation of their tertiary structure and thus, their global folded structure. In this way, we have shown that even this kind of simple hydrophobic potentials, in combination with the hydrogen bond contribution, is able to fold a complete protein with a designed simple sequence and folded structure.

We have been able to design proteins belonging to the most common structural families (all-$\alpha$, all-$\beta$ and $\alpha + \beta$), to make them stable in isolation and to successfully reproduce the competition between folding and aggregation in terms of temperature and concentration changes.

For all these systems, we have performed equilibrium folding simulations at four different concentrations. In all the cases, we have observed that high concentrations lead to the formation of aggregates that are stabilized by hydrophobic interactions and interchain hydrogen bonds; at low concentrations, we have observed the folding and unfolding processes of the isolated chains.

Therefore, we can conclude that the use of simple models has revealed a powerful tool for the study of the folding and aggregation processes of peptides and proteins. We have proved the important role of hydrogen bonds in these systems and how a good description of this interaction, together with a simple hydrophobic potential, has allowed the proper obtention of the different secondary structure elements either in independent pep-tides or in full proteins, where hydrogen bonds and hydrophobic interactions have to work together to build the complete proteins and keep reasonable properties for their folding and aggregation propensities.

# 6

# Summary and conclusions

Along this PhD dissertation we have discussed many relevant aspects of protein folding and aggregation and how hydrogen bonds play a determinant role in these processes. Using molecular simulation tools, we have committed ourselves to the building of simple models to understand these complex processes, keeping in mind that a minimalist approach should be able to grasp their main properties. Therefore, we have just used $\alpha$-carbon representations of the protein chain and functional forms as simple as possible for the interaction potentials.

After the Introduction of Chapter 1, we dedicated Chapter 2 to detailed considerations of the technical aspects of this PhD Project. Among them, the development of the kinetic Monte Carlo method deserves a particular attention, as it corresponds to a new strategy for our research group that can provide useful information on the dynamics of the folding process, as we have proved in Chapter 4. The code parallelization is also a new remarkable technical achievement, as it has allowed the calculation of very costly simulations at a reasonable computational effort.

Regarding the Results themselves, the central aspect of this Project has consisted in the design and development of a hydrogen bond potential.[91] We have focused on this interaction because of its key role in aggregation and folding, stabilizing amyloids and secondary structure elements. In addition, the hydrogen bond interaction constitutes a methodological challenge because of its directional nature, that has seldom been considered in proper terms in simple simulation models. We have started, then, by analyzing several hydrogen bond potentials available in the literature, finding that none of them –to our knowledge– can properly combine a detailed geometrical representation of the hydrogen bond interaction with a reduced representation of the system without additional energetic terms.

This fact, closely related to the partially covalent nature of the hydrogen bond interaction, led us to an extensive investigation of the properties of native hydrogen bonds, as they appear in the PDB database.[88] As a result, we have proposed a hydrogen bond model that is based on a careful geometric representation of this interaction. Its design is composed by two steps, namely the calculation of three geometric restrictions and the application of a step-like potential that distinguishes between local and non-local interactions. In addition, our hydrogen bond potential includes a specific treatment for terminal residues and takes into account the right-handed orientation of natural helices.

To check the performance of this model, we have carried out simulations for the thermal folding and unfolding of sequenceless peptides under equilibrium conditions, evaluating their thermodynamic characteristics. In first place, we have analyzed the effect of the chain length in the model behavior. We have concluded that our model is able to correctly describe the properties of moderately long peptides, far above the average length of the natural fragments of secondary structure elements in globular proteins.

Secondly, we have studied the effect of concentration in these peptide systems, as they are the most simple case where can explore the competition between folding and aggregation under a hydrogen bond potential. We have shown that our interaction model can describe the interplay among different kinds of structures in reasonable terms, being sensitive to changes in the environment such as temperature and concentration. As a result, we have built a "structural phase diagram" for sequenceless peptides.

Once the first target of this PhD dissertation has been completed, we have undertaken the study of more complex systems such as complete proteins, using two different approaches to simulate tertiary interactions.

The first one, presented in Chapter 4, has been the use of a structure-based model, developed in our group some years ago.[62, 122] This type of potentials builds their interactions in terms of the native topology of proteins. For this reason, it is very appropriate to check whether the secondary structure elements generated from our hydrogen bond potential are fully compatible to the structure-based definition and to investigate folding processes of real proteins.

We have selected a set of eight representative proteins whose folding has been analyzed from two different points of view: the thermodynamic and the dynamic one. We have proved that the inclusion of hydrogen bonds improves the description of the folding process in many aspects.

Regarding our thermodynamic results, we have compared the original structure-based model and the combined one, concluding that the latter always improves the properties found for the folding process. The new potential nicely modulates the height of the free energy barriers (either increasing them, like in the case of 2GB1, or reducing them, as it happens in the downhill folder 1HYW). In addition, it improves the structural description during the overall process, as we have checked in 1PGB (where the native state is remarkably better defined), 1R69 (where the denatured state presents some remaining helicity that the original structure-based model could not reproduce) or 1NLO (where our intermediate structures at the transition temperature seem to be linked to the experimental transition state).

In relation to the kinetic results of the combined potential of hydrogen bonds and structure-based interactions, we have firstly explored the performance of the Kinetic Monte Carlo method used in this work and the relationship between the kinetic rate and the height of the free energy barrier, obtaining a reasonable correlation that validates the transition state theory in our simulations of two-state proteins and proves that their folding follows a first order kinetics.

To get a more detailed view on the dynamics of protein folding, we have carried out a study on two helical proteins, comparing the downhill folder 1BBL to the two-state pro-tein 1ENH. The presence or absence of a folding barrier completely modifies the dynamic performance of these proteins, presenting numerous alternatives if there is no barrier and exhibiting similar trends in independent folding simulations if a perceptible folding barrier is detected.

In this latter case, we have obtained representative statistics of the dynamic behavior of the protein, being able to analyze the effect of an accurate representation of hydrogen bonds in the folding pathway. Again, we have proved that the inclusion of hydrogen bonds results in an experimentally-consistent description of the folding process, being able to distinguish the conformational events which indicate how the folded structure is achieved.

However, the use of structure-based potentials presents some strong limitations, as they just consider the interactions that are present in the native state of proteins, and treat them all in the same way. Even if hydrogen bonds are included in the calculation (for native and non native interactions), the resulting potential is still too biased towards native structures. This prevents to carry out a serious study on the competition between folding and aggregation based on this description of the system interactions.

For this reason, the last part of this PhD Dissertation has been devoted to the combination of our original hydrogen bond potential with a hydrophobic one that reproduces the tertiary interactions among residues without imposing any bias towards the native structure. A considerable effort within this piece of work has consisted in the selection of the optimal properties of the hydrophobic interaction.

Strongly based on previous effort from Brown and co-workers,[110] we have modified their hydrophobic potential relying again on the principles of simplicity and computational efficiency that have guided us along this PhD Project. Therefore, we have chosen an $\alpha$-carbon representation and a three-letter alphabet for the description of the interactions (using hydrophilic, hydrophobic and neutral beads). Besides, we have included a "stiffness" contribution that does not impose any local geometry on the residues (as it formerly did in Brown *et al.* potential), but avoids an excessive "stickiness" of the hydrophobic potential.

The use of a simplified hydrophobic potential, where tertiary interactions do not take place among the real side chains of the residues and they have been simplified up to a three letter code, derives in a loss of accuracy in the system description. For this reason, we have not worked with real sequences, but with regularized ones that aim to reproduce the essential features of peptides and simple proteins belonging to different topological families, but lacking the detail of a real sequence.

We have started with small systems formed of $\alpha$-prone and $\beta$-prone peptides, aiming to obtain their related phase diagrams in terms of temperature and concentration. This result has allowed the comparison among these two kinds of peptides and the sequenceless ones that we studied in Chapter 3. We have concluded that the overall behavior of peptides is maintained (i.e. the sensitivity of the system on the concentration conditions that makes high concentrated systems aggregate) and the particular characteristics on each kind of system reflect the differences in the sequences.

However, the greatest interest of this combined potential with hydrogen bonds and hydrophobics lies in the possibility of building complete proteins from very basic principles. We have obtained stable proteins of different structural families, namely an all-$\alpha$ protein, an all-$\beta$ one and an $\alpha+\beta$ protein. After the design of the particular sequences and arrangements of secondary structure elements, we have performed extensive simulations to explore the interplay between folding and aggregation, as we aimed to do since the beginning of this PhD Project.

We have found that the aggregation propensity is a common feature of any kind of protein, regardless its specific sequence, at high concentrations. In this way, we have observed aggregated structures at high concentrations for the three kinds of proteins we have studied, mainly stabilized by long range (or $\beta$-type) hydrogen bonds and hydrophobic interactions. Nevertheless, the particular properties of each structural family have been maintained in the independently folded structures.

As a result, we can state that the original aim of understanding folding and aggregation processes through the study of hydrogen bonds has been successfully achieved. Along the learning process, we have also obtained a much deeper understanding on how the small individual interactions acting on a system can drive much larger processes. The models and algorithms developed in this work can be used now to study further folding and aggregation processes of interest. The comparison of the results from our simple models with experimental data can therefore enhance our understanding of the physical basis of these fundamental transitions.